\documentclass[printer]{tADP2e}

\usepackage[dvips]{color}
   \usepackage{multicol}

\newcommand{\eqn}[1]{\begin{eqnarray} #1 \end{eqnarray}}

\newcommand{\ci}{{\rm i}}

   \newcommand{\ensm}[1]{\ensuremath{#1}}

\renewcommand{\alph}{\ensm{\alpha}}

\newcommand{\braai}{\ensm{\langle\alpha,i|}}

\newcommand{\Heff}{\ensm{H_{eff}}}
\newcommand{\Hint}{\ensm{H_{int}}}
\newcommand{\HO}{\ensm{H_{0}}}

\newcommand{\ketaj}{\ensm{|\alpha,j\rangle}}

\newcommand{\ketbj}{\ensm{|\beta,j\rangle}}

\newcommand{\Sone}{\ensm{S_1}}
\newcommand{\Stwo}{\ensm{S_2}}

\newcommand{\lrb}{\ensm{\left(}}
\newcommand{\rrb}{\ensm{\right)}}
\newcommand{\lsb}{\ensm{\left[}}
\newcommand{\rsb}{\ensm{\right]}}
\newcommand{\lgb}{\ensm{\left\{}}

\newcommand{\Palpha}{\ensm{P_{\alpha}}}

\newcommand{\Pbeta}{\ensm{P_{\beta}}}
\newcommand{\Qalphai}{\ensm{Q_{\alpha i}}}
\newcommand{\Qalphaj}{\ensm{Q_{\alpha j}}}

\begin{document}


\title{
Ultracold  atomic gases in optical lattices:\\ 
Mimicking condensed matter physics and beyond
}

\author{
Maciej Lewenstein,\\ {\tiny{ICREA and ICFO-Institut de Ci\`encies Fot\`oniques, E-08660 Castelldefels (Barcelona) Spain}}\\
 Anna Sanpera,\\ {\tiny{ICREA and Grup de F\'isica Te\`orica, Universitat Aut\`onoma de Barcelona, E-08193 Bellaterra, Spain}}\\
 Veronica Ahufinger,\\ {\tiny{ICREA and Grup d'Optica, Universitat Aut\`onoma de Barcelona, E-08193 Bellaterra, Spain}}\\
 Bogdan Damski,\\ {\tiny{Theory Division, Los Alamos National Laboratory, MS-B213, Los Alamos, NM 87545, USA}}\\
 Aditi Sen(De), 
and Ujjwal Sen\\ {\tiny{ICFO-Institut de Ci\`encies Fot\`oniques, E-08660 Castelldefels (Barcelona) Spain}}\\
}  


\maketitle

\vspace{2cm}

\begin{abstract}
We review recent developments in the physics of ultracold atomic and molecular gases in optical lattices. 
Such systems are nearly perfect realisations of various kinds of Hubbard models, and as such may very well serve to mimic 
condensed matter phenomena.  We show how these systems may be employed as {\it quantum simulators} to answer some challenging
open questions of condensed matter, and even high energy physics. After a short presentation of 
the models and 
the methods of treatment of  such systems, 
we discuss in detail, which challenges of condensed matter physics can be addressed with (i) disordered ultracold lattice gases, (ii) frustrated 
ultracold gases, (iii) spinor lattice gases, (iv) 
lattice gases in ``artificial" magnetic fields, and, last but not least, (v) quantum information processing in lattice gases. For completeness, also some recent progress related to the above topics with trapped cold gases will be discussed.\\

\pagebreak

\centerline{\bfseries \large{Contents} }\medskip

\noindent
{1.}~Introduction \hfill \pageref{introduction}\\
\hspace*{10pt}{1.1.}~Cold atoms from  a historical perspective \hfill \pageref{cold_atoms_history}\\
\hspace*{10pt}{1.2.}~Cold atoms and the challenges of condensed matter physics \hfill \pageref{cold_atoms_challenges}\\
\hspace*{10pt}{1.3.}~Plan of the review \hfill \pageref{plan}\\
{2.}~The Hubbard and spin models with ultracold lattice gases \hfill \pageref{sec-Kalimpong-jabo}\\
\hspace*{10pt}{2.1.}~Optical potentials \hfill \pageref{cold_atoms_in_lattices} \\
\hspace*{10pt}{2.2.}~Hubbard models \hfill \pageref{hubbard_models}\\
\hspace*{10pt}{2.3.}~Spin models \hfill \pageref{spin_models}\\
\hspace*{10pt}{2.4.}~Control of parameters in cold atom systems \hfill \pageref{control}\\
\hspace*{10pt}{2.5.}~Superfluid - Mott insulator quantum phase transition in the Bose Hubbard model \hfill \pageref{SMI}\\
{3.}~The Hubbard model: Methods of treatment \hfill \pageref{sec-sadher-lau}\\
\hspace*{10pt}{3.1.}~Introduction \hfill \pageref{subsec-sadher-lau}\\
\hspace*{10pt}{3.3.}~Weak interactions limit \hfill \pageref{subsec-alaler-ghorer-dulal}\\
\hspace*{10pt}{3.4.}~Strong interactions limit \hfill \pageref{subsec-alaler-ghorer-dulal1}\\
\hspace*{10pt}{3.5.}~The Gutzwiller mean-field approach \hfill \pageref{subsec-alaler-ghorer-dulal2}\\
\hspace*{10pt}{3.6.}~Exact diagonalizations \hfill \pageref{subsec-alaler-ghorer-dulal3}\\
\hspace*{10pt}{3.7.}~Quantum Monte Carlo \hfill \pageref{subsec-alaler-ghorer-dulal4}\\
\hspace*{10pt}{3.8.}~Phase space methods\hfill \pageref{subsec-alaler-ghorer-dulal5}\\
\hspace*{10pt}{3.10.}~1D methods\hfill \pageref{subsec_kachupora1}\\
\hspace*{10pt}{3.11.}~Bethe ansatz\hfill \pageref{subsec_kachupora2}\\
\hspace*{10pt}{3.12.}~A quantum information approach to strongly correlated systems\hfill \pageref{subsec_quantu_raju}\\
\hspace*{24pt}{3.12.1.}~Vidal's algorithm\hfill \pageref{subsec_quantu_raju1}\\
\hspace*{24pt}{3.12.2.}~Matrix product states\hfill \pageref{subsec_quantu_raju2}\\
\hspace*{10pt}{3.13.}~Fermi and Fermi-Bose Hubbard models\hfill \pageref{subsec_kachupora}\\
{4.}~Disordered ultracold atomic gases\hfill \pageref{sec_disorder_Konark}\\
\hspace*{10pt}{4.1.}~Introduction\hfill \pageref{sub_int_disorder}\\
\hspace*{10pt}{4.2.}~Disordered interacting bosonic lattice models in condensed matter\hfill \pageref{sub_bosons}\\
\hspace*{10pt}{4.3.}~Realization of disorder in ultracold atomic gases\hfill \pageref{sub_realization}\\ 
\hspace*{10pt}{4.4.}~Disordered ultracold atomic Bose gases in optical lattices\hfill \pageref{diorderedbose}\\ 
\hspace*{10pt}{4.5.}~Experiments with weakly interacting trapped gases and Anderson localization\hfill \pageref{sub_exp}\\ 
\hspace*{10pt}{4.6}~Disordered interacting fermionic systems\hfill \pageref{sub_fermions}\\
\hspace*{10pt}{4.7}~Disordered Bose-Fermi mixtures\hfill \pageref{sub_bosefermi}\\
\hspace*{10pt}{4.8}~Spin glasses\hfill \pageref{sub_spinglass}\\
{5.}~Frustrated  models in cold atom systems\hfill \pageref{frustration_atom}\\
\hspace*{10pt}{5.1.}~Introduction\hfill \pageref{intr}\\
\hspace*{10pt}{5.2.}~Quantum antiferromagnets\hfill \pageref{frust_qantiferromagnet}\\
\hspace*{24pt}{5.2.1.}~The Heisenberg model \hfill \pageref{subsec-tapa-tepi}\\
\hspace*{24pt}{5.2.2.}~The $J_1-J_2$ model\hfill \pageref{subsec-tapa-tepi1}\\
\hspace*{10pt}{5.3.}~Heisenberg antiferromagnets and atomic Fermi-Fermi mixtures in kagom{\'e} lattices\hfill \pageref{frust_H_antiferro}\\
\hspace*{24pt}{5.3.1.}~Heisenberg kagom{\'e} antiferromagnets\hfill \pageref{subsubsec-radha}\\
\hspace*{24pt}{5.3.2.}~Realization of kagom{\' e} lattice by Fermi-Fermi mixtures\hfill \pageref{subsubsec-radha1}\\
\hspace*{10pt}{5.4.}~Interacting Fermi gas in a kagom{\'e} lattice: Quantum spin-liquid crystals\hfill \pageref{frust_Fermi_spinliquid}\\
\hspace*{24pt}{5.4.1.}~The quantum magnet Hamiltonian\hfill \pageref{subsubsec-kumropotas}\\
\hspace*{24pt}{5.4.2.}~Classical analysis\hfill \pageref{subsubsec-kumropotas1}\\
\hspace*{24pt}{5.4.3.}~Quantum mechanical results\hfill \pageref{subsubsec-kumropotas2}\\
\hspace*{10pt}{5.5.}~Realization of frustrated models in cold atom/ion systems\hfill \pageref{frust_other}\\
\hspace*{24pt}{5.5.1.}~Simulators of spin systems with topological order\hfill \pageref{ultra}\\
\hspace*{24pt}{5.5.2.}~Frustrated models with polar molecules\hfill \pageref{subsubsec-polar}\\
\hspace*{24pt}{5.5.3.}~Ion-based quantum simulators of spin systems\hfill \pageref{subsubsec-ionsimulator}\\
{6.}~Ultracold spinor atomic gases\hfill \pageref{sec-radhamadhab}\\
\hspace*{10pt}{6.1.}~Introduction\hfill \pageref{sec-radhamadhab1}\\
\hspace*{10pt}{6.2.}~Spinor interactions\hfill \pageref{sec-radhamadhab2}\\
\hspace*{10pt}{6.3.}~$F=1$ and $F=2$ spinor gases: Mean field regime\hfill \pageref{sec-radhamadhab3}\\
\hspace*{24pt}{6.3.1.}~F=1 gases in optical lattices \hfill \pageref{sec-Mr.Yusuf_Zamadar}\\
\hspace*{24pt}{6.3.2.}~Bose-Hubbard model for spin 1 particles \hfill \pageref{sec-radhamadhab4}\\
\hspace*{24pt}{6.3.3.}~F=2 gases in optical lattices \hfill \pageref{sec-Mr.Shakti_Sarkar}\\
\hspace*{24pt}{6.3.4.}~Bose-Hubbard model for F=2 particles\hfill \pageref{sec-radhamadhab5}\\
\hspace*{24pt}{6.3.5.}~Spinor Fermi gases in optical gases \hfill \pageref{sec-Mrs.Sabita_Bardhan}\\
{7.}~Ultracold atomic gases in ``artificial" magnetic fields\hfill \pageref{artificial}\\
\hspace*{10pt}{7.1.}~Introduction -- Rapidly rotating ultracold gases\hfill \pageref{rapidly}\\
\hspace*{10pt}{7.2.}~Lattice gases in ``artificial" Abelian magnetic fields\hfill \pageref{abelian}\\
\hspace*{10pt}{7.3.}~Ultracold gases and lattice gauge theories\hfill \pageref{lgt}\\
{8.}~Quantum information with ultracold  gases\hfill \pageref{sec-goru65}\\
\hspace*{10pt}{8.1.}~Introduction\hfill \pageref{sec-goru165}\\
\hspace*{10pt}{8.2.}~Entanglement: A formal definition and some preliminaries\hfill \pageref{subsec-poila-qi}\\
\hspace*{24pt}{8.2.1.}~The partial transposition criterion for detecting entanglement\hfill \pageref{subsubsec-jog-biyog}\\
\hspace*{24pt}{8.2.2.}~Entanglement measures\hfill \pageref{subsubsec-entanglement-measures}\\
\hspace*{10pt}{8.3.}~Entanglement and phase transitions\hfill \pageref{subsec-dusra-qi}\\
\hspace*{24pt}{8.3.1.}~Scaling of entanglement in the reduced density matrix\hfill \pageref{subsubsec-babajibon-qi}\\
\hspace*{24pt}{8.3.2.}~Entanglement entropy: Scaling of spin block entanglement\hfill \pageref{subsubsec-barishal-qi}\\
\hspace*{24pt}{8.3.3.}~Localizable entanglement and its scaling\hfill \pageref{subsubsec-bardhaman-qi}\\
\hspace*{24pt}{8.3.4.}~Critical behaviour in the evolved state\hfill \pageref{subsubsec-sundarban-qi}\\
\hspace*{10pt}{8.4}~Quantum computing with lattice gases~\hfill \pageref{subsec-teesra-qi}\\
\hspace*{10pt}{8.5.}~Generation of entanglement: The one-way quantum computer\hfill \pageref{subsec-chauthha-qi}\\
\hspace*{24pt}{8.5.1.}~The one-way quantum computer\hfill \pageref{subsec-chauthha-qi1}\\
\hspace*{24pt}{8.5.2.}~Disordered lattice\hfill \pageref{subsubsec-hazarchurasi}\\
{9.}~Summary\hfill \pageref{sec-summary-oma}\\
Acknowledgements\hfill \pageref{ackn}\\
Appendix A: Effective Hamiltonian to second order\hfill \pageref{effhamil}\\
Appendix B: Size of the occupation-reduced Hilbert space \hfill \pageref{appendix_S}\\
References\hfill \pageref{Refs}\\ \\
\end{abstract}



\renewcommand*{\DefineNamedColor}[4]{%
      \textcolor[named]{#2}{\rule{7mm}{7mm}}\quad
      \texttt{#2}\strut\\}
   \newcommand{\tr}{{\rm tr}}
\def\intlarge{\mathop{\int}\limits}

\definecolor{Red}{rgb}{.85,0,0}

\pagebreak

\vskip 2 cm

\noindent \small{Motto:} 

\centerline{{\it There are more things in heaven and earth, Horatio,}}
\centerline{{\it Than are dreamt of in your philosophy} \cite{Shakespeare03}.}

\vskip 2 cm


\section{Introduction}
\label{introduction}
\subsection{Cold atoms from  a historical perspective}
\label{cold_atoms_history}

Thirty years ago, atomic physics was a very well established and respectful, but evidently not a   ``hot" area of physics. On the theory side, 
even though one had to deal with complex problems of many electron systems, most of the methods and techniques were developed. The main 
questions concerned, how to optimize these methods, how to calculate more efficiently, etc. These questions were reflecting an evolutionary progress, 
rather than a revolutionary search for totally new phenomena. Quantum optics at this time was entering its Golden Age, but in the first place  
on the experimental side.
Development of laser physics and nonlinear optics led in 1981 to the 
Nobel prize for A.L. Schawlow and N. Bloembergen ``for  their contribution to the 
development of laser spectroscopy". 
Studies of quantum systems at the single particle level culminated in 1989 with the Nobel prize for H.G. Dehmelt and W. Paul ``for the development of the ion trap technique", shared with N.F. Ramsey ``for the invention of the 
separated oscillatory fields method and its use in the hydrogen maser and other atomic clocks". 

Theoretical quantum optics was born in the 60'ties with the works  on quantum coherence theory 
by the 2005 Noble prize winner, R.J. Glauber \cite{Glauber63a,Glauber63b}, and with the development of the laser theory by 
M. Scully and W.E. Lamb (Nobel laureate of  1955) \cite{Scully67}, and H. Haken \cite{Haken66}. 
In the 70'ties and 80'ties, however, theoretical quantum optics was not considered to be a separate, established area of theoretical 
physics. One of the reasons of this state of art, was that indeed the quantum optics of that time was primarily dealing with 
single particle problems. 
Most of the many body problems of quantum optics, such as laser theory, or more generally optical instabilities \cite{Walls06},  
could have been 
solved either using linear models, or employing  relatively simple versions of mean field approach. Perhaps the most sophisticated theoretical 
contributions concerned understanding of quantum fluctuations  and quantum noise \cite{Walls06,Gardiner04}.

\begin{figure}[t]
\centering
\includegraphics[width=10cm,height=8cm]{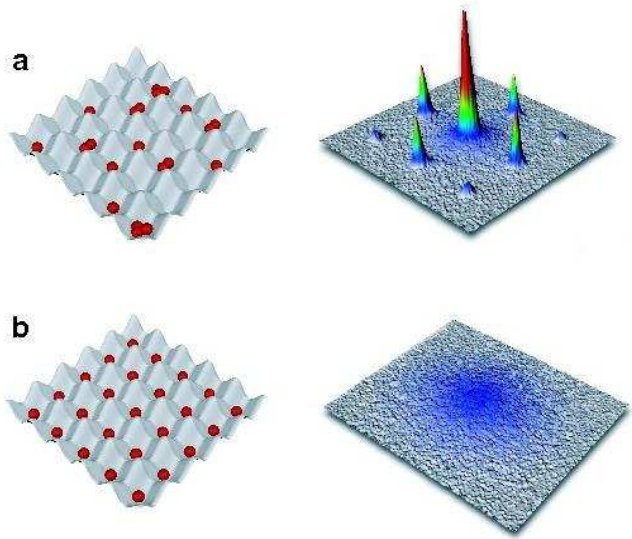}
\caption{
Distribution of atoms among lattice sites vs.   momentum distribution 
of atoms released from an optical lattice. (a): low lattice potential:
the largest interference peak is at zero momentum. The system is a superfluid 
with fluctuating number of atoms  per site. (b): high lattice
potential case  where a perfect Mott state with exactly one atom per site
is achieved -- the momentum distribution is blurred (from \cite{Bloch04}).}
\label{figure1.1}
\end{figure}

This situation has drastically changed in the last ten -- fifteen years, and there are several
 seminal discoveries that have triggered these changes:
\begin{itemize}
\item First of all, atomic physics and quantum optics have developed over the years quite generally an unprecedented level 
of {\it quantum engineering}, i.e. preparation, manipulation, control and detection of quantum systems.
\item Cooling and trapping methods of atoms, ions and molecules have reached regimes of  
low temperatures (today down to nanoKelvin!)  and precision, that 15 years ago were considered unbelievable.  These developments have been recognised by the 
Nobel Foundation in 1997, who awarded the Prize to S. Chu \cite{Chu98}, C. Cohen-Tannoudji\cite{Cohen-Tannoudji98} and W.D. Phillips 
\cite{Phillips98} ``for the development of methods to cool and trap atoms with laser light". Laser cooling and mechanical manipulations of 
particles with light \cite{Metcalf01} was essential for development of completely new areas of atomic physics and quantum optics, 
such as atom optics \cite{Meystre01}, and for reaching new territories of precision metrology and quantum engineering.
\item Laser cooling combined with evaporation cooling technique allowed in 1995 for experimental observation 
of Bose-Einstein condensation (BEC) \cite{Anderson95,Davis95}, a phenomenon predicted by S. Bose and A. Einstein 
more than 
70
years earlier. 
The authors of  these experiments, E.A. Cornell and C.E. Wieman \cite{Cornell02} and W. Ketterle \cite{Ketterle02} received  
the Noble Prize in 2001, ``for the achievement of Bose-Einstein condensation in dilute gases of alkali atoms, and for early fundamental
 studies of the properties of the condensates". This  was a breakthrough moment, in which ``atomic physics and quantum optics 
has met condensed matter physics" \cite{Ketterle02}.  Condensed matter community at this time remained, however, still reserved. 
After all, BEC was observed in weakly interacting dilute gases, where it is very well described by the mean field 
Bogoliubov-de Gennes theory \cite{Pitaevskii03}. Although the finite size of the systems, and spatial inhomogeneity play there a 
crucial role, the basic theory of such systems was developed in the 50'ties. 
\item The seminal theoretical works of the late A. Peres \cite{Peres95},  the proposals of quantum cryptography by
C.H.  Bennett and G. Brassard \cite{Bennett84},  and A.K.  Ekert \cite{Ekert91}, the quantum communication 
proposals by 
C.H. Bennett and S.J. Wiesner \cite{eita-holo-dense} 
and
C.H. Bennett, G. Brassard, C. Cr{\' e}peau, R. Jozsa, A. Peres, and W.K. Wootters
\cite{eita-holo-tele}, the discovery of the quantum factorizing 
algorithm by P. Shor \cite{Shor94}, and the quantum computer proposal   by J.I. Cirac and P. Zoller \cite{Cirac95} 
have given birth to experimental studies of quantum information \cite{Bouwmeester00,Zoller05a}. These studies, 
together with rapid development of the theory have, in particular in the area of
 atomic physics and quantum 
optics, led to enormous progress in our understanding of what quantum correlations and 
quantum entanglement are, and how to prepare and use entangled states as a resource. The impulses from quantum 
information enter nowadays constantly into the physics of cold atoms, molecules and ions, and stimulate new approaches. It is very 
probable that the first quantum computers will be, as suggested already by Feynman \cite{Feynman86}, computers of special purpose -- 
quantum simulators \cite{Cirac04}, that will efficiently simulate quantum many body systems that otherwise cannot be simulated 
using ``classical" computers \cite{Zoller05a}. 
\item The physics of ultracold atoms entered the areas of strongly correlated systems with the  seminal 1998 paper of 
Jaksch et al. \cite{Jaksch98} on the superfluid-Mott insulator transition in cold atoms in an optical lattice. 
Apart from stimulation from the condensed matter physics \cite{Fisher89}, the authors of this paper were in fact motivated by the 
possibility of realising quantum computing with cold atoms in a lattice \cite{Jaksch99}. Transition to the Mott 
insulator state   was supposed to be an efficient way of preparation of a quantum register with a fixed number of atoms per lattice site.
 The experimental observation of the superfluid-Mott insulator transition by the  Bloch--H{\"a}nsch group \cite{Greiner02} (see Fig. \ref{figure1.1})
marks the beginning of age of the experimental studies of strongly correlated systems with ultracold atoms\cite{Bloch04}. Several  other groups 
have observed bosonic superfluid-Mott insulator transitions in pure Bose systems \cite{Kohl05}, in disordered Bose systems \cite{Fallani06}, 
or in Bose-Fermi mixtures \cite{Ospelkaus06,Gunter06}. Very recently MI state of molecules have been created \cite{Volz06}, 
and bound repulsive pairs of atoms (i.e. pairs of atoms at a site that cannot release 
their repulsive energy due to the band structure of the spectrum in  the lattice) have been observed \cite{Winkler06}.
 
\end{itemize}

Since 1998 the physics of ultracold atoms has made enormous progress in the studies of 
strongly correlated systems. Number of the theory 
papers that propose to mimic various  condensed matter
systems of interest is hardly possible to follow, and the number of experiments  in which strongly correlated 
regime has been met grows also very significantly. Moreover, condensed matter physicists, in particular theorists, 
joint the efforts of atomic physicists and quantum opticians. Among those who have ``committed" a paper on cold atoms are 
Noble Prize winners: A.J. Leggett \cite{Leggett01}, F. Wilczek \cite{Liu04}, D. Politzer \cite{Politzer91}, or such personalities 
of condensed matter theory as M.P.A. Fisher \cite{Buechler05}, or mathematical physics as M. Aizenman or  E.H. Lieb\cite{Aizenman04}.

\subsection{Cold atoms and the challenges of condensed matter physics}
\label{cold_atoms_challenges}

The physics of cold atoms touches nowadays the same frontiers of modern physics 
as condensed matter and high energy physics. In particular many of the important challenges of the latter two disciplines can be addressed in the 
context of cold atoms: 
\begin{itemize}
\item {\it 1D systems.} The role of quantum  fluctuations and correlations is particularly important in 1D. 
The theory of 1D systems is very well developed due to existence of exact methods
such as Bethe ansatz and quantum inverse scattering theory (c.f. \cite{Essler05}),
powerful approximate approaches, such as bosonisation, or conformal field theory \cite{Giamarchi04}, 
and efficient computational methods, such as density matrix renormalisation group technique (DMRG. c.f. \cite{Schollwoeck05}). 
There are, however, many open experimental challenges that have not been so far directly and clearly realized in condensed matter, 
and can be addressed with 
cold atoms (for a review see \cite{Cazalilla04}). Examples include atomic Fermi, or Bose analogues of spin-charge 
separation, or more generally observations of microscopic properties of Luttinger liquids \cite{Recati03,Paredes03,Kollath05}.
Recent experimental observations of the 1D gas in the  deep Tonks-Girardeau regime by Paredes et al. \cite{Paredes04} 
(see also \cite{Moritz03,Kinoshita04,Stoferle04,Laburthe04}) are the first steps
in this direction.  
 
\item{\it Spin-boson model}. A two state system coupled to a bosonic reservoir is a paradigm, 
both in quantum optics, as well as in 
condensed matter physics, where it is termed as spin-boson model (for a review see \cite{Leggett87}). 
It has also been proposed \cite{Recati05}
that an atomic quantum dot, i.e.,  a single atom in a tight optical trap 
coupled to a superfluid reservoir via laser transitions, may realise this model. 
In particular, atomic  quantum dots embedded in a 1D Luttinger liquid 
of cold bosonic atoms realizes a spin-boson model with 
Ohmic coupling, which exhibits a dissipative phase transition and allows to directly measure 
atomic Luttinger parameters.

\item{\it 2D systems}. According to the Mermin-Wagner-Hohenberg theorem, 2D 
systems with continuous symmetry do not exhibit long range order at finite temperatures $T$. 
2D systems may, however, undergo Kosterlitz-Thouless-Berezinskii transition (KTB) to a state in 
which correlations decay is algebraic, rather than exponential. Although KTB transition 
has been observed in liquid Helium, 
its microscopic nature (binding of vortex pairs) has never been seen. A recent experiment of 
Dalibard's group \cite{Hadzibacic06} makes an important step in this direction.

\item{\it Hubbard and spin models}. Very many important examples of strongly 
correlated state in condensed matter physics are realised in various types of Hubbard models 
\cite{Essler05,Auerbach94}. While in condensed matter Hubbard models are 
``reasonable caricatures" of real systems, ultracold atomic gases in optical lattices 
allow to achieve  practically perfect realizations of a whole variety of  Hubbard models \cite{Jaksch05}.
 Similarly, in certain limits Hubbard models reduce to various spin models; 
again cold atoms and ions  allow for practically perfect realizations of such spin models (see for instance 
\cite{Dorner03,Duan03,Santos04,Garcia-Ripoll04}). Moreover, one can use such realizations as quantum simulators to mimic 
specific condensed matter models, and to address various very particular and well focused  questions.

\item{\it Disordered systems: Interplay localisation-interactions}. Disorder plays a central role in 
condensed matter physics, and its presence leads to various novel types of effects and phenomena. 
One of most prominent quantum signatures of disorder is Anderson localisation \cite{Anderson58} of the wave function 
of single particles in a random potential. The question of the interplay between disorder and interactions 
has been intensively studied. For attractive interactions, disorder might destroy the 
possibility of superfluid transition (``dirty" superconductors). Weak repulsive interactions play 
a delocalising role, whereas very strong ones lead to Mott type localisation \cite{Mott68a},   
and insulating behaviour. In the intermediate situations there exist a possibility of 
delocalized ``metallic" phases. Cold atom physics starts to investigate these questions. 
Controlled disorder, or pseudo--disorder  might be created   in atomic traps, or optical 
lattices by adding an optical  potential created by speckle radiation, or several lattices with 
incommensurate periods of spatial oscillations \cite{Damski03,Roth03a,Roth03b}. 
In an optical lattice this should allow to study Anderson-Bose glass and crossover between 
Anderson-like (Anderson glass)   to Mott type (Bose glass) localisation. 
Very recently, Bose glass   state has been realized experimentally by the M. Inguscio group \cite{Fallani06}.
The same group, as well as several others have initiated experimental and theoretical studies of 
the role of interactions in Anderson localisation effects for trapped Bose gases \cite{Lye05,Fort05,Clement05,Schulte05}. 
According to theoretical predictions of Ref. \cite{Schulte0609774,Schulte05, Kuhn05, Bilas06, Paul0509446, Bilas0506020},
prospects of detecting signatures of Anderson localisation  in the presence of weak nonlinear 
interactions and quasi-disorder in  BEC are quite promising.  One expects in such system the appearance of a novel Lifshits glass phase
\cite{Lugan_cond-mat0610389}, where 
bosons condense in a finite number of states from the low energy tail (Lifshits tail) of the single particle spectrum. 

\item{\it Disordered systems: spin glasses}. Since the seminal papers of Edwards and Anderson 
\cite{Edwards75}, and Sherrington and Kirkpatrick \cite{Sherrington75} the question about the nature of the spin glass 
ordering has attracted a lot of attention\cite{Mezard87,Fisher86,Bray87}.  The two competing 
pictures: the  replica symmetry breaking picture of G. Parisi, and a droplet model of D.S. Fisher and D.A. Huse 
are probably applicable in some situations, and not applicable in others. Cold atom physics might 
contribute to resolving this controversy, and even add understanding of some quantum aspects, 
like for instance behaviour of Ising spin glasses in transverse (i.e. quantum mechanically non-commuting) fields 
\cite{Sanpera04,Ahufinger05}. 

\item{\it Disordered systems: Large effects by small disorder}. There are many examples of such situations. 
In classical statistical physics a paradigm is the random field Ising model in 2D
(that looses spontaneous magnetization at arbitrarily small disorder). In quantum physics the paradigmatic example 
is Anderson localisation, which occurs at arbitrarily small disorder in 1D, 
and should occur also at arbitrarily small disorder in 2D. Cold atom physics may address 
these questions, and, in fact, much more (cf. the Ref. \cite{Wehr06}, where a disorder breaks the continuous symmetry in 
a spin systems, and thus allows for long range ordering).

\item{\it High $T_c$ superconductivity}. Despite many years of research, opinions on the nature of high $T_c$ superconductivity still 
vary quite appreciably \cite{Claeson03}. It is, however, quite 
established (c.f. contribution of P.W. Anderson in Ref. \cite{Claeson03}) 
that understanding of the 2D Hubbard model in the, so called, $t-J$ limit \cite{Auerbach94} for two component (spin 1/2) 
fermions  provides at least a part of the explanation. 
The simulation of these models are very hard and numerical results are also full of contradictions. Cold fermionic atoms with 
spin (or pseudospin) 1/2 in optical lattices might provide a quantum simulator to resolve these problems \cite{Hofstetter02} (see also
\cite{Koetsier06}).
First  experiments with both ``spinless", i.e. polarized, as well as spin 1/2 unpolarized  ultracold fermions 
\cite{Kohl05,Stoferle06},
 and Fermi-Bose mixtures
in lattices has already been realized \cite{Ospelkaus06,Gunter06}.

\item{\it BCS-BEC cross-over}. Physics of high $T_c$ superconductivity can be also addressed with trapped ultracold gases. 
Weakly attracting spin 1/2 fermions in such situations undergo at (very) low temperatures the Bardeen-Cooper-Schrieffer (BCS) 
transition to a superfluid state of loosely bounded Cooper pairs. Weakly repulsing fermions, on the other hand may form  
bosonic molecules, which in turn may form at very low temperatures a BEC. Strongly interacting fermions undergo also a transition to 
the superfluid state, but at much higher $T$. Several groups  have employed
 the technique of Feshbach resonances \cite{Inouye98, Cornish00, Timmermans99} to observe such BCS-BEC cross-over 
(for the recent status of experiments 
see 
\cite{Regal04,Bartenstein04,Chin04,Zwierlein04,Kinast04,Kinast05,Bourdel04,Partridge05,Zwierlein05,Zwierlein06,Partidge06}).

\item{\it Frustrated antiferromagnets and spin liquids}. The ``rule of thumb" says that everywhere,
in a vicinity of  a high  $T_c$ superconducting phase, there exists a (frustrated) antiferromagnetic phase. 
Frustrated antiferromagnets  have been thus in the centre of interest in condensed matter physics  for decades. 
Particularly challenging here is the possibility of creating novel, exotic quantum phases, such as valence bond solids,
 resonating valence bond states, and  various kinds of quantum spin liquids (spin liquids of I and II  kind, 
according to C. Lhuillier \cite{Misguich04,Lhuillier05}, and topological and critical spin liquids, according to M.P.A. Fisher
\cite{Alet05a}). Cold atoms offer also in this respect opportunities to create various frustrated spin models in triangular, 
or even kagom\'e lattice \cite{Santos04}. In the latter case, it has been proposed by Damski et al. \cite{Damski05a,Damski05b} 
that cold dipolar Fermi gases, or Fermi-Bose mixtures might allow to realize a novel state of quantum matter: {\it quantum spin liquid crystal}, 
characterized by N\'eel like order at low $T$ (see also \cite{Honecker_cond-mat0609312}), 
accompanied by extravagantly high, liquid-like density of low energy excited states.

\item{\it Topological order and quantum computation}. Several very ``exotic" spin systems with topological 
order has been proposed recently \cite{Kitaev06,Doucot05} as candidates for robust quantum computing. 
Despite their unusual form, these models can be realized with cold atoms \cite{Duan03,Micheli06}. 
Particularly interesting \cite{Lewenstein06} is the recent proposal by Micheli et al. \cite{Micheli06},
who propose to  use heteronuclear polar molecules in a lattice, excite them using  microwaves to the lowest rotational level, 
and employ strong dipole-dipole interactions in the resulting spin model. The method provides an universal  ``toolbox"  for spin models with 
  designable range and spatial anisotropy of couplings.

\item{\it Systems with higher spins}. Lattice Hubbard models, or spin systems with higher spins are 
also related to many open challenges; perhaps the most famous being the Haldane conjecture concerning 
existence of a gap, or its lack  for the 1D antiferromagnetic spin chains with integer or half-integer spins, respectively (see for instance \cite{Auerbach94}). 
Ultracold spinor gases \cite{Stamper-Kurn00} might help to study these questions. Again, particularly interesting are in this 
context spinor gases in optical lattices \cite{Demler02,Imambekov03,Yip03a,Yip03b,Zawitkowski06}, 
where in the strongly interacting limit the Hamiltonian reduces to a generalized Heisenberg Hamiltonian. Using Feshbach resonances and varying lattice geometry  one should be able in 
such systems to generate variety of regimes and  quantum phases, including the most interesting antiferromagnetic (AF) regime. 
 Garc{\'i}a-Ripoll et al. \cite{Garcia-Ripoll04} propose to use a duality between the AF and ferromagnetic (F) Hamiltonians, $H_{AF}=-H_F$, 
which implies that minimal energy states of $H_{AF}$ are maximal energy states of $H_F$, and vice versa.  
Since dissipation and decoherence are practically negligible 
in such systems, and affect equally both ends of the spectrum, one can study AF physics with $H_F$, preparing adiabatically AF 
states of interest.

\item{\it Fractional quantum Hall states}. Since the famous work of Laughlin \cite{Laughlin83}, there has been enormous 
progress in our understanding of the fractional quantum Hall effect (FQHE) \cite{Jacak03}. Nevertheless, many challenges 
remain open: direct observation of the anyonic character of excitations, observation of other kinds of strongly 
correlated states, etc. FQHE states might be studied with trapped ultracold rotating  gases \cite{Wilkin00,Cooper01}. 
Rotation induces there effects equivalent to an ``artificial" constant magnetic field directed along the rotation axis.  There 
are proposals  how to detect directly fractional excitations in such systems \cite{Paredes01}. 
Optical lattices might help in this task in two aspects: first, FQHE states of small systems of 
atoms could be observed in a lattice with rotating site potentials, or an array of rotating microtraps (cf. \cite{Popp04,Barberan06} 
and references therein). Second, ``artificial" magnetic field might be directly created in an lattices via appropriate 
control of tunneling (hopping) matrix element in the corresponding Hubbard model \cite{Jaksch03}. 
Such  systems will also   allow to create FQHE type states \cite{Mueller04,Sorensen05,Palmer06}.

\item{\it Lattice gauge fields}. Gauge theories, and in particular lattice gauge theories (LGT) \cite{Montvay97}  
are fundamental for both high energy physics and condensed matter physics, and despite the progress of our 
understanding of  LGT, many questions in this area remain open.  Physics of cold atoms might help here in two aspects: 
``artificial" non-abelian magnetic fields may be created in lattice gases via appropriate control of hopping matrix elements 
\cite{Osterloh05}, or in trapped gases using effects of electromagnetically induced transparency \cite{Ruseckas05}.
One of the most challenging tasks in this context concerns the  possibility of realizing generalizations of Laughlin states 
with possibly non-abelian fractional excitations. Another challenge concerns the 
possibility of ``mimicking" the dynamics of gauge fields. 
In fact, dynamical realizations of $U(1)$ abelian gauge theory, that involves ring exchange interaction in a square lattice \cite{Buechler05}, or 3 particle interactions in a 
triangular lattice \cite{Pachos04,Tewari06}
have been also recently proposed.

\item\emph{Superchemistry}.  This is a challenge of quantum chemistry, rather than condensed matter physics: to perform
a chemical reaction in a controlled way, by using photoassociation or Feshbach resonances from a 
desired initial state to a desired final quantum state. Ref. \cite{Jaksch02} proposed to use 
MI with two identical atoms, to create via photoassociation, first a MI of homonuclear molecules, and then a molecular SF via ``quantum melting''.
In Ref. 
\cite{Damski02}, a similar idea was applied to heteronuclear molecules, in order to achieve molecular SF. 
Bloch's group have indeed
observed photoassociation of \(^{87}\)Rb molecules in MI with two atoms per site \cite{Rom04},
while Rempe's group have realized the first molecular MI using Feshbach resonances \cite{Volz06}.
Formation of three-body Efimov trimer states was observed in trapped Cs atoms by Grimm's group \cite{Kraemer06}.
This process could be even more efficient in optical lattices \cite{Stoll05}.

\item\emph{Ultracold dipolar gases}.  Some of the most facinating experimental and theoretical challenges 
of the modern atomic and molecular physics 
concern ultracold dipolar quantum gases (for a review, see \cite{Baranov02}). The recent experimental realisation of the dipolar Bose 
gas of Chromium \cite{Griesmeyer05}, and the progress in trapping and cooling of dipolar molecules \cite{SpecialIssueEurphysD04} 
have opened the path towards ultracold quantum gases with dominant dipole interactions. Dipolar BECs and BCS states of trapped gases are expected
 to exhibit very interesting dependence on the trap geometry \cite{Baranov02}. Dipolar ultracold gases in optical lattices, described by extended
 Hubbard models, should allow to realize various quantum insulating "solid" phases, such as checkerboard, 
and superfluid phases,  such as supersolid 
phase \cite{Goral02,Trefzger06}. Particularly interesting in this context are the {\it  rotating dipolar gases} (RDG). 
Bose-Einstein condensates
 of RDGs exhibit novel forms of vortex lattices: square, ''stripe crystal", and ''bubble crystal" lattices  \cite{Cooper05}. 
We have demonstrated that pseudo-hole gap survives the large $N$ limit for the Fermi RDGs \cite{Baranov05}, making them perfect 
candidates to achieve the stongly correlated regime, and to realise Laughlin liquid at filling $\nu=1/3$, and quantum Wigner 
crystal at $\nu\le 1/7$ \cite{Fehrmann:2006} with mesoscopic number of atoms
$N\simeq 50-100$. 
 
\end{itemize}
Several of the above mentioned open questions and challenges are addressed in this review. 
However, before we turn to the discussion of how ultracold atomic gases  can mimic condensed matter systems, 
let us discuss shortly the properties of optical potentials in general,  and optical lattices in particular (for a review,
see \cite{Grimm00}).

\subsection{Plan of the review}
\label{plan}

This review is addressed to two kinds of readers. 
First of all,  it gives for condensed matter and, perhaps, high energy physicists 
 an overview of what is being done in atomic physics and quantum optics in the area of ultracold gases in optical lattices.
 The particular emphasis is put here on the problems that are directly related to open problems and challenges of condensed matter,
 or even high energy physics. We discuss how to mimic condensed matter, and even go beyond toward completely new areas and problems.  
Second, the review is directed to atomic and quantum optics community. For these readers it should give some basic 
 information and basic literature about challenging problems of condensed matter physics that can be attacked with atoms, ions or molecules.

The plan of the review is as follows. In Section 2, we review the most general type of Hubbard-type  model
that can be realized with cold gases, and also review some spin models that can be reduced from the Hubbard model in 
specific limits.
In Section 3, we present some basic theoretical methods of treatment of Hubbard models. 
Most of the material here is standard in condensed matter theory, but we include also a subsection about very recent 
developments in numerical 
treatments of many body systems based on quantum information and quantum optics ideas. In the following sections  we address 
some of the 
challenges and open question described in this introduction. Each section has its own short introduction with basic 
condensed matter references to the considered problems, and then focuses on results 
obtained within the atomic physics and quantum optics context. Section 4 treats disordered ultracold gases, 
section 5 -- frustrated ultracold gases, while 
sections 6 and 7  make short overviews of spinor ultracold gases and ultracold gases in 
``artificial" magnetic fields, respectively. The final  section 8 discusses  relations between
 ultracold gases  and quantum information.  

Since the review intends to strength the analogies between condensed matter systems and cold gases in optical lattices, it is mainly focused on the strongly interacting regime. Nevertheless, there is a wide range of interesting phenomena that appears in the weakly interacting regime that are not cover here.

This review has been written by theorists, and as such describes  experiments only in aspects concerning the 
physical results, or the experimentally accessible  ranges of parameters. We do not discuss here experimental techniques
and methods. This review should be considered as  complementary to the excellent review by 
Bloch and Greiner \cite{Bloch05r}.



\section{The Hubbard and spin models with ultracold lattice gases}
\label{sec-Kalimpong-jabo}

In this section, we begin by a short discussion on optical potentials (Subsec. \ref{cold_atoms_in_lattices}), and then 
 review  the most general Hubbard-type model that can be realized with cold gases (Subsec. \ref{hubbard_models}). 
We report then (in Subsec. \ref{spin_models}) the spin models that such Hubbard models reduce to, in different limits.
In Subsec. \ref{control}, we discuss the amount of control that we have in the parameters involved in the Hubbard model, 
when realized with cold atoms.
We then focus (in Subsec. \ref{SMI}) on the 
the paradigmatic   
model of a system that 
exhibits a quantum phase transition,
namely on the homogeneous 
Bose-Hubbard (BH) model \cite{Fisher89,Sachdev99}. 
The model undergoes  the  
superfluid--Mott 
insulator (SF--MI) quantum phase
transition.

\subsection{Optical potentials}
\label{cold_atoms_in_lattices}

The basic tool to create ultracold lattice gases are optical potentials. An electron in an atom  in the presence of 
oscillating electric field ${\bf E}({\bf r},t)$ of a laser attains a time dependent dipole moment 
$\bf d$. When the field oscillations are far off resonance (i.e. they do not cause any real transition in
the atom), the induced dipole  moment follows the laser field oscillations,
\begin{equation}
{d}_i=\sum_{j=x,y,z}\alpha_{ij}(\omega_L)E_j({\bf r},t),
\end{equation}
where $d_i$ is the corresponding component of ${\bf d}$($i=x,y,z$), 
$\omega_L$ is the laser frequency,  
and $\alpha_{ij}(\omega_L)$ denotes the matrix elements of the polarizability tensor.  
The polarizability  depends in general on the laser frequency, and on the energies  
of the non-resonant excited states of the atom.  One of these states
(with excitation energy, say,  $E_1=\hbar\omega_1$)  is usually much closer to 
the resonance than the others; in such case the 
polarizability becomes inversely proportional to the laser detuning from the resonance, 
$\Delta=\omega_L-\omega_1$.

\begin{figure}[t]
\centering
\includegraphics[width=0.70\textwidth]{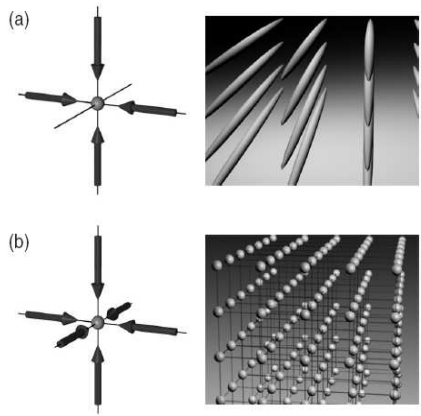}
\vspace*{0.5cm}
\caption{Schematic pictures of optical lattice potentials; a) 2D square lattice of quasi 1D traps; b) 3D simple cubic lattice (from \cite{Bloch04}).}
\label{kal-tumi-aleya}
\end{figure}

Electronic energy undergoes in this situation a shift, $\Delta E$, which is nothing else than a AC-version 
of the standard quadratic Stark effect. The energy shift is proportional to
\begin{equation}
\Delta E({\bf r})=\sum_{i,j=x,y,z}\alpha_{ij}(\omega_L)\langle E_j({\bf r},t)E_j({\bf r},t)\rangle\propto I({\bf r})/\Delta,
\label{optpot}
\end{equation}
where the bra-ket denotes  the averaging  of the product of electric fields over the fast 
optical oscillations, and $I({\bf r})$ is the laser beam intensity.  

The consequences of the above simple formula are enormous. The atom feels an optical potential 
$V_{opt}({\bf r})=\Delta E({\bf r})$, that follows the spatial pattern of the laser field 
intensity. 
This is the basis for optical manipulations and trapping of atoms! 
If the laser is red-detuned (i.e., laser detuning $\Delta<0$),
 the atoms  are attracted toward the regions of high intensity; conversely, blue 
 detuned laser pushed the atoms out of the regions of high intensity.

Adding two, or more laser fields of the same frequency leads in general to interferences and the corresponding interference 
pattern of the intensity. In particular, two counter propagating laser waves of the same polarisation will create a standing 
wave, and thus a spatially  oscillating potential 
for atoms. One can easily avoid interferences when adding more and more laser fields, and add corresponding intensities, i.e. optical 
potentials. To this aim one can use laser fields with 
orthogonal polarisations. Three pairs of counter-propagating laser beams with orthogonal polarizations will form then the 3D optical 
lattice represented schematically in Fig. \ref{kal-tumi-aleya}b.
An alternative way to avoid interferences on demand is to use slightly, but sufficiently  different frequencies. In this case, 
time averaging over the ``sufficient" different frequencies washes out the interference effects. 
Similarly, one can use laser beams with polarisations oscillating at different frequencies  to avoid interferences.

\subsection{Hubbard models}
\label{hubbard_models}

We are interested in the Hubbard-type models that are realizable with cold atoms in an optical lattice.
The simplest optical lattice is a 3D simple cubic lattice, as presented in Fig. \ref{kal-tumi-aleya}b. It is formed by three pairs of laser 
beams creating three orthogonal  standing waves with orthogonal polarisations. As we will discuss below, one can, however, 
create practically arbitrary lattices
on demand using  optical potentials. Also, as the intensity of one of the standing waves increases, the probability of hopping along 
this direction decreases rapidly to zero \cite{Jaksch98}.
In effect we obtain an 1D array of 2D square lattices. 
Consequently, an increase of  the laser intensity of another of the 
standing waves, creates effectively a 2D array  of 1D lattices (Fig. \ref{kal-tumi-aleya}a). 

Optical lattices provide an ideal (contain no defects) and rigid (do not support phonon excitations 
\footnote{This statement has to be revised  when the lattice is created
 inside of  an optical cavity. As we discuss later, the presence of atoms may affect the cavity field.})  
periodic potential in which the atoms move. As it is well known from solid state theory\cite{Ashcroft76}, single particle  
energy spectrum (in the absence of  interactions) consists of bands, and the energy eigenstates of the Hamiltonian  are 
Bloch functions. If the lattice potential is strong, the band-gaps are large, and the bands are very well separated energetically. 
For low temperatures regime, it is easy to achieve a situation in which only the lowest band is occupied (tight binding approximation). 
The Bloch functions of the lowest band can be expanded in Wannier functions, which are not the eigenstates of the single particle 
Hamiltonian, but are localised at each site. In the tight binding approximation, we project all of the atomic quantum field operators 
that describe the systems in question onto the lowest band, and then expand into the Wannier basis. Leaving then just the most relevant
terms  in the Hamiltonian (such as hopping between the nearest neighboring sites) leads directly to Hubbard-type Hamiltonian.

Let us write the most general Hubbard model that may be created in this way. Let us assume that we have 
several bosonic and fermionic species (or bosons/fermions with several internal states), enumerated by $\alpha$.
The basic objects of the theory will be thus annihilation and creation operators of $\alpha$--bosons, $b_{i\alpha}, b^{\dag}_{i\alpha}$ at the site $i$, 
and analogously   annihilation and creation operators of $\beta$--fermions, $f_{i\beta}, f^{\dag}_{i\beta}$ at the site $i$.
Bosonic (fermionic) operators fulfill, of course, the standard canonical commutation (anticommutation) relations: 
\begin{eqnarray}
&&\left[b_{i\alpha}, b_{j\alpha'}\right]=\left\{f_{i\alpha}, f_{j\alpha'}\right\}=0 \\
&&\left[b_{i\alpha}, b^{\dag}_{j\alpha'}\right]=\left\{f_{i\alpha}, f^{\dag}_{j\alpha'}\right\}=
\delta_{\alpha\alpha'}\delta_{ij},
\label{commut}
\end{eqnarray}
where $\delta_{ab}$ denotes the Kronecker delta. 

The most general Hubbard type of hamiltonian that one can realize with cold atoms, assuming  
lowest band occupation only \footnote{Some authors go beyond this assumption. See for instance Ref. \cite{notka2-ekhaney}.}, 
consists of four parts:
\begin{equation}
H_{Hubbard}=H_{hop}+ H_{int} + H_{pot} + H_{Rabi}.
\label{hubbard}
\end{equation}
The hopping part describes hopping (tunneling) of atoms from one site to another. Since the hopping probability 
amplitude decreases exponentially with the distance, hopping is typically assumed to occur between the nearest neighboring sites, 
denoted $\langle ij\rangle$. Hopping, on the other hand, might be laser assisted, and thus might allow for transitions from one 
internal state to another (of course, or perhaps unfortunately, it cannot lead to a change of element, or even isotope!),
\begin{equation}
H_{hop}=-\sum_{\alpha,\beta,\langle ij\rangle}\left[t^B_{ij \alpha\beta}b^{\dag}_{i\alpha}
b_{j\beta}+ h.c.\right]-\sum_{\alpha,\beta\langle ij\rangle}\left[t^F_{ij\alpha\beta}f^{\dag}_{i\alpha}
f_{j\beta}+ h.c.\right].
\label{hoping}
\end{equation}

Atoms interact in the first place via short range Van der Waals forces, which in the low energy limit are very well described 
by various kinds of the zero range pseudopotentials 
\cite{Pitaevskii03,Huang87,Idziaszek06}. That means that the dominant part of the interactions 
is of contact type, i.e. occurs on-site. There are, however, situations in which  interactions do affect 
neighbouring sites, or even have long range (such as dipole-dipole interactions). We thus write the interaction part as 
\begin{equation}
H_{int}=H_{on-site}+ H_{ext},
\label{int}
\end{equation}  
where
\begin{eqnarray}
H_{on-site}&=&\frac{1}{2}\sum_{i, \alpha,\beta,\alpha'\beta'}\left[U^{BB}_{\alpha\beta\alpha'\beta'}(i) b^{\dag}_{i\alpha}b^{\dag}_{i\beta}
b_{i\beta'}b_{i\alpha'} +
U^{FF}_{\alpha\beta\alpha'\beta'}(i) f^{\dag}_{i\alpha}f^{\dag}_{i\beta}f_{i\beta'}f_{i\alpha'} \right.\nonumber\\
&+&\left.
2U^{BF}_{\alpha\beta\alpha'\beta'}(i) b^{\dag}_{i\alpha}f^{\dag}_{i\beta}f_{i\beta'}b_{i\alpha'}\right].
\label{int-onsite}
\end{eqnarray}
In the simplest Hubbard models interactions depend only on the on--site atom numbers, $n^B_{i\alpha}=b^{\dag}_{i\alpha}b_{i\alpha}$, 
and $n^F_{i\alpha}=f^{\dag}_{i\alpha}f_{i\alpha}$.
In general, however,  they may depend in an non-trivial way on internal states, or atomic species; this is for instance the standard case for 
spinor gases \cite{Stamper-Kurn00}. 
The models with non-contact interactions are usually termed as extended Hubbard models, and include
\begin{eqnarray}
H_{ext}&=&\frac{1}{2}\sum_{i,j \alpha,\beta,\alpha',\beta'}\left[V^{BB}_{\alpha\beta\alpha'\beta'}(i,j) b^{\dag}_{i\alpha}b^{\dag}_{j\beta}b_{j\beta'}b_{i\alpha'} +
V^{FF}_{\alpha\beta\alpha'\beta'}(i,j) f^{\dag}_{i\alpha}f^{\dag}_{j\beta}f_{j\beta'}f_{i\alpha'}\right.\nonumber\\
&+&\left. 
2V^{BF}_{\alpha\beta\alpha'\beta'}(i,j) b^{\dag}_{i\alpha}f^{\dag}_{j\beta}f_{j\beta'}b_{i\alpha'}\right].
\label{int-onsite-eta-ageo-chhilo}
\end{eqnarray}
Most typically, the non-contact interactions will depend on the distance between the sites, $|{\bf r}_i-{\bf r}_j|$, and will be of the density-density form
(i.e. they depend only on $n^B_{i\alpha}$, and $n^F_{i\beta}$), but  in general, again  this  does not have to be the case. 
Dipolar interactions, for instance, depend on the angles between the dipole moments and on the vector  
${\bf r}_i-{\bf r}_j$.

The last two parts of the Hamiltonian (Eq. (\ref{hubbard})) describe  on--site single atom processes, and essentially have the same form as the tunneling part, 
except that they occur on-site. We do distinguish them since they are  well defined and controllable in experiments. 
$H_{pot}$  combines effects of all potentials felt by atoms such as external trapping potential, possible additional superlattice
(i.e. additional lattice) potentials, possible disorder potentials, and last,  but not least, chemical potential which is necessary if one uses
 the statistical description based on the grand canonical ensemble: 
\begin{equation}
H_{pot}=-\sum_{\alpha,i}\left[\mu^B_{i\alpha}b^{\dag}_{i\alpha}
b_{i\alpha}+ \mu^F_{i\alpha}f^{\dag}_{i\alpha}
f_{i\alpha}\right].
\label{pot-onsite}
\end{equation}
 The last part of the Hamiltonian, \(H_{Rabi}\), describes possible coherent
 on--site transitions between the internals states of atoms; such transitions may be achieved using laser induced resonant Raman
transitions, or microwave Rabi type transitions (we can write this part of the Hamiltonian as time--independent in the interaction picture with respect to the on site internal states Hamiltonian)    
\begin{equation}
H_{Rabi}=\frac{1}{2}\sum_{\alpha,\beta,i}\left[\Omega^B_{i\alpha\beta}b^{\dag}_{i\alpha}
b_{i\beta}+ \Omega^F_{i\alpha\beta}f^{\dag}_{i\alpha}
f_{i\beta}\right].
\label{pot-onsite-eta-ageo-chhilo!}
\end{equation}

\subsection{Spin models}
\label{spin_models}

As it is well known (see for instance \cite{Auerbach94}), Hubbard models reduce to spin models in certain limits. 
Most of these limits, and even more, are accessible with cold atoms. 
Generally speaking, if  bosonic atoms can occupy only $2S+1$ different states in a lattice site, 
then one can always map these states 
onto the states of pseudo-spin $S$. These states may even correspond to 
different number of bosons, and the dimension of the local, on--site Hilbert space might vary in space; in such 
case we will deal with inhomogeneous models, where at each site there is, in general, a different spin. Similar construction 
might be done for fermionic atoms, with the remark that at a given site, the fermion number differences  might attain only even 
numbers, since otherwise the fermionic character of particles cannot be eliminated. 

When constructing specific spin models two aspects play a role: lattice geometry (which we discuss in the next subsection), 
and the form of interactions, which includes Ising, $XY$, Heisenberg, $XXZ$, and anisotropic $XYZ$ types, as well as 
ring exchange types. Below  we list the most obvious constructions of spin models, that has been discussed in the literature on cold atoms.
We restrict ourselves here to translationally invariant models.  

\begin{itemize}

\item {\it Hard core bosons and $XY$ models}.    Perhaps the simplest way to obtain a non-trivial spin model is to use the simplest Bose-Hubbard 
Hamiltonian for one component (``spinless") bosons
\begin{equation}
H=-t\sum_{\langle ij\rangle}\left[b^{\dag}_{i}
b_{j}+ h.c.\right] + \frac{U}{2}\sum_i n_i(n_i-1) -\mu\sum_i n_i,
\label{bosehub}
\end{equation}
where ${\langle ij\rangle}$ denotes sum over nearest neighbors. In the hard boson limit (i.e. when $U\gg t,\mu$) we may have at most 1 boson per site. We may encode the spin 1/2 states as presence ($\uparrow$), 
or absence ($\downarrow$) of the boson at the site. The Hamiltonian reduces then to that of $XY$ model in a transverse field,
\begin{equation}
H=-t\sum_{\langle ij\rangle}\left[\sigma^{\dag}_{i}
\sigma_{j}+ h.c.\right]  -\frac{\mu}{2}\sum_i (\sigma_{z,i}+1),
\label{xytrans}
\end{equation}
where $\sigma_i=(\sigma_{x,i}+i\sigma_{y,i})/2$ and $\sigma_{x,y,z,i}$ denote the standard Pauli matrices at site $i$. This model has the advantage that in 1D it is exactly solvable via 
Jordan--Wigner transformation \cite{Sachdev99}. One interesting application of this approach concerns the 1D disordered chain studied in 
Ref. \cite{deMartino05}. The same approach was used recently in \cite{Wehr06} to realize $XY$ model in random parallel field. 

\item{\it Spatially delocalised qubits}. Somewhat similar idea considers two neighbouring traps, 
or potential wells (``left" and ``right"), 
assumes 1 atom per double well, and  encodes the spin 1/2 (qubit) 
as the presence of the atom on the ``left", or ``right", respectively \cite{Mompart03}. 
In Ref. \cite{Dorner03} was proposed to encode one qubit by the presence or the absence of a whole string of neutral atoms yielding improved robustness. This system may be used for 
generation of maximally entangled many atom states (Schr\"odinger cat
states) by crossing a quantum phase transition.

\item{\it Multi-component atoms in Mott states}. Whenever we deal with a system of multicomponent atoms, i.e. atoms 
with say $2S+1$ internal states, in the Mott insulator limit, the system will be well described by the appropriate spin model. 
The most  prominent example is a two-component (or spin 1/2) Fermi gas \cite{Auerbach94}, 
which in the Mott state with one atom per site forms a perfect Heisenberg model. Several groups are planning experiments with 
ultracold spin 1/2 Fermi atoms heading toward antiferromagnetism in various kinds of lattices. Prospects for observing
antiferromagnetic transition in such systems are quite good, especially since one expect to be able to employ interaction 
induced cooling (an analogue of the Pomeranchuk effect, known in liquid Helium physics), \cite{Werner05}, or disorder 
induced increase of $T_c$ \cite{Wehr06}.  
One should stress, however,
 that although the Mott transition takes typically place when $t<U$, the low temperature 
physics of the resulting Heisenberg model requires temperatures  of order $t^2/U\simeq k_BT$. Such temperatures 
are often in the nanoKelvin range, i.e. still hard to achieve experimentally. There is, however, a lot of new proposals of 
cooling atoms in the Mott states\cite{Rabl03,Daley05,Popp06a,Popp06b}, and hopefully the temperatures will not set any 
limitations on experiments 
with these  kinds of spin models in the next future.   

The calculation of the effective Hamiltonians in pseudo-spin 1/2 Bose-Bose or Bose-Fermi 
mixtures is quite complicated, and 
was accomplished for the first time recently \cite{Svistunov03,Lewenstein04}.  The Bose-Bose case 
can be reduced 
to an XXZ spin model (for the case of XXZ model in random fields see for instance \cite{Wehr06}). The 
effective Hamiltonian for the Fermi-Bose mixture  in general cannot be reduced to a spin model, 
since it involves Fermi operators describing {\it composite fermions}, consisting of one fermion 
paired with some number of bosons, or bosonic holes  \cite{Lewenstein04}. The Hamiltonian describes
 a ``spinless" interacting Fermi gas of such composites. It can, however, 
 be transformed to an XXZ model in external fields in 1D via Jordan-Wigner transformation. 
 
 \item{\it Spinor gases in Mott states}. Of course, the above statements are particularly valid for spinor gases, 
which for atoms with spin 
 $F>1/2$ have effective Hamiltonians containing generalizations (a power series) of Heisenberg interactions. For $F=1$ 
and the Mott state with 1 atom per site, we deal with the so-called Quadratic-Biquadratic Hamiltonian 
\cite{Demler02,Imambekov03,Yip03a,Yip03b,Garcia-Ripoll04}. In the Mott state with two atoms per site, the pair can  either
compose a singlet state or a state with on--site  spin $S=2$. The resulting Hamiltonian contains then higher powers of the Heisenberg term. 
The situation is obviously more complicated for higher Mott states,  and atoms with higher individual spin $F$. In Refs. 
\cite{Zawitkowski06, barnett, zhoulast}, the case $F=2$ is studied; here, already with one atom per site, the effective Hamiltonian is a 
polynomial of the fourth order in Heisenberg term.  

\item{\it Spin models in polymerized lattices}. Yet another interesting way to obtain spin models with cold atoms, 
is based on the use of polymerized (dimerized, trimerized, quadrumerized etc.) lattices. 
These are lattices that can be easily realised 
with optical potentials, and have no analogue in condensed matter physics. A simple example of dimerized lattice in 2D 
 is a square lattice of pairs of close sites; trimerized kagom\'e lattice, discussed in section 4, is a triangular lattices of 
 trimes of close sites located on a small unilateral triangle; 2D quadrumerized square lattice is a square lattice of 
small squares of close sites, etc. When one considers
ultracold gases in such lattices, one has to  take into account  first the 
lowest energy state in a dimer, trimer etc. If we deal with polarized ``spinless" fermions in trimerized lattice, 
and we consider two fermions per trimer, fermions have to their  disposal zero momentum state 
(which will be necessarily filled at low temperatures), and the two states 
with left and right chirality. The latter two are obviously degenerated, and can thus be used to encode the effective spin 1/2.
This is the model discussed in Refs. \cite{Santos04,Damski05a,Damski05b}.    Note, that similar encoding 
is possible in a quatromerized lattice with 2 atoms per quadrumer.

\end{itemize}

\subsection{Control of parameters in cold atom systems}
\label{control}

Atomic physics and quantum optics offer many new types of methods to quantum engineer systems in question. 
{\it Toutes proportions gard{\'e}es},
there are instances in which atomic physics and quantum optics not only meets, but rather ``beats" condensed matter physics. That is 
one of the reasons why the physics of ultracold atoms attracts so many theorists from other disciplines. 

Let us list shortly what can be controlled in the experiments with cold atoms in optical lattices:

\begin{itemize}

\item{\it Lattice geometry and dimensionality}. As mentioned above, practically any lattice geometry may be achieved with optical potentials. 
The method of superlattices (i.e. adding a new lattice on top of the exiting one) is very well developed. Changing of the lattice 
dimentionality does not pose any problem (compare Fig. \ref{kal-tumi-aleya}).
Also periodic boundary conditions 
can be realized in ring shapeed optical lattices \cite{Amico_PRL95_063201}.

\item{\it Phonons}. 
Optical lattices are rigid and robust: they do not have any phonons. An interesting 
situation arises when 
the lattice is formed in an optical cavity: atom-light coupling might suffice then to shift the cavity resonance. Cavity 
will affect the atoms, but atoms will perform back action, and create ``phonon" like 
excitations. For early works, see  \cite{Maschler05,Nagy06}. For 
more recent studies of superfluid-Mott insulator crossover, see \cite{Larsson06}, and for the first attempts
towards ``refracton'' physics (analogs of ``phonons''), see \cite{Lewenstein_cond-mat0609587}.
Phonons, obviously, play a role in ion traps, 
where they provide the major mechanism for ion-ion interactions.

\item{\it Tunneling}. Tunneling can be controlled to a great extend using combination of pure tunneling, 
laser assisted coherent 
transitions,
and lattice tilting (acceleration) techniques.    
The prominent example of such control describe the proposals for creating artificial 
magnetic fields 
\cite{Jaksch03,Mueller04,Sorensen05,Palmer06,Osterloh05}.

\item{\it On-site interactions}.
These interactions are controlled by scattering lengths, which 
can be modified using Feshbach resonances in magnetic fields, \cite{Inouye98, Cornish00, Timmermans99}, or optical Feshbach resonances 
(for theory see \cite{Fedichev96}, for experiments \cite{Theis04,Thalhammer05}). On--site interactions can be set to zero 
in dipolar gases, by changing the shape of the on--site potential \cite{Goral02}.

\item{\it Next neighbour and long range interactions}. Effective models obtained by calculating effect of tunneling in the 
Mott insulator phases, contain typically short range 
interactions of energies $\propto t^2/U$. Stronger interactions can be achieved using dipolar interactions, such as
those proposed in Refs. 
\cite{Goral02,Micheli06,Barnett06}. Dipolar interactions are of  long range type, 
are anisotropic, and exhibit a very rich variety of phenomena (for a review, see \cite{Baranov02}). 
They can also be achieved in trapped ion systems, where 
they are mediated via phonon vibrations of the equilibrium ionic configuration.   This case is discussed 
in detail in the section \ref{artificial} of this review. The group of T. Pfau has 
recently realized the first experimental observation of ultracold dipolar gas \cite{Griesmeyer05}, by condensing bosonic Chromium. 
Dipolar interaction are mediated here by magnetic dipoles of the Chromium atoms; they are weak, but nevertheless lead to observable 
effects \cite{Stuhler05}.   

\item{\it Multiparticle (plaquette) interactions}. It has been also demonstrated how
to generate  effective three-body interactions in triangular optical lattices\cite{Pachos04}. These interactions result
 from the possibility of atoms tunneling along two different paths. Similarly, ring exchange interactions in square 
 optical lattice can be generated employing the correlated hopping of two bosons \cite{Buechler05}.  

\item{\it Potentials}. Various types of external potentials can be applied to the atoms, depending on the situations. One can use 
magnetic potentials whose shape can be at least cotrolled on the scale of few microns. Magnetic potentials with larger gradients can be created 
on  atom chips (cf. \cite{Jorg1}). The most flexible are, however, optical potentials. Apart from limitations set by the diffraction limit, 
they can have practically any desired shape and can form any kind of optical lattice: regular, disordered, modulated, etc. Recently, 
J. Schmiedmayer demonstrated also great possibilities offered by the so called radio frequency potentials \cite{Jorg2}.

\item{\it Rabi transitions}. Similarly, apart from  limitations set by diffraction, they are in practice highly controlable. 

\item{\it Temperatures}. The typical critical temperatures $T$ of trapped ultracold condensed Bose gases are of order of  nano-Kelvins.  
Using evaporative cooling one can reach, however, lower $T$ (which in fact are not very well known, because of the lack of reliable temperature 
measurement methods. For recent advances see \cite{temperature}). 
Similarly, the temperature of superfluid Fermi gases are in the range of tens 
of nK. One can thus say that temperatures in the  range of tens of nK are becoming nowadays a standard. There are many proposals for reaching even 
lower $T$s employing additional cooling and filtering procedures \cite{Popp06b}. 
SF-MI transition occurs in the regime of $T$s accessible nowadays. 
Many of the strongly correlated phases occur in the regime when the tunneling $t$ is much smaller than $U$ and require temperatures of 
order $k_BT\simeq t^2/U$, i.e. 
10-20 nK, or even less. This is at the border of the current possibilities, but the progress in cooling and quantum engineering techniques 
allow us to believe that these limitations will be overcome very soon (for a detailed discussion see \cite{Fehrmann04a})

\item{\it Time dependences}. The time scales of coherent unitary dynamics of these systems are typically in the millisecond range. 
It implies that, in contrast to condensed matter systems, all of the controls 
discussed above can be made time dependent, adiabatic, or diabatic, on demand. Some of the 
fascinating possibilities include change of lattice geometry, 
or turn-on of the disorder in real time.

\end{itemize}

The huge range of parameters which are experimentally controllable indicates the rich possibilities offered 
by cold gases in optical lattices to implement condensed matter models and beyond. The first proof came with the seminal
paper of Greiner {\it et al.} \cite{Greiner02} reporting the superfluid-Mott insulator transition with cold bosons in an optical lattice.

\subsection{Superfluid - Mott insulator quantum phase transition in the Bose Hubbard model} 
\label{SMI}
Let us now consider the ideal homogeneous Bose-Hubbard (BH) 
model of the form
\begin{equation}
H= -t \sum_{\langle i,j\rangle} (b_i^\dag b_j + h.c.)
+\frac{U}{2} \sum_i n_i(n_i-1),
\label{BHH}
\end{equation}
where $\langle i,j\rangle$ indicates sum over nearest neighbors. 
We denote here the tunneling energy by $t$ 
(both $t$ and $J$ are used in the literature). 
Below we discuss the superfluid (SF) - Mott insulator (MI) quantum 
phase transition separately for the case of 
transitions at fixed density, and for the, so-called, generic
(density-driven) transitions, where the number of atoms  changes:
\begin{itemize}

\item{\it Transitions at fixed integer density.}
We consider here transitions  driven by a change of 
$t/U$ ratio in  a system with a fixed number of bosons.
The phase transition occurs when the lattice 
filling factor $\bar n$ (the number of atoms per site) is exactly integer. 
For $t/U<(t/U)_c \ll 1$ there is a Mott insulator (MI) phase, while 
for $t/U>(t/U)_c$ there is a superfluid (SF) phase. As discussed 
by Fisher {\it et al.} \cite{Fisher89}, such a transition 
in a $d$-dimensional BH model lies in
the universality class of the $(d+1)$-dimensional $XY$ spin model. 
This result implies that the one dimensional Bose-Hubbard 
model undergoes a Kosterlitz-Thouless phase transition. Also it permits
to determine the critical exponents. The quantum phase transition happens,
 ideally, at absolute temperature equal to zero \cite{Sachdev99}, 
and its signatures are reflected in different quantities as discussed
below.

The superfluid fraction \cite{Roth03c,Lieb02}, which is defined through 
the response of the system to an externally imposed velocity field 
(equivalent to a twist in boundary conditions)  vanishes in the Mott phase. 
This was verified in a 1D  system numerically by a DMRG calculation \cite{Pai96}, and analytically in a simplified BH model subjected to the restriction 
of a maximal site occupation of two particles \cite{Krauth91a}. 
Additionally, a jump in the superfluid fraction was observed 
at the transition point\cite{Pai96} of a one dimensional 
system, a result expected at the 
Kosterlitz-Thouless critical point \cite{Kosterlitz73}.
It is worth to stress  here,  that the superfluid fraction 
should not be confused with the condensate fraction \cite{Leggett01}. 
The latter one is equal to the highest eigenvalue of 
the single particle density matrix $\langle b^\dag_i  b_j\rangle$, divided by the number of particles.
These quantities, both equal  
$100\%$ for an untrapped 3D Bose-Einstein condensate in the dilute
limit \cite{Lieb02}, can be very different in the Bose-Hubbard model, 
see e.g. \cite{Damski03} for a simple example where the condensate 
fraction is $100\%$, while the superfluid fraction hits zero once a sufficiently strong on-site disorder is present.

The excitation spectrum is gapless for the SF phase while gapped for the MI. In the MI neighborhood of the transition point, the gap scales 
exponentially in a one dimensional system as
$\propto \exp(\sim[(t/U)_c-t/U]^{-1/2})$, where the proportionality factor 
in the exponent is smaller than zero.
In two and three dimensional models, it exhibits a power law behaviour
$\propto[(t/U)_c-t/U]^{z\nu}$,
where $z$ and $\nu$ are the critical exponents \cite{Sachdev99,Fisher89}.
These exponents for a two dimensional system
are $z=1$ and $\nu\approx0.67$ \cite{Campostrini01}, and for a three dimensional model
they read as $z=1$ and $\nu=1/2$ \cite{Fisher89}.

Another quantity that shows a critical behaviour is the correlation length
$\xi$:
\begin{equation}
\xi^2= 1/2\sum_r r^2  \langle  b^\dag_j b_{j+r}+ h.c.\rangle/\sum_r 
\langle  b^\dag_j b_{j+r} + h.c.\rangle.
\label{xi}
\end{equation} 
It is finite in the Mott phase,
diverges at the critical point and stays divergent in the superfluid phase.
In the neighborhood of the critical point 
it behaves as: $\xi \sim {\rm gap}^{-1/z}$, where
the critical exponent $z=1$ in one, two, and three dimensional
systems \cite{Fisher89}.

\begin{figure}[t]
\centering
\includegraphics[width=8cm]{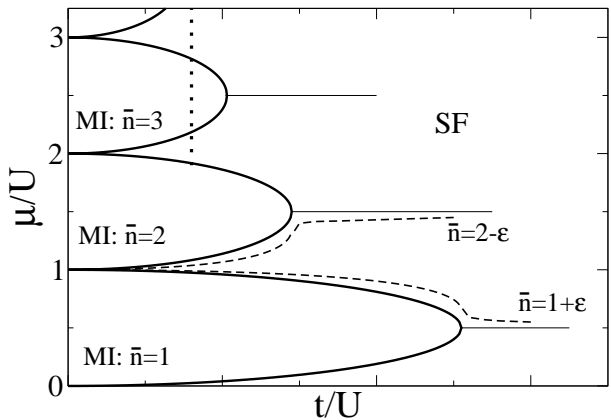} 
\caption{A schematic plot of the phase diagram of the Bose-Hubbard model.
The lobes, surrounded by the superfluid sea, 
correspond to the Mott insulator islands with integer filling factor $\bar n$. 
The thin solid lines represent the lines of constant, integer density. The dashed lines
show  trajectories  of a system with  fixed, non-integer,
filling factor $\bar n$ ($0<\varepsilon\ll1$). The dotted line presents an
example of  a trajectory leading to a generic phase transition when
the system enters either Mott, or superfluid phase by  changing a total number of atoms.
}
\label{loby}
\end{figure}

Additional insight into the superfluid--Mott insulator phase transition 
can be obtained investigating the relation between $\mu/U$, $t/U$, 
and the number of particles in the system. The chemical potential $\mu$
is conveniently introduced during  minimization of 
$\langle H-\mu\sum_i n_i\rangle$ leading to the determination of 
ground state with a $\mu$-dependent total number of atoms. 
When  $t/U=0$, one obtains that for 
$\bar n-1<\mu/U<\bar n$ ($\bar n$ is an integer number) the ground state is 
a single Fock state
\begin{equation}
|\bar n,\bar n, \dots\rangle,
\label{fock}
\end{equation}
When tunneling is nonzero, the range of $\mu/U$ describing the system with 
integer $\bar{n}$ gradually shrinks and finally disappears at some 
$\{(\mu/U)_c,(t/U)_c\}$. In this way, the famous lobes are formed: see Fig. \ref{loby}.
The fixed density transition happens when the system 
moves along the thin solid lines depicted in Fig. \ref{loby}. 
This schematic plot
illustrates also that a system with a non-integer filling factor  
never enters the MI lobes, i.e., stays always superfluid 
as depicted by the dashed lines. It is 
easily understood by considering a state with one particle added
(subtracted) to (from) a system having integer filling factor.
Such a particle (hole) can freely flow through the lattice, 
so that  the system becomes  superfluid for all values of $t/U$ ratio. 

It is important to stress that though inside the lobes 
the filling factor $\bar n$ is integer, the ground state 
is not a single Fock state (\ref{fock}) for $t/U\neq0$.
To illustrate this fact one can look at the expectation values of some
operators at the {\it transition point} of a 1D BH model at $\bar n=1$. 
The ground state wave-function of this
system  has presumably pronounced deviations from (\ref{fock}).
For instance, one obtaines there that (i) the 
nearest neighbor correlation length function 
$\langle b_j^\dag b_{j+1}\rangle$ is approximately $0.8$ and (ii) 
the variance of on-site number operator, 
$[\langle n_j^2 \rangle-\langle n_j \rangle^2]^{1/2}$, equals about $0.6$ 
\cite{Damski06}. These quantities significantly differ from 
the predictions obtained from the ground state for $t/U=0$ (\ref{fock}). 
Notice also that they attain substantial values since they  
are bounded by unity for any $t/U$. 

It is  interesting to ask what are the critical points $(t/U)_c$  
in different dimensions, and at different filling factors. 
Here we list the most accurate estimations up to date 
for the case $\bar n= 1$ which has been systematically studied in the past. 
In one dimension, the critical point was precisely determined by DMRG 
analysis:  $(t/U)_c\approx0.29$ \cite{Kuhner00}.
In two dimensional system, a recent quantum Monte Carlo studies 
estimate the position of the critical point, $(t/U)_c$, to be around 
$0.061$ \cite{Wessel04}. In the three dimensional model, 
the perturbative expansion gives 
$(t/U)_c\approx0.034$ \cite{Freericks96}.
The locations  of the critical points for different small filling factors  in
different dimensional lattices  were 
quite accurately calculated in Ref. \cite{Freericks96}.

\item{\it Generic transitions.}
\label{generic}
A quick look at the dotted line in Fig. \ref{loby} reveals that the system can cross the
superfluid - Mott insulator phase boundary through  trajectories that 
do not correspond to a fixed filling factor. 
In such a case we say that the system undergoes the generic
(density-driven) SF - MI phase transition. This transition  does 
not belong to the universality class of the $XY$ spin model, 
thus it is characterized by different 
critical exponents. In particular, one finds that $z\nu=1$ during generic 
transitions in 1D, 2D and 3D systems \cite{Fisher89}.

Since for this case the number of particles is not conserved, 
one should add the term $-\mu\sum_i
n_i$ to the Hamiltonian (\ref{BHH}), and find a ground state with
a number of particles depending on the chemical potential $\mu$.
The critical behaviour can be observed in at least two quantities: 
the compressibility  $\kappa=\partial\rho/\partial\mu$, 
(where $\rho$ is atom density),
and the superfluid fraction $\rho_s$. 
The first one, diverges as one approaches the Mott
lobes from a superfluid side (Fig. \ref{loby}), while the second  
goes to zero in this limit. 
Both the compressibility and the superfluid fraction 
stay zero inside the lobes. 

To illustrate these statements let us consider a one dimensional system,
at fixed $t/U$, undergoing a phase transition induced by a change 
in the number of atoms.
In this case, the theory of Fisher {\it et al.} \cite{Fisher89}
predicts that (i) $\kappa\sim|\mu-\tilde\mu|^{-\nu(z-1)}$ 
($\tilde\mu$ is a chemical potential at phase boundary); (ii) 
$\rho_s\sim|\rho-\tilde\rho|^{z-1}$ ($\tilde\rho$ is a filling factor
of a Mott lobe approached during transition). 
The  early Quantum Monte Carlo simulations \cite{Batrouni90,Batrouni92}
have verified these predictions giving the following estimations 
of the critical exponents: $z\approx2.04$ and $\nu\approx0.48$.


Recently, there has been a lot of interest in generalizing MI-SF transition and Bose-Hubbard model to more 
``exotic'' situations, such as atoms in 
optical cavities \cite{Lewenstein_cond-mat0609587, Larsson06}, or \(p\) band Hubbard model, where transverse staggered order occurs
\cite{Wu_PRA_74_013607}. In triangular lattices, stripe order was predicted both in SF and MI phases \cite{Wu_PRL_97_190406}. 

\end{itemize}

\section{The Hubbard model: Methods of treatment}
\label{sec-sadher-lau}
\subsection{Introduction}
\label{subsec-sadher-lau}

In this section we review at a rather basic level 
some of the standard theoretical tools used in condensed matter theory 
to treat many body systems of interest. We discuss also novel  
developments which  -- taking advantage of quantum information methods -- 
provide an efficient way to calculate ground state properties  
and dynamical evolution of many condensed matter systems. 
The underlying philosophy of these new methods is closely 
related to the well-established  
Density Matrix Renormalization Group (DMRG) method \cite{White92,Schollwoeck05}, 
and consist in truncating the dimension of the 
Hilbert space,
which diverges exponentially with the size of the system, to a manageable
size, considering entanglement properties 
of different bipartite partitions.

Analytical and numerical methods often rely on the size and 
dimensionality of the system. Powerful techniques like bosonization \cite{Giamarchi04,Cazalilla04,Tsvelik04},
Bethe ansatz  \cite{Korepin94,Korepin97,Essler05}, Jordan--Wigner transformation \cite{Sachdev99,Tsvelik03}, 
or the mentioned DMRG exist and allow to solve some paradigmatic one dimensional
systems, such as for instance Heisenberg spin $1/2$, $XXZ$ spin chains, or 1D Hubbard model of  strongly interacting
electrons. These methods are very well established to treat many body problems in one dimension,
but often fail in higher dimensions. Finite size effects are also crucial 
in the study of strongly correlated systems, because 
quantum phase transitions occur only in the thermodynamic 
limit at zero temperature. It is thus important to know   
how finite size effects affect the statics and 
dynamics of strongly correlated systems, and the signatures of 
 quantum phase transitions.

We begin this section by focussing 
on methods of treatment of the ideal homogeneous Bose Hubbard model. As we have discussed in the previous section, 
this system exhibits a phase transition:
the superfluid--Mott 
insulator (SF--MI) quantum phase
transition. Despite  its simplicity, the interplay between tunneling
and on-site interactions in the BH model is by no means trivial.
Excitations in the limit of small interactions can be described with 
the help of Bogoliubov transformations. 
In the strong interaction limit, 
when the ratio between tunneling and on-site interactions 
is much smaller than one, tunneling can be treated as a perturbation, and the original 
Hamiltonian can be replaced by an effective  Hamiltonian. 
In some cases, it is also possible 
to use a mean field approach, like for instance the Gutzwiller ansatz, which assumes that the 
many body wave functions have a product--over--sites form, and is    
conceptually and numerically relatively easy to implement. 
As we shall see, it predicts  quite correctly the critical points 
separating the Mott phase from the SF phase for 3D, or even 2D  lattices, but
its accuracy decreases dramatically  for 1D systems. 
We discuss then briefly exact diagonalisation, Quantum Monte Carlo, and phase space methods.
Subsequently, we discuss very shortly 1D methods: Bethe ansatz, Jordan--Wigner and bosonisation.  
We analyze in more detail the novel approach to DMRG provided by 
Quantum Information. Finaly, we discuss some methods to treat Fermi and Fermi-bose Hubbard models.

\subsection{Weak interactions limit}
\label{subsec-alaler-ghorer-dulal}

Here we want to illustrate how the Bose-Hubbard model can be solved in the limit of small interactions. Our discussion follows Ref. \cite{Oosten01}, where 
the Bogoliubov approach was developed.

After adding the $-\mu\sum_i n_i$ term to Eq.(\ref{BHH}), the Hamiltonian can be 
conveniently written as: 
\begin{equation}
 H= -t\sum_{\langle i,j\rangle} (b_i^\dag b_j+h.c.)+ \frac{U}{2}\sum_i 
b_i^\dag b_i^\dag b_i b_i-\mu\sum_i  b_i^\dag b_i,
\label{BHa}
\end{equation}
where $\mu$ denotes the chemical potential. We assume a regular $d$-dimensional lattice,
consisting on $M$ sites, and the distance between neighboring sites is ``$a$". 
In the limit of $t/U\to \infty$, interactions between atoms 
are negligible. The system is completely condensed  in the ground
state, and $N_0$ (the number of condensed atoms) equals $N$ (the 
total number of
atoms). When interactions become non-negligible,  atoms gradually leave
the condensate. To describe this process, it is convenient to work in momentum space:
$b_j=\frac{1}{\sqrt{M}}\sum_{\vec k}\exp(-i\vec k\vec x_j) a_{\vec k}$, 
where $\vec x_j$ points into $j$-th lattice site, and $\vec k$ is discretised 
 over the first Brillouin zone.
Using the identity 
$\sum_i e^{i(\vec k_1-\vec k_2)\vec x_i}= M\delta_{\vec k_1,\vec k_2}$ one
obtains: 
\begin{equation}
 H= \sum_{\vec k} \left[-\varepsilon(\vec k)-\mu\right] a^\dag_{\vec k}
 a_{\vec{k}}
+\frac{U}{2M}\sum_{\vec
k_1,\vec k_2, \vec k_3, \vec k_4}\delta_{\vec k_1+\vec k_2,\vec k_3+\vec k_4}
 a_{\vec k_1}^\dag  a_{\vec k_2}^\dag a_{\vec k_3} a_{\vec
k_4},
\label{BHk}
\end{equation}
where $\varepsilon(\vec k)= 2t\sum_{i=1}^d\cos(k_i a)$. As $t/U\to \infty$,
the ground state converges towards $\sim a_0^{\dag N}|0\rangle$. 
The Bogoliubov approach relies on the transformation 
$a_0\to \sqrt{N_0}+  a_0$, 
where the {\it new} operator $ a_0$ is responsible for {\it fluctuations} of 
the number of condensed atoms. Substituting the above expression in Eq.(\ref{BHk}), one
finds, up to the quadratic terms: 
\begin{eqnarray}
H &=& N_0\left(-zt-\mu+\frac{U}{2}n_0\right)+ \sqrt{N_0}(Un_0-tz-\mu)( a_0+ a_0^\dag)
\nonumber\\ &+&
\sum_{\vec k} (-\varepsilon(\vec k)-\mu) a_{\vec k}^\dag a_{\vec k}
+\frac{Un_0}{2}\sum_{\vec k} (4 a_{\vec k}^\dag a_{\vec k}
+ a_{\vec k} a_{-\vec k} +
 a_{\vec k}^\dag a_{-\vec k}^\dag),
\label{BH2}
\end{eqnarray}
where $n_0=N_0/M$ is the condensate density and $z=2d$. 
Setting the chemical potential to $\mu=Un_0-zt$ removes the linear part 
while the quadratic one is diagonalized 
by the Bogoliubov transformation: 
$ c_{\vec k}= u_{\vec k}  a_{\vec k}+ v_{\vec k} a_{-\vec k}^\dag$. Notice
that $|u_{\vec k}|^2-|v_{\vec k}|^2=1$ from the requirement that 
$[ c_{\vec k}, c_{\vec k}^\dag]= 1$.
After a simple algebra, one obtains that within the quadratic approximation
the Hamiltonian reduces to:
\begin{equation}
 H= -\frac{Un_0N_0}{2}+ 
\frac{1}{2}\sum_{\vec{k}}(\hbar\omega_{\vec k}+\varepsilon(\vec k)-zt-Un_0)+
\sum_{\vec k}\hbar\omega_{\vec k} c_{\vec k}^\dag c_{\vec k},
\label{HB}
\end{equation}
if  
\begin{eqnarray}
&&(|u_{\vec k}|^2+|v_{\vec k}|^2)\left[Un_0-\varepsilon(\vec k)+zt\right]-
Un_0(v_{\vec k}u_{\vec k}^*+v_{\vec k}^*u_{\vec k})=\hbar\omega_{\vec k},\nonumber\\
&&(u_{\vec k}^2+v_{\vec k}^2)Un_0- 2\left[Un_0-\varepsilon(\vec k)+zt\right]
v_{\vec k}u_{\vec k}= 0.\nonumber
\end{eqnarray}
Assuming that $u_{\vec k}$ and $v_{\vec k}$ are real, one easily obtains from these
equations: 
\begin{eqnarray}
\label{homega}
&\hbar\omega_{\vec k}&= \sqrt{[zt-\varepsilon(\vec k)]^2+2Un_0[zt-\varepsilon(\vec k)]},\\
&v_{\vec k}^2&=u_{\vec k}^2-1= 
\frac{1}{2}\left(\frac{zt-\varepsilon(\vec k)+Un_0}{\hbar\omega_{\vec k}}-1\right).
\label{v}
\end{eqnarray}
These results reveal that the excitation spectrum is gapless in the thermodynamic limit 
$M,N \rightarrow\infty$ at $M/N$ being fixed. Indeed, for the
long wavelength (phonon) modes ($|\vec k|a\ll1$) we find that:
$$\hbar\omega_{\vec k}\approx|\vec k|a\sqrt{t}\sqrt{t|\vec k|^2a^2+2Un_0}\; ,$$
i.e., the energy of a single excitation can be arbitrarily small: an expected
result showing that the Bogoliubov approach  does not work in the 
Mott phase. 

At zero temperature, there are no excitations in the system, so that 
the ground state is a Bogoliubov vacuum $|{\rm vac}\rangle$, such that 
$ c_{\vec k}|{\rm vac}\rangle= 0$. At finite temperature, say
$T$, excitations are present, and occupations of different 
modes satisfy 
$\langle c_{\vec{k}}^\dag c_{\vec k}\rangle=
\left[\exp\left(\frac{\hbar\omega_{\vec{k}}}{kT}\right)-1\right]^{-1}$,
in accordance with the Bose-Einstein statistics. Using these properties and 
the solutions (\ref{homega}) and (\ref{v}) one can easily calculate different 
quantities (e.g., correlation functions, number of condensed atoms, etc.)
both at zero and finite temperatures.
It has to be remembered, however, that reliable predictions can be obtained 
as long as $N_0(T)\sim N$, which can be self-consistently verified 
within this approach. It is also worth to stress that the Bogoliubov approach
can be applied to {\it time-dependent} problems without further complications. 
Time dependent Bogoliubov-de Gennes method, together with variational approach and
the Kibble-Zurek mechanism has been recently used to show the scaling behaviour of the time-dependent correlations
\cite{Cucchietti06}.

\subsection{Strong interactions limit}
\label{subsec-alaler-ghorer-dulal1}

Let us now study the Bose-Hubbard model in the limit of stong 
interactions, i.e.,  when the system 
is in the Mott phase. A systematic approach for studies of 
the Mott insulator phase is provided by the strong coupling expansion, 
i.e., a perturbative expansion in $t/U$ of Eq.(\ref{BHH}).  

To  perform  strong coupling expansion one splits the Hamiltonian into 
two parts: $ H_0$, whose eigenstates are exactly known at 
$t/U=0$, and treat
$-t/U \sum_{\langle i,j\rangle}(b_i^\dag b_j+h.c.)$ the tunneling part, as a perturbation.
Within this approach, the expectation values of some operators 
are expressed as a series of the form 
$\sum_i a_i\left(t/U\right)^i$.
Expansions up to $14$th-order have been calculated,  which guarantees 
in some cases, a high accuracy of perturbative predictions. 
High order calculations can be performed symbolically on a 
computer, so that the expansion coefficients ($a_i$'s), 
can be obtained exactly (not as the double precision numbers).
The theoretical background for these calculations
was set up in \cite{Gelfand90,Gelfand96}, where this method was 
applied to spin systems. Below we review the relevant results in the
context of the Bose-Hubbard model.

To start with, one can use the strong coupling expansion to determine 
the phase boundaries on the $(\mu/U,t/U)$ plane (Fig. \ref{loby})
\cite{Freericks94,Freericks96,Kuhner98,Elstner99,Kuhner00}.
In this case, the unperturbed bare hamiltonian is
$H_0= \frac{1}{2} \sum_i n_i(n_i-1)-\mu/U\sum_i n_i$,
and one calculates perturbatively 
(i) the energy of the ground state with  exactly $\bar n$ atoms per site; 
(ii) the ground state energy of the system 
with one particle added (subtracted) to (from) the system with filling factor $\bar n$. 
Setting the energy difference between (i) and (ii) cases to zero, one obtains the 
value of the chemical potential at the upper (lower) boundary between insulator 
and superfluid phases. 
These calculations can be performed for any dimensional lattice, and the 
order of expansion can be as large as $13$th \cite{Elstner99}.  For one dimesional
systems \cite{Kuhner98,Kuhner00}, the predicted 
structure of Mott insulator lobes is in a very good agreement 
with the numerical results obtained via DMRG calculations. 
The perturbative expansions can also be used together with different
extrapolation methods leading to the determination of 
the critical exponents $z$ and $\nu$ \cite{Freericks96,Elstner99}.

\begin{figure}[t]
\centering
\includegraphics[width=0.70\textwidth]{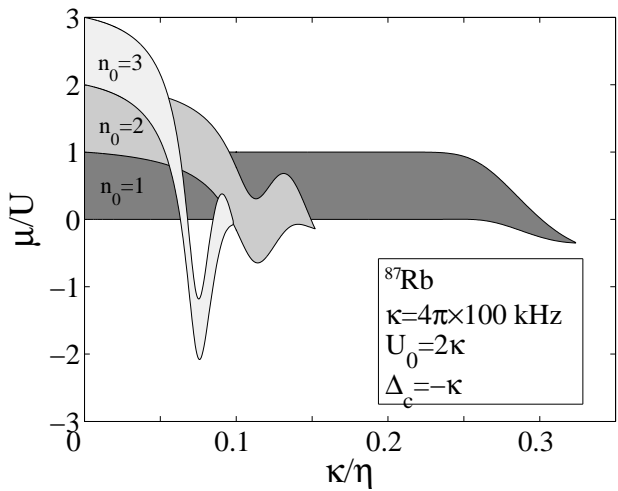}
\vspace*{0.5cm}
\caption{Overlapping MI phases for the Bose-Hubbard model in an optical cavity (from \cite{Larsson06}).}
\label{figure2.2}
\end{figure}

Very recently this method has been applied to calculate the lobes for a modified 1D Bose-Hubbard model describing atoms in an optical lattice created 
by pumping a laser beam into the cavity \cite{Larsson06}. 
The major difference to the standard case is that the intensity 
of the cavity field depends on the number of atoms present, since the atoms shift 
collectively the cavity resonance. In effect the coefficients $t$ and $U$ become very 
complex functions of all of the relevant parameters; cavity detuning, intensity of the 
pumping laser, $N$, etc. Moreover, quantum fluctuations of the resonance shift    induce 
long range interactions between the atoms. The phase diagram, as a function of the dimensionless parameters $\mu/U$ 
and $\kappa/\eta$, where $\kappa$ is the cavity width and  $\eta $ is the pumping laser strength, 
is shown in Fig. \ref{figure2.2}. (\(1/\sqrt{n_{ph}}=\kappa/\eta\), where \(n_{ph}\) is the number of 
photons in the cavity.) 
The striking effect is the overlap of  different Mott 
phases, which is the consequence of the fact that the expressions for $t$ and $U$ for 
$n_0=1, 2,3, \ldots$ Mott phases are different. 

The strong coupling expansion has also been used to calculate 
the correlation functions  $\langle b_i^\dag b_{i+j}\rangle$ and the structure factor
\cite{Elstner99,Damski06}.
A typical prediction of the strong coupling expansion reads as
\begin{eqnarray}
\langle b_i^\dag b_{i+1}\rangle &=& 4\frac{t}{U}-8\left(\frac{t}{U}\right)^3
-\frac{272}{3}\left(\frac{t}{U}\right)^5+\frac{20272}{81}\left(\frac{t}{U}\right)^7
- \frac{441710}{729} \left(\frac{t}{U}\right)^9 \nonumber\\&+& 
\frac{39220768}{2187} \left(\frac{t}{U}\right)^{11} + 
\frac{8020902135607}{94478400}\left(\frac{t}{U}\right)^{13}+   
{\cal O}(\left(t/U\right)^{15}),\nonumber
\end{eqnarray}
which was obtained in a one dimensional system at unit filling factor in
\cite{Damski06}. The differences between such analytical result and the numerical
calculation turn out to be hardly visible for $t/U<0.3$, i.e., 
in entire Mott phase. The strong coupling expansion has been also
employed to determine density-density correlations,
$\langle n_i n_{i+j}\rangle$, and the variance of on-site atom occupation
$[\langle n_i^2\rangle-\langle n_i\rangle^2]^{1/2}$ \cite{Damski06}. 
All these quantities should be directly measurable in an ongoing experiment 
in a homogeneous, one dimensional lattice \cite{Meyrath05}.

\subsection{The Gutzwiller mean-field approach}
\label{subsec-alaler-ghorer-dulal2}

The Gutzwiller mean-field approach \cite{Rokhsar91,Krauth92,Jaksch98}
has been  used in numerous papers devoted to the Bose-Hubbard model. 
In its simplest version, it is  based on the approximation of 
the many-body wave function by the product over single site contributions
\begin{equation}
|\Psi\rangle= \prod_i \sum_{n=0}^{n_{\rm max}} f^{(i)}_n |n\rangle_i\,,
\label{gwf}
\end{equation}
where $|n\rangle_i$ denotes the Fock state of $n$ atoms in the $i$-th lattice site,
$n_{\rm max}$ is a system size-independent cut off in the number of atoms per site, and 
$f^{(i)}_n$ corresponds to the amplitude of having $n$ atoms in the $i$-th lattice site. The 
amplitudes are normalized to $\sum_n |f^{(i)}_n|^2=1$.

To see what Gutzwiller approach predicts for quantum phase transitions,
we focus now on the case of an homogeneous system having 
an integer number of particles per site $\bar n$. In this case, 
one obtains that for $t/U<(t/U)_g$, where $(t/U)_g$ denotes a critical value, 
the Gutzwiller amplitudes are 
$f^{(i)}_n=\delta_{n\bar{n}}$, so that the wave function becomes
a single Fock state (\ref{fock}).  
Therefore, the Gutzwiller wave function 
exactly reproduces the system wave function  when $t/U\to 0$. 
It is also
possible to argue that  the differences between exact result and (\ref{gwf})
are negligible in the limit of a large lattice and $t/U\to \infty$ \cite{Zwerger03}. 
Therefore, the common expectation is that (\ref{gwf}) reasonably interpolates between 
superfluid and Mott insulator limits, which we discuss below pointing
out the advantages and disadvantages  of the Gutzwiller method.

The critical point according to the Gutzwiller approach, $(t/U)_g$, is located at $\frac{1}{5.8z}$ for $\bar n=1$ \cite{Zwerger03}, 
where $z$ is the number of nearest neighbors. Comparing this result to 
the more reliable findings  from 
Sec. \ref{SMI}, one observes that the agreement improves with the system
dimensionality: the Gutzwiller result is poor for a 1D system ($z=2$), and
satisfactory for a 3D one ($z=6$). 

The Gutzwiller method is quite straightforward in the numerical implementation.
In the static case the amplitudes $f^{(i)}_n$ are real numbers and can be
found by minimization of $\langle\Psi| H -\mu\sum_i n_i|\Psi\rangle$,
where $ H$ is given by (\ref{BHH}), and 
$\mu$ is a chemical potential used to enforce a desired number of atoms
in the ground state. The minimization can be done in a
standard way, e.g. with  the conjugate gradient method
\cite{NumericalRecipes87}, and faces no problems as long as the system 
is homogeneous, or the external potential imposed on it is quite regular, e.g. harmonic.
In the later case, a calculation in an experimentally  realistic 3D
configuration consisting of $65^3$ lattice sites has been recently done \cite{Zakrzewski05}.

The extension of the Gutzwiller approach to time-dependent problems is  simple
\cite{Jaksch02}.  (Alternative dynamical mean field approach has been formulated in Refs. \cite{Amico98}.) Indeed, the stationarity of 
$\langle\Psi|i\frac{d}{dt}- H|\Psi\rangle$, where $ H$ is given by (\ref{BHH}), leads to
the equation
\begin{eqnarray}
i\frac{d}{dt}f^{(i)}_n = \frac{U}{2}n(n-1)f^{(i)}_n-
t\Phi^{\star}_i \sqrt{n+1} f^{(i)}_{n+1}- t\Phi_{i} \sqrt{n} f^{(i)}_{n-1},
\label{gwft}
\end{eqnarray}
where
$\Phi_{i}=\sum_{\langle i,j\rangle}\langle b_j\rangle=
\sum_{\langle i,j\rangle}\sum_n \sqrt{n}f^{(j)\star}_{n-1} f^{(j)}_{n}$ 
(the first sum goes over all $j$ being nearest neighbors of $i$).
Numerical integration of (\ref{gwft}) is straightforward.

One should also appreciate the simplicity 
of Gutzwiller approach extensions to different systems, e.g.,
mixtures of  bosonic gases \cite{Jaksch02,Damski02}, 
Bose-Fermi mixtures \cite{Fehrmann04a,Fehrmann04b,Ahufinger05}, etc.
Particularly interesting are 
extended Hubbard models such as those involving dipolar interactions, 
where one expects the appearance of supersolid 
and checkerboard-like phases at low filling factors \cite{Goral02, Barnett06, Scarola06a}. 
Despite all these nice features, there are also problems
associated with the Gutzwiller ansatz. 

The picture that the Mott insulator corresponds to 
a single Fock state (\ref{fock}) for the filling factor $\bar n$
is obviously incorrect -- see a more detailed discussion of the 
Mott phase in Sec. \ref{SMI}. Another drawback of (\ref{gwf}) 
is that the correlation functions between 
different sites factorize into products of single site contributions, e.g.,
$\langle\Psi| b_i^\dag b_j|\Psi\rangle= 
\langle\Psi| b_i^\dag|\Psi\rangle\langle\Psi| b_j|\Psi\rangle$
for $i\neq j$. As a result, there is lack of dependence of correlations 
on the distance between lattice sites. Also, formula  (\ref{gwf}) does not correspond to a  well defined 
number of particles. This problem 
can be solved by a proper projection of the wave function (\ref{gwf}) onto
the
subspace with fixed number of atoms \cite{Jaksch02,Krauth92}, but
the subsequent calculations become complicated. 
Finally, the Gutzwiller approach underestimates
finite size effects: it predicts a  ``quantum phase
transition'' in  systems of any size due to decoupling into the product
of single site contributions (\ref{gwf}). The true quantum phase transition, 
however, requires a large system.

The performance of the Gutzwiller ansatz can be, to a limited  extent,
perturbatively improved \cite{Schroll04}. 
These  corrections  significantly modify the Gutzwiller wave function  
for $t/U$ smaller than $(t/U)_g$. As a result, both the variance of an on-site 
atom occupation and the correlation functions 
$\langle b_i^\dag b_{j\neq i}\rangle$
become nonzero for $t/U>0$, which is a progress with respect to the 
traditional Gutzwiller approach. 

Finally, we mention that the Gutzwiller approach can be supplemented by other 
mean-field-like calculations exploring the properties of Green functions  
\cite{Sheshadri93,Dickerscheid03,Sengupta05b,Konabe06,Ohashi06, Menotti06}. 
They  predict virtually the same  transition points in different dimensions
as the Gutzwiller method, but allow for determination of the excitation spectra
and finite temperature calculations.

\subsection{Exact diagonalizations}
\label{subsec-alaler-ghorer-dulal3}

Exact diagonalizations of the Bose-Hubbard model can be done for small systems only. 
The problems result from the enormous size of the Hilbert space, given by
\begin{equation}
{\cal HS}(N,M)= \frac{(N+M-1)!}{N! (M-1)!},
\label{HS}
\end{equation}
where $N$ and $M$  stand  for the number  of atoms and the number of 
lattice sites, respectively.
To illustrate predictions of (\ref{HS}) we  consider the simplest system
 undergoing a quantum phase transition in the 
thermodynamical limit, i.e., the $M=N$ case. For instance, for 
$N=8$, $10$, $12$ one obtains
${\cal HS}(N,N)= 6435$, $92378$, $1352078$ respectively. 
Moreover, one can easily verify that 
$$\frac{{\cal HS}(N+1,N+1)}{{\cal HS}(N,N)} \stackrel{N\gg1} \approx  4,$$
which quantifies the fast increase of 
the Hilbert space size with system size.
This shows that a standard  diagonalization where 
all matrix elements are stored, and no symmetries are employed, 
can face problems already for $N=M>8$. Additionally, from the exponential 
increase of the Hilbert space size with the system size, any 
significant progress due to improvement of computer resources is unlikely.  
To overcome to some extend the above limitations, one can take into account 
the following. First, one can use  
numerical routines that store non-zero matrix elements only, e.g., ARPACK
\cite{Arpack}.
Then, diagonalization of $N=M=12$ system faces no problems on a  
computer with about $1$Gb of memory provided  that one looks for  a limited 
number of eigenstates instead of the full spectrum. Second, one can cut the Hilbert
space by restricting maximal site occupation to $K$ atoms.
Such a choice can be justified by a quadratic increase of the interaction energy
with the site occupation number and is present in DMRG and Quantum
Monte Carlo schemes. The size of the Hilbert space (see Appendix \ref{appendix_S}) is then
\begin{equation}
{\cal HS}(N,M)|_K=
\sum_{j=0}^{\left[\frac{N}{K+1}\right]}  (-1)^j
\left(
\begin{array}{c}
M+N-1-j(K+1) \\
M-1
\end{array}
\right)
\left(
\begin{array}{c}
M \\
j
\end{array}
\right),
\label{HSK}
\end{equation}
where $[x]$ stands for the largest integer number not greater than $x$.
For $K=N$, (\ref{HSK}) reduces to (\ref{HS}). A typical choice leading to 
well converged results in a wide range of Bose-Hubbard model parameters 
is  $K=4$ \cite{Pai96,Kuhner98}. 
It results in the size of  Hilbert space equal to $5475$, $72403$,
$975338$ for $N=8$, $10$, $12$, respectively. These numbers suggest that rather a slight
progress can be achieved by cutting the Hilbert space size this way. 
The most powerful simplifications are possible when 
the problem under consideration has some symmetries, i.e., when
there exists a set of operators $\{ O_i\}$ commuting with 
the Hamiltonian. Then, one can split a Hilbert space into subspaces 
composed of states with well defined 
eigenvalues of the $ O_i$ operators. In this way, the
problem  of diagonalization of the full Hamiltonian reduces to a few independent 
diagonalizations of smaller matrices. It allows for reduction of memory 
requirements and helps in getting a relatively large number of
excited eigenstates. An example relevant for this review 
can be found in \cite{Damski05a,Damski05b}, where
a system of cold fermions placed in a kagom\'e lattice is considered. There exists
a translational symmetry generated by an operator having $S$ different eigenvalues.
The Hilbert space splits into $S$ subspaces, and 
diagonalization of the full Hamiltonian,
 say  $L\times L$ matrix, reduces to $S$ independent diagonalizations of 
$\sim L/S\times L/S$ matrices. A more detailed description of 
exact diagonalization procedures  for many-body quantum systems 
can be found in \cite{Noack05}.

\subsection{Quantum Monte Carlo}
\label{subsec-alaler-ghorer-dulal4}

The Quantum Monte Carlo (QMC) method was successfully used for studies of 
the Bose-Hubbard model  
\cite{Batrouni90,Krauth91b,Krauth91c,Scalettar91, Batrouni92},
after publication of the seminal paper of 
Fisher {\it et al.} \cite{Fisher89}.  The computer resources 
and QMC algorithms allowed, at that time, to consider systems of the size of
a few tens of lattice sites/atoms in one and two dimensional models. 
Nowadays, systems composed of  $10^3$ sites and atoms in any dimension can be
routinely studied. It allows to consider 3D configurations,  
which are quite realistic from an experimental perspective  
\cite{Kashurnikov02,Wessel04}. 
The progress comes obviously from better 
computer resources and more efficient algorithms (see e.g., \cite{Prokofev98,Alet05a, Alet05b}). 
Interestingly, efficient numerical codes for 
QMC simulations are now publicaly available \cite{alps05}.

The  QMC approach allows to calculate different properties 
of a system being in equilibrium at finite temperature. 
During studies  of the Bose-Hubbard
model these temperatures can be chosen so low that the simulation 
describes essentially the  zero-temperature physics of the system.
The quantities usually calculated within the QMC approach  
are: the superfluid fraction, the chemical potential ($\mu$), 
the density of atoms ($\rho$), the variance of the on-site occupation, and the
compressibility ($\kappa=\partial \rho/\partial \mu$).

Early QMC calculations \cite{Batrouni90,Krauth91b,Krauth91c,Scalettar91, Batrouni92} focused on determination of the phase diagram 
in the $(\mu/U,t/U)$ plane (Fig. \ref{loby}), and on the influence of 
disorder on it. In particular, they  provided 
estimations of the position of the critical points, and they have 
verified critical behaviour of different quantities, 
 e.g., scaling properties of compressibility and superfluid fraction, in the 
neighbourhood of the phase boundary, predicted in \cite{Fisher89}  (see Sec. \ref{generic}). 

Among the recent QMC studies of the Bose-Hubbard related to physics of cold 
atoms in optical lattices we would like to focus on investigations of
harmonically trapped systems
\cite{Batrouni02,Wessel04,Sengupta05a,Batrouni05}. 
Below we briefly review findings of these works, stressing the differences between 
the harmonically trapped model  and the homogeneous one.

The Bose-Hubbard  Hamiltonian in the presence of the harmonic trap 
takes the form (\ref{BHH}) with  an additional 
term $-\sum_i \mu_i^{\rm local} n_i$. The   
$\mu_i^{\rm local}=\mu-Vr_i^2$ is a local chemical potential,
$r_i$ is the distance from the trap centre, and
$V$ is the strength of a harmonic  trap. 
The state diagram for such  a problem in one dimension 
was calculated in \cite{Batrouni02}. In a generic situation 
one finds that 
there are plateaus, characterized by integer density of atoms
and small atom number fluctuations, surrounded by regions of space where atom
number fluctuations are large  and density is    non-integer. The
former (latter) are identified as  Mott insulator (superfluid) domains. Amazingly, 
images of these domain shells have been recently observed experimentally \cite{Ketterle06, Bloch06}.
It should be stressed, however, that there are significant differences 
between superfluid and Mott insulator ``phases'' in harmonically trapped 
and  homogeneous models.

The harmonically trapped system is gapless even in the presence of Mott 
domains \cite{Batrouni05}. It is also always globally compressible \cite{Batrouni02}. 
Both these properties are in a striking difference to the homogeneous case.
Inhomogeneity of a trapped system suggests that  
the critical behaviour might be recovered in local quantities, e.g.,  
the local compressibility
$\kappa_i^{\rm local}=\partial \langle\sum_j n_j\rangle/\partial \mu_i^{\rm local}$
(see \cite{Wessel04,Gygi06} for a systematic discussion in both harmonic
and quartic trapping potentials).
This quantity, however, does not  show the critical behaviour
at the border between Mott and superfluid domains. 
All this leads to the conclusions that: i) in a harmonically trapped system, instead of a true 
quantum phase transition, there is rather a crossover, and ii) it will be very interesting to perform 
the experiments in homogeneous systems in a ``box" potential. Such potentials are currently being realized, 
for instance, by the group of M. Raizen \cite{Meyrath05}. 

Another difference between homogeneous and trapped models is 
observed in the  visibility of interference patterns 
measured after releasing the atoms from the external potentials 
\cite{Gerbier05,Sengupta05a}. 
 While in a homogeneous system the visibility of interference fringes
is a smooth monotonic function of $t/U$ ratio, in the trapped model 
one finds both kinks (sudden changes of the slope) and non-monotonic 
behaviour caused by the presence of 
correlations between disconnected superfluid domains. 

\subsection{Phase space methods}
\label{subsec-alaler-ghorer-dulal5}

Phase space methods have been introduced to quantum mechanics relatively early by Wigner \cite{Wigner31,Wigner32} and Moyal \cite{Moyal49}, 
but their rapid development started in the 60-ties with the applications to quantum optics by Glauber \cite{Glauber63b} 
and Sudarshan \cite{Sudarshan63}. To a great extent, quantum optical studies of quasi-probability distributions, 
such as Wigner functions, or $P$- or $Q$- distributions, contributed enormously to our modern understanding of  
quantum noise \cite{Gardiner04,Walls06}. These methods usually map the density matrix of the considered system onto a 
function fulfilling a generalized Fokker-Planck type of equations, and try to replace this equation by systems of 
Langevin--like equations that can be simulated using classical Monte Carlo methods. 
The necessary condition is that the corresponding quasi-probability, or phase space  quasi-distribution must be a reasonable probability measure, which not always is the case. Glauber-Sudarshan 
$P$-representation often is a highly singular distribution, and Wigner function might take negative values. 

All these problems can be sometimes overcome, and several groups have started to use phase space methods 
to simulate many body problems. Pioneering work in this direction for bosonic gases has been done by P. 
Drummond's group \cite{Drummond98}. These methods were extended to fermions in Ref.\cite{Corney04}. The authors study 
various signatures of strongly correlated ultra-cold fermions in optical lattices, performing collective mode 
calculations, where a sharp decrease in collective mode frequency is predicted at the onset of the Mott metal-insulator transition. They have  also  looked at correlation functions at finite temperatures, using  a new exact method that applies the stochastic gauge technique with a Gaussian operator basis.

Somewhat similar approach has been developed by Y. Castin group \cite{Sinatra00}. This approach allowed, for instance, for precise determination of the fluctuations of the number of condensed 
atoms in an interacting Bose gas. It has been recently extended to describe lattice Hubbard problems \cite{Carusotto03},
and to Fermi gases \cite{Castin04,Montina06}. 

It is worth stressing that stochastic, or phase space methods become very efficient when the fluctuations become 
classical, i.e. at high, or at least moderate high temperatures, which for the BEC might mean 0.1-0.2$T_c$. 
Truncated Wigner approach, or simulations of the Gross-Pitaevskii equations with random initial conditions, that mimic 
the initial thermal equilibrium are easy to implement and very accurate. Several groups have used such approach with great 
success. Hannover group used such method to describe phase fluctuations in quasi-1D BEC's \cite{Dettmer01,Hellweg01,Kreutzmann02,Mebrahtu06}. 
Barcelona-Hamburg collaboration applied these approach to study dynamics of spinor condensates \cite{Mur-Petit06}. 

The so-called classical field method was developed by K. Burnett \cite{Burnett04} and 
K. Rz\c a\.zewski \cite{Kazik04}  groups. Both groups put some emphasis on understanding 
the concept of temperature in the microcanonical ensemble, whereas the latter recently realized the modeling of the decay of unstable vortex states.

\subsection{1D methods}
\label{subsec_kachupora1}

In this section we give a very short guide of 1D methods, illustrating the subsequent steps from Bose-Hubbard model to  Luttinger liquid theory. 
We start by analyzing the Bose-Hubbard Hamiltonian in 1D (Eq. (\ref{bosehub})), i.e. a chain of sites with open ends
and we perform then the following steps:

\paragraph{a. Hard core bosons}
In this  limit (i.e. when $U\gg t,\mu$), as descibed in Section \ref{spin_models}, the Bose Hubbard hamiltonian (Eq. (\ref{bosehub})) reduces to the XY model in a transverse field (Eq. (\ref{xytrans})).

\paragraph{b. Jordan-Wigner transformation} Jordan-Wigner transformation is a way to ``fermionize" the, otherwise, bosonic system \cite{Sachdev99}. One defines:
\begin{eqnarray}
f_i&=&\prod_{j<i}\sigma_{z,j}\sigma_i, \\
f^{\dag}_i&=&\prod_{j<i}\sigma_{z,j}\sigma^{\dag}_i, 
\label{jw}
\end{eqnarray}
so that $f^{\dag}_if_i=\sigma^{\dag}_i\sigma_i= (1+\sigma_{z,i})/2$. It is easy to check that such nonlocal operators are fermionic, i.e.
\begin{equation}
\{f_i,f_j\}=\{f^{\dag}_if^{\dag}_j\}=0, \ \ \ \{f_if^{\dag}_j\}=\delta_{ij}.
\end{equation}
The inverse relations are also simple:
\begin{eqnarray}
\sigma_i&=&\prod_{j<i}(2f^{\dag}_if_i-1)f_i, \\
\sigma^{\dag}_i&=&\prod_{j<i}(2f^{\dag}_if_i-1)f^{\dag}_i. 
\label{jwi}
\end{eqnarray}
With this transformation the Hamiltonian becomes the one of ``spinless" fermions:
\begin{equation}
H=+t\sum_{\langle i\rangle}\left[f^{\dag}_{i}
f_{i+1}+ h.c.\right]  -\mu\sum_i f^{\dag}_if_i ,
\label{xytrans2}
\end{equation}
and the model becomes exactly solvable. Had we started from an extended Hubbard model with, 
say, nearest neighbour interactions of the form $V\sum_i \sigma^{\dag}_i\sigma^{\dag}_{i+1}\sigma_{i+1}\sigma_i$, 
we would end up with ``spinless" interacting fermions with the term $V\sum_i f^{\dag}_if^{\dag}_{i+1}f_{i+1}f_i$. 

Jordan-Wigner transformation is essentially a 1D transformation. There were very many attempts of generalizing it to 
higher dimension (for a review see \cite{Tsvelik03}, for a discussion in the context of applications for atoms see
 \cite{Fehrmann04a}). Only recently, new ideas related to the concept of PEPS, allowed Verstraete and Cirac to 
propose an interesting efficient generalization of the Jordan-Wigner transformation to 2D \cite{Vertraete05}.

\paragraph{c. Bosonisation, and Luttinger liquid theory}  Let us consider now the interacting ``spinless" Fermi gas, 
and try to formulate the low energy effective theory for this model. We assume that fermions do not fill the lowest band, so that in the 
absence of interactions, the Fermi level is somewhere in the middle of the band. We expect that at low temperatures interesting physics 
will occur close to the Fermi energy. It is thus reasonable to linearize the
fermionic dispersion relation at the Fermi energy, $\epsilon(k)=\epsilon(\pm k_F) \pm c_F(k\pm k_F)$, with 
the "sound velocity" $c_F=d\epsilon(k)/dk|_{k_F}$. In 1D there are two values of momenta $\pm k_F$ where the Fermi energy 
is reached: one corresponds to left, and another to right going fermions. Note that all of the interesting low energy physics happens close to the Fermi surface (points). That implies that the states with momenta far from $\pm k_F$ will practically never participate in any relevant physical process, and will remain deep in the filled Fermi sea. It is thus reasonable treat the left and right going fermions close to $\pm k_F$ as independent and introduce two fermionic species, described in the momentum representation  by  $L(k), L^{\dag}(k)$, 
and $R(k)$, $R^{\dag}(k)$, respectively;  $k$ is here momentum  relative to $\pm k_F$, so that it attains values from $-\infty$ to $+\infty$. The Hamiltonian,  becomes 
\begin{eqnarray}
H&=&+\sum_{k}[-c_F(k)L^{\dag}(k)L(k) + c_F(k)R^{\dag}(k)R(k)] \nonumber \\ 
&+& \frac{1}{L}\sum_{k_1,k_2,q} V(q)L^{\dag}(k_1-q)R^{\dag}(k_2+q)R(k_2)L(k_1).
\label{xytrans3}
\end{eqnarray}
This is a Luttinger model \cite{Mahan93,Giamarchi04}. One introduces now the operators
\begin{eqnarray}
&&\rho_L(q)=\sum_k L^{\dag}(k+q)L(k),\\
&&\rho_L(-q)=\sum_k L^{\dag}(k)L(k+q)=\rho_L^{\dag} (q),\\
&&\rho_R(q)=\sum_k R^{\dag}(k+q)R(k),\\
&&\rho_R(-q)=\sum_k R^{\dag}(k)R(k+q)=\rho_L^{\dag} (q),
\end{eqnarray}
and observes that $[\rho_R(-q),\rho_R(+q')]=\sum_{k_F-q'}^{k_F}R^{\dag}(k+q-q')R(k)$. In the Luttinger-Tomonaga
approximation \cite{Mahan93} one replaces now these and similar commutators by their averaged values over  
the ideal Fermi sea.
The result is
\begin{equation}
[\rho_R(-q),\rho_R(+q')]= \frac{qL}{2\pi}\delta(q-q'),
\end{equation}
implying that for $q>0$, we can introduce 
$b^{\dag}_q=\rho_R(q)\sqrt{2\pi/qL}$, $b_q=\rho_R(-q)\sqrt{2\pi/qL}$, which may
be regarded as bosonic annihilation and creation operators, respectively, 
since $b_q,b^{\dag}_{q'}=\delta(q-q')$.
Similar construction is done for the operators 
$c^{\dag}_{-q}=\rho_L(-q)\sqrt{2\pi/qL}$, and $c_{-q}=\rho_L(q)\sqrt{2\pi/qL}$.
These operators have the interpretation that $b^{\dag}_q$  takes a particle 
from state $k$ and puts it into $k+q$. It creates thus 
fermion-hole pairs when $k<k_F$ and $k+q>k_F$. 
Analogously, the operator $c^{\dag}_{-q}$ takes a fermion from an unoccupied state 
$-k_F<k+p$ to the unoccupied state $k<-k_F$.  The whole Hamiltonian becomes
\begin{equation}
H=\sum_{q>0}qc_F[ b^{\dag}_q b_q + c^{\dag}_{-q} c_{-q} ] + \frac{1}{2\pi}\sum_{q>0} [V(q)c^{\dag}_{-q}b^{\dag}_q +  V(q)^* b_q c_{-q}].
\label{xytrans4}
\end{equation}
The problem has been thus reduced to an exactly solvable system of interacting harmonic oscillators, 
describing linear 1D hydrodynamics of a, so called, Luttinger liquid. Bosonisation theory \cite{Giamarchi04} assures that such description
can be found for most of the 1D bosonic and fermionic systems with local interactions. For the very recent  
and very complete description    of the  interacting Bose gases in quasi-one dimensional optical lattices see the review article of 
Cazalilla et al. \cite{Cazalilla06}.

\subsection{Bethe ansatz}
\label{subsec_kachupora2}

Bethe ansatz is an analytical method 
to find exact eigenstates and eigenvalues of some strongly correlated 
one-dimensional models (although some times it has to be 
complemented with numerical analysis). 
By exploiting appropriately the 
symmetries involved in the Hamiltonian it is possible to diagonalise the
Hamiltonian exactly.  However, its power relies
in the fact that Bethe ansatz characterises all eigenstates by a set
of quantum numbers which enumerate the states, according to their physical
properties. This method was originally developed by
Bethe in 1931 \cite{Bethe31} to solve a 1D  array of electrons with uniform 
next neighbour interactions, i.e.,  spin 1/2 Heisenberg model: 
\begin{equation}
H=-J\sum_{i=1}^N {\bf {\sigma_i}} \cdot {\bf \sigma_{i+1}},
\label{Heisenber1/2}
\end{equation}
where ${\bf \sigma}=(\sigma_x,\sigma_y,\sigma_z)$ are the standard Pauli matrices and N denotes the number of sites. 
The parametrisation of the eigenvectors, i.e. {\it the Bethe ansatz} has become 
a fundamental tool with which  many other 1D quantum systems have been shown to be
solvable. From the eigenvectors 
one can compute easily the quantities of interest 
by calculating the expectation values of the desired 
operators. 
In what follows, we summarize the original 
formulation of Bethe following
the pedagogical work  of Karbach et. al. \cite{Karbach98a,Karbach98b}.  
Since, the Heisenberg  Hamiltonian commutes with the total 
spin along the z-direction $\sigma_T^z=\sum_i \sigma_i^z$, $[H,\sigma_T^z]=0$, then:
\begin{equation}
H=-J\sum_{i=1}^N\left[\frac{1}{2}( \sigma^+_i \sigma^-_{i+1} +  \sigma^-_i \sigma^+_{i+1})+  
\sigma^z_i \sigma^z_{i+1}\right].
\end{equation}
Thus, eigenstates of $\sigma^z$ are also eigenstates of $H$.
Bethe ansatz is, as discussed below, a basis transformation.

We denote by $|0\rangle=|\uparrow\rangle$ the eigenstate
of $\sigma^z$ with value +1/2 and by $|1\rangle=|\downarrow\rangle $ the eigenstate with
eigenvalue -1/2. 
Sorting the basis according to the quantum number $N/2-r$, 
where $r$ is the number of  flipped spins, is all 
that is needed to block diagonalise the Hamiltonian.
The block $r=0$ corresponds to a single state with all spins up which we call 
ferromagnetic: $H|F\rangle = H|\uparrow...\uparrow\rangle=-JN/4$. 
The case $r=1$ (one spin down) has $N$ invariant vectors, which  are
labeled by the position of the flipped spin: $|n\rangle=\sigma_n^-|F\rangle$.  
To diagonalise this block,
which has size $N\times N$, we take into account that the Hamiltonian 
possesses translational symmetry. Therefore a translational invariant
basis can be constructed in the subspace with $r=1$,
\begin{equation}
|\psi\rangle=\frac{1}{\sqrt N}\sum_{n=1}^N e^{ikn}|n\rangle,
\label{BETHEr=1}
\end{equation}
with wave numbers $k=2\pi m/N,m=0,1,...N-1$.  The vectors  $|\psi\rangle$ correspond
to a complete spin alignment of the ferromagnetic 
ground state  $|F\rangle$ which is periodically disturbed 
by a spin wave with wavelength $\lambda=2 \pi/k$. These states are called magnons, or spin-waves.

The subspace $r=2$ cannot be solved by applying any further symmetry and
here is where the full power of the Bethe ansatz appears. 
Solutions (\ref{BETHEr=1}), for \(r=1\), can be also obtained starting from the
following ansatz 
\begin{equation}
|\psi\rangle=\sum_{n=1}^N a(n)|n\rangle
\end{equation}
and finding the values of the  coefficients $a(n)$ that satisfy the eigenvalue
equation $H|\psi\rangle=E |\psi\rangle$ using periodic boundary conditions.
 If one applies the same procedure to
the subspace $r=2$, i.e. one  looks for the coefficients $a(n_1,n_2)$ that determine
the eigenstates, one finds:
\begin{equation}
|\psi\rangle=\sum_{1\le n_1< n_2\le N}^N a(n_1,n_2)|n_1,n_2\rangle.
\end{equation}
There are $N(N-1)/2$ eigenstates. These are characterised by
pairs of Bethe (integer) quantum numbers $\lambda_1,\lambda_2$, which define the ``momenta" via the relations
$Nk_1= 2\pi \lambda_1+\theta$, $Nk_2= 2\pi \lambda_2-\theta$, such that
\begin{equation}
a(n_1,n_2)=e^{i(k_1n_1+k_2n_2+\theta_{12}/2)}+e^{i(k_1n_2+k_2n_1+\theta_{21}/2)},
\label{wzor}
\end{equation}
whereas the angles $\theta_{12}=-\theta_{21}=\theta$ fulfill
$2\cot(\theta/2)= \cot(k_1/2) - \cot(k_2/2)$.
Depending on the 
relative values of the pair $\lambda_1,\lambda_2$, different type of solutions exist. 
In general, the Bethe ansatz for an unrestricted number $r$ of flipped spins
reads:
\begin{equation}
|\psi\rangle=\sum_{1\le n_1\le...\le n_r\le N}^N a(n_1...n_r)|n_1...n_r\rangle
\end{equation}
and the eigenvectors span a subspace of dimension $N!/(N-r)!r!$. 
The coefficients $a(n_1...n_r)$ have an analogous form
to (\ref{wzor}), 
and a generalised solution of the resulting transcendental equations can be easily found  
numerically. 

It is also worth to mention that the Bethe ansatz can be applied to the 
one dimensional Bose-Hubbard model -- see \cite{Krauth91a} for a complete discussion.
In this case, however, the method faces fundamental problems since one has 
to assume that the maximal occupation of each lattice site is not larger than
2. As a result, the ansatz is never exact though it gives some valuable and 
correct, predictions, e.g., vanishing superfluid fraction in the Mott phase. 


\subsection{A quantum information approach to strongly correlated systems}
\label{subsec_quantu_raju}

The density matrix renormalisation group (DMRG)\cite{White92,Schollwoeck05} is a variational 
method that has had an enormous success in describing ground states 
of some strongly interacting 1D systems with rather 
modest computational effort. 
The underlying philosophy of all DMRG oriented algorithms 
is that many body systems can be treated almost ``exactly" if one is
able to truncate the full Hilbert space by removing the degrees of freedom that
are not involved neither in the ground state, nor in the 
dynamical evolution of the system. The difficulty and glory of the method
relies on how reliable the truncation is done.
Very recently \cite{Vidal03a,Vidal03b,Verstraete04b,Orus04,Verstraete04c}, 
quantum information theory has provided a new perspective on the following questions: 
(i) how to perform an efficient truncation of the Hilbert space, 
(ii) which quantum systems can be efficiently simulated, 
(iii) how to simulate dynamical evolutions of strongly correlated systems, 
(iv) how and when DMRG-oriented methods can be 
implemented to investigate ground states of 2D and
3D  systems,
(v) how classical concepts like correlation length, which diverge
on the critical points is linked to entanglement \cite{Verstraete04a, Verstraete04c}, etc.
In general, this  approach is 
shedding new light in our understanding of complex many body physics. Anders \emph{et al.}
have recently proposed an approximate method to calculate ground states of quantum many-body systems, 
based on quantum information concepts of ``weighted graph states'' \cite{Anders06natun}.

Real space renormalisation methods are iterative methods to describe
accurately mainly 1D many body systems.
The starting point is a small 1D chain system 
which can be exactly diagonalised. Then, the size of the chain is iteratively increased by adding a new lattice site at each step. The Hamiltonian that coupled the new site to the chain is renormalized by disregarding all physically irrelevant couplings and the resulting Hamiltonian diagonalised.



There are, at least, 3 different approaches to perform 
the truncation\cite{Schollwoeck05}:
\itemize
\item By optimizing expectation values \cite{White98}. In this approach the value of some bounded operator (energy, magnetization or density) 
determines which are the most relevant
states contributing to it.
\item By optimizing the wave function\cite{White92}. The truncation is done
such that the renormalized wave function minimizes the Schmidt norm distance
to the exact one, i.e. $|||\Psi\rangle-|\Psi\rangle_{DMRG}||^2\rightarrow 0$ 

\item By optimizing entanglement properties. 
It was shown by G. Vidal \cite{Vidal03a} that
the efficiency in simulating many body systems is directly related to
their entanglement behaviour. Efficient simulation is possible if the entanglement of a subsystem with respect to the whole is bounded, or grows at most 
logarithmically with its size. Also, if entanglement grows linearly in time
and block size, simulation of time evolution may be not efficient even in
some non-critical 1D systems \cite{Orus04}.

\subsubsection{Vidal's algorithm}
\label{subsec_quantu_raju1}

In the following we briefly review the algorithm of
G. Vidal \cite{Vidal03a} (also called Time Evolving Block Decimation (TEBD)) to calculate ground states and dynamics using the Schmidt decomposition to truncate the Hilbert space. Implementing operations on individual
sites and on 2 neighboring sites requires only local updating 
of the expansion, rending the calculations of ground states, expectation 
values, and 2-body correlations efficient.
We also comment how the expansion is linked to the Matrix Product states (MPS) ansatz
used in DMRG, and why MPS describe so well ground states of quantum spin
systems \cite{Verstraete04b}.

Assume that our system consists of a set of atoms loaded in a 1D optical lattice with $M$-sites, whose Hamiltonian is well described by a 
Bose-Hubbard Hamiltonian (see Eq. (\ref{BHH})) and we are seeking for
the ground state $|\Psi\rangle$ of the system. We choose as ansatz
the most general description of a state in such Hilbert space which 
is given by
\begin{equation}
|\Psi\rangle=\sum_{i_0=0}^{d}...\sum_{i_M=0}^d C_{i_1...i_M},
|i_1\rangle\otimes...|i_M\rangle
\label{vidal}
\end{equation}
where $\{|i_j\rangle\}$ denotes an orthonormal basis, $M$ is the number of sites, $N$ is the number of atoms, and $d$ indicates
the maximal occupation number of site $j$ (maximal $\langle a_j^{\dagger}a_j\rangle$). 
Note that by taking $d=N$, we can be sure that
Eq. (\ref{vidal}) gives an {\it exact} representation of the state.   
Let us now split the 1D chain into two blocks: 
block A, which consists of a single  site, and block B, consisting of the remaining 
 $M-1$ sites. For this (or any other bipartite)
splitting ($A,B$), there exist always 
a bi-orthonormal basis denoted by  $|\Phi_{\alpha}^{[A]}\rangle$ and 
$|\Phi_{\alpha}^{[B]}\rangle$, respectively such that:
\begin{equation}
|\Psi\rangle=\sum_{\alpha}^{\chi_A}\lambda_{\alpha}
|\Phi_{\alpha}^{[A]}\rangle
|\Phi_{\alpha}^{[B]}\rangle,
\end{equation}
where $\chi_A\le d$, i.e. the number of coefficients 
(Schmidt coefficients) of the expansion is 
bounded by the dimension of the smallest of the two subspaces 
of the partition. This simply indicates that the 
number of degrees of freedom that can be entangled between $A$ and $B$, is the maximum number of degrees of freedom
of the smaller subsystem. If there is a single coefficient different from
zero in the expansion, then the two parties are in a product state, i.e. not
entangled. The coefficients of the expansion are unique (up to the degeneracy
of the reduced density matrices), real and 
$\sum_{\alpha }\lambda_{\alpha}^2=1$. Notice also that 
$\rho_A=Tr_B (|\Psi\rangle\langle\Psi|)=
\sum_{\alpha}\lambda_{\alpha}^2|\Phi_{\alpha}^{[A]}\rangle\langle\Phi_{\alpha}^{[A]}|$
and $\rho_B=Tr_A (|\Psi\rangle\langle\Psi|)=
\sum_{\alpha}\lambda_{\alpha}^2|\Phi_{\alpha}^{[B]}\rangle
\langle\Phi_{\alpha}^{[B]}|$. 
The Schmidt decomposition is nothing else than the usual Single Value
Decomposition (SVD) of the coefficient matrix \(C\) corresponding to the 
decomposition of $|\Psi\rangle$ in an 
arbitrary orthogonal basis $|i\rangle_A$ ($|j\rangle_B$):
\begin{equation}
|\Psi\rangle=\sum_{i,j}C_{i,j}|i\rangle_A  |j\rangle_B.
\end{equation}
 $C_{i,j}$ is a $d\times d^{M-1}$ matrix. Applying the Single Value 
Decomposition to C one obtains that  $C=U\, D\,V$, where $U$ ($V$) 
is a unitary matrix of
dimensions $d\times d$ ($d^{M-1}\times d^{M-1}$) and $D$ is a 
diagonal matrix of dimensions $d\times d^{M-1}$. Therefore,
\begin{eqnarray}
|\Psi\rangle&=&\sum_{i,j}C_{i,j}|i\rangle_A  |j\rangle_B=
\sum_{\alpha}[\sum_l U_{l,\alpha}|i\rangle_A]\,D_{l,l}\,[\sum_l 
V_{\alpha,l}|j\rangle_B] \nonumber
\\
&=&\sum_{\alpha=1}^{d}\lambda_{\alpha}|\phi_{\alpha}^{[A]}\rangle 
|\phi_{\alpha}^{[B]}\rangle.
\end{eqnarray}
  
An iterative treatment of the above procedure (i.e. 
splitting now part $B$ into a block of 1 lattice site and block of $M-2$ sites 
and finding its Schmidt decomposition, and again for the block of $M-3$ sites)
leads to the following expression:

\begin{equation}
C_{i_1...i_M}=\sum_{\alpha_1...\alpha_{M-1}}\Gamma^{[1]i_1}_{\alpha_1}
\lambda_{\alpha_1}^{[1]}{\Gamma}^{[2]i_2}_{\alpha_1,\alpha_2}
\lambda_{\alpha_2}^{[2]}\Gamma^{[3]i_3}_{\alpha_2,\alpha_3}...
\Gamma^{[M]i_M}_{\alpha_{M-1}}.
\end{equation}
The tensors $\Gamma$ take into account the correlations arising from
the splitting, and  are straightforwardly obtained by performing at each bipartite partition the corresponding SVD. Notice
that except for the first and the last lattice site, for all  the other lattice sites, 
the corresponding tensor $\Gamma$ depends on two $\alpha$ indices, corresponding
to the two involved partitions, and takes into account the correlations  
with the block at its left, and the block at its right.
It can be shown that for translationally invariant 1D systems with 
short range interactions, the coefficients of the 
Schmidt decomposition behave as $\lambda_{\alpha}^{[l]}\sim \exp(-\alpha)$,
and the truncation of the Hilbert space is performed by removing 
all small coefficients.

By grouping the $\Gamma$ tensors with the coefficients 
$\lambda$ the expression of the ansatz takes the more compact form:
\begin{equation}
|\Psi\rangle=
\sum_{i_1,...i_M=0}^d {\rm Tr}[A^{[1],i_1}... A^{[M],i_M}]|i_1,...i_M\rangle,
\end{equation}  
where $A^{[k],1} \ldots A^{[k],d}$ are \(D_k \times \tilde{D}_k \) complex matrices with \(D_{k +1} = \tilde{D}_k \leq D\),
and \(d\) is the occupation number on site $i$.
Operations on a single site $i$ involve only updating 
the value of  $A^{[k],i_k}$, which basically implies of the order of
$D^2$ operations, $A'^{[k],i_k}=U A^{[k],i_k}$. In a similar way, operations
involving two neighbouring sites correspond to updating the matrices of the
corresponding sites and the correlations between them and again can 
be efficiently performed.  
Thus, the ground state of the Bose-Hubbard Hamiltonian, which includes only
next neighbour interactions plus on site collisions can be  
calculated variationally, whereas  time evolutions can be simulated by means of a Trotter expansion.
(For the recent work of dynamical response of BH model at the MI-SF transition, see \cite{Clark06}.)

\subsubsection{Matrix product states}
\label{subsec_quantu_raju2}

\begin{figure}[t]
\centering
\includegraphics[width=12cm,height=7.5cm]{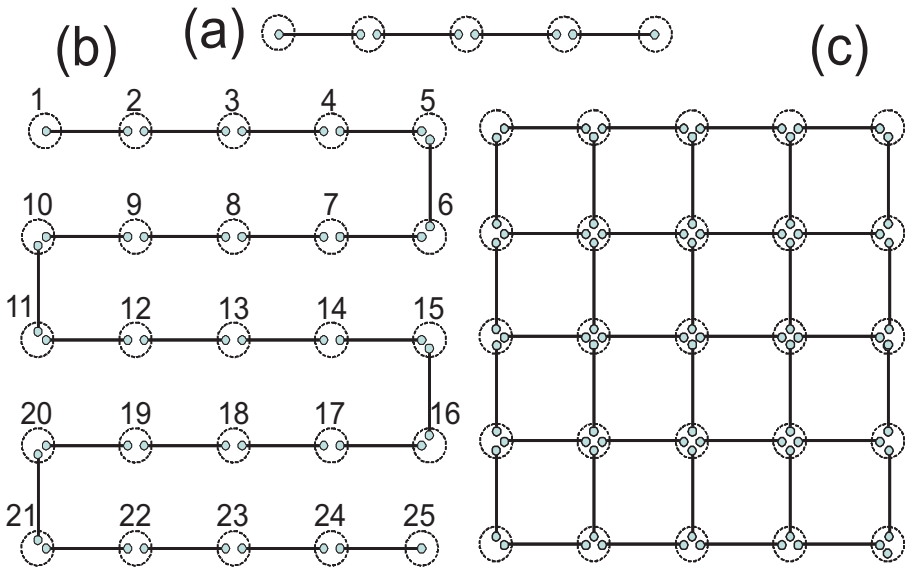}
\caption{Schematic representation of a) MPS for a 1D system, b) ``inefficient" MPS in 2D, and c) PEPS 
for a 2D system. Each pair of nearest neighbour sites is connected via maximally entangled state from an auxiliary Hilbert space. 
In a) and b) matrices, $A$ projects onto the physical space. In c) matrices have to be replaced by 4th rank tensors in the bulk 
(from \cite{Verstrate04a}).}
\label{figure2.4}          
\end{figure}

Already {\"O}stlund and Rommer \cite{Ostlund95,Rommer97} realized that 
DMRG leads to an special type of ansatz known as 
Matrix Product States \cite{Fannes89, Fannes92}. Matrix Product States
(MPS) and their generalization to higher dimensions, Projected entangled-pair states (PEPS) (see Fig. \ref{figure2.4}) corresponds
to  translationally invariant states, 
whose general expression is of the form \cite{Rommer97} :
\begin{equation}
|\Psi\rangle=
\sum_{i_1,...i_M=1}^d {\rm Tr}[A^{[1],i_1}... A^{[M],i_M}]|i_1,...i_M\rangle,
\end{equation}
which describes a chain of $M$ spins of dimension $d$. The $A^{[k]i_k}$
are complex matrices with dimension $ \leq D\times D$ matrices. The reason why this 
type of states (MPS) can reproduce very well  the ground states of quantum spin
systems with nearest neighbor interactions, is because the ground states (that have also the translational invariance of 
the Hamiltonian) and low energy excitations are completely
determined by their two-body reduced density operators as shown by Verstraete
and Cirac \cite{Verstraete06a}. An intuitive but handwaving argument to understand 
the reason why MPS parametrize ground states of gapped quantum spin systems, comes
from Hastings theorem \cite{Hasting04}, which states that for gapped systems 
all correlations decay exponentially. This means that since  
$\langle O_A O_B\rangle -\langle O_A\rangle \langle O_B \rangle\simeq \exp(-l_{AB}/\xi_{corr})$, 
blocks $A$ and $B$, separated by \(l_{AB}\), can be described by a product state, when $l_{AB}$ is 
much larger than the correlation length.

MPS and PEPS have already proven to be an enormously useful tool. In particular, they can be easily extended to calculate time evolution. They have also been used to characterize quantum phase transitions in spin chains \cite{Verstraete04b}, simulate 
infinite systems \cite{Vidal06}, and establish a closer relation between criticality, area law, and block entanglement \cite{Verstraete06b}.




\subsection{Fermi and Fermi-Bose Hubbard models}
\label{subsec_kachupora}

Most of the methods that are used for Bose-Hubbard models 
(some of which were discussed in the previous subsections of this section) 
can be carried over to fermionic systems, although typically, 
treatment of fermions is more difficult. For example, mean field BCS method, valid for weakly interacting fermions, requires much more 
effort for the trapped gases than the bosonic Bogoliubov-de Gennes approach. Quantum Monte Carlo methods (QMC) 
suffer notoriously the, so-called, ``sign problem" 
(negative probabilities). Gutzwiller ansatz in dimensions higher than 1, does not take 
the fermionic anticommutation relations into account properly, due to non-existence of standard Jordan-Wigner
transformation (for a discussion, see for instance  \cite{Fehrmann04a}). 

When we deal with Fermi-Fermi, or Fermi-Bose mixtures, it is always useful to study some limiting cases, 
that moreover, can be realized in experiments. One limit would be the weakly interacting regime, 
where Hartree-Fock, BCS or Landau Fermi liquid theory could be applied. 
In the strong coupling limit of Mott insulating states, a 
very useful method is to construct 
an effective Hamiltonian. Such Hamiltonian  will often correspond  to a spin model, if we deal with commensurate filling factors 
(as we discussed in the Introduction). If the Mott states involve states differing by odd number of fermions (for instance
 presence or absence of a fermion), then the effective model will necessarily involve composite fermions. 

\paragraph{Fermionic Hubbard models}

Perhaps the most well-known are the effective models obtained in the strong coupling limit of the electronic Hubbard model 
for spin 1/2 fermions, described by the Hamiltonian, 
\begin{equation}
 H= -t\sum_{\langle i,j\rangle,\sigma}(f_{i\sigma}^\dag f_{j\sigma}+ h.c.)+ \frac{U}{2}\sum_i 
f_{i\uparrow}^\dag f_{i\downarrow}^\dag f_{i\downarrow} f_{i\uparrow}-\mu\sum_{i,\sigma}  f_{i\sigma}^\dag f_{i\sigma}.
\label{FHa}
\end{equation}
In the limit of large repulsive interactions, $U\gg t,\mu$, there can be one, or zero 
particles at a site. In this limit the model is termed in the literature ``$t-J$" model (cf. \cite{Auerbach94}). 
When the system is at full filling, i.e. 
there is 1 fermion at each site (in the Mott limit), the effective Hamiltonian reduces to that of Heisenberg 
quantum antiferromagnet with the magnetization fixed by the corresponding numbers of spin--up and spin--down fermions 
(or external ``magnetic" field). In particular, if these numbers are equal, the system has a zero net magnetization  
(vanishing ``magnetic" field).   Allowing for certain number of unoccupied sites, is termed doping. Doped Mott insulators 
provide a key to understand high $T_c$ superconductivity of cuprates 
(for a recent review, see \cite{Lee06};
for the recent contribution in the atomic context, see \cite{Klein06}).
P. Torma's group has investigated in recent years, the role of spin-density imbalance, both in trapped \cite{paivi1}, 
and superfluid lattice 
Fermi gases \cite{paivi2, paivi3}.
Asymmetric Hubbard model with light and heavy fermions has been studied by Ziegler's group,
who has, in particular, shown quantum phases with self-induced disorder
\cite{Ziegler1,Ziegler2,Ziegler3}.

Hubbard models for particles with higher spin, as well as Fermi-Fermi mixtures can be realized with ultracold atoms. 
 There has been thus considerable interest in studying such systems (for a review of $SU(N)$ symmetric Hubbard models 
of fermions and bosons see \cite{Hofstetter06}). In Ref. \cite{Zawitkowski06} effective spin Hamiltonians for spin 3/2 
ultracold spinor gases, described by the spinor version of the Hamiltonian (\ref{FHa}) have been derived and investigated. 
In 1D the same system has been studied using the bosonisation approach \cite{Lecheminant05}, and exact Bethe ansatz method \cite{Controzzi06}.

\paragraph{Fermi-Bose Hubbard models}

The spectacular advances in loading atomic samples in optical lattices
have allowed for the realization of  systems which are well described by
a Fermi Hubbard model, as well as  mixtures of fermions and bosons (FB) 
described by Fermi-Bose Hubbard models. In the latter case, and 
in the limit of strong atom-atom interactions, 
such systems can be described in terms of
composite fermions, consisting of a bare fermion, 
or a fermion paired with one boson (bosonic hole), or two bosons
(bosonic holes), etc. \cite{Kagan0405}. The quantum phase diagram displayed
by those systems is amazingly rich and complex.  
The physics of Fermi-Bose mixtures in this regime has been 
studied recently in \cite{Lewenstein04,Fehrmann04a,Fehrmann04b}, using perturbation theory 
up to second order to derive an effective Hamiltonian.  
There are a number of recent studies of FB mixtures in optical lattices 
\footnote{The studies of trapped FB gases concerned in particular
FB phase separation  \cite{molmer,stoofpu}, the phase diagram \cite{Roth2002}, 
novel types of 
collective modes \cite{stoofpu,capuzziliu}, Fermi-Fermi interactions 
mediated by bosons \cite{stoofpu,albusviveritheisel}, 
the  collapse of the Fermi cloud in the presence of attractive FB interactions
\cite{bfcooling4}, 
or the effects characteristic for the 1D FB mixtures \cite{dascazalilla}.},
%
%
and also of 
strongly correlated  FB mixtures in traps
\cite{Svistunov03,other}. 
In particular, the validity of the effective Hamiltonian for fermionic
composites in 1D was studied using exact diagonalization 
and  Density Matrix Renormalization Group method  in
Ref. \cite{Mehring05}, whereas trap effects were analyzed in Ref. \cite{Cramer04}. 
Very recently, exact Bethe ansatz solution for the special case of FB Hubbard model has been found
and intensively studied \cite{Imambekov06+pra,Frahm05}. 
Also, intensive studies of quantum phases in the multiband Fermi-Hubbard model (that models a Fermi gas subject to 
Feshbach resonance) have been undertaken \cite{Carr2, Zhou, diener}.
Recently, particular interest has been stimulated by the experiments of the 
Hamburg and ETH groups \cite{Klaus, Tilman}, in which presence of fermions attracting bosons was seen to decrease 
the lattice potential threshold 
for MI state. This is quite paradoxical, since DMRG calculations predict the opposite effect \cite{pollet}
 The physics of disordered FB  mixtures will be
 reported in subsection \ref{sub_bosefermi}. 

Here, we focus on the homogeneous case
and explain how to derive the corresponding effective Hamiltonian. We consider a homogeneous mixture 
of ultracold bosons (b) and spinless (or spin-polarized)
fermions (f), for example $^7$Li-$^6$Li or $^{87}$Rb-$^{40}$K, trapped in an optical lattice. 
As for the bosonic case, 
in the tight-binding regime, it is convenient to project the wave functions onto the
Wannier basis of the fundamental Bloch band, 
corresponding to wave functions well
localized in each lattice site \cite{Ashcroft76,Kittel04}. 
This leads to the Fermi-Bose Hubbard (FBH) Hamiltonian 
\cite{Auerbach94,Sachdev99,Jaksch98,other}:

\begin{equation}
H_{FBH} = -\sum_{\langle ij \rangle}\left[ t_b b^{\dagger}_i b_j+t_f 
f^{\dagger}_i f_j + h.c. \right]
+ \sum_i \left[\frac{U}{2} n_i(n_i-1) + V n_i m_i -\mu^b n_i 
-\mu^f m_i \right],
\label{hamiltonian} 
\end{equation}
where $b^{\dagger}_{i}$, $b_{i}$, $f^{\dagger}_{i}$ and $f_{i}$ are bosonic and fermionic 
creation- annihilation operators of a particle in the $i$-th localized Wannier state of 
the fundamental band, and $n_i= b^{\dagger}_i b_i$, $m_i= f^{\dagger}_i f_i$ 
are the corresponding on-site number operators.
The FBH model describes:
(i) nearest neighbor (n.n.) boson (fermion) hopping, with an associated negative energy, $-t_b$ ($-t_f$);
(ii) on-site boson-boson interactions with an energy $U$, which 
we will assume to be repulsive (i.e. positive),(iii) on-site boson-fermion interactions with an energy $V$, 
which is positive (negative) for repulsive (attractive) interactions and finally (iv) on-site energy 
due to chemical potentials in grand canonical description.
Following \cite{Cohen92,Ahufinger05,Fehrmann04a, Fehrmann04b}, let us review 
how to derive an effective Hamiltonian to second (or higher) order in
$t=t_f=t_b$ 
(we assume here the same tunneling for bosons and fermions for the sake of simplicity). 
Generalization to the case \(t_f \ne t_b\) 
is straightforward.
We follow here the notation of Ref. \cite{Lewenstein04}.

\begin{figure}[t]
\centering
\includegraphics[width=1.1\linewidth]{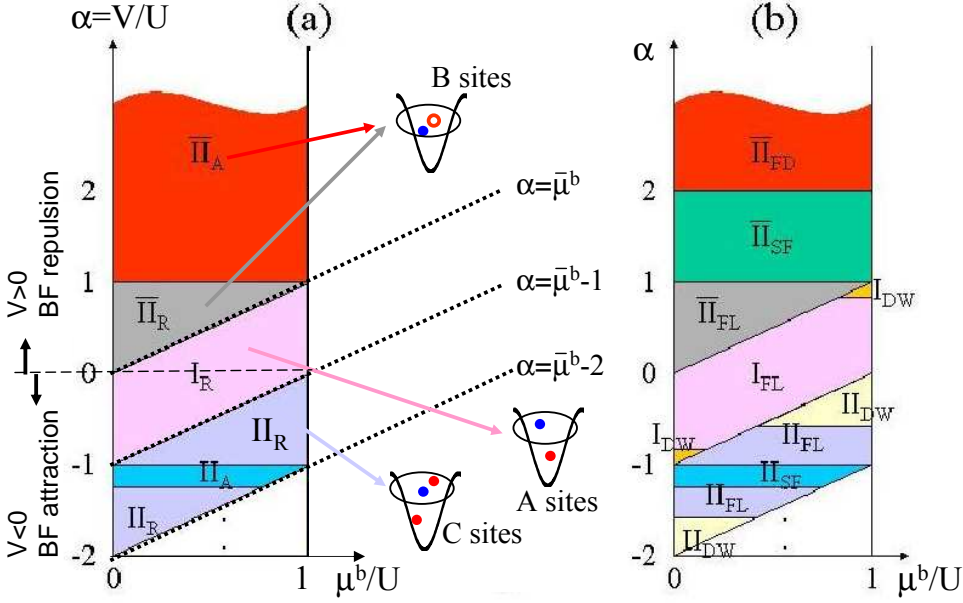}
\caption{Quantum phase diagrams of Fermi-Bose mixtures in an homogeneous optical 
lattice as functions of $\tilde{\mu}^{b}$ and $\alpha=V/U$, for 
$\rho_f=0.4$ and $t/U=0.02$. Roman numbers denote the total number of particles that 
form the composite and a bar means that the composite is formed by bosonic holes rather than bosons. 
(a) Diagram of composites where the filled small (blue) dots symbolize fermions, large (red) dots symbolize 
bosons and empty (red) dots, bosonic holes. The subindex A (R) indicates attractive (repulsive) composites 
interactions. (b) Detailed quantum phase diagram of fermionic composites. The subindices denote here different
 phases: DW (density wave), FL (fermi liquid), SF (superfluid) and FD (fermionic domains). The strongly correlated 
phases for small but finite $t$ are surrounded by characteristic lobes 
\cite{Fehrmann04a}, beyond which bosons become superfluid. Therefore, there are thin regions of bosonic superfluid 
between the various composite phases (from \cite{Lewenstein04}).}
\label{fig:phase}
\end{figure}

In the limit of vanishing tunneling ($t=0$) with finite repulsive boson-boson interaction $U$,
and in the absence of interactions between bosons and fermions ($V=0$), the bosons are in a Mott
insulator (MI) phase with exactly $\tilde{n}=\lceil\tilde{\mu}^b\rceil+1$ bosons per site,
where $\tilde{\mu}^{b}=\mu^{b}/U$ and $\lceil x \rceil$ denotes the
integer part of $x$. In contrast, the fermions can be in any set of Wannier states, since for vanishing tunneling,
the energy is independent of their configuration.
The situation changes when the interparticle interactions between bosons and fermions, $V$, are turned on.
In the following, we define \(\alpha = V/U\), and consider the case of bosonic MI phase with \(\tilde{n}\) bosons per site.
We also assume that the temperature is small enough (\(T< V\)) in order to neglect the particle-hole excitations. 
 It follows that tunneling of a fermion is
necessarily accompanied by the tunneling of $-s$ bosons (if $s<0$) or
opposed-tunneling of $s$ bosons (if $s\geq 0$). The dynamics of the Fermi-Bose
mixture can thus be regarded as the one of composite fermions made of one fermion plus $-s$ bosons
(if $s<0$) or one fermion plus $s$ bosonic holes (if $s\geq 0$). The annihilation operators of the
composite fermions are \cite{Lewenstein04}:
\begin{eqnarray}
F_i & = & \sqrt{\frac{(\tilde n -s)!}{\tilde n !}} \left( b_i^\dagger \right)^s f_i \quad\mbox{for $s$ bosonic holes}\label{eq:composites1} \\
F_i & = & \sqrt{\frac{\tilde n!}{(\tilde n -s)!}} \left( b_i \right)^{-s} f_i \quad\mbox{for $-s$ bosons} .\label{eq:composites2}
\end{eqnarray}
These operators are fermionic in the sub-Hilbert space generated by $|n-ms,m\rangle$ with $m=0,1$ in
each lattice site.
Note that within the picture of fermionic composites, the vacuum state corresponds to MI phase with
$\tilde n$ boson per site.
At this point, different composite fermions appear, depending on the values of $\alpha$, $\tilde n$ and
$\tilde \mu^{b}$, as detailed in Fig. \ref{fig:phase}.

Because all sites are equivalent for the fermions, the ground state is highly
$[N! / N_f! (N - N_f)!]$-degenerated, where $N$ denotes the total number of atoms and $N_f$ the number of fermions. Hence, the manifold of ground
states is strongly coupled by fermion or boson tunneling.
We assume now that the tunneling rate $t$ is small but finite. 
In the homogeneous case, one uses standard second order perturbation theory, to derive 
%
an effective Hamiltonian \cite{Auerbach94} for the fermionic composites (for the disordered case, see Appendix \ref{effhamil} 
\cite{Cohen92,Ahufinger05}):
\begin{equation}
H_{eff}=-d_{eff}\sum_{\langle i,j \rangle}(F^{\dagger}_iF_j+ h.c.)+
K_{eff}\sum_{\langle i,j \rangle}M_iM_j - \overline{\mu}_{eff} \sum_i M_i,
\label{Heff}
\end{equation}
where $M_i= F^{\dagger}_i F_i$ and $\overline{\mu}_{eff}$ is the chemical potential,
which value is fixed by the total number of composite fermions. The nearest neighbor hopping for
the composites is described by $-d_{eff}$ and the nearest neighbor composite-composite interactions
is given by $K_{eff}$, which may be repulsive ($>0$) or attractive ($<0$). 
This effective model is
equivalent to that of spinless interacting fermions. The interaction coefficient $K_{eff}$ originates
from $2$nd order terms in perturbative theory and can be written in the general form \cite{Fehrmann04a}:
\begin{equation}
K_{eff} = \frac{-2t^2}{U} \left[
{(2\tilde{n}-s)(\tilde{n}+1)-s(\tilde{n}-s)}
-
\frac{(\tilde{n}-s)(\tilde{n}+1)}{1+s-\alpha}
-
\frac{(\tilde{n}-s+1)\tilde{n}}{1-s+\alpha}
-
\frac{1}{s\alpha}
\right] .
\label{Keff_hom}
\end{equation}

This expression is valid in all the cases, but when $s=0$, the last term ($1/s\alpha$) should not be taken into account. $d_{eff}$ originates from $(|s|+1)$-th order terms in perturbative theory and thus presents
different forms in different regions of the phase diagram of Fig.~\ref{fig:phase}. For instance in region I,
$d_{eff}=t$, in region $\overline{II}$ $d_{eff}=2 t^2 / V$ and
in region II, $d_{eff}=4 t^2/|V|$.

The physics of the system is determined by the ratio $K_{eff}/d_{eff}$ and the sign
of $K_{eff}$. In Fig.~\ref{fig:phase}(a), the subindex A/R denotes attractive
($K_{eff}>0$) / repulsive ($K_{eff}<0$) composites interactions. Fig.~\ref{fig:phase}(b)
shows the quantum phase diagram of composites for fermionic filling factor $\rho_{f}=0.4$ and tunneling $t/U=0.02$.
For large values of \(t/U\), a transition to the SF state takes place (for the corresponding 3D lobes over the 
different regions in Fig. \ref{fig:phase}, see Ref. \cite{Fehrmann04a}).

\section{Disordered ultracold atomic gases}
\label{sec_disorder_Konark}

\subsection{Introduction}
\label{sub_int_disorder}

In 1958, Anderson \cite{Anderson58} reported for first time the quantum localization phenomenon. 
He pointed out that the extended wave functions of electrons (Bloch waves) in a crystal, for strong enough disorder, 
become localized with an exponentially decaying envelope $\vert \psi \vert \sim exp(\vert\vec{r}-\vec{r_i}\vert/\xi)$, where $\xi$ is 
the localization length. This occurs at the single particle level by coherent back scattering from random impurities. 
In one dimension, it has been rigorously proven that infinitesimally small disorder leads to exponential localization 
of all the eigenfunctions by repeated backscattering \cite{Mott61, Borland63, Mott68a} and it is known that the localization length 
is of the order of the backscattering mean free path. Ref. \cite{Mott68b}  proposed the concept of mobility edge, which 
separates the localized states from the extended ones. 
The scaling theory has been one of the basic tools to deepen the understanding of the Anderson localization phenomenon by 
considering the conductance as a function of the system size or of other scale variables \cite{Edwards72, Thouless74, Licciardello75, Wegner76, Abrahams79, Gorkov79}.  In 1D the localization
length is a function of the ratio between the potential and the kinetic (tunneling) energies of the eigenstate and the 
disorder strength. For the case of discrete systems with constant tunneling rates and local disorder distributed according 
to a Lorentzian distribution (Lloyd's model, cf. \cite{Haake04}) the exact expression for the localization length is known. 
In general, an exact relation between the density of states and the range of localization
in 1D has been provided by Thouless \cite{Thouless74}. In 2D, following the scaling theory, it is believed that localization 
occurs also for arbitrarily small disorder, but its character interpolates smoothly between algebraic for weak disorder, 
and exponential for strong disorder. There are, however, no rigorous arguments to support this belief, and several 
controversies aroused about this subject over the years. In 3D, scaling theory predicts a critical value of disorder, above which 
every eigenfunction exponentially localizes, and this fact has found strong evidence in numerical simulations. 

Since the discovery of Anderson \cite{Anderson58}, the field of quantum disordered systems has been a very active research field in condensed matter physics (for reviews see \cite{Nagaoka82, Lee85, Ando88, Kramer93}). Disordered systems appear in some of the most challenging open questions concerning many body systems (see for instance \cite{Cusack87, Balian79, Chowdhury86}). 
In particular, quenched (i.e., frozen on the typical time scale of the considered systems) disorder
determines the physics of a large variety of phenomena: transport, conductivity, localization effects and metal-insulator transition (cf. \cite{Nagaoka82, Lee85, Ando88, Kramer93}), spin glasses (cf. \cite{Mezard87, Newman03, Sachdev99}), neural networks (cf. \cite{Amit89}), percolation \cite{Aharony94, Barbosa03}, high $T_c$ superconductivity (cf. \cite{Auerbach94}), or quantum chaos (cf.\cite{Haake04}). 

The theoretical studies  of disordered systems imply severe difficulties. One of the main ones is that in order to characterize the systems
 independently of the particular disorder realisations, one needs to average over the disorder. 
Such quenched (frozen over the typical time scales of the system) average of physical quantities, such as for instance free energy, requires usually the use of special methods, 
such as the replica trick (cf. \cite{Mezard87}) or supersymmetry method \cite{Efetov97}. Interestingly,
quantum information approach might help to calculate averages using quantum superpositions, see \cite{Parades05}.
Regarding numerical approaches,  the characterization of the system demands 
either simulations of very large
samples to achieve ``self-averaging'' or 
numerous repetitions of simulations of small samples with different configurations of the disorder. 
Obviously, this difficulty is particularly important for quantum disordered systems.
Another difficulty that arises in the study of disordered systems is the possible existence of a large number of low energy excitations. 
Similar limitations arise in frustrated systems (cf. \cite{Misguich04})(see section \ref{frustration_atom}). Last but not least, the interplay of the disorder and the interactions has been and still is a challenging problem, as it is reported in the following subsections in the case of bosons and fermions.  

Disorder effects are present in many condensed matter systems, and in particular in electronic systems. Nevertheless, these  systems do not provide the best possible  scenario to test the theoretical predictions mainly because of two factors: (i) the disorder is not controllable, but fixed by the specific realization of the sample; and (ii) the electrons interact via the long-range Coulomb interaction, which is fixed by Nature. It is thus desirable to ask whether atomic, molecular physics and quantum optics may help to understand quantum disordered systems, for instance by realizing in experiments disordered models using cold atoms in optical lattices. The advantages of atomic disordered systems are clear: (i) the control of the generation of random potentials that induce disorder in the system (see subsections \ref{sub_realization} and 
\ref{sub_exp}); (ii) the presence of interactions that can also be controlled by means of Feshbach resonances; (iii) the type of quantum statistics, that can be chosen 
by using ultracold bosons or fermions, and (iv) the full control of the trapping potentials and therefore the effective dimensionality of the system. Obviously, standard control parameters, such  as for example the temperature of the system, can also be controlled, but in these aspects atomic system do not differ much form the condensed matter one. Of course, there are also disadvantages: atomic systems are 
``small". Typical experiments with ultracold lattice gases involve lattices of sizes up to  $100\times 100\times 100$
and up to few times $N=10^6$ atoms. 

\subsection{Disordered interacting bosonic lattice models in condensed matter}
\label{sub_bosons}

The theory of disordered interacting bosons is complex and there are essentially  no exact solutions, not even in one-dimension. In this section we review different numerical and approximate results existing in the condensed matter literature. 

A system of bosons in a lattice with short-ranged repulsive interactions in the presence of external random potentials was considered in \cite{Fisher89} where, by treating the tunneling as a perturbation, the phase diagram of the system was worked out. The three possible ground states predicted were: (i) an incompressible Mott insulator with a gap for particle-hole excitations; (ii) a gapless Bose-glass insulator with finite compressibility, exponentially decaying superfluid correlations in space and infinite superfluid susceptibility; and (iii) a superfluid phase with the usual off-diagonal long ranged order. It was predicted that the gapless Bose-glass intervenes between the Mott and the superfluid phases, so that the superfluid transition always occurs from the Bose-glass. The critical properties of this transition were characterized by three exponents: a dynamical exponent that equals the dimensionality of the system, a correlation length exponent that is bounded from above and an order parameter exponent that is bounded from below. 

Previously, the onset of superfluidity in a random potential was studied in \cite{Ma86} and \cite{Giamarchi88}. In \cite{Ma86}, a system of strongly disordered hard-core bosons was considered in the framework of a mean field theory including quantum fluctuations. A renormalization group approach was developed to study a one-dimensional system of interacting bosons in a random potential \cite{Giamarchi88}. In this work it was shown the existence of a localized-superfluid transition and universal power laws for correlation functions on the transition line were found. 
The case of zero temperature interacting bosons at commensurate density (one atom per site) on disordered lattices in one and two dimensions was addressed in \cite{Singh92a} by using a real-space renormalization group. The results showed that when weak disorder is introduced, a transition directly from the Mott insulator to the superfluid occurs, so that infinitesimally weak disorder does not stabilize a Bose-glass 
at commensurate filling. The Bose-glass is found beyond a threshold disorder, in contradiction with the arguments in \cite{Fisher89}. The critical exponents for the superfluid-Bose-glass phase transitions at zero temperature for hard-core bosons in one-, two-dimensional \cite{Zhang92} and three-dimensional \cite{Zhang93a} disordered lattices were calculated using a quantum real-space renormalization group method once the system was mapped onto a quantum spin-1/2 XY model with transverse random field. From these calculations \cite{Zhang92,Zhang93a}, it was concluded that randomness is always relevant in one-dimensional systems while in two and three dimensions, there is a critical amount of disorder below which the superfluid phase is stable. Moreover it is also stated that there is only one universality class for the superfluid-Bose-glass transition. Starting from the mapping of a system of hard core disordered bosons onto a quantum spin-1/2 XY model with transverse random field and generalizing it to a system of spins with arbitrary magnitude, a perturbative study was applied to get more insight into the low-energy excitations in a weakly disordered bosonic system \cite{Zhang93b}.

In the early 90's, disordered Bose condensates were studied using the Bogoliubov approximation 
\cite{Lee90,Huang92,Singh94}. In \cite{Lee90}, the screening of the random potential in a two-dimensional dense Bose gas in a lattice due to the short-range repulsive interactions was addressed. This screening is effective when the healing length is short enough so that the condensate can adjust to the variations of the random potential. The effects of the fluctuations deplete the condensate nonuniformly and lead to a spectrum of collective, phonon-like excitations. In \cite{Singh94} was reported that weak disorder hardly affects the condensate fraction or the superfluid density although the condensate distorts to screen the imposed random potential. In the strong disorder limit, the condensate fraction and the superfluid density tend to vanish as the disorder increases and a constant density of states appears at low energy. In case of three-dimensional  bosonic systems in random external potentials the depletion of the condensate and the superfluid density were 
explored in \cite{Huang92}. The conclusion obtained from this analysis was that disorder is more active in reducing superfluidity than in depleting the condensate. In this scenario, a formalism based on dispersive quantum hydrodynamics at zero temperature was applied to investigate the propagation of phonons in the system \cite{Giorgini94}. In particular, the shift of the sound velocity and its damping were calculated. In \cite{Lopatin02}, building on the continuum model of \cite{Huang92} and \cite{Giorgini94}, a systematic diagrammatic perturbation theory for a dilute Bose gas with weak disorder at finite temperature below the superfluid transition temperature was developed and the disorder-induced shift of the superfluid transition temperature was derived. 
  
Monte Carlo techniques (world-line algorithm) have been also applied to study the interplay between interactions and disorder in bosonic systems. In \cite{Scalettar91} was reported the first convincing evidence of a second insulating phase, the Anderson-glass phase, in a one-dimensional lattice system. The Anderson-glass phase appears at weak couplings where the interactions compete with the disorder and tends to delocalize the bosons, contrary of what happens in the Bose-glass phase, that appears at strong couplings where disorder and interactions cooperate. Moreover, these two phases differ substantially on the nature of the boson density distribution: in the Bose-glass, the density is reasonably uniform while in the Anderson-glass, boson density correlations are expected to decay exponentially \cite{Scalettar91}. The existence of two such separate insulating phases was conjectured previously in \cite{Giamarchi88}. The same year 1991, in \cite{Krauth91}, path-integral Monte Carlo techniques were used to study also the superfluid-insulator transition but in a two-dimensional square lattice. Using this technique in two dimensions, three phases were predicted: superfluid, Bose-glass and Mott-insulator and it was stated that at commensurate density, the system seems to undergo a direct transition from the superfluid phase to Mott insulator contradicting the picture of disorder of \cite{Fisher89} and \cite{Giamarchi88}, but in agreement with the results obtained using real-space renormalization group by \cite{Singh92a}. The direct transition from superfluid to Mott insulator without intervening Bose-glass at weak disorder was also reported for instance in \cite{Pai96} by using density-matrix renormalization group in one dimension, in \cite{Kisker97} using quantum Monte Carlo simulations in two dimensions, or in \cite{Pazmandi95} using a mean field theory. Contrarily, the transition via the Bose-glass phase was predicted, for instance in \cite{Wallin94, Prokofev04} by using Monte Carlo simulations, in \cite{Kim94} by using the renormalization group method, in \cite{Freericks96} using a strong-coupling expansion for the phase boundary of the Mott-insulator or in \cite{Rapsch99} by an improved  application of the density matrix renormalization group with respect to \cite{Pai96}. It is worth noticing that in the results of \cite{Rapsch99} there is no indication of a qualitative difference between the glass phase at small or large values of the repulsion i.e., between Anderson and Bose-glass. Thus, a complete understanding of the phase diagram of interacting bosons in the 
presence of disorder and in various dimensions is still under debate. In particular, the possibility of a direct Mott-insulator to superfluid transition in the presence of disorder remains a controversial issue.

\subsection{Realization of disorder in ultracold atomic gases}
\label{sub_realization}

In order to perform a detailed analysis of the properties of the disordered interacting Bose lattice gases it 
would obviously be very useful to have  an experimentally accessible system,
 that could be studied in a controlled way. As we will see below, ultracold atomic lattice gases provide such opportunity.

As has been discussed in section 1, since the experimental realization of Bose-Einstein 
condensation there
has been an enormous progress in the studies of the ultracold gases: first in the weakly interacting regime and more recently also in the strongly interacting regime. Nowadays there exist a complete control of the generation and manipulation of ultracold bosonic, or fermionic gases, as well as their mixtures. Among all the techniques of control, the transfer of these ultracold samples to optical lattices offers an unprecedented possibility to study disorder related phenomena. In fact, by superimposing laser beams from different directions and with different frequencies, it is possible to generate a huge variety of lattice geometries  in a very controlled way. 
For instance, it has been proposed the use of two colour superlattices \cite{Damski03,Roth03a,Roth03b}, i.e., the superposition of two standing-wave lattices with comparable amplitudes and with different wavelengths, as a form of quasidisorder. The so-called quasicrystal optical lattices in two and three dimensions have also been explored in \cite{Guidoni97,Guidoni98,Guidoni99,Sanchez-Palencia05}. These systems present long range order but not translational invariance. An example of a laser configuration \cite{Sanchez-Palencia05} that gives rise to a quasicrystal lattice consists on $N_b$ laser beams arranged on the $xy$ plane with $N_b$-fold symmetry rotation. The polarization $\vec{\epsilon}_j$ of laser $j$ with wavevector $\vec{k}_j$ is linear and makes an angle $\alpha_j$ with the $xy$ plane. The optical potential is in this case:
\begin{equation}
V(\vec{r})=\frac{V_0}{\vert \sum_{j}\varepsilon_j \vert^2}
\left|\sum_{j=0}^{N_b-1} \varepsilon_j \vec{\epsilon}_j exp^{-i(\vec{k}_j\vec{r}+\varphi_j)}\right|^2,
\label{quasicristal}
\end{equation}   
where $\vec{r}=[x,y]$, $0\le\varepsilon_j\le1$ stand for eventually different laser intensities and $\varphi_j$ are the corresponding phases. 
In  \cite{Damski03} a two-dimensional local quasidisordered potential was numerically simulated by using a main optical lattice, $V_l(\vec{r})$, to which a secondary lattice, $V_r(\vec{r})$, with much smaller amplitude and with frequency incommensurable with respect to the main lattice, is superimposed. The quasidisorder potential in this case reads:
\begin{equation}
V(\vec{r})=V_l(\vec{r})+V_r(\vec{r})=V_0[\cos^2(kx)+\cos^2(ky)]+V_1[\cos^2(\vec{k_1}\vec{r})+
\cos^2(\vec{k_2}\vec{r})]
\label{superlattice}
\end{equation}
where $\vec{r}=[x,y]$. The secondary lattice is responsible for the introduction of the (quenched) pseudodisorder which is determined by the ratio between the wavelengths of the main and additional lattices, $k_1/k$ and $k_2/k$. The same kind of potential has been also studied in one-dimensional geometries \cite{Diener01}.

\begin{figure}[t]
\centering
\includegraphics[width=0.75\textwidth]{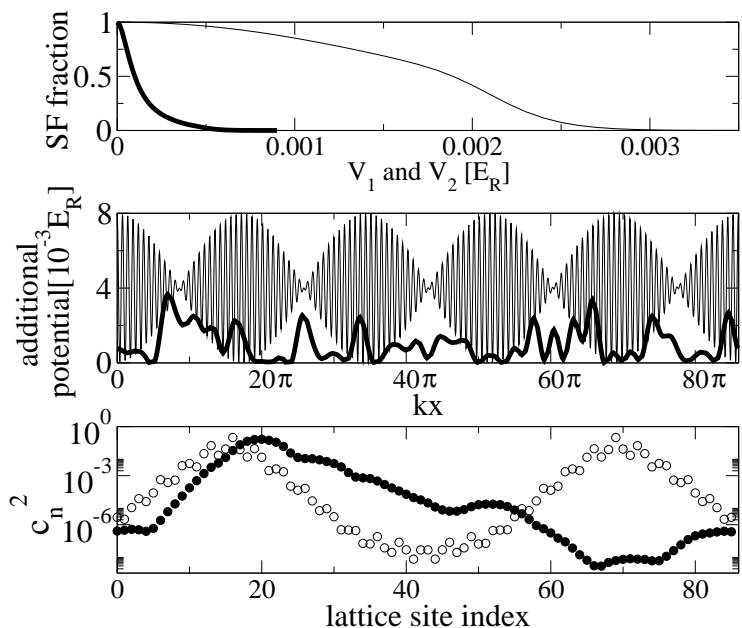}
\vspace*{.5cm}
\caption{Theoretical prediction of Bose-Anderson glass in a 1D optical lattice; 
top: superfluid fraction as a function of disorder $V_1$ (thickline), respectively 
quasi-disorder $V_2$ (thinline); 
middle: disordered speckle potential (thickline), 
and quasi-disordered (thinline) potential formed by incommensurable superlattices; 
bottom: occupation numbers of the ground state in the presence of disorder (black dots), and quasi-disorder (open dots) (from \cite{Damski03}).} 
\label{figure3.1}
\end{figure}

All the potentials discussed so far in this section are not strictly speaking disordered, but quasidisordered. 
Truly random potentials can be achieved by using a speckle pattern 
\cite{Goodman75, Horak98, Boiron99, Damski03, Lye05, Fort05, Schulte05, Clement05}. The speckle field is a light field 
with highly disordered intensity and phase distributions but stationary and coherent. Such a speckle field 
can be easily generated experimentally, for instance, by 
introducing a diffusor in the path of a laser plane wave or by reflecting a laser beam in a surface that is rough on the scale of the laser wavelength. Specifically, the random intensity of an speckle field follows an exponential statistical distribution with a standard deviation given by the average intensity, $P(I)=\exp(-I/\langle I\rangle)$. 
In addition, the intensity correlation length or disorder correlation length, given by the half-width of the autocorrelation function, is limited by optical resolution being of the order of at least few $\mu$m (see Fig. \ref{figure3.1}
for examples of 1D random, and quasi-random potentials leading to Bose-Anderson glass). 

One can also create a disordered potential by using atoms of a second species or internal state quenched randomly  
at the nodes of an optical lattice \cite{Gavish05, Zoller05b}. These randomly distributed atoms act as point-like scatterers 
for the atoms of the first species, that are not trapped by the optical lattice. One possibility to randomly trap the atoms of 
the second species is by rapidly quenching them from the superfluid to the Mott insulating phase. In this approach, 
the correlation length is very short since the distance between the lattice sites is of the order of $0.5\mu $m.
It is possible that such quenched random scatterers of fermionic composites (consisting of fermions bounded to 6-7 bosons) have been recently realised in an experiment with Fermi-Bose mixtures by the Hamburg group \cite{Ospelkaus06}.
It is also worth pointing out that with this  kind of disorder, the effective Hamiltonian for composites will 
involve disorder with bimodal, or at most trimodal distribution. This kind of disorder is particularly interesting to study spin 
glass transition in the 2D Ising model \cite{Kawashima04}.

Recently, also the generation of disorder in ultracold atoms in optical lattices through on-site interparticle interactions has been addressed \cite{Kantian04, Gimperlein05}. The idea is to exploit the significant modifications of the scattering properties of two atoms that can be induced near Feshbach resonances \cite{Inouye98, Cornish00} by slight modifications of the magnetic field. 
In this scenario, spatial
 random variations of the local interatomic interactions arise. The system is placed at the verge of a Feshbach resonance 
by means of an offset magnetic field in the presence of a spatially random magnetic field. This spatially random magnetic field 
appears for instance in magnetic microtraps and atom chips as a result of the roughness of the underlying 
surface \cite{Folman02}. The intrinsic disorder that appears in magnetic microtraps and atom chips has also 
been addressed in the absence of an optical lattice \cite{Henkel03, Wang04}. Controllable disorder on the level
 of next-neighbors interactions can be generated by means of tunneling induced interactions in systems with local 
disorder \cite{Sanpera04}. It is also worth mentioning a very recent attempt to create controlled disorder using 
optical tweezers methods \cite{Raizen05}.

\subsection{Disordered ultracold atomic Bose gases in optical lattices}
\label{diorderedbose}


Previous section \ref{sub_realization} reports on experimental feasible realizations of random and pseudorandom potentials for cold atoms in optical lattices. All these possibilities lead to experimental realizations of the disordered Bose-Hubbard model \cite{Damski03, Roth03a, Roth03b}, where the interplay between interactions and disorder could be explored. The Bose-Hubbard hamiltonian in the presence of disorder reads like Eq. (\ref{bosehub}) but with site dependent tunneling rate,$t_{ij}$, and local chemical potential,$\mu_i$:
\begin{eqnarray}
H=-\sum_{<ij>} \left[ t_{ij} b^{\dagger}_i b_j+ h.c.\right] +\sum_i \frac{U}{2} n_i(n_i-1) +\sum_i \mu_i n_i,  
\label{bos_hub_dis}
\end{eqnarray}
Although tunneling coefficients in the most general disordered case should be site dependent, it has been shown \cite{Damski03} that in optical lattices the hopping disorder is suppressed against on-site disorder included in the term $\mu_i$. The last term in (\ref{bos_hub_dis}) gives the on-site single-particle energy which originates from the external potentials and the on-site part of the kinetic energy. Therefore it accounts for an external harmonic trapping potential plus the inhomogeneities produced by the speckle pattern, or by the superlattice.  Eventually, $\mu_i$ also contains the chemical potential in the grand canonical description. 
In \cite{Damski03}, the pseudorandom on-site energies are calculated by using:
\begin{eqnarray}
\mu_i=\int d^3r w^*(\vec{r}-\vec{r_i})V_r(\vec{r})w(\vec{r}-\vec{r_i}),
\label{mus}
\end{eqnarray}
where $w(\vec{r}-\vec{r_i})$ are the Wannier functions in the lowest Bloch band and $V_r(\vec{r})$ is the superlattice potential 
introduced in (\ref{superlattice}) or the potential induced by a numerically generated speckle pattern characterized by its mean
 value and the average speckle size. In \cite{Roth03a, Roth03b} there was considered a sinusoidal variation of the $\mu_i$ in the 
range $[-\Delta,0]$ recovering the 
regular case for $\Delta=0$.
\begin{figure}[t]
\centering
\includegraphics[width=0.75\textwidth]{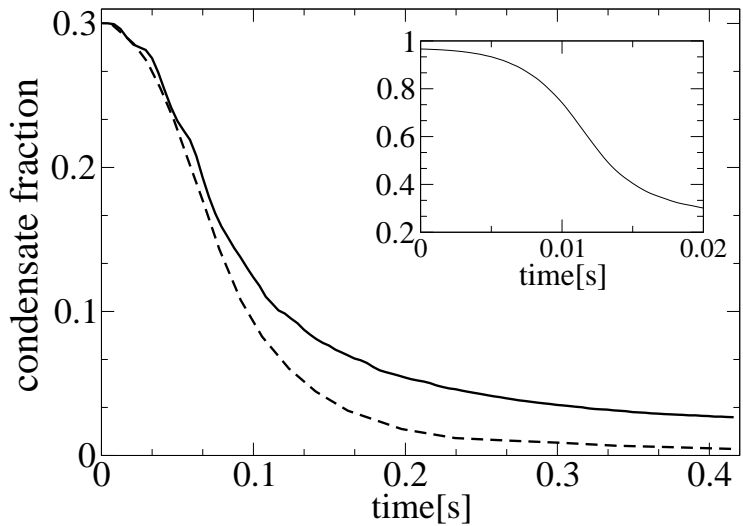}
\vspace*{.5cm}
\caption{Numerically simulated dynamical transition to the Bose glass state; first (in inset) a superfluid  
at high value of lattice potential is formed, and then (in the main figure) the disorder is turned on gradually. Condensate (solid line) and superfluid (dashed line)
fractions tend to zero (from \cite{Damski03}).}
\label{figure3.3}
\end{figure}
In \cite{Damski03}, the dynamical generation of the Bose-glass 
(strong interactions and non-integer filling factor) and the Anderson
 glass phases (weak interaction regime) in a two-dimensional ultracold
 bosonic gas was calculated using the Gutzwiller ansatz method (see Fig. \ref{figure3.3}), while in 
\cite{Roth03a, Roth03b} the ground state of the system was determined by solving 
the eigenvalue problem of the one-dimensional Bose-Hubbard hamiltonian numerically. 
In this last case, also both the Anderson-glass and the Bose-glass phases were found (see Fig. \ref{figure3.2}).

\begin{figure}[t]
\centering
\includegraphics[width=0.75\textwidth]{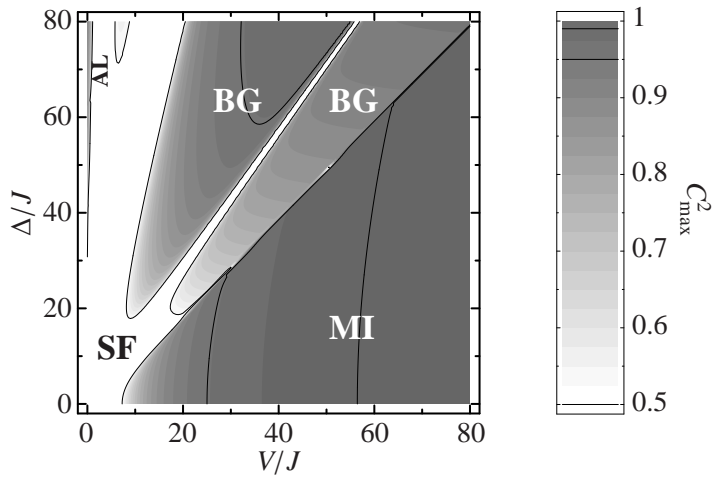}
\vspace*{.5cm}
\caption{Contour plot of the square of the largest coefficient, \(C_{\max}^2\), of the expansion of the
state of the system described by a Bose-Hubbard model in the number basis,          as
a function of disorder \(\Delta\) and on-site interaction \(V\), in units of the hopping energy \(J\).
 The results correspond to exact diagonalization of a one-dimensional system with 8 sites
and 8 bosons (from \cite{Roth03a}. Note that $V$ accounts for $U$ and $J$ for $t$ in the figure).
The labels identify the
predicted different phases: superfluid, Mott insulator, Bose glass (BG), and Anderson localization (AL).} 
\label{figure3.2}
\end{figure}

Very recently 
\cite{deMartino05},  the problem of disordered one-dimensional bosonic systems with hard core boson interactions has been
  exactly solved via a Jordan-Wigner transformation (see subsection \ref{subsec_kachupora1}). This mapping establishes a connection to non-interacting disordered 
fermions and allows to take advantage of many known results on Anderson localization. In this scenario, the correlation functions of the
 particle density and the local density of states for the interacting bosonic case coincide with the non-interacting fermionic ones, because 
only depend on the modulus square of the wave-function. This is not the case for the momentum distribution, but the mapping allows for a simple 
calculation of the disorder-averaged boson momentum distribution \cite{deMartino05}. These calculations show the complete destruction of quasi-long 
range order by disorder and the flattening of the momentum distribution for sufficiently strong disorder. Note that in the homogeneous case, the 
momentum distribution is well known to possess a singularity \cite{Lenard64, Vaidya79, Gangardt04}. In \cite{deMartino05} it was also shown that 
the Bose-Fermi mapping can also be established via the effective low-energy theory \cite{Giamarchi88}. Moreover, it was pointed out  that a similar 
mapping is also available for arbitrary interaction strength, but involves interacting fermions with nonstandard contact interactions \cite{Cheon99}. 
For strong (but finite) repulsive bosonic interactions, the weak fermionic 
interactions can be treated perturbatively. In Ref. \cite{Scarola06}, specific effects of quasi-disorder,
in contrast to real disorder, in 1D lattices were discussed.

Just before submission of this review,
 Krutitsky et al. \cite{Krutitsky06} calculated the mean field phase diagram of disordered bosons at finite $T$. They made the important observation that the distinction of MI and Bose glass at finite $T$ in atomic lattice gases should be possible by looking at the density of states of low energy excitations. 

\begin{figure}[t]
\centering
\includegraphics[width=0.75\textwidth]{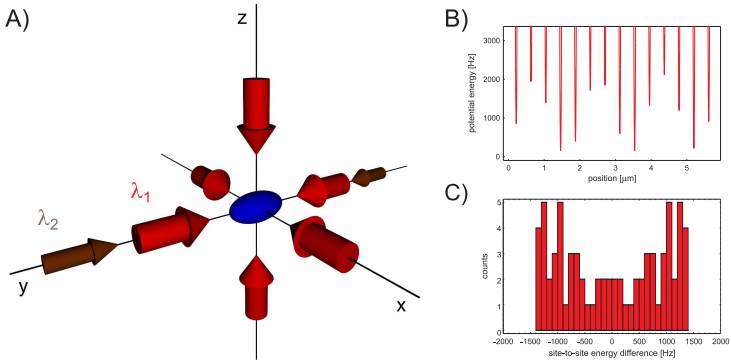}
\vspace*{.5cm}
\caption{Quasidisorder used in the experiments reported in \cite{Fallani06}.}
\label{figure3.4}
\end{figure}

\begin{figure}[t]
\centering
\includegraphics[width=0.75\textwidth]{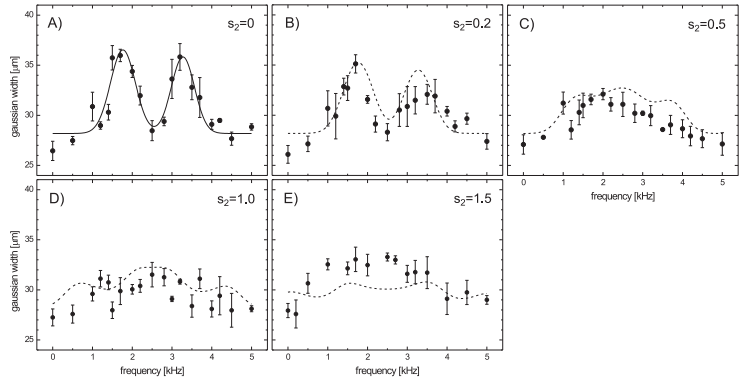}
\vspace*{.5cm}
\caption{Experimental signatures of the MI--Bose glass transition: Broadening of the excitation spectrum as disorder ($s_2$) grows. 
One observes discrete linbes in the gapped MI, and 
significantly broadened and structureless spectrum in the BG phase  (from \cite{Fallani06}).}
\label{figure3.5}
\end{figure}

The experiments on disordered ultracold gases in optical lattices has just started,  and most of them deal with trapped gases and weak interaction limit (we discuss them in the next subsection). Nevertheless,  spectacular results have been
achieved in the area of disordered lattice gases, already. The Florence group has recently applied to a basic lattice formed by light with wavelength $\lambda_1=830$nm another superlattice created by $\lambda_2=1076$nm light, creating a quasi--disordered potential (see Fig. \ref{figure3.4}). They have applied the method of Bragg spectroscopy 
developed in \cite{Stoferle04} and measured effects of the lattice modulation. This allows essentially to measure the low energy excitation spectrum; broadening of this spectrum was identified in Ref. \cite{Fallani06} to be the signature of the Bose glass phase (see Fig. \ref{figure3.5}).

In the experiment of the Hamburg group \cite{Ospelkaus06} (see also \cite{Gunter06}), an unexpected  shift of the SF--MI transition
point towards weaker lattice potentials has been observed in a Fermi-Bose mixture with attractive interspecies interactions. 
It has been speculated that this effect might have to do with the formation of composite fermions. 
Composite fermions, consisting of a fermion and few bosons, are hardly mobile and should remain in the conditions of the mentioned experiments immobile and play a role of random scatterers for (still relatively movable) bosons.

\subsection{Experiments with weakly interacting trapped gases and Anderson localization}
\label{sub_exp}

During the last year the first experimental results on Bose-Einstein condensates in random 
potentials were reported \cite{Lye05, Fort05, Schulte05, Clement05}. In the first experiments \cite{Lye05}, 
static and dynamic properties of an harmonically trapped $^{87}$Rb condensate in the presence of a stationary,
but spatially random potential created by imaging an speckle pattern onto the condensate were addressed. 
In this experiment the smallest length scale of the speckle was $L_{dis}=10\mu$m while the size of the quasi one 
dimensional condensate was $110\mu$m in the axial direction and $11\mu$m in the radial. Absorption images after free 
expansion for different values of the speckle height, $V_s$, reveal different regimes of behavior: (i) for strengths of 
the disorder much smaller than the chemical potential of the condensate in the harmonic trap, practically no change with 
respect to the standard Thomas-Fermi profile is observed; (ii) for intermediate values of $V_s$, stripes appear in the expanded 
density profile, which can be a signature of the development of different phase domains across the condensate, like the ones 
associated with phase fluctuations in highly elongated condensates \cite{Dettmer01, Richard01}, or they can be explained through 
the presence of different momentum components due to the growth of instabilities \cite{Fallani04}; and  (iii) when $V_s$ is larger 
than the chemical potential of the harmonically trapped condensate, the tunneling between the different minima of the random 
potential is strongly suppressed and the system enters the tight binding regime, where a broad Gaussian profile is seen in the expansion 
as expected for the expansion of randomly spaced condensates isolated in the individual speckle wells. The transport properties of the 
system were also studied in \cite{Lye05} through collective excitations, dipolar and quadrupolar modes. The dipolar mode was excited by 
abruptly displacing the magnetic trap in the axial direction obtaining: (i) for  $V_s$ much smaller than the chemical potential, slightly 
damped oscillations were obtained with the same frequency as in the undamped case in the absence of the speckle; (ii) for intermediate $V_s$, 
strong damping of the oscillations is observed; and (iii) for high values of $V_s$, the atomic cloud does not oscillate and remains localized 
on the side of the magnetic trap. The quadrupolar mode was excited with a resonant modulation of the radial trapping frequency observing that 
the anharmonicities of the random potential result in frequency shifts that are not correlated with the dipolar frequency like it happens in the 
absence of speckle. Moreover, it is shown that both the sign and the amplitude of the shift depend on the exact realization of the speckle 
potential.

Following these first experiments, the suppression of the one-dimensional transport of an interacting elongated condensate 
in a random potential was reported nearly simultaneously by \cite{Fort05, Clement05}. In these experiments, one-dimensional 
expansion of a $^{87}$ Rb condensate along an optical \cite{Fort05} or magnetic \cite{Clement05} guide in the presence of a speckle 
potential was studied. Without the speckles, the condensate freely expand and the growth of the root mean squared (rms) radius (rms$=\sqrt{<x^2>+<y^2>}$) is self-similar and linear in 
agreement with the theory \cite{Kagan94, Castin96}. In the presence of a random potential with high enough amplitude, but without 
entering in the tight binding regime, the expansion dynamics changes completely and both the expansion and the centre of mass motion are 
inhibited. Although this strong suppression of expansion corresponds to disorder-induced trapping of the BEC, it does not correspond to
 Anderson like localization. The reasons are mainly: (i) the screening of the disorder potential played by the interaction energy; (ii) 
the resolution of the optics of the disordered potential that fixes the correlation length of the disorder, $L_{dis}$, which 
is much larger than the healing length, $l_{heal}=1/ \sqrt{8 \pi n a}$ where $n$ is the density and $a$ the atomic scattering
 length. The healing length provides the typical distance over which the order parameter of the condensate recovers its bulk value 
when it is forced to vanish at a given point by an impurity for instance. Therefore, the density profile tends to follow 
the modulations of the disordered potential; and (iii) the fixed order of magnitude of the distance between the speckle
 sites imposes that the typical axial size of the system, $L$, is only ten times larger than the correlation length of the disorder.
 In order to achieve Anderson localization, the correlation length of the disorder has to be smaller than the size of the system, 
which seems to be difficult to achieve with speckles. Anderson localisation of elementary excitations in 1D BEC seemingly sets a 
little less rigid requirement, but is still not easy to achieve with the presently achievable speckle patterns 
\cite{Kuhn05,Bilas06}. Recently, a very detailed study shows that by expanding a quasi 2D cloud to a size of \(1 cm^2\), it should be possible 
to see effects of weak localization and perhaps even strong ones, using available speckle potentials \cite{Kuhn05}.

\begin{figure}[t]
\centering
\includegraphics[width=0.75\textwidth]{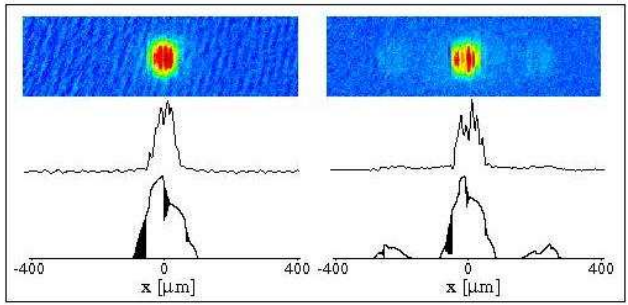}
\vspace*{.5cm}
\caption{Density profiles (absorption images) of quasi 1D BEC released from the combined harmonic trap plus
random speckle (left column), and from the combined harmonic trap, optical lattice, and random speckle (right column). The second row shows the column density and the third row shows the result of the numerical simulation 
(from \cite{Schulte05}). }
\label{figure3.6}
\end{figure}

\begin{figure}[t]
\centering
\includegraphics[width=0.75\textwidth]{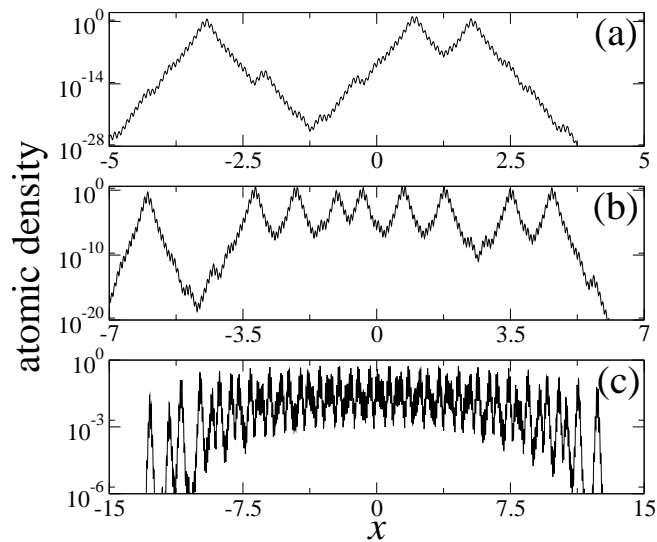}
\vspace*{0.5cm}
\caption{Calculated condensate wavefunction for a BEC in a combined potential formed by harmonic  trap, 
and two incommensurate standing waves.  The sequence (a)--(c) 
shows the effects of the increasing nonlinearity (i.e.,   increasing the number of atoms) (from \cite{Schulte05}). }
\label{figure3.7}
\end{figure}

In Ref. \cite{Schulte05}, the first realization of an ultracold disordered lattice gas was reported, so the 
effect of a speckle pattern was superimposed to a quasi-1D condensate in a regular 1D optical lattice. 
In this work, it was shown that the fragmentation, already reported in \cite{Lye05}, 
also appears in the presence of an optical lattice (see Fig. \ref{figure3.6}). More interestingly, in \cite{Schulte05} 
it has been explored the crossover from the Anderson localization in the absence of interactions, 
where the ground state wave function is characterized by an exponential localization, to the screening 
regime, where the number of localization centres is so high that one can no longer distinguish the individual 
localized states and the signature of nontrivial localization vanishes (see Fig. \ref{figure3.7}). On one hand, the interactions can be reduced 
by reducing the number of atoms, lowering the trap frequencies or tuning the scattering length via Feshbach resonances 
and on the other hand, the limitations in the ratio between the correlation length of the disorder and the healing length 
and between the correlation length of the disorder and the size of the system, that arise when the disorder potential is
created using an speckle pattern, could be overcome using the quasidisorder created by several lasers with 
incommensurable frequencies. Measurement of the superfluid fraction in an accelerated optical lattice is proposed 
\cite{Schulte05} as a way of detection of the localization rather than the usual measurement of the density distribution after 
ballistic expansion.

\subsection{Disordered interacting fermionic systems}
\label{sub_fermions}

So far we have discussed disordered interacting bosonic systems and the experimental possibilities of testing the still 
controversial points of the theory using ultracold bosonic systems in optical lattices. Fermionic gases have also been cooled to the quantum degeneracy regime using sympathetic cooling of two fermion species, 
or boson-fermion mixtures (for the first experiments see \cite{DeMarco99, Truscott01, Schreck01, Hadzibabic02, Roati02}). Moreover,  
fermionic $^{40}$K atoms have been also loaded into a 
three-dimensional optical lattice \cite{Kohl05}. Several proposals of using ultracold fermionic atoms in
 optical lattices to study various condensed matter  models have already been formulated. 
In \cite{Hofstetter02} for instance, it was  discussed how  fermionic atoms in optical lattices 
allow for the realization of antiferromagnetism or high-temperature superconductivity. More recently, it has been shown 
\cite{Honerkamp04} that the fermionic SU(N) Hubbard model on a 2D square lattice can be realized with 
ultracold fermions in an optical lattice. 
Regarding the disordered case, very recently it has been reported \cite{Paredes05} that fermionic atoms in an optical superlattice exhibit strongly correlated phases from Kondo singlet formation to magnetism of localized spins. 
                          
Let us now briefly review the literature on the physics of disordered interacting electronic  systems 
with the aim of identifying the still open questions that could be addressed exploiting the ultracold fermionic atoms in disordered optical lattices.   

As we have discussed in subsection \ref{sub_int_disorder}, originally disorder in non-interacting electronic systems attracted a lot of interest, but the effects of electron-electron interactions received full attention only recently.
 It was believed that weak disorder should not modify essentially the Fermi liquid picture of Landau. Later a Fermi liquid theory for electrons in disordered solids was formulated, and termed the Fermi glass theory \cite{Anderson70,Freedman77}. According to this approach the Landau's quasiparticle description is valid but now the quasiparticle wave-functions are not extended, as in the translationally invariant case, but they have no long-range coherence and may be even localized. Each quasiparticle is viewed as a single entity moving in the self-consistent field of all the other quasiparticles, and the resulting theory has the form of a mean-field theory for the quasiparticles. In \cite{Freedman77}, the phase transition into a {\it Fermi glass} state was discussed. In \cite{Altshuler79a, Altshuler79b, Altshuler80}, and independently in \cite{Fukuyama80}, it has been shown, applying perturbation theory to the lowest order in the interaction strength, that even weak disorder leads to surprisingly singular corrections to the electronic density of states near the Fermi surface, and to transport properties. 
In 1994, Shepelyansky \cite{Shepelyansky94} stimulated further discussion about the role of interactions by considering two interacting particles in a random potential, arguing that there exist an interaction-induced enhancement of the two-particle localization length compared to the noninteracting case. The lines of this work were developed further \cite{Imry95, Frahm95, Weinmann95, vonOppen96, Martin96, Jacquod97}. A Fermi liquid approach for finite densities was suggested by \cite{Imry95}, and later developed by \cite{vonOppen96, Jacquod97} reducing the problem to the study of the delocalization of few quasiparticles above the Fermi sea. 
                 
Several groups have tried to study  effects of interplay between disorder and (repulsive) interactions
in more detail in the regime, when Fermi liquid becomes unstable as the Mott insulator state is approached by increasing the interactions. 
In other words, the crossover between Fermi glass (Anderson localized states populated with the restriction of the Pauli principle) and Mott insulator (state where repulsion dominates over kinetic energy and disorder) has been studied. Density matrix renormalization group studies were performed for spinless fermions with nearest neighbor (n.n.) interactions in a disordered mesoscopic ring \cite{Schmitteckert98}, and for spin 1/2 electrons in a ring described by the half-filled Hubbard-Anderson model \cite{Gambeti-Cesare02}. Spinless fermions with Coulomb repulsion (reduced to n.n. repulsion) in 2D \cite{Benenti99, Waintal00} were also studied. This collection of works shows that as interactions become comparable with disorder, delocalization does takes place. In a 1D ring it leads to the appearance of
 persistent currents.  In 2D, the delocalized state exhibits also an ordered flow of persistent currents, 
which is believed to constitute a novel quantum phase corresponding to the metallic phase observed in experiments for instance with a gas of holes in GaAs heterostructures for the similar range of parameters.

Another intensive subject of investigation concerns metal (Fermi liquid)-insulator transition driven by disorder in 3D. Theoretical description of this phenomenon goes back to the seminal works of Efros and Shklovskii 
\cite{Efros75, Efros76, Efros81} and Mac Millan \cite{MacMillan81}. In this context, recent results of experiments on  disordered alloys \cite{Lee98, Lee00} allowed to determine the critical exponents that govern the conductivity dynamics on both sides of a quantum phase transition in a disordered electronic system. Weakly doped semiconductors provide a good model of a disordered solid,  and their critical behavior at the metal-insulator transition has been intensively studied (cf. \cite{Paalanen83, Bogdanovich99}). Results concerning various forms of electronic glass: from Fermi glass (with negligible effects of Coulomb repulsion) to Coulomb glass 
\cite{Davies82, Davies84, Pollak85}, dominated by the electronic correlations were obtained in the group of M. Dressel \cite{Hering05, Scheffler05}.
  
Very recently, the ground state phase diagram at half filling, for arbitrary interaction and disorder strength, has been calculated \cite{Byczuk05a} by applying the Dynamical Mean-Field Theory (DMFT) with the geometrically averaged local density of states. It was shown that the presence of disorder increases the critical interaction at which the Mott-metal transition occurs and turns the sharp transition into an smooth crossover. Regarding the critical disorder strength for the Anderson-localization transition, it was reported that it increases for weak interactions, and it is suppressed by strong interactions \cite{Byczuk05a}.  In the phase diagram obtained with the same method, but with an arithmetic average of the local density of states the Anderson transition is missing \cite{Byczuk05b}. 

Again, we would like to point out that many of the questions discussed in this section can be addressed using ultracold atomic lattice gases.

\subsection{Disordered Bose-Fermi mixtures}
\label{sub_bosefermi}

As mentioned in the subsection \ref{sub_fermions}, in order to cool fermionic gases to the quantum degeneracy regime it is necessary to use sympathetic cooling of two spin species or boson-fermion (B-F) mixtures \cite{DeMarco99, Truscott01, Schreck01, Hadzibabic02, Roati02}. In the latter case, the final phase of the system is a quantum degenerate B-F mixture. The loading of a B-F mixture in an optical lattice has been also recently reported \cite{Ott04,Ospelkaus06,Gunter06}. 

In the absence of disorder, and in the limit of strong atom-atom interactions such lattice B-F systems can be 
described in terms of composite fermions consisting of a bare fermion, or a fermion paired with 1 boson (bosonic hole), 
or 2 bosons (bosonic holes), etc. \cite{Lewenstein04}. The physics of Fermi-Bose mixtures in this regime has been studied recently in a series of papers  
\cite{Lewenstein04,Fehrmann04a,Fehrmann04b,other,Roth04a, Roth04b}, where it has been shown that the low temperature dynamics of the fermionic composites is
 described by an effective Hamiltonian (see also Sec. \ref{subsec_kachupora} and Appendix \ref{effhamil}), 
describing a spinless interacting Fermi gas. The validity of the effective Hamiltonian for fermionic
composites in 1D was studied using exact diagonalization and the DMRG method \cite{Mehring05}. The effects of inhomogeneous trapping potential on lattice mixtures was for the first time discussed by 
Cramer {\it et al.} \cite{Cramer04}, while the disordered case in the strong coupling limit was studied in 
\cite{Sanpera04, Ahufinger05}. In the presence of disorder, degenerate second order
perturbation theory cannot be applied to derive the effective Hamiltonian, as it was used in 
\cite{Svistunov03,Lewenstein04}, since even for zero hopping rates there exists a well defined single ground state determined 
by the values of the local chemical potentials. Nevertheless, in general, there will be a manifold of many states with similar energies. 
The differences of energy inside a manifold are of the order of the difference of chemical potential in different sites, 
whose random distribution is bounded. Moreover, the lower energy manifold is separated from the exited states by a gap given 
by the boson-boson interaction. Therefore, one can apply a form of quasidegenerate perturbation theory by projecting onto the 
manifold of near-ground states, as described in Appendix \ref{effhamil} \cite{Cohen92, Ahufinger05}. 
As in the homogeneous case  \cite{Lewenstein04}, composite fermions behave as a 
spinless interacting Fermi gas, but in the presence of local disorder they interact 
via random couplings and feel effective random local potential. The effective Hamiltonian that describes their physics can be written as follows:
\begin{equation}
H_{eff}=\sum_{\langle i,j \rangle} \left[ -d_{ij} F^{\dagger}_i F_j + h.c.\right] + \sum_{\langle i,j \rangle} K_{ij} M_i M_j + \sum_i \overline{\mu}_i M_i,
\label{Heffinhom}
\end{equation}
where $F_i$ are the annihilation operators of the
composite fermions (Eqs. (\ref{eq:composites1}) and (\ref{eq:composites2})), and $M_i= F^{\dagger}_i F_i$. The nearest neighbor hopping for
the composites is described by $-d_{ij}$, the nearest neighbor composite-composite interactions
is given by $K_{ij}$, which may be repulsive ($>0$) or attractive ($<0$), and $\overline{\mu}_i$ are the on-site energies. The explicit calculation of the
coefficients $d_{ij}$, $K_{ij}$ and $\overline{\mu}_i$ depends on the concrete type of composites fermions \cite{Sanpera04, Ahufinger05}. For fermion-bosonic hole composite and for fermion-boson composite, it has been shown that the hopping amplitudes $d_{ij}$ are always positive. Depending on the ratio $\alpha=V/U$ between the boson-fermion and the boson-boson interactions, the effective interactions between composites $K_{ij}$ may be either repulsive or attractive for all the values of disorder,  or for certain values of $\alpha$ the qualitative character of the interactions is controlled by the inhomogeneity. These two types of composites have been studied in two limits: (i) the small disorder limit, where the contributions of the disorder to the interactions and hopping are neglected ($K_{ij}=K$ and $d_{ij}=d$), and only the leading contributions in the on-site energies are kept; and (ii) the large disorder limit or spin glass limit, where the tunneling becomes non-resonant and can be neglected in Eq.~(\ref{Heffinhom}), while the couplings $K_{ij}$ fluctuate strongly. This situation corresponds to the (fermionic) Ising spin glass model (see Section \ref{sub_spinglass} for details).

For the case of  disorder applied only for the bosonic component of the mixture, and in the small disorder limit (when disorder does not affect the composite formation), the following quantum phases can be achieved:
\begin{figure}[t]
\centering
\includegraphics[width=0.75\textwidth]{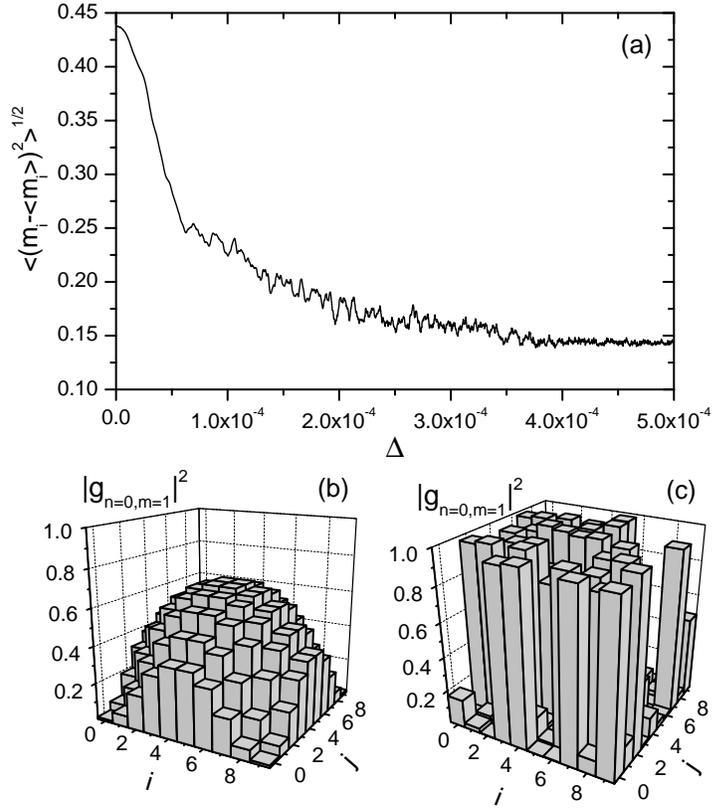}
\caption{Numerically calculated dynamical transition from the Fermi liquid to the Fermi glass of fermionic composites
(fermion $+$ bosonic hole composites) in a 2D lattice with \(N=10 \times 10\) sites. In (a), the decrease of
the variance of the number of fermions per lattice site is shown as a function of the
amplitude of the disorder. In (b), the  probability of having one composite at each lattice site
in the absence of disorder is given, and (c) gives the same as (b) after adiabatically ramping up the disorder
(from \cite{Ahufinger05}).}
\vspace*{0.5cm}
\label{figure3.8}
\end{figure}

\begin{itemize}
\item Fermion $+$ bosonic hole composites 

When $K/d\ll 1$, i.e., when the interactions are negligible, the system is in the Fermi glass phase,
i.e. Anderson localized (and many-body corrected) single particle states are occupied according to the Fermi-Dirac rules. For large repulsive interactions, $K/d\gg 1$ and $K>0$, the ground state will be a Mott insulator and the composite fermions will be pinned for large filling factors. The value of $K/d$ is, however,  bounded from above by $2$, which seems not enough to achieve the Mott insulator state. For intermediate values of $K/d$, with $K>0$, delocalized metallic phases with enhanced persistent currents are possible. For attractive interactions ($K<0$) and  $|K|/d< 1$ one expects competition between pairing of fermions and disorder, i.e., a ``dirty" superfluid phase while for $|K|/d\gg 1$, the fermions will form a domain insulator, that is a state in which fermionic composites will stick together to form a rigid immobile cluster. In Ref. \cite{Ahufinger05} the crossover from the Fermi gas to the Fermi glass phases has been studied numerically by means of the dynamical Gutzwiller ansatz method (Fig. \ref{figure3.8}); similarly, dynamics of a transition from the fermionic domain insulator to a disordered insulating phase was  investigated (Fig. \ref{figure3.9}).

\begin{figure}[t]
\centering
\includegraphics[width=0.75\textwidth]{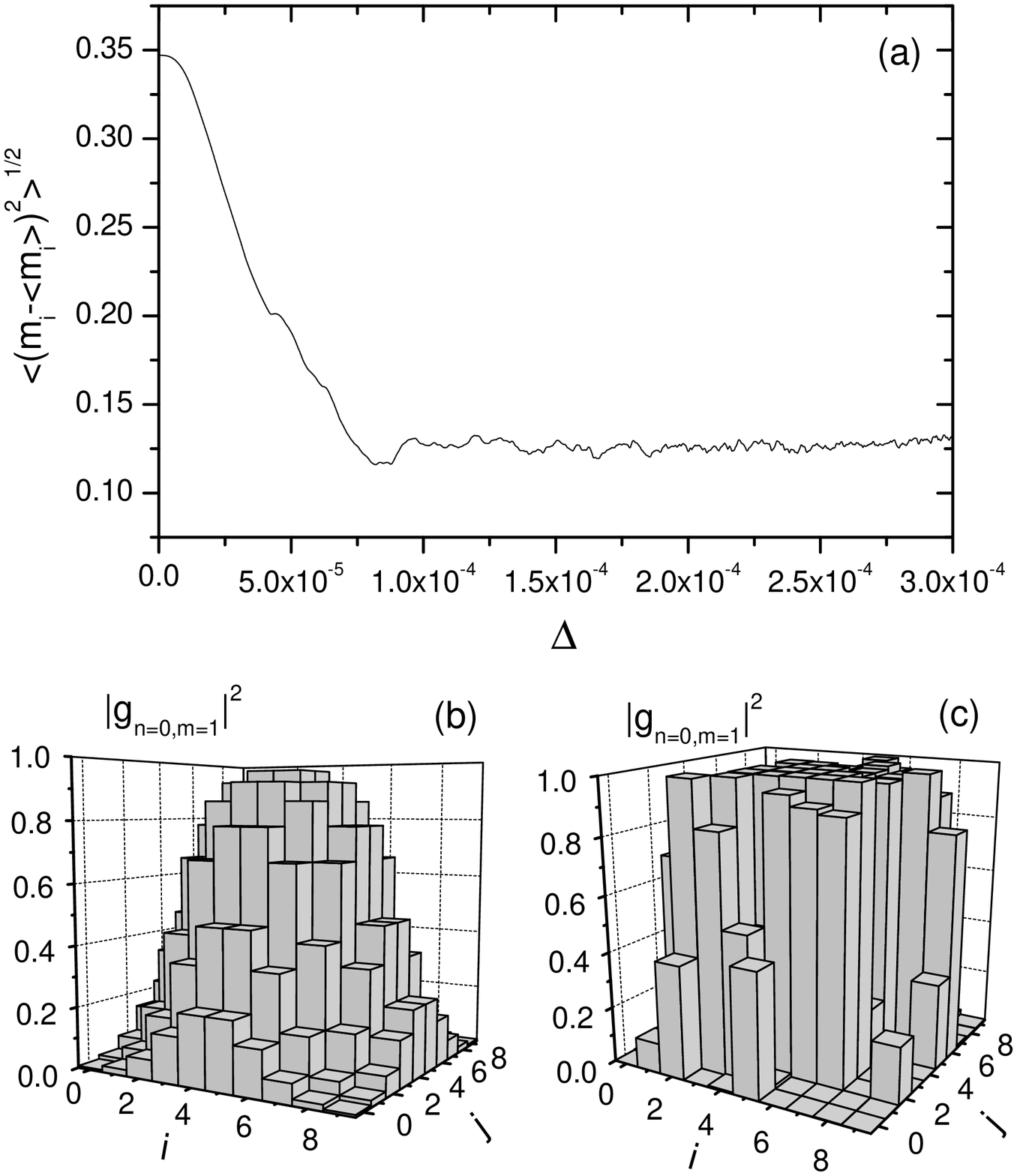}
\caption{Numerically calculated dynamical transition from the Fermi domain to the Fermi glass of fermionic composites
(one fermion and zero boson) in a 2D lattice with \(N=10 \times 10\) sites. In (a), the decrease of
the variance of the number of fermions per lattice site is plotted as a function of the
amplitude of the disorder. In (b), the probability of having one composite at each lattice site
in the absence of disorder is given, and (c) gives the same after adiabatically ramping up the disorder
(from \cite{Ahufinger05}).}
\vspace*{0.5cm}
\label{figure3.9}
\end{figure}

\item Fermion $+$ boson composites

The regimes where $K \ll d$ lead to a non-interacting Fermi glass while the regimes of strong effective repulsive interactions, where $K \gg d$ and $K>0$, would correspond to Mott insulator or checkerboard phase if the filling factor is $1/2$. In this case, no strong attractive interactions regime occurs since $K/d$ reaches a minimum of $\simeq -0.07$. Therefore, the domain insulator phase does not appear, and even the 
``dirty" superfluid phase may be washed out.
\end{itemize}

\begin{figure}[t]
\centering
\includegraphics[width=0.75\textwidth]{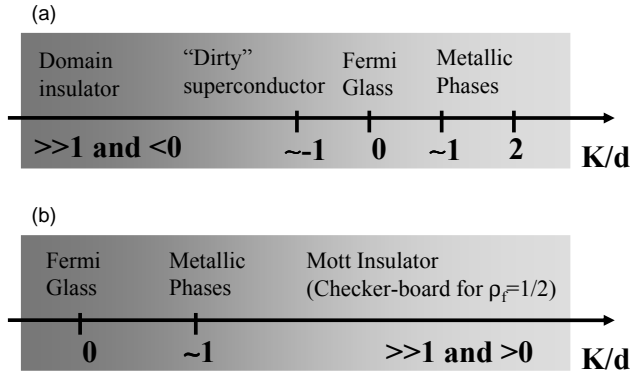}
\caption{Schematic predicted phase diagram of (a) fermion plus bosonic hole composites and  of (b) fermion plus boson composites \cite{Ahufinger05}).}
\vspace*{0.5cm}
\label{figure3.10}
\end{figure}

 The summary of the possible phases, described in the case of fermions plus bosonic hole composites,
and in the case of fermion plus boson composites, is presented in the schematic diagram of Fig.\ref{figure3.10}.
In the case of bare fermion composites, it has been shown that for finite boson-fermion interactions, the fluctuations of the effective composite interactions may be large, and the dynamics of this type of composites resembles {\it quantum bond percolation}. One can assume, in a somehow simplified view, that the interaction parameter $K_{ij}$ takes either very large, or zero values. The lattice decomposes then into two sub-lattices: a ``weak" bond sub-lattice (corresponding to $K_{ij} \ll d_{ij}$) in which fermions flow as
in an almost ideal Fermi liquid, and a ``strong" sub-lattice (corresponding to $K_{ij} \gg d_{ij}$),
where only one fermion per bond is allowed. 
Additionally, for the case of lattices with different types of sites (i.e. sites in which disorder affects the formation and character of composites), it has been predicted that physics of quantum site percolation will become relevant. 

\subsection{Spin glasses}
\label{sub_spinglass}

Spin glass is a phase that appears in spin systems interacting via random couplings that can be positive (anti-ferromagnetic), or negative (ferromagnetic). Such variations of the couplings lead typically to frustration i.e., if there are only two possible spin orientations and the interactions are random, no spin configuration can simultaneously satisfy all the couplings (see section \ref{frustration_atom} for more details). Thus, spin glass behavior requires two essential ingredients: quenched disorder and frustration. 

In a somewhat oversimplified picture, this kind of systems are characterized by two order parameters: (i) the order parameter for magnetic ordering, the magnetization given by $M:=<\overline{\sigma_i}>$ and (ii) the Edwards-Anderson order parameter for spin glass ordering, $Q_{EA}:=\overline{<\sigma_i>^2}$, where $\overline{(\cdot)}$ denotes the average over disorder and $<(\cdot)>$ the thermal average. Experimental studies have identified the phase diagram of the spin glasses consisting on three phases \cite{Sherrington98}: (i) At high temperature and small average spin exchange, $\overline{K}$, the system exhibits a paramagnetic phase with $M=0$ and $Q_{EA}=0$; (ii) for $\overline{K}<0$ and large, the system exhibits a ferromagnetic phase with $M\ne0$ and $Q_{EA}\ne 0$; and (iii) for weak $\overline{K}$ and small temperatures, a spin glass phase appears with $M=0$ and $Q_{EA}\ne 0$. $Q_{EA}\ne 0$ signals that the local magnetization is frozen, but may vary from site to site, so that the disorder prevents long range magnetic order. Spin glass phase is an example of ``order in the presence of disorder" \cite{Sherrington98}.

The physics of spin glasses is up to now not well understood and remains as one of the challenges of statistical physics. In particular, the question of the nature of their order is still open. There exist
 two competing theories: the {\sl M\'ezard-Parisi} (MP) model, and the {\sl droplet} model. The MP picture is a mean-field theory based on the replica method \cite{Mezard87} that predicts that the spin glass phase consists of a large number of low-energy states with very similar energies. The applicability of the MP picture for short range spin glasses (like the Edwards-Anderson model \cite{Edwards75}) is very questionable. The rival theory, the {\sl droplet} model \cite{Fisher86,Bray87,Fisher88a,Fisher88b,Newman03} is a phenomenological theory based on scaling arguments and numerical results. The {\sl droplet} model predicts that there are two ground states related by spin-flip symmetry, and that excitations over the ground state are regions with fractal boundary, the droplets, in which the spins are inversed with respect to the ground state.   

One of the most remarkable results of Refs. \cite{Sanpera04,Ahufinger05} is the possibility of the realization of a fermionic Ising spin glass model by using a disordered Bose-Fermi mixture. The spin glass limit is obtained from the composite fermionic model (Eq.(\ref{Heffinhom})) when the interactions between fermions and bosons are of the same order, but slightly smaller than the interactions between bosons, and in the limit of large disorder. In this situation, the hopping vanishes due to strong site-to-site energy fluctuations and the nearest neighbor interactions, $K_{ij}$, fluctuate around mean zero with random positive and negative values. Replacing the composite number operators with a classical Ising spin variable $\sigma_i:=2M_i-1=\pm 1$, one ends up with the Hamiltonian:

\begin{equation}
H_{E-A}=\frac{1}{4}\sum_{\left\langle ij\right\rangle }
K_{ij}\sigma_i \sigma_j + \frac{1}{2}\sum_{i}\overline\mu_i \sigma_i .
\label{HamiltonianSG}
\end{equation}
It describes an (fermionic) Ising spin glass \cite{Oppermann99,Oppermann03}, which differs from the standard Edwards-Anderson model 
\cite{Binder86,Mezard87} in that the 
model (Eq.(\ref{HamiltonianSG})) includes an additional random magnetic field $\overline{\mu}_i$, and, moreover, it has to satisfy the constraint of fixed magnetization value, $m=2N_{f}/N-1$, as the number of fermions, $N_f$, in the underlying BFH-model is conserved. $N$ is the number of sites. It shares, however,  the basic characteristics with the Edwards-Anderson model as being a spin Hamiltonian with random spin exchange terms $K_{ij}$. In particular,
this provides bond frustration, which in this model is essential for the appearance of a spin glassy phase. Due to the mentioned differences, it is necessary to reformulate  slightly the M\'ezard-Parisi mean field description of the system by adapting the Sherrington-Kirkpatrick-Parisi calculations \cite{Sherrington75,Mezard87} to this specific case 
\cite{Ahufinger05}.

The experimental study of this limit in ultra-cold Bose-Fermi mixtures could present a way to address various open questions of spin
 glass physics concerning the nature and the ordering of its ground- and possibly metastable states
(the M\'ezard-Parisi picture \cite{Mezard87} versus the droplet picture \cite{Fisher86, Bray87, Fisher88a, Fisher88b, Newman03}), 
broken symmetry and dynamics in classical (in absence of hopping) and quantum (with small, but nevertheless present hopping) 
spin glasses \cite{Sachdev99, Georges01}. Particularly interesting are prospects for studying questions that have been addressed 
only recently: the existence of the celebrated  \emph{de Almeida-Thouless line} (i.e. the line separating the replica symmetry broken 
spin glass phase from the replica symmetric phase in the magnetic field in the Parisi picture) in short range spin glasses 
\cite{Young04,Katzgraber05a}, disorder chaos in spin glasses (i.e. ground state sensitivity to small changes of 
disorder, \cite{Krzakala05}), 
spin glass transition in 2D and its dependence on the bond distribution \cite{Katzgraber05b}, 
universality classes \cite{Katzgraber06} 
etc. (for a recent review see \cite{Kawashima04}).

Usage of  atoms offers unprecedented possibilities for detection of the spin glass properties. 
Recently for instance, it has been proposed \cite{Buchler04} how to create replicas of disordered systems, i.e.
 systems with identical disorder landscape and to the measure of correlations between the replicas. 
These correlations can provide information about the ground state phases of the system. The replicas can be 
created in a three dimensional lattice with incommensurate frequencies by quenching the hopping between 
different planes obtaining then a set of two-dimensional incommensurate lattices, with the same realization of 
disorder in each plane. 
Another possibility is to use localized impurity atoms. In this case, the procedure would 
be first to prepare a disordered 
distribution of localized impurity atoms in such a way that only two or no atoms per lattice site are allowed. 
This can be achieved, for instance, using repulsive 
bound pairs, observed recently  by the  Innsbruck group \cite{Winkler06}. Then, 
periodicity of the lattice in one direction should be  adiabatically doubled by stretching the lattice in that direction. 
One would in this way obtain an array of replicated pairs of 2D random landscapes.  That would be, obviously, 
 an ideal tool to study the questions concerning existence of finite $T$ spin glass transition in Ising model
with bimodal bond distribution \cite{Kawashima04}.   

Refs. \cite{Sanpera04, Ahufinger05} demonstrate that ultracold disordered Bose-Fermi mixtures in optical 
lattices may serve as a paradigm fermionic system to study a variety of disordered phases and phenomena: 
from Fermi glass to quantum spin glass and quantum percolation. 
No doubts, one can  mimic condensed matter and even go beyond!

\section{Frustrated  models in cold atom systems}
\label{frustration_atom}

\subsection{Introduction}
\label{intr}

Frustration appears when all the constraints imposed by the Hamiltonian cannot be simultaneously fullfilled and it is an inherent property of some strongly correlated systems. It
introduces various interesting features, such as 
a rich phase diagram. 
Particularly fascinating effects appear when a system is both disordered and frustrated 
(see, e.g., \cite{Mezard87,Sachdev99}). In this section, however, we will
study several frustrated models in  \emph{regular} lattices and discuss their possible realizations. 
Antiferromagnetic (AF) models in a regular lattice are frustrated 
if the geometry of the lattice is sufficiently complicated 
(such as triangular or kagom{\' e} lattice).

\begin{figure}[t]
\centering
\includegraphics[width=7cm]{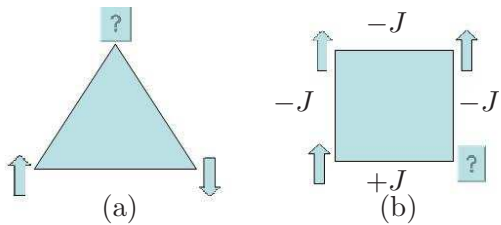}
\put(-160, 5) {(a)}
\put(-55, 5) {(b)}
\put(-95, 45) {\(- J\)}
\put(-25, 45) {\(- J\)}
\put(-60, 75) {\(- J\)}
\put(-60, 15) {\(+ J\)}
\caption{(a) Antiferromagnetic Ising model (\(J> 0\)) on a triangular lattice:
 Orientation of a spin cannot be determined by satifying all the couplings simultaneously. 
(b) Square lattice with all ferromagnetic Ising
interactions, except one, in which the interaction is antiferromagnetic. The interactions
are indicated on the sides of the lattice.
The model is again frustrated.} 
 \label{fig_frust_triangle}
\end{figure}

A physical system described by the Ising Hamiltonian 
\begin{equation}
\label{eq_Ising}
H_I = J \sum_{\langle ij \rangle} \sigma_i^z \sigma_j^z
\end{equation}
is AF, if  \(J > 0\), where \( \langle ij \rangle\)  
denotes the summation over nearest neighbors.  
This model, when considered on a two-dimensional triangular lattice, is an example 
of a frustrated system \cite{Moesner01}.  It is clear from Fig. \ref{fig_frust_triangle}(a)
that it is not possible to minimize the energy on such a triangular lattice, satisfying 
all the bonds simultaneously. In this case, we obtain a degeneracy in the ground state
 due to the frustration. It is not due to the 
invariance under global spin flips of the Hamiltonian, 
which is the case for the ferromagnetic Ising model ($J <0$) on the same lattice.

Frustration can  be met even in a  lattice, which has a relatively simple structure 
(e.g., a 2D square lattice), if some interactions are ferromagnetic while others are AF.  
In the ferromagnetic Ising model, e.g., if one replaces 
an odd number of ferromagnetic bonds by AF ones, 
one obtains a frustrated model from a nonfrustrated one.  
The square lattice in Fig. \ref{fig_frust_triangle}(b) is an example of such a model.
In general, we say that the model is frustrated, when 
the orientations of its  spins cannot be obtained by satisfying all the couplings
simultaneously to the aim of minimizing the energy in the model \cite{Mezard87}.
Further discussion on this subject can be found in \cite{Misguich04}. Detailed studies of quantum antiferromagnets can be found 
in the books of Auerbach \cite{Auerbach94} and Sachdev \cite{Sachdev99}. 
Frustrated models have been reviewed  very recently
by Misguich and Lhuillier \cite{Misguich04} (see also \cite{Lhuillier05,Alet05a}). 
They have been also discussed in the context of spin glasses (see \cite{Binder86}).
The concept of frustration has been also
studied from the perspective of high temperature superconductors 
(e.g., see the reviews \cite{ Rasolt92} and  \cite{Sigrist95}).

The plan of this section is as follows.
We will present a short review on quantum antiferromagnets in the succeeding subsection. 
The next two subsections (\ref{frust_H_antiferro} and \ref{frust_Fermi_spinliquid}) will be
devoted to the kagom{\' e} lattices: we will review some recent results 
and discuss possibilities of experimental verifications of the theoretical predictions. 
In the last subsection,
we will discuss the possible realization of  quantum magnets with cold
atoms, ions, and molecules.

\subsection{Quantum antiferromagnets}
\label{frust_qantiferromagnet}

In this section we discuss some basic properties of the quantum
antiferromagnets. The interest in this subject is motivated mainly by 
 two things:
 (a) frustrated quantum antiferromagnets are related to most challenging open questions of condensed matter physics;
 (b) there exist already several theoretical proposals for experimental realization of quantum 
magnets in cold atom/ion systems. We expect that these experiments will 
verify existing theoretical predictions, and point out new directions 
in theoretical investigations of the quantum magnetism.

According to reviews \cite{Misguich04, Lhuillier05, Alet05a}, quantum antiferromagnets exhibit at low temperatures 
one of four generic behaviors (see below). 
We add to the list a fifth possibility \cite{Damski05a, Damski05b}.
\begin{enumerate}
\item N{\'e}el order. This is a standard up-down-up-down type of ordering that occurs often in regular lattices such as square in 2D
or cubic in 3D. Generalized (planar) N{\'e}el order might, however, occur also in  frustrated antiferromagnets, for instance
in the triangular lattice \cite{Bernu92}. 
N{\'e}el order breaks rotational and translational symmetry, exhibits long range order,  and leads to gapless 
spin wave excitations. 

\item Valence bond solids (VBS). This order breaks the lattice translational symmetry and 
consists of an ordered covering of the lattice
by singlets, called also dimers. There is a long range order in dimer correlations, but the system is gapped and correlation functions 
decay exponentially. A famous example of a VBS is the Affleck-Kennedy-Lieb-Tasaki (AKLT) state, although it does not break the 
lattice translational symmetry \cite{Auerbach94}. Excitations in 2D VBS are gapped (confined spinons) \cite{Alet05a}.

\item Spin liquid of the I type. This order has no apparent symmetry breaking, and
decaying correlations. Particularly interesting are  topological spin liquids \cite{Alet05a, Wen04}, for which
degeneracy of the ground state depends on the topology of the underlying lattice.

\item Spin liquids of the II type. This order has no symmetry breaking, is gapless, and
has large density of states of low energy excitations. 
Heisenberg antiferromagnet in trimerized kagom{\'e} 
lattice, discussed below is  an example of such behavior, representing in a resonating valence bond state (RVB). The concept
of resonating valence bond liquid was first introduced by Anderson \cite{Anderson87}, 
in the context of superconductivity. 
When a valence bond solid starts melting due to quantum fluctuations introduced in the 
system, the situation gives rise to a new phase called resonating valence bond liquid. Numerical
Monte Carlo simulations suggest that the liquid does not show
any long-range correlations \cite{Liang88}, and is a products of singlets \cite{Alet05a}.

\item Quantum spin-liquid crystal. This order combines N{\'e}el order with spin liquid type II behavior. 
Numerical studies reveal \cite{Damski05a, Damski05b, Honecker_cond-mat0609312}
that it has a large number of low energy excitations as in the spin-liquid type II, and presumably no gap.

\end{enumerate}

\subsubsection{The Heisenberg model}
\label{subsec-tapa-tepi}

Let us first focus on spin systems that do not show  frustration. 
Introduction of such models will help us to explain many results in 
the frustrated models.  Let us consider the Heisenberg model 
\begin{equation}
\label{eq_Heisenberg}
H_{H} = J \sum_{ \langle ij \rangle} \vec{\sigma_i} \cdot \vec{\sigma_j},
\end{equation}  
 where \(\vec{\sigma}\) is the vector of spin-1/2 Pauli operators.  

\paragraph{The ferromagnetic case} For the sake of completeness, we first 
look  at the 2D ferromagnetic model ($J <0$) in a square lattice. 
The exact ground state is 
given by the state \cite{Ashcroft76} 
\begin{equation}
\label{eq_gr_ferro}
|\psi^{G}_{F}\rangle = \prod_i |0\rangle_i,
\end{equation} 
where \(|0\rangle_i\) and  \(|1\rangle_i\) represent respectively  the up and down 
eigenstates of $\sigma^z$ at the $i$-th site.

\paragraph{The antiferromagnetic case}
In the case of 2D antiferromagnets ($J>0$) on the square lattice, the exact ground
state is hard to find though the system is not frustrated.
Nonetheless, the classical ground states, obtained after 
replacement of the $\vec{\sigma}$ operators  by vectors, can be explicitly 
written down. To this aim the lattice is divided into two sublattices A and B in such 
a way that all nearest neighbors of sublattice A belong to the sublattice B, and vice-versa. 
Such a splitted lattice is called a bipartite lattice. 
The classical ground state can be obtained by taking some orientation 
(e.g., in the z-direction)
of all spins in one sublattice and opposite orientation 
of all spins in the other sublattice. 
In this way, one can find the classical ground state of a square lattice 
\cite{Sachdev99, Manousakis91}:
\begin{equation}
\label{eq_cl_gr_antiferro}
|\psi^{G}_{cl, AF}\rangle = \prod_{i\in A, j\in B} |0\rangle_i |1\rangle_j.
\end{equation} 
This classical ground state is known as the
N{\'e}el state.  
Note here that any pair of spins in this classical
ground state is either parallel or antiparallel, which means that the ordering 
is collinear. The ordering of classical ground states in frustrated systems, however, 
is noncollinear, even in bipartite lattices \cite{Sachdev99}.

In  the case of one-dimensional AF model,  
 the exact ground state of the spin-$1/2$
antiferromagnet  in the thermodynamic limit is  known (Bethe solution) \cite{Bethe31, Yang66, Hulthen38, Auerbach94}.
For comparing the properties of integer
 and half-integer spin systems \cite{Haldane83} (cf. \cite{Bonner87}), and the famous Haldane conjecture, the systems may be realized with
atoms either by  using Fermi-Fermi mixtures, or by employing the approach of Ref.  \cite{Garcia-Ripoll04}.

For  quantum antiferromagnets, made up of spin-1/2 particles, and placed 
in  the infinite square lattice, the ground state energy can be approximated 
by that of the N{\'e}el state (Eq.(\ref{eq_cl_gr_antiferro})),
despite the fact that it is 
not an eigenstate of the quantum Hamiltonian 
(Eq.(\ref{eq_Heisenberg})).

\subsubsection{The $J_1-J_2$ model}
\label{subsec-tapa-tepi1}

Let us now move on to the $J_1-J_2$ model that can be frustrated for some
values of interaction couplings, and is a paradigmatic example of VBS. The Hamiltonian of this model can be written as
\begin{equation}
\label{eq_Ham_J1_J2}
H_{J_1J_2} = 2 J_1 \sum_{\langle ij \rangle} \vec{\sigma_i}\cdot \vec{\sigma_j} 
+ 2 J_2  \sum_{ \langle \langle ij \rangle \rangle } \vec{\sigma_i}\cdot \vec{\sigma_j}, 
\end{equation}
where \(\langle \langle ij \rangle \rangle\) denotes the next nearest neighbors. 
This is a frustrated model 
(even in the case of a linear chain), when both 
\(J_{1,2} \neq 0\) and at least one of them is positive. 
As will be discussed in Sec. \ref{frust_H_antiferro}, this Hamiltonian can also be obtained 
using Fermi-Fermi mixtures in an optical lattice.
Its importance comes from its usefulness  in 
explanation of magnetic properties of 
Li$_2$VOSiO$_4$ and Li$_2$VOGeO$_4$ compounds \cite{Melzi00}.
Below, we will discuss the different phases of this model.

\paragraph{The one-dimensional system}
The Hamiltonian (Eq.(\ref{eq_Ham_J1_J2})) for the case of the 1D lattice,  
with an even number of sites, periodic boundary conditions, and 
$J_1 = 2 J_2 > 0$ is known as
the Majumdar-Ghosh model \cite{Majumdar69}.         
The ground state space of the model is spanned by 
the two dimers \footnote{The Majumdar-Ghosh model is one of the few frustrated spin models for which the exact ground states are known. 
See also, e.g. \cite{IndraniBose91a, IndraniBose91b, Surendran02, Kumar02}.}          
 \begin{equation}
 \label{eq_MG_ground}
 |\psi^{G}_{MG}\rangle_{\pm} = \prod_{i =1}^{N/2} \left(|0\rangle_{2i} |1\rangle_{2i \pm 1} - 
|1\rangle_{2i} |0\rangle_{2i \pm 1}\right)/\sqrt{2},
  \end{equation}      
where \(N\) is the number of sites in the lattice. We also call these singlet states as 
``valence bond states''. Note that unlike the AKLT state, this model breaks translational symmetry. A ``valence bond
solid'' is formed from these valence bond states after they order between themselves.  

\paragraph{The two dimensional model}
The 2D case is far more complicated. The quantum phase diagram  of this model is not 
clear even for the square lattice, 
while the classical phase diagram is quite well-understood \cite{Misguich04}.  
In the latter case, it is known that when \(J_2/J_1\) is very small,
 the system is N{\'e}el ordered, and in the 
opposite extreme (i.e., $J_2\gg J_1$), it has 
collinear ordering (different than N{\'e}el) characterized by
\begin{equation}
\label{eq_collinear}
|\Psi_{cl, J_2\gg J_1}^{G} \rangle = \prod_{i} \prod_{\mbox{\footnotesize{odd} }j} |1\rangle_{ij} \prod_{\mbox{\footnotesize{even} } j} |0\rangle_{ij}, 
\end{equation}
where \(i\) and \(j\) are respectively the indices of rows and columns of the 2D lattice.
 However, when 
\(0.4 <J_2/J_1 < 0.6\) (the strongly frustrated
regime),  no such orderings exist. 

These classical predictions are in qualitative agreement with 
semi-classical and fully quantum calculations based on:
series expansion  \cite{Dagotto89,Singh90,Ceccatto92,Richter93,Oitmaa96} 
(see  \cite{Singh88, Gelfand90a} for an introduction to this method), 
spin wave theory \cite{Shender94, Henley89}
(first introduced by Anderson \cite{Anderson51}, and then extended to higher order by
Kubo \cite{Kubo52} and Oguchi \cite{Oguchi60}),
exact diagonalizations \cite{Poilblanc91, Schulz92, Schulz95}, 
and Quantum Monte Carlo \cite{Sandvik97}.

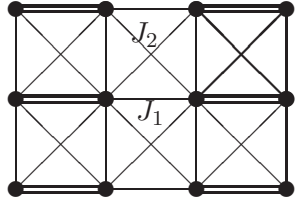
\begin{figure}
\begin{center}
\unitlength=0.4mm
\begin{picture}(250,60)(0,0)


\thicklines
\put(90, 4){\line(1,0){30}}
\put(90, 6){\line(1,0){30}}

\put(90,34){\line(1, 0){30}}
\put(90,36){\line(1, 0){30}}
\thinlines

\put(90,5){\line(0, 1 ){30}}
\put(120, 35){\line(0, -1){30}}
\put(90,5){\line(1,1){30}}
\put(120,5){\line(-1,1){30}}
\put(90, 5) {\circle*{5}}
\put(120, 5) {\circle*{5}}
\put(120, 35) {\circle*{5}}
\put(90, 35) {\circle*{5}}


\thinlines
\put(120, 5){\line(1,0){30}}
\put(120,35){\line(1, 0){30}}
\put(150, 35){\line(0, -1){30}}
\put(150, 35) {\circle*{5}}
\put(150, 5) {\circle*{5}}
\put(120,5){\line(1,1){30}}
\put(150,5){\line(-1,1){30}}


\thicklines
\put(150, 6){\line(1,0){30}}
\put(150,36){\line(1, 0){30}}
\put(150, 4){\line(1,0){30}}
\put(150,34){\line(1, 0){30}}

\thinlines
\put(180, 35){\line(0, -1){30}}
\put(180, 35) {\circle*{5}}
\put(180, 5) {\circle*{5}}
\put(150,5){\line(1,1){30}}
\put(180,5){\line(-1,1){30}}

\thicklines
\put(90,64){\line(1, 0){30}}
\put(90,66){\line(1, 0){30}}

\thinlines
\put(120, 65){\line(0, -1){30}}
\put(90, 35){\line(0,1){30}}
\put(120, 65) {\circle*{5}}
\put(120, 35) {\circle*{5}}
\put(90, 65) {\circle*{5}}
\put(90,35){\line(1,1){30}}
\put(120,35){\line(-1,1){30}}
\put(130,28){$J_1$}
\put(128,54){$J_2$}


\thinlines
\put(120, 35){\line(0,1){30}}
\put(120,65){\line(1, 0){30}}
\put(150, 65){\line(0, -1){30}}
\put(150, 65) {\circle*{5}}
\put(150, 35) {\circle*{5}}
\put(120, 65) {\circle*{5}}
\put(120,35){\line(1,1){30}}
\put(150,35){\line(-1,1){30}}

\thinlines
\put(150, 35){\line(0,1){30}}
\put(180, 65){\line(0, -1){30}}

\thicklines
\put(150,64){\line(1, 0){30}}
\put(150,66){\line(1, 0){30}}
\put(180, 65) {\circle*{5}}
\put(180, 35) {\circle*{5}}
\put(150, 65) {\circle*{5}}
\put(150,35){\line(1,1){30}}
\put(180,35){\line(-1,1){30}}
\end{picture}
\end{center}
\caption{The $J_1 - J_2$ model with dimer configurations in a square lattice. 
The double lines indicate  singlets between the two corresponding spins. 
The interaction strength corresponding to the diagonals of the squares are \(J_2\), 
while that of the  sides  are \(J_1\).
} 
 \label{fig_frus_dimer}
\end{figure}

In the semi-classical limit,
spin wave theory predicts a first-order phase transition from  the N{\'e}el  ordered phase to the collinear 
ordered phase
 (in the highly frustrated region at \(J_2/J_1 = 0.5\)). However, 
when quantum fluctuations are introduced in this region, a new phase will 
appear to separate these two 
phases.
From the exact diagonalizations \cite{Dagotto89,Figueirido90,Schulz96} one obtains that 
there are two phase transitions. One is at \(J_2/J_1\approx 0.38\) 
and another one is at  \(J_2/J_1\approx 0.6\). 
The first one is a second order phase transitions from the N{\'e}el state to a spin 
liquid valence bond state.
The second transition is a first order phase
transition from the spin liquid state to a collinear state. In the regime \(0.38 < J_2/J_1 < 0.6\),
 many calculations 
\cite{Read89,Read91, Murthy90, Kotov99, Gelfand89, Gelfand90b, Singh99} suggest that 
the ground state may have the VBS dimer configuration 
 with long range ordering, as shown in Fig. \ref{fig_frus_dimer}.  

Despite all these efforts, the phase diagram of this model is still not completely 
understood. In particular, many candidates for the ground states are proposed for this highly frustrated
regime. For example, the variational approach suggests \cite{Capriotti01}
that  a ground state is a spin liquid resonating valence bond state   
for a spin-1/2 system.
Sushkov \emph{et al.} \cite{Sushkov01} (see also \cite{Poilblanc91}) 
found two new
second-order phase transitions by using series expansion method:
one of them around   \(J_2/J_1 =   0.34 \pm 0.04\),
while the other  in the strongly frustrated regime (at \(J_2/J_1 = 0.50 \pm 0.02\)).
      
\subsection{Heisenberg antiferromagnets and atomic Fermi-Fermi mixtures in kagom{\'e} lattices}
\label{frust_H_antiferro}

In the preceding subsection, we have discussed quantum magnets 
either in a linear chain or in a square lattice. In this section, 
we move to a more complicated lattice: the kagom{\' e} lattice. 
The Heisenberg antiferromagnet in this lattice is frustrated.  













\begin{figure}[t]
\centering
\includegraphics[width=4cm,height=4cm]{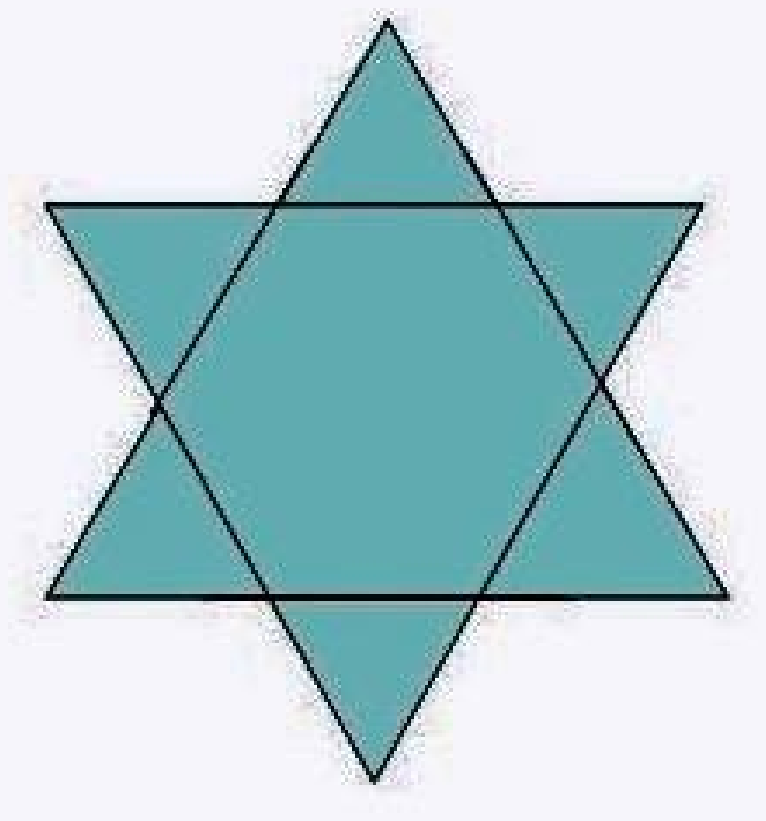}
\includegraphics[width=4.5cm,height=4cm]{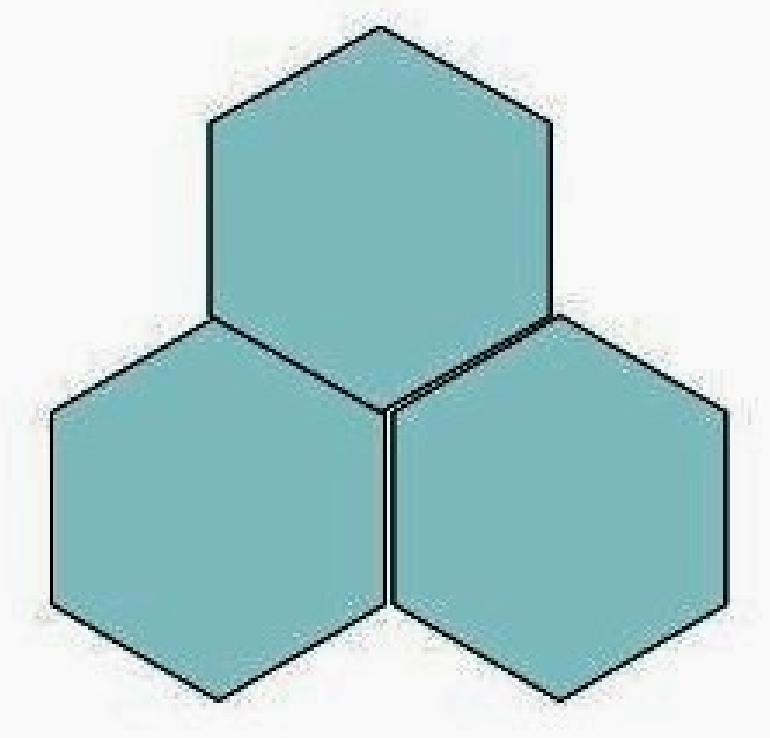}
\thinlines
\put(-76,20){x}
\put(-50,18){y}
\put(-60,34){z}
\put(-60,57){y}
\put(-76,57){x}
\put(-63.5,24){\line(0,1 ){25}}
\caption{(a) (Left) The kagom{\' e} lattice. 
Two parent triangles superimposed on each other. Each of the parent triangles (one of which is inverted)
contains three triangles at its three corners.
The triangles in one of the parent triangles (say, in the non-inverted one) have interaction couplings \(J\), while those 
in the other have \(J'\), between all their vertices. 
The kagom{\'e} lattice consists of a 2D array of such superimposed parent triangles.
(b) (Right) The honeycomb lattice for a spin-1/2 system, whose Hamiltonian is given in Eq. (\ref{eq_Honeycomb}). The three types of links are depicted in the 
figure.
}
\label{fig_kagome}
\end{figure}
A schematic diagram of the kagom{\' e} lattice is depicted in Fig. \ref{fig_kagome}.
During the last 15 years, extensive work have been done  on this model with Ising-type
 nearest neighbor interactions \cite{Moesner00, Moesner01}, or 
with Heisenberg-type nearest and next-nearest neighbor interactions 
\cite{Waldtmann98, Zeng90, Singh92b, Leung93, Zeng95, Singzingre94, Nakamura95, 
Lecheminant97, Mila98, Mambrini01, Budnik04}. All that 
 give us a lot of  information about the ground and excited states of these models. 
 There are  still, however, many questions to be answered.

\subsubsection{Heisenberg kagom{\'e} antiferromagnets}
\label{subsubsec-radha}

The Hamiltonian, in this case, is given by 
\begin{equation}
\label{eq_kagome}
H_{kag} = J \sum_{\langle i j\rangle}\vec{\sigma_i}\cdot\vec{\sigma_j} +  
J' \sum_{\langle i j\rangle}  \vec{\sigma_i}\cdot\vec{\sigma_j},
\end{equation}
where \(J\) and \(J'\) are the  couplings (see caption of Fig.
\ref{fig_kagome}). Both $J$ and $J'$ are positive. 

Numerical simulations of the spin-1/2 system on the kagom{\' e} lattice 
suggest that the energy gap between the ground state and the lowest triplet
state, if any, is very small (of order \(J/20\)). 
This gap is filled with low-lying singlets, whose number 
scales with the number of spins, $N$,  as $1.15^N$ \cite{Waldtmann98}.   
All these results suggest that
this model may be described by the resonating valence bond states.

In the trimerized limit, i.e., when the ratio \(J{'}/J\) is very small, 
Mila and Mambrini \cite{Mila98, Mambrini01} have found the number, the form, and the
spectrum of singlets by using mean field approximation. It corresponds to  short-ranged resonating
valence bond states. 
 This method also predicts a gap of \(2 J{'}/3\) between singlet and triplets, and the gap remains
in the thermodynamic limit. 
In the fully trimerized limit (i.e., \(J{'}/J = 0\)), the ground state in the subspace of
short-ranged resonating valence bond states,  is one singlet in 
each of the triangles.
Although there are many theoretical predictions about this model, there are no clear 
experimental confirmations.
We will now consider a possible way of verifying these theoretical results by using 
ultracold atoms \cite{Santos04}.

\subsubsection{Realization of a kagom{\' e} lattice by Fermi-Fermi mixtures}
\label{subsubsec-radha1}
 
Consider a Fermi-Fermi mixture at  1/2 filling for each species
\cite{Santos04,Damski05b}.
 The Hamiltonian in this case is the spin-1/2 Hubbard model, given by 
 \begin{equation}
 \label{eq_Fermi-Fermi}
H_{FF}=-\sum_{\left\langle ij\right\rangle} t_{ij}
(f_{i}^{\dagger }f_{j}+ \tilde f_{i}^{\dagger }\tilde f_{j} + {\rm h.c.}) +
\sum_i V  n_{i}\tilde n_{i}, 
  \end{equation}
where $n_i=f_i^\dag f_i$ ($\tilde n_i=\tilde f_i^\dag \tilde f_i$), and 
the operators $f_i$ and $f_i^\dag$ ($\tilde f_i$ and $\tilde f_i^\dag$) are 
the annihilation and creation operators for the two species. 
Here  $t_{ij}$  takes the value $t_0$ for intratrimer,
  and \(t{'}_0\) for intertrimer hopping.
 In the strong coupling limit, $t_0,t'_0 \ll V$ 
($t-J$ model) \cite{Auerbach94}, $H_{FF}$ reduces to the Heisenberg 
 antiferromagnet Hamiltonian \(H_{kag}\) (Eq.(\ref{eq_kagome})), 
where $J=4t_0^2/V$, and $J'=4{t'_0}^2/V$,	and 
$\vec \sigma=(\sigma^x,\sigma^y,\sigma^z)$, with 
$$
n-\tilde{n} = 2\sigma^z, \ \ f^\dag \tilde f = \sigma^x+i\sigma^y, \ \ \tilde f^\dag f = \sigma^x-i\sigma^y.  
$$
The total spin in the trimer takes the minimal value $1/2$, 
and there are four degenerate states having $\sigma^z=\pm 1/2$ and  
left or right  chirality.  
The spectrum of the system in the singlet 
sector consists of a narrow band of low energy states 
of a width of the order  $J'$, 
separated from the higher singlet (triplet) 
bands by a gap of the order $3J/4$ ($2J'/3$). 

\subsection{Interacting Fermi gas in a kagom{\'e} lattice: Quantum spin-liquid crystals}
\label{frust_Fermi_spinliquid}

In this section we show that an interacting Fermi gas placed in a
trimerized kagom\'e lattice behaves as a very special quantum magnet,
a {\it quantum spin-liquid crystal},
possessing both antiferromagnetic order and an exceptionally large number of
low-energy excitations \cite{Damski05a,Damski05b}. 

\begin{figure}[t]
\centerline{\includegraphics[width=8cm]{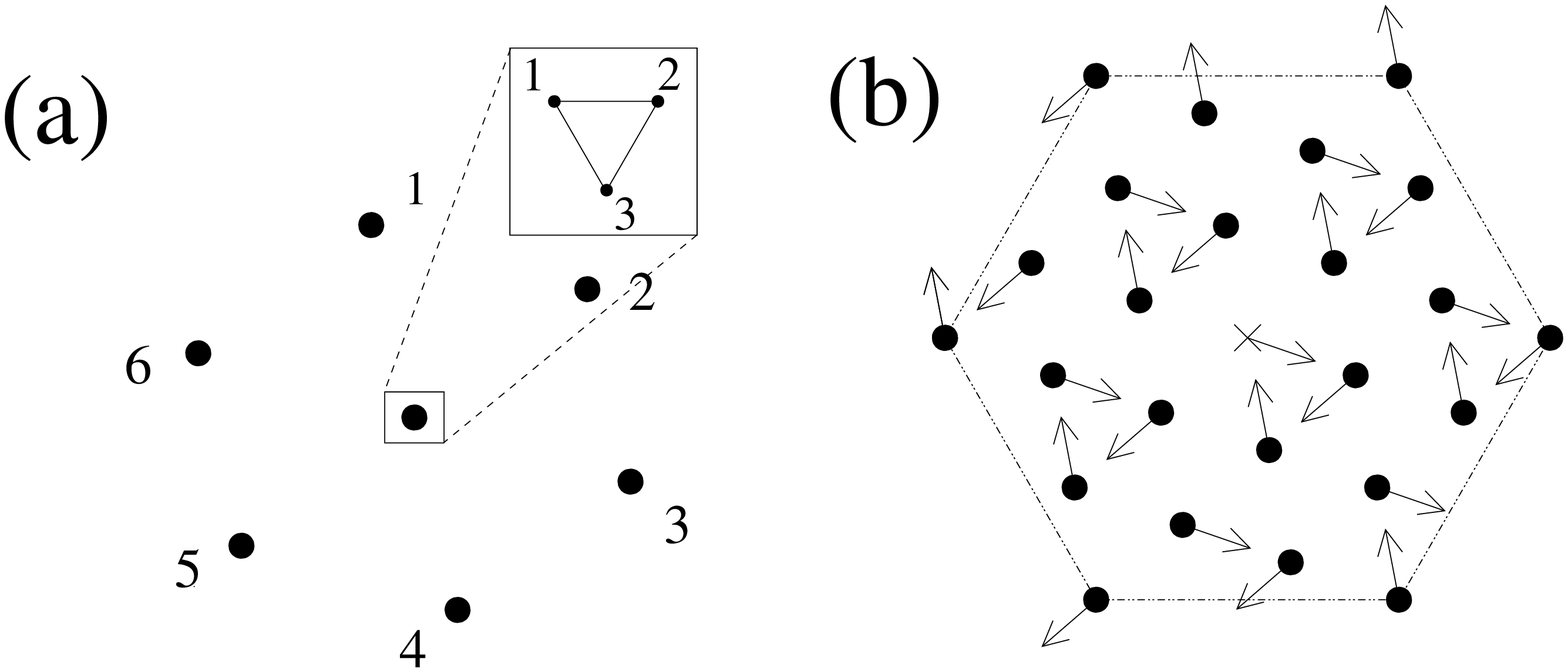}}
\caption{(a) Enumeration of intertrimer (intratrimer) nearest
neighbours; (b) Classical $120^\circ$ N\'eel state  with {\it left} chirality.}
\label{trimer}
\end{figure}

The trimerized kagom\'e lattice can be created by 
a proper superposition of standing laser beams \cite{Santos04,Damski05b}.
Such a  lattice consists of trimers (i.e., sets of 
three closely packed potential minima) arranged in a perfect  triangular 
pattern (see Fig. \ref{trimer}(a)). 
The ultracold fermions that are
loaded into the lattice   stay localized around 
potential minima, and the Fermi-Hubbard model can be shown to capture all
essential properties of this system. The density of fermions is assumed 
to be such that there are  two fermions per each trimer. It is also assumed that  
there is a long-range interaction between fermions at nearest neighbouring sites 
in nearest neighbour trimers. This long-range interaction can be achieved, e.g., 
in a gas of polarized dipolar fermions.

\subsubsection{The quantum magnet Hamiltonian}
\label{subsubsec-kumropotas}

The spinless interacting Fermi gas in the trimerised kagom\'e lattice
is described by the following extended Fermi-Hubbard Hamiltonian
$$
 H_{\mathrm{FH}}=-\sum_{\left\langle a,b\right\rangle }(t_{ab}  f_a^{\dagger } f_b + h.c.)
+\sum_{\left\langle a,b\right\rangle}
U_{ab}  n_{a} n_{b},
\label{FH}
$$
where $\left\langle a,b\right\rangle$ denotes nearest neighbors,
$a=\{\alpha,i\}$ with $\alpha$ referring to intra-trimer
indices and $i$ numbering the trimers.
The $t_{ab}$ and $U_{ab}$ take the values $t$ and $U$ for intratrimer,
and $t'$ and $U'$ for intertrimer couplings, $n_a=f_a^{\dagger } f_a$   
and $f_a$ is the fermionic annihilation operator.
The sites in each trimer are enumerated as in Fig. \ref{trimer}(a). We
denote the 3 different intra-trimer modes by
$f^{(i)}=(f_{1,i}+ f_{2,i} + f_{3,i})/\sqrt{3}$
(zero momentum mode), and
$ f_\pm^{(i)}=(f_{1,i}+z_{\pm} f_{2,i} +z_{\pm}^2 f_{3,i})/\sqrt{3}$
(left and right chirality modes), where $z_{\pm}=\exp(\pm 2\pi i/3)$.

In the limit of weak coupling between trimers, the extended Hubbard Hamiltonian 
for the problem of two fermions per trimer becomes equivalent to a
quantum magnet on a triangular lattice with couplings
that depend on the bond directions \cite{Santos04,Damski05a,Damski05b} 
\begin{equation}
  H_{magnet} =
\frac{J}{2} \sum_{i=1}^{N} \sum_{j = 1}^6  \sigma_i (\phi_{i \to j})
 \sigma_j (\tilde{\phi}_{j \to i}),
\label{subra}
\end{equation}
where $N$ denotes number of trimers,
$J=4U'/9$, and the nearest neighbours are enumerated as
in Fig. \ref{trimer}(a).
In  Eq. (\ref{subra}) we have
$ \sigma_i(\phi) = \cos(\phi)   \sigma_x^{(i)} + \sin(\phi)  \sigma_y^{(i)}$,
where the spin-$1/2$ operators
are defined as:
$ \sigma_x^{(i)} = ( f_+^{(i)\dag} f_-^{(i)} + f_-^{(i)\dag} f_+^{(i)})/2$,
$ \sigma_y^{(i)} = -i ( f_+^{(i)\dag}  f_-^{(i)}
-   f_-^{(i)\dag} f_+^{(i)})/2$. The angles $\phi$ are:
$\phi_{i \to 1} = \phi_{i \to 6} = 0$, $\phi_{i \to 2} = \phi_{i \to 3} = 2\pi/3$,
$\phi_{i \to 4} = \phi_{i \to 5} = -2\pi/3$,
$\tilde{\phi}_{1 \to i} = \tilde{\phi}_{2 \to i} = -2\pi/3$,
$\tilde{\phi}_{3 \to i} = \tilde{\phi}_{4 \to i} = 0$,
$\tilde{\phi}_{5 \to i} = \tilde{\phi}_{6 \to i} = 2\pi/3$. The physical
picture behind mapping the Fermi-Hubbard onto spin model is the following. There
are two fermions in each trimer. In the ground state configuration one of them 
 occupies the zero momentum mode, while the second one has to
choose between either right or left chirality modes, so it stays in the
superposition of these two modes that are further identified as spin-$1/2$
states.

\subsubsection{Classical analysis}
\label{subsubsec-kumropotas1}
\begin{figure}[t]
\centerline{\includegraphics[width=8cm]{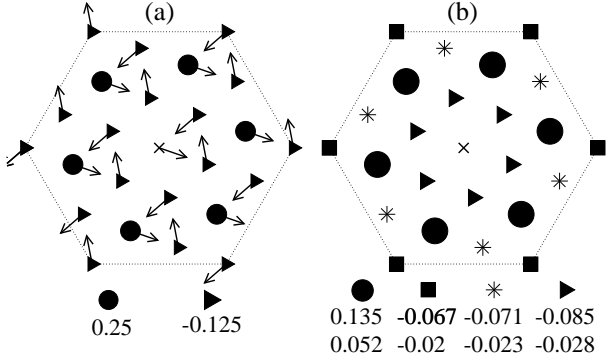}}
\caption{
(a) Classical $120^\circ$ N\'eel state with {\it right} chirality.
Dots and triangles show $\sigma_x^{(i)}\sigma_x^{(10)}+\sigma_y^{(i)}\sigma_y^{(10)}$, 
where $|\vec{\sigma}\;|=1/2$ and 
the central spin defines the x axis;
(b) Spin-spin correlations: 
$\langle  \sigma_x^{(i)} \sigma_x^{(10)}+ \sigma_y^{(i)} \sigma_y^{(10)}\rangle$.
The upper (lower) set of values corresponds to $kT=0$ ($kT=10^{-2}J/2$).
In both plots $N=21$ and  $i= 10$ at the central site.}
\label{chirality}
\end{figure}

We first discuss the classical theory of the model (Eq.(\ref{subra})),
i.e., the large $S$ (spin) limit.
In addition to being translationally invariant, the model  (Eq.(\ref{subra})) is invariant
under the point group $Z_6=Z_3\cdot Z_2$. The generator of $Z_3$ 
is the combined rotation of the lattice by the angle $4\pi/3$, and of the spins
by the angle $2\pi/3$, while the generator of $Z_2$  is the spin
inversion in the lattice plane.
There exist three ordered classical states with small unit cells that
are compatible with this
point-group symmetry of the  model:
a ferromagnetic state and two $120^{\circ}$ N{\'e}el type
structures with left (Fig. \ref{trimer}(b)) and right
(Fig. \ref{chirality}(a)) chiralities formally defined in \cite{Damski05a,Damski05b}.
The energies per site of these states are (i)
 $-3 S^2 J/4$ for ferromagnetic  and right-handed N\'eel states;
 (ii) $3 S^2 J/2$ for left-handed N\'eel state. Hence, for
$J<0$ the state with left-handed chirality will be the
ground state. For $J>0$ the situation is more
complicated since the state with right-handed chirality and the ferromagnetic state are degenerate
ground states.

The  analysis of classical ground states can be supplemented  by a numerical study 
of the $12$-spin cell done by fixing the direction of every spin to  $n\pi/3$ ($n=0\cdots5$),
and checking  the energies of the resulting $6^{12}$ configurations. This analysis
has revealed that for $J<0$ there are $6$ ground states
($Z_6$ symmetry of (\ref{subra}))
each of them exhibiting the left chirality N\'{e}el order.
For $J>0$ the results are dramatically different:
there are $240$ degenerate classical ground states in this case. 
Among them 6  right chirality  N\'{e}el states and 6  ferromagnetic states.
As will be seen below,
the large number of degenerate {\em classical} ground states for $J>0$ finds its analogue in a
large density of low-lying excitations of the {\em quantum} version of (Eq.(\ref{subra})).

\subsubsection{Quantum mechanical results}
\label{subsubsec-kumropotas2}

The insight into quantum mechanical properties of the system  can be
obtained through exact diagonalization of the Hamiltonian (Eq.(\ref{subra})).

Numerical findings are presented in Fig. \ref{chirality}(b)
and Tables \ref{tab_fermi_1} and \ref{tab_fermi_2}. For $J>0$
the ground state exhibits the $120^{\circ}$
N\'eel order with right chirality. This
is illustrated in Fig. \ref{chirality}(b), where
the planar spin-spin correlations are presented.
Direct comparison with the correlations
of the classical state, Fig. \ref{chirality}(a),
shows that  the exact quantum
correlations, although smaller, have the
same order of magnitude and sign as the classical ones. Especially,
the relative values of correlations compare nicely to the classical
result.
Interestingly, the $120^{\circ}$ N\'eel order survives at finite temperatures,
as is indicated by  the results obtained for $kT=10^{-2}J/2$ (Fig. \ref{chirality}(b)).
At such temperatures about 800 low energy eigenstates contribute
to the  correlations. To quantify how exceptionally dense the excitation spectrum is,
we note that  for $N=21$ there are about $2000$ ($800$) excited states
with energies less than $0.09J$ ($0.05J$) above the ground state. 
Most of them support the
spin order of the ground state  so that antiferromagnetic order persists at finite temperatures.
Moreover, numerical simulations show that the spectrum becomes more dense as $N$
increases.

\begin{table}
\begin{tabular}{c c c c c c}
\hline\hline
  &  1  & $\sqrt{3}$ & 2 & $\sqrt{7}$ & 3 \\ \hline
$120^{\circ}$& -0.125 & 0.25  & -0.125 & -0.125 & 0.25 \\ 
N=24         & -0.096 & 0.162 & -0.083 & -0.080 & 0.156\\ 
N=21         & -0.085 & 0.135 & -0.071 & -0.067 &      \\ \hline\hline
\end{tabular}
\caption{Spin-spin correlations, 
$\langle  \sigma_x^{(i)} \sigma_x^{(j)}+ \sigma_y^{(i)} \sigma_y^{(j)}\rangle$,
for $J>0$ as a function of distance (expressed in lattice units) 
between  sites $i$ and $j$. The first row presents classical predictions
for the $120^\circ$ N\'eel state (compare to Fig. \ref{chirality}).}
\label{tab_fermi_1}
\end{table}

The exact diagonalizations do not give a definite answer of whether the
gap vanishes in the limit of infinite lattice. What can be found out from 
exact diagonalizations is that the gap, if any,  should be  smaller than
about $10^{-2}J/2$.
The appearance of this very small energy scale is surprising. 
The smallness of the gap and
the large density of low-energy states resemble very much the behaviour 
of a quantum spin liquid of type II. 
The spin liquids of type II, however, possess extremely short range
correlations, which is in striking opposition to the behaviour of the considered
quantum magnet. For these reasons it was proposed to name this system a
{\it quantum  spin-liquid crystal}.

The above results for $J>0$ are in a strong contrast to 
those for $J<0$. In the latter case the ground state is 
the standard quantum antiferromagnet with  $120^{\circ}$
N\'eel order and left chirality.
The spectrum is gapped, and the
classical spin-spin correlations approximate well
the quantum ones (Table \ref{tab_fermi_2}). In fact, 
the semiclassical theory works remarkably well
even for system sizes as small as $N=12$.
The gap is of the order of $|J|/2$ in this case, so that 
there are at most few states with energies substantially below $|J|/2$
for $J<0$, as opposed to the huge number for $J>0$.

\begin{table}
\begin{tabular}{c c c c c}
\hline\hline
  & $1$ & $\sqrt{3}$ & 2 & $\sqrt{7}$  \\ \hline
$120^{\circ}$& -0.125 & 0.25 & -0.125& -0.125 \\ 
N=21 & -0.134 & 0.237 & -0.117  & -0.116 \\ 
N=12 & -0.137 & 0.251 &  -0.125 & \\ \hline\hline
\end{tabular}
\caption{The same as in Table \ref{tab_fermi_1} but for $J<0$.}
\label{tab_fermi_2}
\end{table}

As it is evidenced from the above discussion, the theoretical studies
of the quantum spin liquid crystals are limited to relatively small systems.
We regard future experiments as the best possible verification of
these predictions.

\subsection{Realization of frustrated models in cold atom/ion systems}
\label{frust_other}

\begin{figure}[ht]
\begin{center}
\includegraphics[width=4cm]{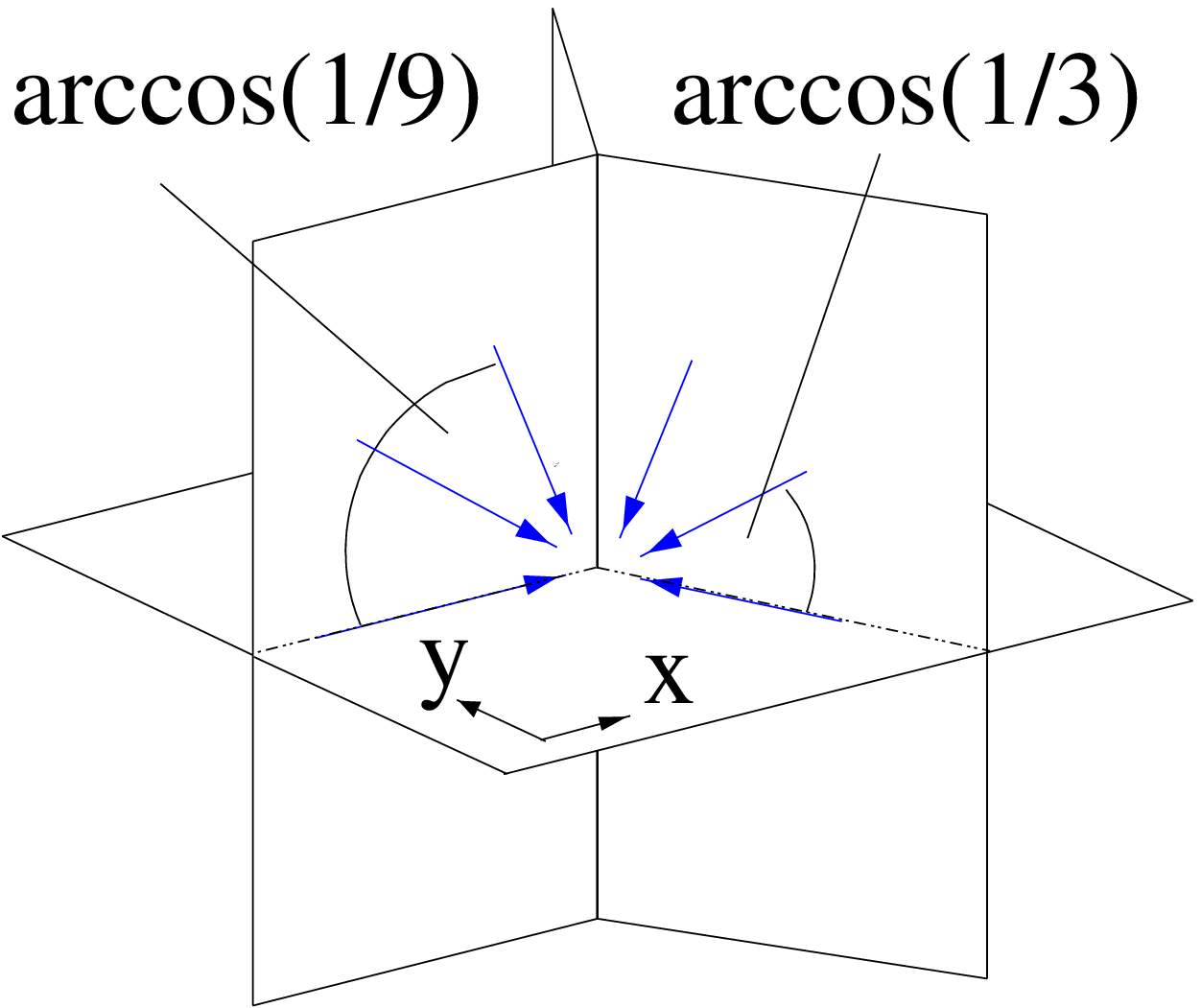}
\includegraphics[width=3cm]{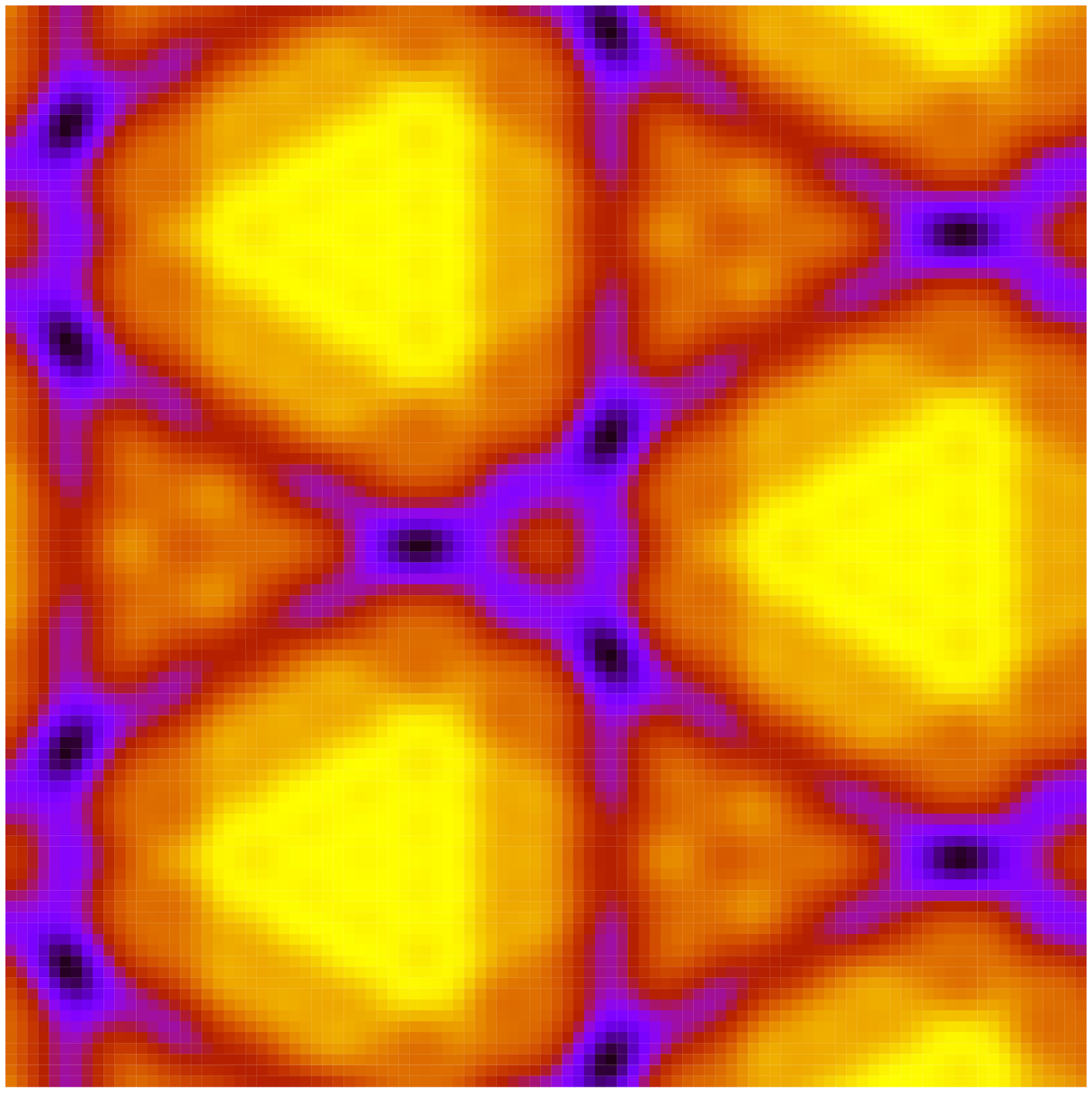}
\end{center}
\caption{Scheme of the proposed experimental set-up.
Each arrow depicts a wave vector of  a standing wave laser. The three vertical 
planes intersect at an angle of $120^{\circ}$.  
Dark (dark blue in the online version) spots in the 
right kagom\'e figure indicate the potential 
lattice minima (from \cite{Damski05b}).
}
\label{fig_Setup}
\end{figure}

Here we focus on control of interactions and  
recent proposals based on ultracold bosonic
or fermionic atoms \cite{Duan03}, cold gases of polar molecules \cite{Micheli06}, and 
 trapped ions \cite{Porras04,Deng05,Porras06}.
Realizations of frustrated models in ultracold atomic systems requires either the creation of a lattice with appropriate geometry, 
or the control of the effective atomic interactions. Geometry of the lattice, as we already mentioned, can be engineered using superlattice techniques 
(see Fig. \ref{fig_Setup} and Ref. \cite{Damski05b} for the details in the case of kagom{\'e} lattices). 

\subsubsection{Simulators of spin systems with topological order}
\label{ultra}
We discuss here  the proposal of Duan, Demler and Lukin, 
for realization of quantum magnets in a system of cold atoms  placed in
an optical lattice \cite{Duan03}. For simplicity, we assume that the lattice is filled with bosons. It is 
made of standing laser beams producing the periodic potential
\(V_{\mu \sigma} \sin^2(\vec{k}_\mu \cdot \vec{r})\) in the direction \(\mu\)
($\vec k$ is the beam wave vector). The index $\sigma= \{\uparrow, \downarrow\}$
accounts for the fact that bosons are trapped in two different internal states
by independent lattice potentials.

The system is governed by the following Hubbard Hamiltonian 
\begin{equation}
\label{eq_H_Duan}
H_{atoms} = - \sum_{\langle ij \rangle \sigma} (t_{\mu \sigma} b_{i\sigma}^{\dagger} b_{j\sigma} 
+ {\rm h. c.}) 
+ \frac{1}{2} \sum_{i \sigma} U_{\sigma} n_{i\sigma} (n_{i\sigma} -1) 
+ U_{\uparrow \downarrow} \sum_i  n_{i\uparrow}  n_{i\downarrow},
\end{equation}
where \(\langle ij \rangle\) denotes nearest neighbors,
$b_{i\sigma}$ ($b_{i\sigma}^\dagger$) 
is the annihilation (creation) operator of bosons in the $i$-th lattice site.
The Hamiltonian (Eq.(\ref{eq_H_Duan})),  describes tunneling of
bosons in different internal states between neighbouring lattice sites, 
and their on-site interactions.

The Hamiltonian (Eq.(\ref{eq_H_Duan})) can be cast into the quantum magnet form if the
following conditions are satisfied (i) 
$t_{\mu \sigma} \ll U_{\sigma}, U_{\uparrow \downarrow}$,
which typically requires large enough $V_{\mu\sigma}$;
(ii) $\langle n_{i\uparrow} \rangle + \langle n_{i\downarrow} \rangle \simeq  1$.
Then, in the lowest order nontrivial approximation in 
$t_{\mu\sigma}/U_{\uparrow\downarrow}$, one gets 
\begin{equation}
H_{atoms}\approx\sum_{\langle i,j\rangle}\left[\lambda_{\mu z}\sigma^z_i\sigma^z_j-
\lambda_{\mu\perp}(\sigma^x_i\sigma^x_j+\sigma^y_i\sigma^y_j)\right],
\label{model}
\end{equation}
where $\sigma^z_i=n_{i\uparrow}-n_{i\downarrow}$, 
$\sigma^x_i=b_{i\uparrow}^\dag b_{i\downarrow}+ b_{i\downarrow}^\dag b_{i\uparrow}$,
and $\sigma^y_i=-i(b_{i\uparrow}^\dag b_{i\downarrow}- b_{i\downarrow}^\dag b_{i\uparrow})$,
and
$$
\lambda_{\mu z}=
\frac{t^2_{\mu\uparrow}+t^2_{\mu\downarrow}}{2U_{\uparrow\downarrow}}-
\frac{t^2_{\mu\uparrow}}{U_\uparrow}-
\frac{t^2_{\mu\downarrow}}{U_\downarrow}, \ \ \lambda_{\mu\perp}=
\frac{t_{\mu\uparrow}t_{\mu\downarrow}}{U_{\uparrow\downarrow}}.
$$
The Hamiltonian (Eq.(\ref{model})) corresponds to anisotropic Heisenberg spin-1/2
model, whose different variations can be studied due to the possibility of
adjustment of $t_{\mu\sigma}$, $U_\sigma$ and $U_{\uparrow\downarrow}$
couplings. This can be  achieved, e.g., via manipulation of the lattice potentials 
imposed on atoms: $V_{\mu\sigma}$.

To illustrate the possibilities offered by cold atomic systems, let us consider
the limit of $t_{\mu\downarrow}/t_{\mu\uparrow}\to0$, i.e., 
$V_{\mu \uparrow}\ll V_{\mu \downarrow}$. Then, obviously the $H_{atoms}$ Hamiltonian
approximates the nearest-neighbor Ising model (Eq.(\ref{eq_Ising})). Depending on 
$V_{\mu\uparrow}$ the couplings $\lambda_{\mu z}$
can be   either isotropic or anisotropic. Moreover,
they can be all either positive or negative depending on the
$U_{\uparrow\downarrow}/U_\uparrow$ ratio. Therefore, as discussed in 
Sec. \ref{intr} both ferromagnetic and antiferromagnetic Ising model can 
be simulated with cold bosons in a lattice. Additionally, 
as discussed in \cite{Duan03}, the successful experimental 
realization of the Ising  model in a system of cold atoms
can be very useful in the experimental implementation of  the one-way quantum computer, 
introduced by Raussendorf and Briegel  \cite{Briegel01, Raussendorf01}. 

Perhaps the most unusual quantum magnet, that can be realized using the ideas
presented above, corresponds to the anisotropic 2D spin-1/2 model on a hexagonal
lattice. A hexagonal lattice can be created in a carefully designed setup
of laser beams \cite{Duan03}. Assuming the same as above for derivation 
of  Eq.(\ref{model}), and additionally  
$U_{\uparrow\downarrow}\approx U_\uparrow\approx U_\downarrow\approx U$ the
Hamiltonian (Eq.(\ref{eq_H_Duan})) can be transformed into  
\begin{equation}
\label{eq_Honeycomb}
H_{atoms} \approx \lambda_x \sum_{x-links} \sigma_i^x \sigma_j^x +
         \lambda_y \sum_{y-links} \sigma_i^y \sigma_j^y + 
	 \lambda_z \sum_{z-links} \sigma_i^z \sigma_j^z, 
\end{equation}
where the summation goes over nearest neighbours $i$ and $j$ placed 
on the $x, y, z$-links -- see Fig. \ref{fig_kagome}. The couplings
read $\lambda_\mu=-t^2_{\mu+}/(2U)$, 
where $t_{\mu+}$ is the rate of tunneling of the atom being in the eigenstate of 
the Pauli operator $\sigma_\mu$ with eigenvalue $+1$.




The Hamiltonian (Eq.(\ref{eq_Honeycomb})) describes 
the Kitaev model \cite{Kitaev05} with topological order. 
Excitations in this model are   both Abelian and non-Abelian anyons \cite{Wilczek82a, Wilczek82b}, 
 having exotic fractional statistics. 
It is  
exactly solvable,  and possesses other exciting features
 \cite{Kitaev05}. Its realization should provide
an exceptional possibility for experimental observation of Abelian and
non-Abelian anyons, and could serve for applications in quantum
information as protected qubits.

\subsubsection{Frustrated models with polar molecules}
\label{subsubsec-polar}
 
Micheli \emph{et al.} \cite{Micheli06} have proposed a scheme to realize 
spin Hamiltonians with polar molecules. 
In fact, recently various schemes have been proposed
for trapping different states of cold polar molecules. 
New developments in this area can be found 
in Ref. \cite{SpecialIssueEurphysD04} (for experiments, see \cite{Rieger05}). 

In Ref. \cite{Micheli06}, the authors start with 
two polar molecules, trapped in an optical lattice. The 
outermost shell of an electron of a heteronuclear molecule represents the spin-1/2 system. 
The total Hamiltonian of a pair of heteronuclear molecules
trapped in an optical lattice is given by
\begin{equation}
\label{eq_H_mol}
H = \underbrace{\left(H_{dd} + \sum_{i=1}^{2} H_m^i \right)}_{H_{int}} 
+ \underbrace{\left(\sum_{i=1}^2 \frac{P_i^2}{2m} + V_i(x - \overline{x_i}) \right)}_{H_{ext}}, 
\end{equation}
where \(H_{int}\) and \(H_{ext}\) represent respectively the internal and external dynamics of molecules. 
\(H_{dd}\) is the Hamiltonian for dipole-dipole interactions between two molecules. 
The rotational excitation of each molecule is described 
by the Hamiltonian
\begin{equation}
\label{eq_H_dd}
H_m = B N^2 + \gamma N\cdot S,
\end{equation}
where \(N\) is the dimensionless orbital angular momentum of the nuclei and \(S\) is the electronic spin. Here 
\(B\) is the rotational constant and 
\(\gamma \) is the spin-rotation coupling constant. The rotational motion of molecules is coupled by the dipole-dipole interactions that are 
already present in the heteronuclear molecule. To obtain strong dipole-dipole interactions, 
a microwave field is introduced. If the field polarization is set to the \(z\)-axis of the two molecules, 
and the frequency is fixed as near resonant with the excited state potential, 
it leads to some spin pattern of ground states. 
By changing the frequency and field polarization, it is possible to obtain various spin models, 
e.g., Ising, Heisenberg, and Kitaev models, 
that have been discussed earlier (see Table I of \cite{Micheli06}). 
The advantage of this model is 
the strong dipole-dipole interactions which is due to the inherent properties of these molecules,
and is also due to the
introduction of  microwave fields. 
The couplings are strong relative to decoherence rates.

\subsubsection{Ion-based quantum simulators of spin systems}
\label{subsubsec-ionsimulator}

It was recently proposed by Porras and Cirac that different spin systems can also be  simulated 
with cold ions \cite{Porras04,Deng05}. This protocol, if implemented
experimentally, should allow for studies of different  quantum magnets in  
ion traps. As a result, the ions could then be used  for investigation of
condensed matter problems. In fact, it seems that ions are perfectly
suitable for such experimental studies since (i) they can be trapped and cooled 
very efficiently; (ii) the position and internal state
of every single ion can be  measured and manipulated almost at will; 
(iii) the external parameters of that system can be well controlled and changed
in real time.

Below we will briefly present the Porras and Cirac proposal. The spin $1/2$
states are encoded in  two internal hyperfine states of each ion. 
The local dynamics of these  states is governed by the Hamiltonian
$$
 H_m= \sum_{j=1...N}^{\alpha=x,y,z} B^\alpha\sigma^\alpha_j,
$$
where $N$ is the number of ions,  $B^x$ and $B^y$ are 
``analogs'' of the magnetic field induced by lasers resonant with the internal transition,
whereas  $B^z$ is the energy gap between the two internal states. 

The ions are affected by both an external harmonic trapping potential with frequencies ($\Omega_x$,$\Omega_y$,$\Omega_z$)and Coulomb interactions, 
which is described by the potential: 
$$
V=\frac{m}{2}\sum_{j=1\dots N} (\Omega_x^2x_j^2+\Omega_y^2y_j^2+\Omega_z^2z_j^2)
+ \sum_{j>i}\frac{e^2}{\sqrt{(x_i-x_j)^2+(y_i-y_j)^2+(z_i-z_j)^2}},
$$
where $m$ is the mass of the ion and $e$ is the electron charge.
It is further assumed that the trap has a cigar-shaped geometry with 
axial $z$ direction and radial $x$ and $y$ ones: 
$\Omega_z\ll\Omega_x,\Omega_y$.
The  competition between harmonic squeezing and Coulomb
repulsion  results in oscillations of ions around equilibrium positions. 
Dynamics of these oscillations is governed by the vibrational 
Hamiltonian: 
$ H_v= \sum_{j=1\dots N}^{\alpha=x,y,z}\frac{{p_j^\alpha}^2}{2m}+ V$, where
$p_j^\alpha$ are momentum operators of the $j$-th ion. After expansion of $V$ up to a
quadratic 
order in ions displacement from equilibrium positions, the vibrational  Hamiltonian 
can be diagonalized in a standard way in terms of collective modes (phonons):
$$
 H_v= \sum_n^{\alpha=x,y,z} \hbar\omega_{\alpha,n}
a_{\alpha,n}^\dag  a_{\alpha,n},
$$
where $ a_{\alpha,n}$ and $ a_{\alpha,n}^\dag$ are the 
phonon annihilation and creation operators and $\omega_{\alpha,n}$ are the collective mode frequencies.
 
Finally, one needs a coupling between $ H_m$ and $ H_v$, i.e.,
between ``effective'' spins  and phonons. This
is achieved by placing the ions in off-resonant standing wave beams. 
For instance, imposing standing light waves on ions one can realize the 
following Hamiltonian \cite{Deng05}:
$$
 H_f=-F_x\sum_j x_j|\uparrow\rangle\langle \uparrow|_{z,j} 
-F_y\sum_j y_j|\uparrow\rangle\langle\uparrow|_{y,j},$$ 
where $\sigma^\alpha|\uparrow\rangle_\alpha=|\uparrow\rangle_\alpha$. 
This Hamiltonian assumes that the laser beams
push ions in the upper  state only, which can be achieved by
a proper adjustment of the relative phases of the beams.

The total Hamiltonian of the system becomes then $ H= H_m+ 
H_v+ H_f$. The final step to get the clear picture of the phonon-mediated
interactions between ``effective'' spins  relies on the unitary transformation 
$ H\to U H U^\dag$ with $U$ specified in \cite{Mintert2001,Porras04,Deng05}.
The results of this transformation 
will be illustrated on particular examples.

Assuming that  $F_x\neq 0$ and $F_y=0$, one gets to the
lowest order in
$\eta_\alpha=F_\alpha\sqrt{\hbar/2m\Omega_\alpha}/\hbar\Omega_\alpha$ 
\begin{equation}
 H \approx \frac{1}{2}\sum_{i,j}J_{i,j}^{[x]}\sigma_i^z\sigma_j^z+\sum_i B^x\sigma^x_i+
 H_v+  H_E,
\label{HIL}
\end{equation}
where 
\begin{equation}
J_{i,j}^{[\alpha]}\sim \frac{1}{|\langle z_i\rangle-\langle z_j\rangle|^3},
\label{Jalpha}
\end{equation}
with $\langle z_i\rangle$ being the equilibrium position of the $i$-th ion.
The first two terms in  Eq.(\ref{HIL}) act in the ``spin'' space and 
exhibit the  quantum Ising model type of spin-spin and spin-field 
interactions.
The fact that the interactions are long-range makes quantum properties
even richer then those of a standard Ising model with only nearest neighbor
 terms. Therefore, one gets the possibility of studying the
fascinating Ising-type  model \cite{Deng05}
in an ion chain if the corrections coming from 
the perturbation $ H_E$ \cite{Porras04,Deng05} are negligible. 
To quantify influence of the perturbation $ H_E$ on spin dynamics,
we note that the deviations of expectation values of $M$-spin(site) observables,
induced by skipping the $ H_E$ term, scale as $M\eta_\alpha^2$. 
Since the observables of interest 
during analysis of quantum phase transitions  
correspond usually to $M=1$ or $2$ classes, the corrections 
coming from $ H_E$ should not cause major problems.

Another possibility shows up when there are two forces acting in radial
directions $x$ and $y$ ($F_x, F_y\neq0$). Then one gets
that the $U$ transformed Hamiltonian is 
\begin{equation}
 H\approx \frac{1}{2}\sum_{i,j} (J^{[x]}_{i,j}\sigma^z_i\sigma_j^z
                              +J^{[y]}_{i,j}\sigma^y_i\sigma_j^y)
			      +\sum_i B^x\sigma^x_i+ H_v+   H_E,
\end{equation}
where $J^{[\alpha]}_{i,j}$ are given by Eq.(\ref{Jalpha}).
This time, the first two terms correspond to the XY model with long-range 
interactions. The perturbation affecting spin dynamics, $ H_E$,
is now  a little different than the one above, but  
as long as $\omega_x\neq\omega_y$  the errors induced by leaving the 
$ H_E$ out are of the same order. Hence, we conclude
that ion chains can be used for simulation of yet another important 
spin model.

Though the above discussion was focused on the one-dimensional geometry
of the ion trap, a similar approach can be used for studies of ions in 
two-dimensional geometries, where they form 
Coulomb crystals \cite{Itano1998,Mitchell1998}. This time, two dimensional 
``effective'' spin models should show up. Their experimental realization 
would greatly  facilitate studies  of 
frustrated quantum magnets.

Finally, we remark that the scheme discussed in this section offers an 
access to measurements and manipulations of arbitrarily selected ion(s): 
an exciting opportunity unavailable in traditional condensed matter systems.
Interestingly, since the ``spin'' configuration is encoded in ions internal
states, the  fluorescence signal of the sample can contain enough information 
for detection of different phases.



\section{Ultracold spinor atomic gases}
\label{sec-radhamadhab}

\subsection{Introduction}
\label{sec-radhamadhab1}
Interactions in bosonic systems with spin degrees of freedom host a wide
variety of exotic phases at zero temperature and a dynamics clearly 
differentiated from the one displayed by scalar gases. 
Spin effects are enhanced in the limit of small occupation number and
strong interactions. This fact, makes the study of ultracold 
spinors in optical lattices of primordial importance and a very 
valuable tool to deepen our understanding of magnetic ordering 
and condensed matter related issues.

On the other hand, there exists also an increasing interest in the study of spinor fermions in optical lattices since these systems could serve as quantum simulators of fermionic Hubbard model and shed some light on the problem of high Tc-superconductivity. we shall address this point at the end of the chapter.

The spinor degree of freedom on alkaline gases 
corresponds to the manifold of
degenerate (in the absence of an external magnetic field) 
Zeeman hyperfine energy levels.
The energy levels are described by 
the total angular momentum $\bf {F = I + J}$,  
where ${\bf I}$ refers to the nuclear spin and ${\bf J=L+S}$ 
describes the total electronic angular momentum. 
In the ground state, alkaline atoms have a single electron 
in an $ns$ orbital, 
their electronic angular momentum $J=S=1/2$ and, therefore, 
they have two possible hyperfine of Zeeman states. 
The nuclear spin ${\bf I}$ depends on each atomic species. 
$^{133}$Cs atoms  have a nuclear spin $I=7/2$, while 
$^{85}$Rb atoms have $I=5/2$, and $^{87}$Rb and $^{23}$Na have $I=3/2$.
In the case of Cesium, the hyperfine manifolds  correspond to $F=4,3$; 
for $^{85}$Rb they correspond to $F=3,2$ and $^{87}$Rb and $^{23}$Na have manifolds $F=2,1$. Each ground state manifold consists
of  all  Zeeman states associated to a given total
angular momentum, i.e. $\{|F,m_F\rangle\}$ where $m_F=-F,..,F$. 
In this context, we identify the total angular momentum of the atom $F$ 
as the atomic spin.

If atoms are magnetically trapped, the degeneracy is broken and the
atoms minimize their energy in the so-called ``weak field'' seeking states, 
characterised by a fixed hyperfine level $|F,m_F\rangle$.
The simultaneous confining of different ``low field seeking'' states 
is usually unstable against spin-spin collisions. 
This is why magnetic trapping results in a frozen atomic spin ``$F$'' and spin
projection ``$m_F$'' \cite{StamperKurn99}.  
In that case, the bosonic quantum field operator $ \Psi$ 
describing the creation of a boson in position ${\bf {r}}$
is a scalar in the spinor space 
with no dependence whatsoever on the hyperfine 
magnetic level. In turn, the mean field description of the ultracold 
bosonic gas is done via  a scalar order parameter 
which has not explicit dependence on the ground state 
manifold $|F,m_F\rangle$ in which the atoms were trapped.
In some cases, despite the fact that the levels are not
anymore degenerate, it is possible to magnetically trap 
simultaneously more than one hyperfine component. 
This is the case of $^{87}$Rb, where usually the  
singlet and the triplet scattering lengths are practically equal and thus
spin exchange collisions are highly suppressed. It is then 
possible to magnetically trap simultaneously the 
$|F=2,m_F=2\rangle$ and  $|F=1,m_F=-1\rangle$ states.   
In such cases, one speaks of  ``multi-component'' ultracold 
gases,  but still the spin of the atoms remains ``frozen'', 
although with two different possibilities \cite{Cornellgroup_1}.

On the contrary, if atoms are optically trapped with a 
far-off resonance laser, then  all atoms, regardless  of their hyperfine level
are simultaneously trapped. 
The bosonic quantum field operator in such case is no longer
an scalar and has to incorporate this new degree of
freedom $ \Psi_{m_F}$. In turn, the corresponding order parameter
describing mean field approach, 
becomes a vector whose components 
correspond to the different accessible hyperfine levels \cite{StamperKurn99,Ho98}
and which transforms under rotations in the spin space as
a vector preserving the symmetries 
present in the corresponding
spin space.

\subsection{Spinor interactions}
\label{sec-radhamadhab2}

Like in the scalar case, ultracold atomic spinor interactions 
can be parametrised by  two-body short range (s-wave) collisions.
In the most general scenario, symmetry arguments 
impose that the collisions between two {\it identical} bosons 
in a hyperfine spin level $|F,m_F\rangle$  are restricted to 
total spin $S=2F,2F-2,...,0$. The contact potential can be written as:

\begin{equation}
V=\sum_{S=0,2,...,2F}g_S {P_S},
\label{spinpot}
\end{equation}
where ${P}_S$ ($S=0,2,...,2F$) denotes  
the projector onto the subspace with total spin $S$. 
The interaction strength $g_S$ characterising 
the contact potential interaction of the  different $S$ channels  
are given by  $g_S=\frac{4\pi\hbar^2 a_S}{m}$, where $a_S$ is the
corresponding scattering length and $m$ the atomic mass.
The different values of the various
$a_S$ will lead to distinct magnetic ordering.

To understand the  ground state properties as well as the dynamics,
it is convenient to express the 
interaction potential $V$ in terms of spin operators by 
using different operator identities, e.g. 
$I=\sum_{S=0,2,...,2F}{P}_S$ and 
${\bf {F}}_1\cdot {\bf{{F}}}_2=
\sum \lambda_S {P}_S$ where $\lambda_S= (1/2)[S(S+1)-F(F+1)]$.
For spin $F=1$, the total spin is $S=2,0$ 
and using the above identities  
the contact potential can be written as:
\begin{equation}
     V_{(F=1)}=c_0 + c_2 {\bf {F}}_1\cdot{\bf{{F}}}_2.
\label{contactpotentialF=1}
\end{equation}
The terms with coefficients $c_0$ and $c_2$
describe spin-independent and spin-dependent 
binary elastic collisions respectively
in the combined symmetric channels of total spin 0 and 2, 
and are expressed in
terms of the s-wave scattering lengths $a_0$ and $a_2$ as:
$c_0=4 \pi \hbar^2 (a_0+2a_2)/3m$ and
$c_2=4 \pi \hbar^2 (a_2-a_0)/3m$ \cite{Ho98,Koashi00}.

It is straightforward to generalize some spin identities 
to larger spins\cite{Ho98} by noting that 
$({\textbf{\textit{F}}}_1\cdot {\textbf{\textit{F}}}_2)^n=
\sum \lambda_S^nP_S$. Then the interaction potential can be 
rewritten as $V=\sum_{n=0}^{2F} c_n ({\bf{F}}_1\cdot{\bf{F}}_2)^n$
where the $c_n$ are linear combinations 
of the different scattering lengths $a_S$.

For $F=2$, there are 3 possible channels with total spin $S=0,2,4$. 
The contact potential can be expressed as

\begin{equation}
     V_{(F=2)}=c_0 + c_1 P_0+c_2 {\bf {F}}_1\cdot {\bf{{F}}}_2,
\label{contactpotentialF2}
\end{equation}
where the coefficients are given by  
$c_0=4 \pi \hbar^2 (3a_4+4a_2)/7m$ and
$c_1=4 \pi \hbar^2 (3a_2-10a_2+7a_0)/7m$ and 
$c_2=4 \pi \hbar^2 (a_4-a_2)/7m$. The projector 
$P_0$ can be further expressed in terms of
``singlet'' pair operator as we shall see later.

\subsection{$F=1$ and $F=2$ spinor gases: Mean field regime}
\label{sec-radhamadhab3}

The experimental achievement of Stenger {\it et al.} \cite{Stenger98} 
in trapping  $^{23}$Na by optical means in 1998, triggered
the study of spinor ultracold gases. A mean field approach to describe
a $F=1$ condensate was developed by Ho \cite{Ho98}, and independently by 
Ohmi and Mashida \cite{Ohmi98} in the same year. 
Koashi and Ueda \cite{Koashi00}, and Ciobanu, Yip, and Ho \cite{Ciobanu01} have calculated the mean field 
phase diagram of $F=2$ spinor condensates.
More recently, Ueda and Koashi \cite{Ueda02} have studied 
mean field theory for $F=2$ atoms in presence also 
of a magnetic external field. 
The recent success on condensing Chromium atoms \cite{Griesmeyer05} have also
initiated the study  of mean field phases for 
spin $F=3$ \cite{Ho05,Santos05}. 
Recent experiments concern studies of dynamics of spinor BECs in 
traps \cite{Chang04,Kuwamoto04, Schmaljohan04, 
Barrett01,Higbie05,Erhard04} (for the theory, see \cite{Mur-Petit06}), 
and in lattices \cite{Widera05}.

In scalar bosonic weakly interacting systems, where a mean field approach can be used, 
the ground state is found by approximating the bosonic operator 
of the corresponding Hamiltonian by its mean, and
minimizing the energy functional under 
the constrain that the number of particles is 
fixed (grand canonical ensemble) $\partial\langle H-\mu N\rangle/\partial\mu=0$. 
For spinor gases, one follows the same approach\cite{Ho98} 
and minimization of the energy leads to 
different ground states depending on the values
of the spin-spin coupling. 

The Hamiltonian of a trapped cloud of ultracold atoms with spin $F=1$ 
in second quantization reads
\begin{eqnarray}
  H& = &\int d^3r \left\{
    \Psi^\dagger_m \left(-\frac{\hbar^2}{2M}{\bm \nabla}^2
                   + V_{ext}\right) \Psi_m \right.\nonumber \\
    &&
\left.
    +\frac{c_0}{2} \Psi^\dagger_m \Psi^\dagger_{j} \Psi_{j} \Psi_m
    +\frac{c_2}{2}
      \Psi^\dagger_m\Psi^\dagger_{j} {\bf F}_{mk} \cdot 
                            {\bf F}_{jl} \Psi_{l}\Psi_k 
\right\}, 
\label{spinorhamiltonian}
\end{eqnarray}
where $\Psi_m({\bf r})$ $(\Psi_m^\dagger)$ is the field 
operator that annihilates (creates) an atom in the $m$-th 
hyperfine state $|F=1, m_F \rangle$ at point ${\bf r}$. 
The external trapping potential, \(V_{ext}\), is normally assumed 
to be spin-independent.

There are two distinct ground state mean field phases for
spin $F=1$: 
\begin{itemize}
\item 
\noindent Ferromagnetic phase for $F=1$.\\
The system presents ferromagnetic order if $c_2 < 0$ (i.e. $g_0> g_2$).
This configuration minimizes the energy by imposing that 
$\langle F \rangle ^2=1$. 
For a spin pointing along 
 $\vec{n}=(\cos\alpha\sin\beta,\, \sin\alpha \sin\beta\,,
\cos\beta)$ the condensate order parameter is given by: 
\begin{equation}
\xi=e^{i\varphi}U(\alpha,\beta)\left( \begin{array}{c}
1\\ 
0 \\ 
0 \\ 
\end{array} \right)=e^{i\varphi}\left( \begin{array}{c}
e^{-i\alpha}\cos^2(\beta/2)\\ 
\sqrt(2)\cos(\beta/2)\sin(\beta/2) \\ 
e^{i\alpha}\sin^2(\beta/2) \\ 
\end{array} \right).
\end{equation}
This is, for example, the ground state of $^{87}$Rb in the $F=1$ manifold.

\item
\noindent Polar phase for $F=1$.\\
The system presents polar (sometimes called antiferromagnetic) ordering 
if $c_2 > 0$. This is the case for $^{23}$Na or $^{85}$Rb. 
Minimisation is achieved by demanding that 
the expectation values of the spin component are
zero along any direction, i.e. $\langle F\rangle=0$. 
However, variances are not equal to zero, 
indicating that the system does not posses rotational invariance.
The general expression for the spinor in this case reads
\begin{equation}
\xi=e^{i\varphi}\left( \begin{array}{c}
0\\ 
1 \\ 
0 \\ 
\end{array} \right)=e^{i\varphi}\left( \begin{array}{c}
-e^{-i\alpha}\sin(\beta)/\sqrt(2)\\ 
\cos(\beta)\\ 
e^{i\alpha}\sin(\beta)\sqrt(2) \\ 
\end{array} \right).
\end{equation}
\end{itemize}
Notice that this ``antiferromagnetic'' ordering do not refer to orient
the spins antiparallely as it happens classically.

For $F=2$ (for experiments see \cite{Schmaljohan04}) 
collisions can occur in one more channel corresponding 
to total spin $S=4$. As a result there is one more possible magnetic
ground state, the so-called cyclic ground state. 
Now, to characterise the different ground states 
one should consider not only  magnetization, i.e. $\langle F\rangle$,
but also the ``spin singlet pair creation'' expectation value \cite{Ueda02,
Ciobanu01}. The term proportional to $P_0$ 
in the contact potential (see Eq.(\ref{contactpotentialF2}))
can be expressed by means of a ``spin singlet pair operator'' 
as $P_0=(2/5) \overline{S}_+\overline{S}_-$, 
where  $\overline{S}_+=a^{\dagger}_0 a^{\dagger}_0/2-a^{\dagger}_1 a^{\dagger}_{-1}+ 
a^{\dagger}_2 a^{\dagger}_{-2}$, and
$a_{\sigma}^{\dagger}$ ($a_{\sigma}$) creates (annihilates) a particle with spin projection $\sigma$ ($\overline{S}_-=\overline{S}^{\dagger}_+$).
The operator $\overline{S}_-$ applied on the vacuum creates, except for
normalization, two bosons in a spin singlet state.
Ground states can be classified according to their expectation values of
the ``magnetisation'' and the ``singlet pair creation'':
\begin{eqnarray}
\langle F \rangle&=&\sum_{m=-2,n=-2}^{m=2,n=2}F_{mn}\xi_m^{*}\xi_n\nonumber\\
\langle {\overline{S}_+} \rangle&=&\frac{1}{2}\sum_m (-1)^{m}\xi_m\xi_{-m}.
\end{eqnarray}

\begin{itemize}
\item 
\noindent  Ferromagnetic phase for $F=2$.\\
The  ferromagnetic phase is achieved for $\langle F \rangle\neq 0$
and $\langle {\overline{S}_+} \rangle=0$.
The system prefers ferromagnetic order if $c_1<0$ and $c_1-(c_2/20)<0$.
The expectation value of the magnetisation can either be 
$\langle F \rangle=2$ or $\langle F \rangle=1$ 
but for both cases $\langle{\overline{S}_+}\rangle=0$. 
A representative of maximal spin projection, 
ground state is given by:
\begin{equation}
\xi=e^{i\varphi}\left( \begin{array}{c}
1\\ 
0\\ 
0\\ 
0\\
0\\
\end{array} \right).
\end{equation}
This corresponds, for instance, to the  ground state $F=2$ of $^{87}$Rb.

\item Polar condensate for $F=2$.\\
Polar (antiferromagnetic) ordering is described by  
a non zero singlet amplitude $\langle {\overline{S}_+}\rangle\neq 0$ 
and zero magnetisation $\langle F \rangle=0$. 
A representative of the corresponding
spinor order parameter is given by:

\begin{equation}
\xi=e^{i\varphi}\left( \begin{array}{c}
0\\ 
0\\ 
1\\ 
0\\
0\\
\end{array} \right).
\end{equation}
The phase space boundaries for polar phase are given by  
$c_2<0$ and $c_1-(c_2/20)>0$.

\item Cyclic phase for $F=2$.\\
Finally, cyclic ordering appears for $c_1,c_2>0$
there $\langle F \rangle=0 $ and $\langle {\overline{S}_+} \rangle=0$.
\end{itemize}

\subsubsection{F=1 gases in optical lattices}
\label{sec-Mr.Yusuf_Zamadar}

Studies of  $F=1$ systems have already been carried 
out by Demler's group \cite{Demler02, Imambekov03}. They have derived an approximate phase diagram in the case of 
antiferromagnetic interactions.  As in the standard Bose-Hubbard model, an $F=1$ spinor gas undergoes superfluid 
to Mott insulator transition as tunneling is decreased. In the antiferromagnetic case in 2D and 3D, the SF phase is 
{\it polar}, and so are the Mott states with an odd number $N$ of atoms per site (those states are also termed {\it nematic}). In the case of even $N$, for small tunneling the Mott states are singlets, and for moderate tunneling there occurs a first order transition 
to the nematic state.

In 1D there is furthermore the possibility of a dimerized valence bond solid state, as in the Majumdar-Ghosh model 
\cite{Auerbach94} (discussed also in subsection \ref{subsec-tapa-tepi1}). This possibility was studied by Yip \cite{Yip03a}, who derived an effective spin Hamiltonian for the MI state with $N=1$, 
and used the variational ansatz interpolating between dimer and nematic states to argue that in a wide range of parameters the spinor $^{23}$Na 
lattice gas should have the dimer ground state in 1D, 2D, and 3D. This is a very interesting result, since dimer states have not been so far
 observed in experiments. This result has been supported by rigorous studies in Ref. \cite{Yip03b} for spin systems with an even number of spins 
described by the same effective Hamiltonian: it was shown that  while the ground state of the system has total spin $S_{tot}=0$, the first 
excited state has $S_{tot}=2$. Yip's results were recently confirmed by Rizzi {\it et al.} \cite{rizzi}, who numerically studied the SF -- MI 
transition in the $F=1$ Bose-Hubbard model in 1D, and found that the system is always in the dimerized state in low tunneling regime of the first 
MI lobe, where the effective spin model of Ref. \cite{Demler02, Imambekov03} 
works. Similar results were obtained by Porras  {\it et al.} \cite{diego}. 
Thus, nematic order seems to be strictly speaking absent in 1D in the thermodynamic limit. However, susceptibility to nematic ordering 
grows close to the border of the ferromagnetic phases, indicating that it may persist in finite systems.   A completely new insight can 
be gained by looking at entanglement transport properties of $F=1$ chains \cite{Romero06}, which seem to confirm existence of nematic 
and trimer  regions for finite systems. 

Another interesting aspect, namely the possibility of controlling the order of the SF -- MI transition by using appropriately polarized 
(lin--$\theta$—-lin) laser fields to form the optical lattice was investigated in 
Refs. \cite{graham,Kimura}. Such a laser configuration couples the states with $m_F=\pm 1$, so that the system becomes effectively two--component.

\subsubsection{Bose-Hubbard model for spin 1 particles}
\label{sec-radhamadhab4}

The derivation of the Bose-Hubbard Hamiltonian for ultracold
spinor gases is performed in the same way as in the scalar case.
One has to add to the scalar Bose-Hubbard model
the spin dependent part of the interaction. 
Following the identities given in the
previous part of the section, the Bose-Hubbard Hamiltonian for 
spin 1 particles is obtained straightforwardly \cite{Imambekov03}:

\begin{eqnarray}
H&=&-t\sum_{<ij>,\sigma}(b^{\dagger}_{\sigma i} b_{\sigma j}+b^{\dagger}_{\sigma j} b_{\sigma i})
+\frac{c_0}{2}\ \sum_{i}n_i(n_i-1)\nonumber \\
&&+\frac{c_2}{2} \sum_{i}({\vec F}_i^2-2n_i)-\mu\sum_{i}n_i,
\end{eqnarray}
where $b_{\sigma i}$ 
annihilates a boson in a hyperfine state $m_F=\sigma$ at site $i$, 
$n_i$ denotes the number of particles at site $i$ and 
$ {\bf {F}}_i=\sum_{\sigma \sigma^{\prime}} b^{\dagger}_{\sigma i} 
\textbf{\textit{T}}_{\sigma \sigma^{\prime}} b_{\sigma^{\prime} i}$ 
is the spin operator at site $i$  ($\textbf{\textit{T}}_{\sigma \sigma^{\prime}}$ 
being the usual spin matrices for a spin-1 particle) and $<ij>$ denotes 
pairs of nearest neighbours in the lattice.
The first two terms in the Hamiltonian represent 
tunneling between nearest-neighbor sites 
and Hubbard interactions between bosons on the same site, respectively, 
as in the standard Bose-Hubbard model. The third term represents 
the energy associated with spin configurations within lattice sites which 
penalizes non zero spin configurations in each individual lattice. 
This term will induce distinct Mott phases 
that differ from each other in their spin correlations \cite{Imambekov03}. 
The appearance of spin mediated tunneling transitions 
in the optical lattice depends clearly on the ratio between
the different energy scales appearing on the Bose Hubbard Hamiltonian. 
We shall consider in what follows  
the polar superfluid phase characterised by $c_2>0$, and 
assume that spin-independent interactions  $c_0$ are larger 
than the spin-dependent ones $c_2$ \cite{Imambekov03}. 
The superfluid-Mott transition depends then only on the ratio $t/c_0$. 
However, once the Mott regime is achieved, if $c_2>t$ 
one should expect different magnetic ordering in 
the insulator phase due to the spin interactions. 
On the contrary, if $t\gg c_2$ tunneling will manifest
equal for all spin components and the gas 
will behave as a strongly correlated scalar gas.
For small but finite tunneling $t/c_0\ll 1$ it is possible to perform 
perturbation theory. To derive an effective Hamiltonian to second order
in $t/c_0$. We split, as usually, the full Hamiltonian as $H=H_0+H_t$ where 
$H_0=(c_0/2)n(n-1) +(c_2/2){\vec F}^2-2n-\mu n$
describes the one site unperturbed Hamiltonian and $H_t$ describes tunneling
between two adjacent sites, i.e  we consider a two site problem.
To derive the effective Hamiltonian one looks 
how the energy of the unperturbed ground states $|g,S\rangle$ 
is lowered due to tunneling: 
$\epsilon_S=-\sum_{\nu} \frac{|\langle \nu|H_t|g,S\rangle|^2}{E_{\nu}-E_g}$ 
where $\nu$ labels all the (virtual) intermediate states and $E_{\nu},\,E_g$ 
denote the unperturbed energies of the two-site state 
$|\nu\rangle$, $|g,S\rangle$ (which are non-degenerate). 
The dependence of the energy shifts on the total spin $S$  
introduces nearest-neighbor spin-spin interactions in the lattice.
It is sufficient to evaluate these shifts for only one value of $m_S$ 
because tunneling cannot mix states with different $m_S$ and the overlaps 
$|\langle \nu|H_t|g,S\rangle|$ are rotationally invariant. 
The effective Hamiltonian can be written then as: 
\begin{equation}
H_{ij}=\sum_{0}^{F_1+F_2=S} \epsilon_S P_S,
\end{equation}
where, now, the sum extends to all $F_1+F_2$ values. 
The effective Hamiltonian for $S=1$ 
in second order on $t$ can be expressed as a 
generalized quadratic Heisenberg Hamiltonian \cite{Imambekov03}:
\begin{equation}
H_{ij}=-J_0-J_1\sum_{\langle i,j\rangle }{\bf F_i F_j}-J_2\sum_
{\langle i,j\rangle}({\bf F_i}\cdot {\bf F_j})^2,
\label{effectiveF=1}
\end{equation}
where 
the explicit expressions for \(J_0\), \(J_1\), and \(J_2\) can be found in Ref. \cite{Imambekov03}.

\begin{figure}[t]
\centerline{\epsfbox{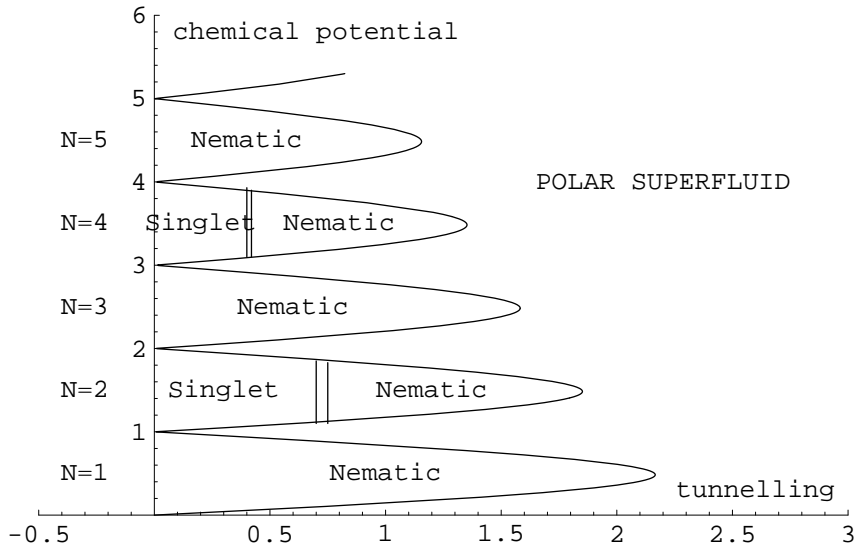}} 
\caption{General phase diagram for $F=1$ bosons in 2D and 3D optical lattices 
(from \cite{Imambekov03}).}
\label{chapter4fig1}
\end{figure}

This Hamiltonian differs from the familiar Heisenberg Hamiltonian
for spin 1/2 particles due to the presence of the quadratic 
term on the spin interaction. Higher order terms on perturbation theory 
can give rise to terms with higher powers in 
$ ({\bf F_i F_j})$ but with much smaller coefficients \cite{Yip03a}. 
We will see, however, in the next subsection, 
that higher orders in the Heisenberg interactions appear
in a perturbative treatment to second order if larger spins are used. 
We should ignore now higher order perturbative terms (i.e. $\propto t^4$ ) 
since their contribution are highly suppressed. 
We summarise the possible quantum phases in such a case. As reported by
Imambekov {\it et al.} \cite{Imambekov03}, for an optical lattice in
2D and 3D, the phase diagram can be summarized as follows. The term \(J_1\) favors ferromagnetic order,
while \(J_2\) enhances the singlet spin configuration on each bond.
To solve this competition, there is 
the possibility of having nematic states that mix states with total spin 
$S=0$ and $S=2$ (and $m_F=0$) at each
bond, but are product states:
\begin{equation}
\label{nematicF=1}
|N\rangle= \prod_i |F_i=1,m_i=0\rangle.
\end{equation}
This is a nematic state, with zero expectation value of each spin but not rotationally
invariant. 
For an odd filling
factor, the Mott insulating phase is always nematic. For even 
filling factors, there is always a spin-singlet phase, in which pairs of atoms at a site form 
singlets. 
See Fig.
\ref{chapter4fig1}.

Apart from nematic and singlet states, in 1D, an additional solution is the dimerized VBS state
\begin{equation}
|D\rangle= \prod_{i=2n} |F_i=1,F_{i+1}=1,F_i+F_{i+1}=0\rangle,
\end{equation}
in which the translational invariance is broken.


\subsubsection{F=2 gases in optical lattices}
\label{sec-Mr.Shakti_Sarkar}

The mean field states of spinor $F=2$ gases have been for the first time investigated in 
Refs. \cite{Ciobanu01,Koashi00,Ueda02}. It is worth noticing that mean field states are also valid for 
insulating Mott states with one atom per lattice site, provided all atoms are described by the same single-particle 
wave function attached to a given site. Refs. \cite{Koashi00,Ueda02} go one step further, and apart from 
the mean field theory also consider the extreme case of quenched (immobile) $F=2$ bosons in an optical lattice. 
In other words, these references characterize possible on--site states for $N$ bosons with total spin $S$ in the absence of tunneling. 

After submission of the first version of this paper, Barnett, Turner and Demler have presented a beautiful and complete 
classification of the mean field phases for arbitrary $F$, based on the 19th  century F. Klein's method of solving quintic 
polynomials by analysis of rotations of regular icosahedra \cite{barnett}. We discuss their results in more details 
in the following. Also, very recently the effective spin Hamiltonians (in the first MI lobe), and quantum insulating phases  
of $F=2$ bosons have been studied by Zhou amd Semenoff \cite{zhoulast}, using variational principle applied to product 
(Guztwiller ansatz, cf. \cite{Jaksch98}), dimer and trimer states.


\subsubsection{Bose-Hubbard model for F=2 particles}
\label{sec-radhamadhab5}

The derivation of the  Bose-Hubbard Hamiltonian for spin $F=2$ 
can be conveniently expressed \cite{Ueda02} as:
\begin{eqnarray}
H&=&-t\sum_{<ij>,\sigma}(b^{\dagger}_{\sigma i} b_{\sigma j}+b^{\dagger}_{\sigma j} b_{\sigma i})
+\frac{c_0}{2}\ \sum_{i}n_i(n_i-1)\nonumber \\
&&+\frac{c_1}{2} \sum_{i}:\textbf{\textit{F}}_i
\cdot \textbf{\textit{F}}_i:
+\frac{2 c_2}{5}\sum_i \overline{S}_{+i} \overline{S}_{-i},
\end{eqnarray}
where the $:\,\,:$ denotes normal ordering of the operators.

The ratios between the various interactions, $c_1/c_0$ and $c_2/c_0$, are fixed by the scattering lengths whereas the ratio $t/c_0$ between tunneling and Hubbard interactions can be tuned by changing the lattice parameters 
\cite{Jaksch98}. The possible ground state phases for $F=2$ have been studied 
in \cite{Zawitkowski06} using second order perturbation 
theory with $t/c_0\ll 1$. At zeroth order, the Hamiltonian is a sum of independent single-site Hamiltonians:
\begin{equation}\label{H_0}
H_{0}=\frac{c_0}{2}\ n(n-1)+\frac{c_1}{2}\,:\textbf{\textit{F}}\cdot \textbf{\textit{F}}: 
+ \frac{2 c_2}{5} \overline{S}_+ \overline{S}_-.
\end{equation}
Exact eigenstates of this Hamiltonian have been obtained in 
Ref. \cite{Ueda02}. Since $S_{\pm}$ commute with the 
total spin operator, the energy eigenstates 
can be labeled with four quantum numbers as $|N,N_S,F\rangle_{m_F}$. They 
have a $2F+1$-fold degeneracy associated with the quantum number $m_F$ and 
their energies are given by:
\begin{equation}
\label{energies}
E=\frac{c_0}{2} N(N-1) + \frac{c_1}{2}(F(F+1)-6N)+
\frac{c_2}{5}N_S(2N-2N_S+3).
\end{equation}
Assuming one atom per site, the 
effective Hamiltonian to second order in the perturbation parameter 
$t/c_0$  has the following form:
\begin{equation}\label{1particleH}
H_I^{(ij)}=-4t^2 \left[ \frac{P_0^{(ij)}}{g_0}+\frac{P_2^{(ij)}}{g_2}+
\frac{P_4^{(ij)}}{g_4}\right].
\end{equation}
Notice that $\epsilon_S=-4t^2/g_S$ indicates 
that the control and engineering of the  magnetic properties 
of the system could be easily achieved by means of e.g. optical 
Feshbach resonances.
The Hamiltonian (Eq.(\ref{1particleH})) can be easily generalized to 
the whole lattice,
$H= \sum_i H_{0,i}+\sum_{<ij>} H^{(ij)}_I$.
It can also be transformed into a polynomial of 
fourth order in the Heisenberg interaction 
$\textbf{F}_i \cdot \textbf{F}_j$:
\begin{eqnarray}
H&= &\sum_i H_{0,i}+
\sum_{<ij>}
\frac{39\epsilon_0-80\epsilon_2}{51}( {\bf{F}}_i \cdot {\bf{F}}_j)+
\frac{9\epsilon_0-8\epsilon_2}{102}( {\bf{F}}_i \cdot {\bf{F}}_j)^2+
\nonumber \\
&&\left(-\frac{7\epsilon_0}{204}+\frac{10\epsilon_2}{204}+\frac{\epsilon_4}{72}\right)
({\bf{F}}_i \cdot {\bf{F}}_j)^3+
\frac{7\epsilon_0+10\epsilon_4}{1020}({\bf{F}}_i \cdot {\bf{F}}_j)^4 .
\end{eqnarray}
 
Very recently, effective spin Hamiltonians (in the first MI lobe) and quantum insulating phases
of $F=2$ have been studied in \cite{zhoulast}.

To study the ground state phase diagram of a one-dimensional $F=2$ spinor gas
assuming that there is the freedom to modify the value of (the zero magnetic field) 
scattering lengths \cite{Zawitkowski06}, one can use Matrix Product States (MPS) methods. 
MPS represent
the ground state of a translationally invariant short range 
Hamiltonian exactly \cite{Verstraete06a}, or 
nearly exactly (see subsection \ref{subsec_quantu_raju2}). A straightforward way to do that \cite{Wolf05} is 
to add to the bond Hamiltonian (\ref{1particleH}) $c$ (\(> 0\)) times 
the identity operator $I^{(ij)}=\sum_S P_S^{(ij)}$ on the bond, 
so that the new Hamiltonian  $H'=H_I^{(ij)}+c I$ 
becomes positive definite,
\begin{equation}
\label{1particH}
{H'}_I^{(ij)}=\sum_{S=0}^4\lambda_S P_S^{(ij)},
\end{equation}
{\it i.e.}, all $\lambda_S$ are non-negative. In particular, 
$\lambda_1=\lambda_3=c$ and $\lambda_0=c-4t^2/g_0$,
$\lambda_2=c-4t^2/g_2$ and $\lambda_4=c-4t^2/g_4$. 
Since by definition,
$\lambda_1=\lambda_3$ are the largest parameters, 
the ground state will obviously belong to the symmetric subspace.



Before we proceed, it is worth to discuss the mean field phase 
diagram obtained under the assumption that the ground state is a product state, $|\Psi\rangle=|e,e,\ldots\rangle$ (see \cite{Ciobanu01,barnett}). 
There are 3 possible mean field (i.e., product) ground states, with $|e\rangle$ given (up to $SO(3)$ rotations) by:
\begin{itemize}
\item Ferromagnetic state, $|e\rangle=(1,0,0,0,0)$; possesses only the $U(1)$ symmetry of rotations around the $z$--axis, and has 
maximal projection of the spin onto $z$ axis.
\item nematic state, has the $\eta$--degeneracy, $|e\rangle=(\sin(\eta)/\sqrt{2},0,
\cos(\eta),0,\sin(\eta)/\sqrt{2})$. This state is a MI version of the polar state in BEC; it has mean value of all components of the 
spin equal zero, but non vanishing singlet projection $\langle\rm singlet|e,e\rangle\ne 0$;
\item Tetrahedratic (cyclic) state, $|e\rangle=(1/\sqrt{3},0,0,\sqrt{2/3},0)$; this is a MI version of the cyclic state. The state 
may be uni- or biaxial, depending on whether the nematic  tensor does, or does not have a pair of degenerated eigenvalues; 
it has vanishing of both, of mean values of all of the spin components, and of the singlet projection. 
\end{itemize}
The mean field phase diagram corresponds to:  
\begin{itemize} 
\item a ferromagnetic state for $\lambda_4=0$, $\lambda_2,\lambda_0>0$, and for $\lambda_0=0$, provided $\lambda_2\ge 17\lambda_4/10$;
\item a nematic state  for $\lambda_0=0$, provided $3\lambda_4/10\lambda_2\le 17\lambda_4/10$;
\item a cyclic state for $\lambda_0=0$, provided $\lambda_2\le 3\lambda_4/10$, and for $\lambda_2=0$.
\end{itemize}

In 1D, since the Hamiltonian is a sum of nearest neighbor bond Hamiltonians, we have
$\sum_{k,k'}\langle e_k,e_k\ldots|\hat{H_I}|e_{k'},e_{k'}\ldots\rangle\propto\langle e_k|e_{k'}\rangle^{N-2}$ and thus in the limit of an infinite chain
the ground states are equally well 
described by product states 
(that will typically break the rotational symmetry). 
This means in this case we expect mean field product states to provide a very good approximation of the ground states with translational symmetry. 
Below we present a schematic classification 
of the possible ground states for the specific $\lambda_i$ values \cite{Zawitkowski06}.

\noindent (\(A\)) For $\lambda_4=\lambda_2=\lambda_0=0$, all symmetric states are ground states, i.e., in particular all product 
states $|e,e\ldots\rangle$ 
with arbitrary $|e\rangle$.

\noindent ($B$) For $\lambda_4= \lambda_2=0,\ \lambda_0 >0$, 
the ground states $|e,e\ldots\rangle$ remind the cyclic states 
states of Ref.\cite{Demler02} (i.e., they correspond to translationally, but nor rotationally invariant product states), which now mix 
$S=2$ and $S=4$ contributions on each bond, and they have to fulfill the condition $\langle {\rm singlet}|e,e\rangle=0$. 
Denoting by $|e\rangle=(e_{2}, e_1,e_0, e_{-1}, e_{-2})$, this implies $e_0^2-2e_{1}e_{-1}+2e_{2}e_{-2}=0$. 
These states form a much greater class than the cyclic ones, since they may have non-vanishing (and even maximal) components of the spin. 
Interestingly, the transition between the cyclic phase for $\lambda_2=0$, and the ferromagnetic phase for $\lambda_4=0$, occurs via such states, 
i.e., at the transition point the degeneracy of the ground states manifold explodes. 

\noindent ($C$) For $\lambda_4=0$ and 
$\lambda_2,\,\lambda_0 >0$, the ground states are ferromagnetic states  $|2\rangle_{{\bf n}}|2\rangle_{{\bf n}}\cdots|2\rangle_{{\bf n}}$, 
corresponding to
a maximal projection of the local spin onto a given direction 
${\bf n=(\sin(\theta)\cos(\phi),\sin(\theta)\sin(\phi), \cos(\theta))}$.
Such vectors for $F=2$ may be parametrized (in the basis of $\hat{\bf F}_{\bf n}$ with descending
$m_F$) as
$\propto(z^{-2},2z^{-1},\sqrt6,2z,z^2)$ with $z=|z|e^{i\phi}, \,\, 
|z|\in(- \infty,\infty)$. It should be stressed that ferromagnetic states are {\it exact} ground state in the entire 
part of the phase diagram whenever $\lambda_4=0$. 

\begin{figure}[t]
\centering
\includegraphics[width=0.8\textwidth]{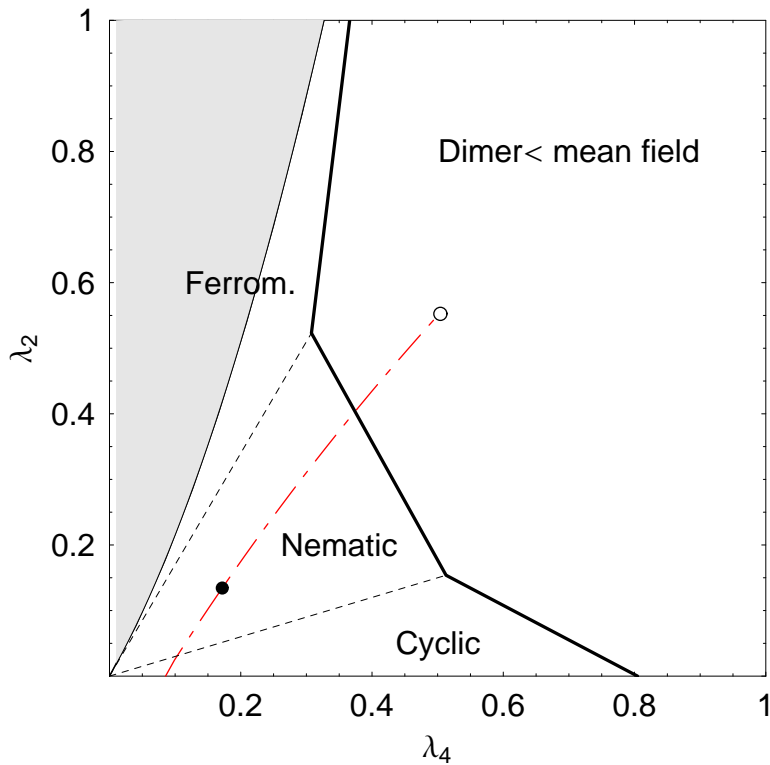}
\vspace*{0.5cm}
\caption{(Color online.) Sketch of the phase diagram,
obtained by applying the variational principle 
in the $\lambda_2, \,\lambda_4$ phase space 
(for $\lambda_0=0$) to mean field, N\'eel, and dimer states with one atom per site. 
The scale is set by letting $\lambda_1=\lambda_3=1$.
N\'eel-type states are never favorable over nematic states.  The {\it ferromagnetic} region (gray) was obtained  
numerically by imaginary time evolution of MPS (and comparing 
the results from runs with $D=1$  and $D=5$)  
in a chain of $50$ sites with open boundary conditions \cite{Verstraete04b}. 
(Note that here \(D\) is 
the dimension of the ancillary system that is used in the MPS method.) 
Of course,  
on the line $(\lambda_4=0,\lambda_2)$ ferromagnetic states give always ground states. 
Dashed lines
indicate the regions where the type of mean field state with lowest energy changes qualitatively.
The red (dashed-dotted) line gives the values of $(\lambda_2,\lambda_4)$ which can be obtained by changing the
spin-independent scattering length $c_0=(3  g_4+4  g_2)/7$ through optical Feshbach resonances.
\cite{Fedichev96,Theis04}.
Black and white circles indicate a change of $ c_0$ of $10$\% and $100$\%, respectively.
}
\label{chapter4fig2}
\end{figure}

\noindent ($D$) For $\lambda_0=0$ and
$\lambda_4,\,\lambda_2 >0$, the ground states apparently favor 
antiferromagnetic order. This, however, can be misleading  if $\lambda_4\ll\lambda_2$. In that case, as the mean field diagram suggests, the 
ferromagnetic order might prevail. We have applied in 1D a more general variational approach, going beyond mean field. We have looked for ground 
states by 
applying the variational principle to mean field (product)  states $|e,e\ldots\rangle$, 
N\'eel-type states $|e,f,e,f\ldots\rangle$, 
and  valence bond solid states with singlet states for distinct pairs (dimers) of neighboring atoms and translational dimer symmetry.
For the mean field case as discussed earlier
the energy is either minimized by the ferromagnetic state $|e\rangle=|2\rangle_{\bf n}$ (for $\lambda_2\ge 17\lambda_4/10$), by a nematic state
 $|e\rangle=|0\rangle_{\bf n}$ (for $3\lambda_4/10 \lambda_2\le 17\lambda_4/10$; in this case the state is a combination of total spin $0,\,2$ and 
$4$), or, for $\lambda_2\leq3\lambda4/10$, by a cyclic state, $|e\rangle=(e_2,e_1,e_0,e_{-1},e_{-2})$ with 
$e_0=1/\sqrt2,\,e_2=-e_{-2}=1/2,\,e_1=e_{-1}=0$. 
Imposing N\'eel order with $\langle e|f\rangle\neq1$ always results in a larger energy, as $\lambda_{1,3}>\lambda_{2,4}$, and the overlap
with the singlet can be maximized already by restricting to product states. On the other hand, for the dimer state 
the energy per bond is given by $\frac12{\rm Tr}(H_I\frac1{25} 1\otimes 1)$. 
One can apply MPS codes \cite{Zawitkowski06} to  
search numerically for the exact ground states in 1D using the method of \cite{Verstraete06a} to confirm the existence and phase borders of the
ferromagnetic ground state, nematic and dimer regions. This particular phase diagram in the $\lambda_2, \lambda_4$ phase space 
is displayed in  Fig. \ref{chapter4fig2}.

\noindent ($E$) For $\lambda_2=0$ and $\lambda_4,\,\lambda_0 >0$, 
as in the ($D$) case, mean field cyclic  states are favorable over N\'eel states. 
One can compare them  variationally to the analogues of the dimer states in the present case, i.e., 
 configurations which have a state with total spin $S=2$
on distinct bonds. We call these state {\it para--dimers}. 
Now, contrary to the dimerized states in (\(D\)), where the states on the bond are unique, 
here states with different $m_{S=2}$ can form superpositions. It is easy to see, however, 
that the best superposition should have maximal possible entanglement, 
in order to minimize interaction on the bonds that are not covered by the para--dimers\cite{Zawitkowski06}.

\subsubsection{Spinor Fermi gases in optical lattices} 
\label{sec-Mrs.Sabita_Bardhan}
We move now to fermionic spinor gases. For $F=1/2$, the quantum simulator of the fermionic Hubbard model has been proposed in \cite{Hofstetter02} and recently studied with 3-component fermions in \cite{paananen}.


Liu {\it et al.} \cite{Liu04} proposed to use fermions with high $F$ to realize spin-dependent Hubbard models, 
in which hopping parameters are spin-dependent. Such models lead to exotic kinds of superfluidity, such as to a phase 
in which SF and normal component coexist at zero temperature. W. Hofstetter and collaborators have written a series of papers, reviewed in \cite{hofstetter2}, on fermionic atoms with $SU(N)$ symmetry 
in optical lattices. Such systems also  have exotic superfluid and flavor-ordered ground states, and exhibit a very rich behavior in the presence of disorder. It is, of course, inevitable to ask which atoms can be used to realize high $F$ fermonic spinor gases in optical lattices. The most commonly used alkali $^6$Li has hyperfine manifolds with $F=1/2$ and $F=3/2$. The latter is, obviously, 
a subject of two body losses, but as in the case of the $F=2$ manifold of Rubidium, one can expect reasonably long life time 
in the lattice (especially in MI states with $N=1$). Another commonly used fermion is a heavy alkali $^{40}$K, which has manifolds 
$F=7/2$ and $F=9/2$. These fermions are particularly useful for spin-dependent Hubbard models 
\cite{Liu04}. There are several atoms whose lowest hyperfine manifold is $F=3/2$, i.e., in those ground states two body 
losses can be avoided: $^9$Be, $^{132}$Cs, or $^{135}$Ba, but so far only the bosonic Cesium BEC has been achieved 
\cite{rudi}. On the other hand, recently a BEC of $^{174}$Yb atoms \cite{ybbec}, as well as degenerate gas of $^{173}$Yb 
fermions with $F=5/2$, has been realized. Finally, fermionic Chromium has 4 hyperfine manifolds with
$F=9/2,7/2,5/2,3/2$, in the ascending order of energies, and after achieving BEC of the bosonic Chromium \cite{Griesmeyer05}, 
the prospect for achieving ultracold degenerate spinor Fermi gases are very good.

Recently, there has been a lot of progress in understanding the special properties of $F=3/2$  and $F=5/2$ lattice
Fermi gases \cite{Zawitkowski06,tu,wu3}. C. Wu {\it et al.} \cite{wu1} realized that the spin-3/2 fermion models with contact 
interactions have a
generic $SO(5)$  symmetry without any fine-tuning of parameters, and employed this fact to form  a quantum Monte Carlo algorithm 
free of the sign-problem and to study  novel competing orders in 1-dimensional optical traps and lattices \cite{wu2}. In particular,
the quartetting phase, a four-fermion counterpart of Cooper pairing,
exists in a large portion of the phase diagram.  The bosonisation approach was applied to 1D systems with 
$F=3/2,\,5/2,\,\ldots$ by Lecheminant {\it et al.} \cite{Lecheminant05}, and exact Bethe ansatz in \cite{Controzzi06}. An overview of
 hidden symmetries and competing orders in spin 3/2 gases can be found in the excellent paper of C. Wu \cite{wureview}.





\section{Ultracold atomic gases in ``artificial" magnetic fields}
\label{artificial}

\subsection{Introduction -- Rapidly rotating ultracold gases}
\label{rapidly}

Quantum systems in magnetic fields exhibit particularly interesting behavior, with
fractional quantum Hall effect (FQHE) being a paradigm \cite{Prange87}. 
It is very well known that  rapidly 
rotating trapped ultracold gases provide a possibility to realize ``artificial" 
(standard, or better to say, Abelian) magnetic fields.

In order to achieve such magnetic fields, one typically 
considers a quasi single component (``spinless") 2D gas of $N$ atoms in the  
$XY$ plane rotating
around the $Z$ axis with frequency $\Omega$, and confined in a
harmonic trap of frequency $\omega_{\perp}$. If the rotation is moderate $\Omega <\omega_{\perp}$, in macroscopic atomic clouds, the Abrikosov vortex lattice is formed \cite{Madison00,Abo-Shaeer01}. As $\Omega$
approaches $\omega_{\perp}$, the vortex lattice melts,
and the system evolves through a sequence of states that have
been identified in literature \cite{Wilkin98,Wilkin00,Paredes01} as
highly correlated quantum liquids.

The various regimes of rapidly
rotating gases can be described in the terminology of fractional
quantum Hall effect (FQHE) theory \cite{Prange87}.
The crucial role is played by the direct analog of the Landau
level filling factor in the FQHE, which can be related to the number
of vortices $N_v$ by $\nu=N/N_v$ as defined in the BEC mean field
description, which is valid for large systems and moderate rotations.

In the first literature \cite{Wilkin00,Paredes01}, the authors considered only 
the lowest Landau level (LLL) for strong enough rotation. Later, attention was paid to 
edge excitations and topological order \cite{Cazalilla05}.
Recently,
correlated liquids at $\nu=k/2$ for $k=1,2,3,\ldots$ for
$\nu\le \nu_c\simeq 6-10$ have been discussed \cite{Cooper01,Sinova02,Regnault04a,Regnault04b}.
These states resemble, to a great extent, the states  from the
Rezayi-Read (RR) hierarchy \cite{Read99}: $k=1$ is the Laughlin state,
$k=2$ is the Moore-Read paired state \cite{Moore91,Cazalilla05}, etc.
Rezayi et al. \cite{Rezayi05} have recently shown that the presence of a small amount of
dipole-dipole interactions unambiguously makes the RR state with $k=3$
the ground state.
This state is particularly interesting, since its excitations are
both fractional, and non-Abelian.
The validity of the LLL approximation for rotating gases is
also discussed in \cite{Morris06}.
Recently, a composite fermion (boson + one flux quantum) theory of rapidly rotating bosons has been formulated \cite{Regnault06}.

\medskip

Most of the literature on ultracold rapidly rotating gases aims at
considering relatively large systems and even the thermodynamic
limit. Thus, in numerical simulations, either periodic (torus) or
spherical boundary conditions are used. Stability of the cloud in an harmonic trap 
requires $\Omega<\omega_{\perp}$, since otherwise centrifugal forces drag the atoms away from the trap. 
Observation of the
Laughlin states requires, on the other hand,
to remain in  the LLL. This in turn can be assured only if $\Omega-\omega_{\perp}=O(1/N)$, i.e., 
it requires a very precise control of the delicate
 balance between $\Omega$ and $\omega_{\perp}$. Unfortunately, in the case of contact (short range Van der Waals) 
interactions pseudo-hole excitations of the Laughlin state
have vanishing interaction energy, similarly as the Laughlin state itself; they can differ only by angular momentum 
contribution $\propto \Omega-\omega_{\perp}=O(1/N)$, i.e. they vanish at large $N$. 
Despite the progress in experimental studies of
vortex lattices \cite{Coddington04,Schweikhard04}, and first steps toward
reaching the LLL physics \cite{Schweikhard02,Simula05,Stock05}, experiments have
not yet reached this regime. 

\medskip

The problems related to the short range nature of the Van der Waals
forces can be overcome by using optical traps more stiff than harmonic, which is technically difficult, but
in principle possible.
 Yet another promising idea is to use  dipolar gases, i.e. gases that interact
via magnetic or electric dipole moments
(for a review see \cite{Baranov02}). Rotating dipolar bosonic gases are
expected to exhibit exotic behaviour
in the weakly interacting regime \cite{Cooper05}, whereas
fermionic dipolar gases  have a finite gap for
the $\nu=1/3$ Laughlin state \cite{Baranov05}. The first observation
of BEC of a dipolar gas of  Chromium atoms
with large magnetic dipole has been recently reported \cite{Stuhler05},
and several groups are trying to achieve an ultracold
gas of heteronuclear molecules with large electric dipole moments
\cite{Eur04}. The recent proposal of  using polar molecules excited to the first rotational state is 
also very promising \cite{Micheli06}.

Completely different approach to create ``artificial" magnetic fields employ the effect of electromagnetically
induced transparency (EIT) and ``slow light"; it  has been proposed by Juzeliunas 
and \"Ohberg \cite{Juzeliunas04,Juzeliunas05a,Juzeliunas05b,Juzeliunas05c}. 
This method is very flexible, and can be used to reach
the integer quantum Hall effect \cite{Ohberg05}, and even regimes of FQHE in planar 
geometries \cite{Juzeliunas05c},
or generate non-Abelian gauge fields \cite{Ruseckas05}.

Nevertheless, atomic gases in optical lattices, or arrays of optical traps provide 
perhaps the best opportunity for  FQHE. There are two ways of reaching the regime of Laughlin liquids. 
A very  promising one, consists in  using atoms in an array of rotating optical
microtraps, either in an optical lattice \cite{Bloch05}, or
created by an array of rotating traps generated by an array of microlenses
 \cite{Birkl01}, or an array of laser beams \cite{Yavuz06}. 

\begin{figure}[t]
\centerline{\epsfbox{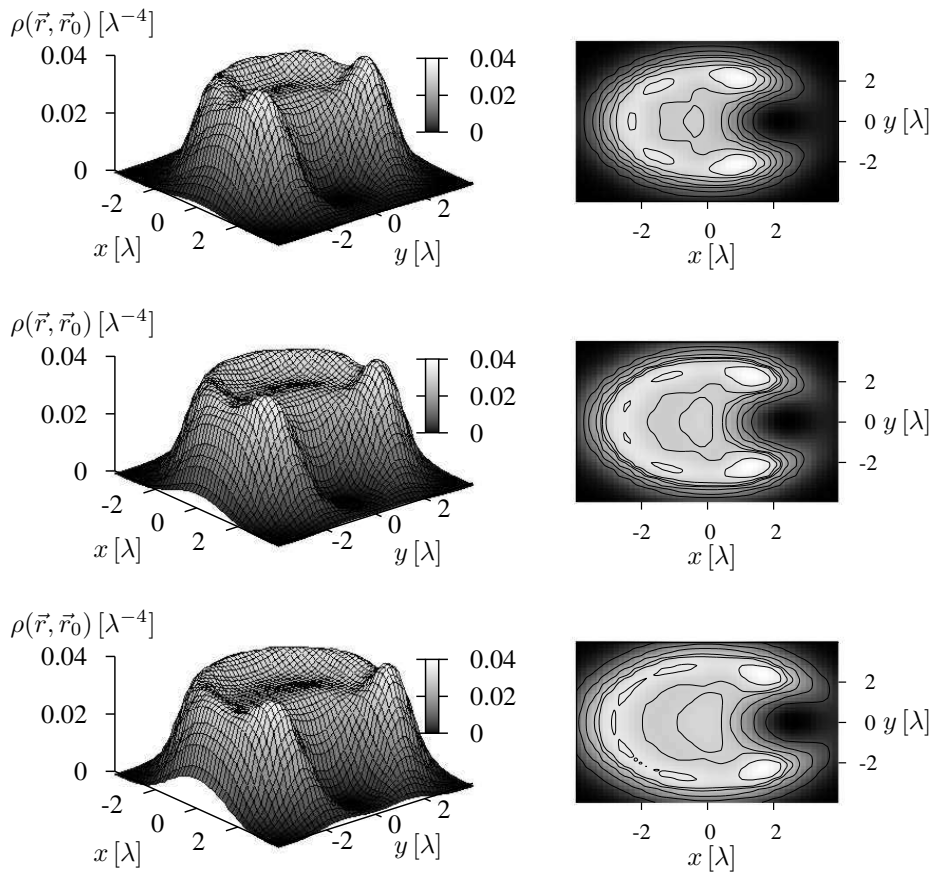}} 
\caption{3D (left) and contour (right) plot of the pair correlation function $\rho(r,r_0)$ for the Laughlin 
states of $N=$6,7 and 8 atoms of mass $m$ in a lattice site; $r_0=\sqrt{N}$ in harmonic oscillator units and $\lambda=\sqrt{\hbar/2m\omega{\perp}}$ (from \cite{Barberan06}). }
\label{figure7.1}
\end{figure}

In such arrangements, it will be natural to study independent mesoscopic or even
microscopic systems of few atoms at each lattice site, i.e., avoid the problems of large $N$. 
The lattice will play, however, an important role. First of all, one can prepare equal number 
of atoms at each site using the Mott transition, and then turn the intersite hopping off. Second, 
the measurement and detection procedures could  automatically be  applied to many copies of the same 
meso--, or micro--system.   Such experiments demand careful theoretical studies of few atom
systems using possibly exact methods, such as exact
diagonalisations of the Hamiltonian with open boundary conditions
in  the presence of the harmonic trap, or even a deformed trap.
Such studies have been initiated recently. Popp et al. \cite{Popp04} analyzed in detail the 
possibilities of an adiabatic path to fractional quantum Hall states
of  few bosonic atoms. The papers of Barberan et al. \cite{Barberan06} discuss the appearance of ordered structures 
(vortices, vortex arrays, Wigner crystals) in systems of few bosonic atoms by looking at density and pair-correlation functions. 
In Fig. \ref{figure7.1}, an example of such structures pair correlation function in the Laughlin states is shown.  
Similar studies (going even beyond LLL approximation) for bosonic and fermionic atoms have been reported in Ref. \cite{Consta}.

Another way to create highly-correlated liquids could be to
mimic effects of magnetic fields not by rotation,
but by appropriately designed control of tunnelling in optical
lattices. This is discussed in the next subsection. 

\subsection{Lattice gases in ``artificial" Abelian magnetic fields}
\label{abelian}

In order to create an effect of constant magnetic field in a 2D square lattice one needs to be able to control the 
phases of the hopping matrix elements. Single atom stationary Schr\"odinger equation, analyzed in the famous paper 
by Hofstadter \cite{Hofstadter76},
 reads:
\eqn{\label{spwaveequation}
      -t e^{-\ci\frac{{\rm e}a}{\hbar c}A_x}\psi(x\!+\!a,y)
     -t e^{\ci\frac{{\rm e}a}{\hbar c}A_x}\psi(x\!-\!a,y)&&\\
     - t e^{-\ci\frac{{\rm e}a}{\hbar c}A_y}\psi(x,y\!+\!a)
     - t e^{\ci\frac{{\rm e}a}{\hbar c}A_y}\psi(x,y\!-\!a)
     &\!\!=\!\!&E\psi(x,\,y)\nonumber,}
where $\psi(x,\,y)$ is the wave function, $a$ is the lattice constant, and $\vec{A}$ is the vector potential. 
The choice of $\vec{A}$  determines the magnetic field $\vec B$, which in turn
determines the behavior of the system.
For uniform field in the $z$ direction, $\vec B=B\vec{\rm e}_z$,
one may choose  the potential $\vec{A}=(0,\,Bx,\,0)$, and  
tunnelings in the $y$-direction acquire phases.
This makes the problem effectively one-dimensional
and Eq. (\ref{spwaveequation}) can be transformed into the Harper's
equation \cite{Harper55}:
\eqn{g(m\!+\!1)-g(m\!-\!1)+2\cos(2\pi m\alpha-\nu)g(m)
=\varepsilon g(m)\label{HarperEquation}}
by using the ansatz $\psi(ma,\,na)={\rm e}^{\ci\nu n}g(m)$, where
$x=ma$, $y=na$, and $\varepsilon=-E/t$.

\begin{figure}[t]
\centering
\includegraphics[width=0.75\textwidth]{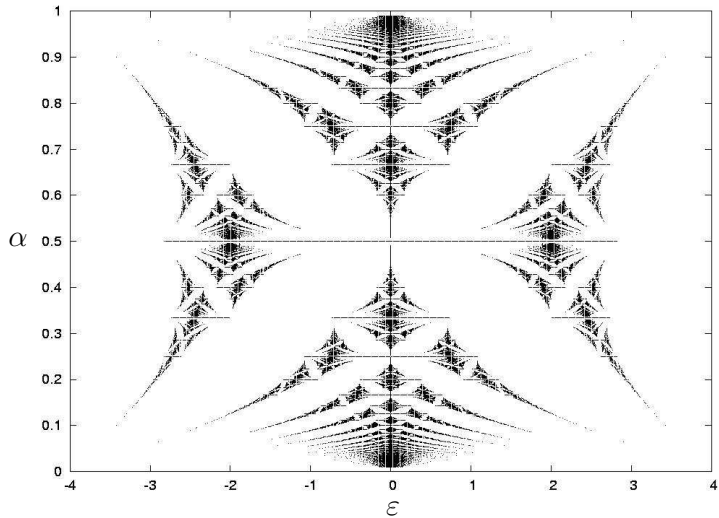}
\caption{The famous Hofstadter butterfly: energy bands  plotted for rational values of the magnetic flux $\alpha$.}
\vspace*{0.5cm}
\label{figure7.2}
\end{figure}

The eigenvalue problem for rational values of the magnetic 
flux $\alpha=\frac{{\rm e}a^2B}{hc}$
per elementary plaquette becomes periodic.
This results in a band spectrum whose bands   form the famous
Hofstadter butterfly (see Fig. \ref{figure7.2}).  Note that the regime
of this spectrum requires finite values of $\alpha$, i.e.,
magnetic fields $B\sim 1/a^2$, which in the continuum limit $a\to 0$
become ultra-intense.

Jaksch and Zoller \cite{Jaksch03}, were the first who 
have recently proposed methods to realize such ``artificial'' magnetic
field effects in lattice gases. Their method   employs  atoms with two internal
states trapped in two lattices, laser assisted tunneling, lattice tilting (achieved by acceleration, or application of 
external static electric fields), and other
experimentally accessible techniques. We discuss the generalization of this proposal to the case of non-Abelian fields 
in the next subsection. 

The physical features in ``artificial'' magnetic fields are, as expected 
extremely rich.  For instance, one may load the lattice with a cold BEC, or add disorder, 
 and observe how the single atom properties of the  
 spectrum changes 
in weakly interacting or
weakly disordered  systems.  The modifications of the
butterfly due to interactions and disorder, or both respectively,
can be measured, as proposed in \cite{Jaksch03}.

Several other groups made alternative proposals for ``artificial electromagnetism for neutral atoms". 
E.J. Mueller \cite{Mueller04} generalized the method of \cite{Jaksch03} to atoms with 3 internal states in  three lattices. 
The advantage of this approach is that it does not use lattice tilting, but on expense of more complicated laser configuration. 
Since the fields are ``artificial" they do not even need to fulfill Maxwell equations, 
and can thus lead to physical realisations of otherwise unphysical effects. Mueller proposes for instance time dependent 
realisations of hopping matrix element on a 1D ring that leads to ``Escher staircase": a single particle in such ring 
undergoes acceleration limited only by the Umklapp process: when the de Broglie wavelength of the particle becomes 
equal to the lattice constant, the matter wave is Bragg reflected off the lattice and reverses its direction. 

Sorensen et al. \cite{Sorensen05} have proposed yet another method that employs also time dependent hopping matrix elements along with a large oscillating quadrupolar potential. These authors performed exact diagonalisations of the 
resulting Bose--Hubbard model in the limit of hard core bosons for $N$ up to 5. In the limit of small $\alpha$ and low densities, 
the continuum limit may be applied and the model reduces then to a system of bosons in a magnetic field
with contact interactions. For filling factor $1/2$ (total angular momentum $L=2N$), the Laughlin wave function is then  the exact ground   
state \cite{Haldane83}.  The numerical calculations show that the exact ground state remains 
very similar to the Laughlin state for $\alpha<0.3$. Preparation of such Laughlin state in the lattice may be 
achieved by creating first a Mott state with one atom per site in a superlattice with the larger period. In this way  
low  density is achieved, while the atoms are quenched and not affected by the ``magnetic" field. By turning off the 
superlattice, the Mott state melts into a Laughlin liquid.

Very recently, Palmer and Jaksch studied high field FQHE in optical lattices. They considered the value of $\alpha\simeq\alpha_c=l/n$, where $l,n$ are small integers, and derived the corresponding effective Hamiltonian 
in the continuum limit, which
for $\alpha_c=1/2$ reduces to a model similar to a bilayer FQHE system \cite{Ezawa00}. The corresponding ground state 
 in homogeneous systems (in the absence of trapping potential) is,  the,  so called, 221 state, 
constructed as the Laughlin  state for $\nu=1/2$ for particles from the same layer, and for $\nu=1$ for particles from different layers.
Denoting the coordinates  by $z_i, s_j$, the wave function for $N=2M$ atoms is
\begin{equation}
\Psi(z_1,\ldots, z_M,s_1,\ldots,s_M)\propto \prod_{i\ne j}(z_i-z_j)^2\prod_{i\ne j}(s_i-s_j)^2\prod_{i, j}(z_i-s_j).
\end{equation}
The angular momentum of this state is $L=2M(M-1)+M^2$, and filling factor $\nu=2/3$. 
Interestingly, the Hall current in such states in the presence of the trap might exhibit unexpected sign changes. 

Bukov and Demler \cite{Burkov06} have recently studied ``vortices'' in a dice lattice, induced by ``magnetic field'' and have 
shown fascinating possibilities of creation of vortex-Peierls state (a bosonic analog of valence bond solids).
Rotating optical lattices has been considered by several authors: Polini \emph{et al.} \cite{polini} proposed a realization of a fully frustrated 
XY model with cold atoms in optical lattices, while Bhat \emph{et al.} \cite{Bhat} studied ground states of various symmetries for rotating 
\(4 \times 4\) lattices. 
The somewhat related problem of vortex configuration in systems of rotating ultracold atoms in optical lattices has been studied
by Wu \emph{et al.} \cite{sombhoboto_prothom_kaj}, who finds that near the superfluid-Mott insulator 
transition, the vortex core has a tendency to approach the Mott insulator state.

\subsection{Lattice gases in ``artificial" non-Abelian magnetic fields}
\label{non-abelian}

In a recent Letter \cite{Osterloh05}, it has been shown that by using atoms with more internal states,
the application  of state dependent, laser assisted tunneling
(cf. \cite{Liu04}), and coherent transfer between
internal states, one can  generalize results of Jaksch and Zoller  and
create ``artificial external magnetic fields''
 corresponding to non-Abelian $U(n)$, $SU(n)$, or even $GL(n)$ gauge fields.
In this case, the tunneling amplitudes are replaced by
(unitary) matrices whose product
around a plaquette is non-trivial and its mean trace (Wilson loop)
is not equal to $n$ \cite{Montvay97,Rothe98}.

We consider an atomic gas in a  3D optical lattice and
assume that tunneling is completely suppressed in the $z$-direction,
so that, effectively, we deal with an array of 2D lattice gases
and we are able to restrict ourselves to one copy.
The atoms occupy two
internal hyperfine states $|g\rangle$, $|e\rangle$, and the
optical potential traps them  in the state $|g\rangle$ and $|e\rangle$
 in every second column, i.e., for the $y$ coordinate equal
to $\ldots,n-1,n+1,\ldots$ ($\ldots,n,n+2,\ldots)$. The resulting
 2D-lattice has
thus the spacing $\lambda/2$ ($\lambda/4$) in the $x$- ($y$-) direction.
The tunneling rates in the $x$ direction are due
to kinetic energy; they are spatially homogeneous and assumed to be equal for
both hyperfine states. The lattice is tilted in the $y$-direction,
which introduces an energy shift $\Delta$ between neighboring columns.
Tilting can be achieved by accelerating the lattice, or by placing it in a
static electric field.
By doing this, standard tunneling rates due to kinetic energy are suppressed
in the $y$ direction. Instead, tunneling is laser assisted, and driven
by two pairs of lasers resonant for Raman transitions between
$|g\rangle$ and $|e\rangle$, i.e., $n\leftrightarrow n \pm 1$.
This can be achieved because the offset energy for both
transitions is different and equals $\pm \Delta$.
Detunings of the lasers are chosen in such a way that the effect of tilting is
cancelled in the rotating frame of reference.
The lasers  generate  running waves
in the $\pm x$-direction, so that the corresponding tunneling rates
acquire local phases $\exp(\pm iqx)$.

In order to realize ``artificial'' non-Abelian fields in a similar scheme,
one may  use atoms with  degenerate Zeeman sublevels
in the hyperfine ground
state manifolds,  $|g_i\rangle$, and $|e_i\rangle$ with $i=1,\ldots, n$,
whose degeneracy is  lifted in external magnetic fields.
These states may be thought of as ``colors'' of the gauge fields.
Promising fermionic candidates with these properties are heavy Alkali atoms,
for instance,  $^{40}$K atoms in states $F=9/2, m_F=9/2,7/2, \ldots$,
and $F=7/2, m_F=-7/2, -5/2, \ldots$; in particular, they allow for
realizing ``spin'' dependent lattice potentials
and hopping \cite{Liu04}.

Having identified the ``colors'', one modifies
the scheme of Ref. \cite{Jaksch03}: laser assisted tunneling rates
along the $y$-axis should depend on the internal state, although not
necessarily  in the sense of  Ref. \cite{Liu04}.  
For a given link $|g_i\rangle$ to
$|e_i\rangle$, tunneling should be  described by a non-trivial
unitary matrix $U_y(x)$ being a member of the ``color'' group
($U(n)$, $SU(n)$, $GL(n)$ etc.). For unitary groups,
the tunneling matrix $U_y(x)$ can be represented as
$\exp(i\tilde\alpha A_y(x))$.
Here, $\tilde\alpha$ is real, and $A_y(x)$ is a
Hermitian matrix from the gauge algebra,
e.g., $U(n)$ or $SU(n)$. Since transitions 
from $|g_i\rangle$ to $|e_i\rangle$ correspond to different frequencies
for each $i$, they are driven by different running wave lasers, and may
attain different phase factors $\exp(\pm i q_ix)$.

\begin{figure}[t]
\centering
\includegraphics[width=0.75\textwidth]{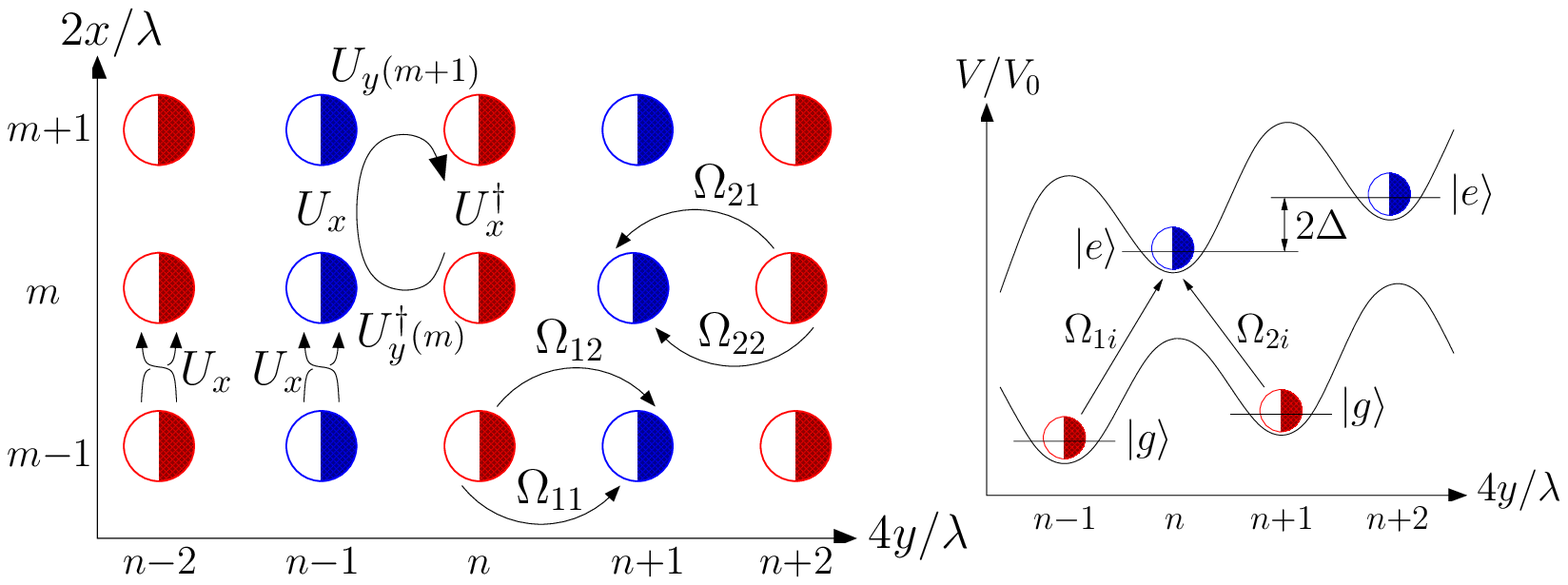}
\caption{{(Color online, see \cite{Osterloh05}).\it Optical lattice setup for U(2) gauge fields:}
Red and blue open semi-circles (closed semi-circles) denote atoms in states
$|g_1\rangle$
and $|g_2\rangle$, respectively ($|e_1\rangle$ and $|e_2\rangle$).
Left) Hopping in the $x$-direction is laser assisted and allows for
unitary exchange of colors; it is described by the same unitary hopping matrix
$U_x$ for both $|g_i\rangle$
and $|e_i\rangle$ states. Hopping along the $y$-direction is also
laser assisted and attains ``spin dependent'' phase factors.
Right) Trapping potential in $y$-direction.
Adjacent sites are set off by an energy $\Delta$ due to the lattice
acceleration, or a static inhomogeneous electric field. The
lasers $\Omega_{1i}$ are resonant for transitions $|g_{1i}\rangle
\leftrightarrow|e_{2i}\rangle$, while $\Omega_{2i}$ are resonant for
transitions between $|e_{1i}\rangle
\leftrightarrow|g_{2i}\rangle$ due to the offset of the lattice sites.
Because of the spatial dependence of $\Omega_{1,2}$ (running waves in
$\pm x$ direction)  the atoms hopping around the plaquette get the
unitary transformation $ U=U^{\dag}_y(m)U_xU_y(m+1)U^{\dag}_x$,
where $U_y(m)=\exp(2\pi i m \ {\rm diag}[\alpha_1,\alpha_2])$,
as indicated in the left figure (from \cite{Osterloh05}).}
\vspace*{0.5cm}
\label{figure7.3}
\end{figure}

In order to create gauge potentials that cannot simply be reduced
to two independent Abelian components, tunneling in the
$x$-direction should be described by a tunneling  matrix $U_x$, which fulfills  t $[U_x, U_y(x)]\ne 0$,
so that a genuine non-Abelian character of the fields is assured.  
  We stress that all elements of this scheme,
as  shown in Fig.~\ref{figure7.3}, are 
experimentally accessible.

For the specific gauge fields considered in Ref. \cite{Osterloh05} one obtains the 
following $U(2)$ generalization of the Harper's equation:
\begin{equation}
\sigma_x g(m\!+\!1)-\sigma_xg(m\!-\!1)+2\cos(2\pi m \alpha-\nu)g(m)
=\varepsilon g(m)
\label{GenHarperEquation},
\end{equation}
where $g(m)$ is the two component wave function obtained by using the ansatz 
 $\psi(ma,\,na)={\rm e}^{\ci\nu n}g(m)$, with
$x=ma$, $y=na$, and $\varepsilon=-E/t$; $\sigma_x$ is the Pauli matrix, and the ``magnetic flux" 
matrix $\alpha={\rm diag}[\alpha_1,\alpha_2]$.

\begin{figure}[t]
\centering
\includegraphics[width=0.75\textwidth]{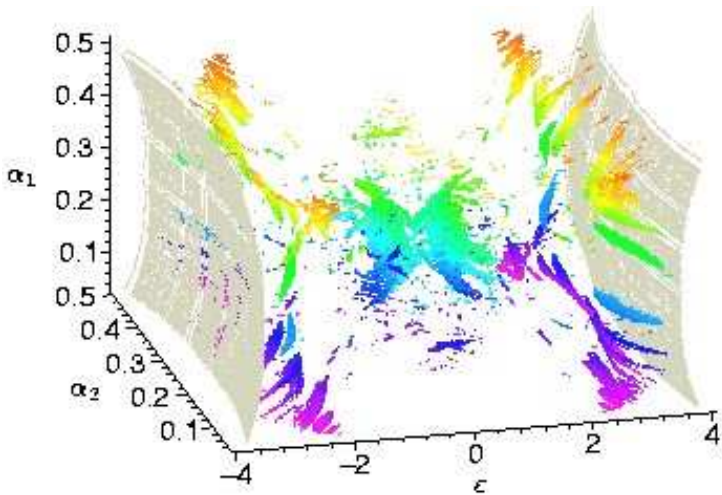}
\caption{(Color online). The Hofstadter ``moth'' spectrum.
Forbidden eigenenergies $\varepsilon$ are plotted versus
$\alpha_i=p_i/q_i,\in[0,\,=0.5]$ ($i=1,2$), where $q_i\leq 41$
and $\alpha_1\neq\alpha_2$ (from \cite{Osterloh05}).}
\vspace*{0.5cm}
\label{figure7.4}
\end{figure}

Given each $\alpha_i=p_i/q_i$ rational, the problem is $Q$-periodic
(where $Q$ equals the smallest common multiple of $q_1$ and $q_2$).
The spectrum shows a band structure,
 and is bounded by two hyperplanes (Fig.~\ref{figure7.4}).
It exhibits a very complex formation of holes of finite
measure and various sizes, which we name the Hofstadter ``moth''. 
Although a rigorous proof can not be provided, the ``moth'' reminds a
fractal structure. Obviously, this fractal structure will
be very sensitive to any sort of perturbation
(finite size of the system, external trapping potential etc.)
on very small scales.
But, since the holes are true 3D objects with finite volume, 
the spectrum will be more robust on a larger scale
to perturbations than in case of the Hofstadter ``butterfly''.
Very recently, the same gauge fields were used to investigate metal-insulator transition for cold atoms \cite{Satija} and 
integer quantum Hall effect \cite{gaspard}.

The method presented in Ref. \cite{Osterloh05} can be easily generalized to  external gauge
potentials of the form 
$$ A_i(x,y)=  a_i +b_i(x/a) +c_i(y/a),$$
 with $i=x,y$, and  $a_i,b_i,c_i$ being essentially arbitrary $n\times n$ matrices. 
Furthermore, local disorder may be introduced in a controlled way
that allows for small fluctuations of the matrices $a_i,b_i,c_i$. In particular,
disorder can be made annealed, i.e. changing on a time scale comparable
with the relevant time scales of the system, and, thus, mimics thermal
fluctuations. It can be of significant amplitude,  provided it does not
drive the assisting lasers system out of resonance.
Also,  more complicated spatial dependences, e.g., piecewise linearity,
of $\vec A$ are feasible by using static electric fields, laser induced
potentials, etc. Additional lasers may introduce local,
and in general time dependent  unitaries.
Such transformations would generate arbitrary local temporal
components of the gauge potential, $A_0(x,\,y)$.
Although for Yang-Mills fields in (2+1)D, this component may be gauged out
adapting the Weyl, or strict temporal gauge \cite{Schulz00}, the corresponding
gauge transformations may introduce more complex spatial and temporal forms
of the remaining two components of $\vec A$.

In the limit 
of weak fields, the continuum limit may be used, and the single particle Hamiltonian reduces to that 
of a particle in the gauge field corresponding to the  above potential, i.e.: 
\begin{equation}
H=\frac{1}{2m}\left[(p_x- A_x(x,y))^2 + (p_y- A_y(x,y))^2\right].
\label{singleham}
\end{equation}

The non-Abelian gauge fields and Hofstadter ``moth" are interesting not only for fundamental reasons: 
they offer also  a fascinating possibility of observing the non-Abelian Aharonov-Bohm effect and realizing  
non-Abelian atom interferometry and non-Abelian nonlinear atom optics  \cite{Osterloh05,Jacob06}. Phase shifts in non-Abelian interferometers will correspond to matrices, and will depend not only on the perturbations of the 
 trajectories of the interfering particles, but also on the specific locations of these perturbations. 

Another fascinating possibility concerns the possibility of realizing novel FQHE states, that would be 
associated with fractional non-Abelian excitations. For instance, moving of a pseudo-hole around origin should 
induce a non-Abelian Berry ``phase" described by a non-trivial matrix \cite{Shapere89}. Particularly interesting 
in this context would be generalizations of methods used in \cite{Palmer06}. 
The studies of non-Abelian FQHE in such fields has 
just began. The resulting model \cite{Lewenstein_cond-mat0609587} is related to, yet very different from,
the bilayer of FQHE with tunneling between the layers \cite{Wen04, Ezawa00}. 

We expect that pseudo-hole in such model will transform according to the non-Abelian representations of the permutation group, 
or more precisely Artin's braid group, which in 2D considers particle exchange along the topologically distinct paths that avoid 
other particles (for a pedagogical review see 
\cite{read}, for recent discussion of wave functions of non-Abelions see \cite{schoutens}). 
There has been a long lasting quest for non-Abelian anyons 
in the condensed matter literature, but no clear experimental observation so far. The most prominent candidates are electronic (fermionic) 
$\nu=5/2$ FQHE state. This state has been observed in experiments, and Moore and Read \cite{Moore91} (see also 
\cite{Read_PRB_54_16864}), and independently Greiter, Wen and Wilczek 
\cite{greiter}  proposed  to explain it in terms of the ``Pfaffian state'' 
(see also Das Sarma and Pinczuk \cite{Das}). Recently, however, T\"oke and Jain \cite{toke}  
proposed an alternative ``composite fermions'' model of the $\nu=5/2$ state, which does not relate to non-Abelian statistics in any obvious manner. 
For bosons, a promising candidate is the $\nu=3/2$ state from the, so called, Read-Rezayi sequence of 
incompressible correlated liquids. This state 
seem to be a true ground state for the rapidly rotating gas of bosons interacting via contact (Van der Waals) forces with a moderate amount of 
dipolar interactions \cite{Rezayi05}. Such situation may be achieved, for instance, with Bose condensed Chromium, as in experiments of
 T. Pfau's group \cite{Griesmeyer05}. We hope that non-Abelian FQHE, due its profound and direct non-commutative character,  will provide further,
 experimentally feasible examples of non-Abelian anyons. An alternative way to realize Pfaffian-like states and non-Abelions in 1D ultracold 
gases has been very recently  proposed in Ref. \cite{paredes}.

\subsection{Ultracold gases and lattice gauge theories}
\label{lgt}

Another fascinating question concerns the  possibility of using ultracold atoms for simulations 
of lattice gauge theories in (2+1)D. The main difference is that in lattice gauge theories (LGT) 
gauge fields are dynamical variables,
whereas they  are obviously not in the scheme of Ref. \cite{Osterloh05}.
Moreover, the scheme is realized in real rather than in imaginary time.
Nevertheless, the big advantage of the proposal is that given a gauge
field configuration, the dynamics of matter fields in real time are
given for free. 
By generating  various configurations of gauge fields, we may try to
``mimic" the Monte
 Carlo sampling of LGT in the limit in which gauge fields affect the matter fields, but not vice versa.
 Averaging over both, annealed
disorder and quantum fluctuations should approximate the statistical
average in LGT. Such approach  requires that
generated configurations represent the characteristic
or statistically relevant ones of corresponding LGT phases. For instance, configurations in the confinement sector
should exhibit an area law fulfilled by Wilson loops,  appropriate distributions
of centre vortices, Abelian magnetic monopoles, instantons, merons, calorons,
etc. (for a recent reviews, see \cite{Engelhardt04,Greensite03}; see also \cite{Baig86}). 
Although the gauge fields accessible in the
scheme of Ref. \cite{Osterloh05}  are limited, at least some of them share 
characteristics with LGT phases.

The situation is much better when we turn to Abelian lattice gauge theories, since they reduce 
in some limits to models with ring exchange interactions, that involve product of operators over an elementary plaquette of the underlying lattice.  The first proposal for such interactions has been formulated in Ref. \cite{Pachos04}. The authors 
derived a spin model for a multicomponent Bose, or Fermi gas in the Mott limit with one particle per site. 
Due to the fact that in a triangular lattice tunneling from one site to another may occur directly, or via the remaining third site, 
they obtain 3-spin interactions in the third order of $t/U$ expansion. This proposal has the disadvantage  
that it involves very small energy scales and long time scales, since in the Mott limit $(t/U)^3\ll 1$. 

\begin{figure}[t]
\centering
\includegraphics[width=0.75\textwidth]{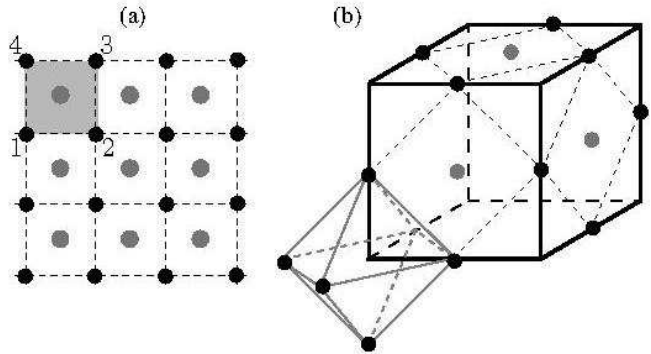}
\vspace*{0.5cm}
\caption{The scheme of realizing ring exchange interactions (redrawn from \cite{Buechler05})}
\label{figure7.5}
\end{figure}

B\"uchler et al. \cite{Buechler05} consider a standard single component Bose-Hubbard model in a square (cubic) lattice coupled to 
lattice of diatomic ``molecules", trapped in the centre of each plaquette (see Fig. \ref{figure7.5})
The coupling to the molecular state (of $d$-symmetry)  takes the form:
\begin{equation}
H_m=\nu\sum_{plaquettes\ \square}m_\square^{\dag}
m_{\square} + g\sum_{plaquettes\ \square}\left[m^{\dag}_{\square}(b_1b_3-b_2b_4) + h.c. \right].
\label{bosemolec}
\end{equation}
Two atoms may perform a Raman transition to a molecular state with the coupling $g$; $\nu$ denotes detuning of this transition. 
Perturbative elimination of the molecules leads to the effective ring exchange Hamiltonian for bosons:
\begin{equation}
H_{RE}=K\sum_{plaquettes\ \square}\left(b^{\dag}_1b_2b^{\dag}_3b_4 + b^{\dag}_4b_3b^{\dag}_2b_1 -n_1n_2-n_3n_4\right).
\label{ring}
\end{equation}
More precisely, the model can be realised using atoms with two internal states: one state 
trapped in the square (cubic) lattice and described by the simple Bose-Hubbard model, and  a second one 
trapped in the centres of the plaquettes in a site potential that is not symmetric, but rather has a point 
symmetry of the lattice. ``Molecules" are build  from two atoms in that second internal state, and have 
the $d$--symmetry. The model described by the Bose-Hubbard Hamiltonian (Eq.(\ref{BHH})) plus $H_m$ is a  promising candidate 
for a deconfined quantum critical point  in 2D \cite{Senthil04}. This is because it undergoes (most probably) 
a quantum phase transition between the bosonic superfluid phase (which occurs when $\nu\gg g$, and $t\gg g^2/\nu$, 
and which breaks the $U(1)$-symmetry), and a molecular ``density wave" (stripes) phase (which occurs in the opposite limit, 
when $\nu<0$ and $|\nu|\gg t,g$, and breaks the translational symmetry). Both in 2D and 3D this model describes a $U(1)$ gauge 
theory, and is likely to exhibit a variety of quantum phases, 
such as $U(1)$ deconfined insulator, $U(1)$ deconfined phase, etc.  \cite{Alet05a,Hermele04}. 

Just before this review was submitted, Tewari et al. \cite{Tewari06}
proposed to use a dipolar Bose gas to simulate the compact $U(1)$ lattice gauge theory. Such model is well described by an 
extended Bose-Hubbard Hamiltonian, as discussed in \cite{Goral02}. Tewari et al. propose to use 2D kagom\'e, or 3D 
pyrochlore lattices, assuring 
isotropy of dipolar interactions  between nearest neighbors on the lattice. In the limit in which the on-site 
interactions $U$ are comparable to the n.n. interactions $V$, 
and both are strong, the physics is dominated by the configurations where the fluctuations of the 
number of atoms in all elementary plaquettes is zero.  In the lowest relevant order of perturbation theory ($(t/U)^3$) one obtains a hamiltonian 
with ``ring exchange" term on a dual hexagonal lattice. This model reduces directly to the $U(1)$ lattice gauge theory, and allows to realize various fractionalized topological phases. 
Moreover, the authors propose methods of detecting signatures of the emergent $U(1)$ Coulomb phase, terming it as ``emergence of artificial 
light in an optical lattice". It is finally worth stressing that by combining, vortex lines with fermions, one
can realize perhaps the first connections of ultracold atoms with string theory \cite{Snoek06}.


\section{Quantum information with ultracold  gases}
\label{sec-goru65}
\subsection{Introduction}
\label{sec-goru165}
There has recently been a real  explosion of  interest in the studies of the interface between  quantum information (QI) and many
body systems, both in condensed matter, as well as in the physics of ultracold atomic gases. First, in Ref. \cite{Jaksch99} 
a proposal 
of using ultracold atomic gases in optical lattices for QI was formulated. 
This has stimulated interest in {\it distributed quantum information processing}. 
Second, in Refs. \cite{Osborne01, Osborne02,Osterloh02},  the first connections between entanglement 
and quantum phase transitions (QPT) were discussed. This discussion  has opened fundamental questions of deeper 
understanding of QPT's as well as practical questions of employing QI ideas in simulation codes for many body quantum systems 
(see subsec. \ref{subsec_quantu_raju}).

In this section we will (partially) review these two major themes of such studies. One of them 
uses realizable many body  systems as viable systems for quantum computation and other quantum 
information tasks.   In the other theme, 
the aim is 
to use quantum information concepts 
such as entanglement, to  understand condensed matter and many body phenomena, like for instance quantum phase transitions.

This section has a slightly different character than the rest of the paper,
 in that it does not make distinction between condensed matter and atomic systems, but rather considers 
general properties of quantum states. It is also organized differently. 
In order to make the review self-contained, we begin with subsec. \ref{subsec-poila-qi}  by presenting necessary basic
notions and tools of quantum information theory in general,  and entanglement theory in particular.  
In the next subsection (subsec. \ref{subsec-dusra-qi}), we consider quantum phase transitions from the quantum information 
perspective. In subsec. \ref{subsec-teesra-qi}, we discuss possibilities of 
performing quantum computation using atoms in optical traps. 
The last subsection \ref{subsec-chauthha-qi} is devoted to the concept of measurement-based 
quantum computation, both in ordered and disordered lattices.

\subsection{Entanglement: A formal definition and some preliminaries}
\label{subsec-poila-qi}

The concept of entanglement can be traced back to manuscripts of E. Schr{\"o}dinger  around 1932, when 
it was realised that the quantum mechanical formalism allows the existence of \emph{pure} 
states of two systems, for which the information about the whole system is not the sum of the 
information about the separate 
systems. With the advent of quantum information, a more operational definition of 
entanglement was required. Such a definition was formulated by Werner in 1989 \cite{Werner89}. Suppose that there are two systems belonging to two observers called Alice (A) and Bob (B). 
They are, in principle, at distant locations, which, operationally means that 
they can only act on their local parts. Under such conditions, the most general state that they are able 
to prepare is of the form
\[
\varrho_A \otimes \varrho_B. 
\]
However, they may additionally be allowed to communicate classical information between them. In that case, the most general 
state that they are able to prepare is of the form
\begin{equation}
\label{alada-qi}
\sum_i p_i \varrho^i_A \otimes \varrho^i_B,
\end{equation}
where the \(\{p_i\}\) are a set of probabilities. 
The set of operations that Alice and Bob are allowed to perform is then called local operations and 
classical communication (LOCC). The states that are of the form in Eq. (\ref{alada-qi}) are called 
{\it separable states}, and they (and only they) can be prepared under LOCC only. However, there are states that cannot be 
written in the form in Eq. (\ref{alada-qi}), and consequently it is not possible 
to prepare them by LOCC, an example being the singlet state. These states are called entangled states. 

A pure state \(\left|\psi\right\rangle_{AB}\)  is  separable  if  and  only  if  its  local subsystems are also pure, that is when it 
can  be written as \(\left|\psi_1\right\rangle_A \otimes \left|\psi_2\right\rangle_B\). 
The case of mixed states is, however,  nontrivial. In fact, so far  an operational 
 necessary and sufficient criterion for detecting 
entanglement  in mixed states has not been found (see \cite{SenDe05a} for a tutorial). 
Using a semidefinite programming, it has been shown that separability can be tested in a finite number of steps, although there 
is no limit on the number of steps needed \cite{kajey-lagteo-parey}. See also Ref. \cite{Dagmar}. 
However, there exist many efficient criteria, that are either necessary or sufficient. One of them is described below.

\subsubsection{The partial transposition criterion for detecting entanglement}
\label{subsubsec-jog-biyog}

The partial transposition criterion is a very useful
criterion for detecting entanglement. We will require the notion of 
partial transposition of a density matrix \(\varrho_{AB}\). If partial transposition
of \(\varrho_{AB}\) is taken over the subsystem \(A\), denoted 
by \(\varrho_{AB}^{T_A}\), then its elements are  defined as 
\begin{equation}
\label{eq_partial_trans}
(\rho_{AB}^{T_A})_{i\mu, j\nu} = (\rho_{AB})_{j\mu, i\nu},
\end{equation}
 where the subscripts \(i,j\) are for Alice's subsystem and the subscripts \(\mu, \nu\) are for 
Bob's subsystem.  Note that although the partial transposition depends upon the choice of the basis in
which \(\varrho_{AB}\) is written, the eigenvalues do not depend on the basis.
We  say that a state has  positive partial transposition (PPT) 
whenever \(\varrho^{T_A} \geq 0\), i.e. all the eigenvalues of \(\varrho^{T_{A}}\) are nonnegative. 
Otherwise the state has negative 
partial transposition (NPT). Peres  \cite{Peres96} noticed  that for separable states, the 
eigenvalues of the  partially transposed density matrix are always positive. 
That is, all 
separable states are PPT, and there exist entangled states which are NPT. Failing to be PPT
is thus a signature of entanglement. An NPT state is often said to violate the  PPT criterion. 
One of the most important results of the theory of entanglement has been proven by the Horodecki family, who 
showed that in \(\mathbb{C}^2 \otimes \mathbb{C}^2\) and \(\mathbb{C}^2 \otimes \mathbb{C}^3\), 
a state is separable if and only
if it is PPT \cite{Horodecki96}.  This is no longer true in higher dimensions.

\subsubsection{Entanglement measures}
\label{subsubsec-entanglement-measures}

The partial transposition criterion provides a tool to check 
whether the given state is entangled or not. 
Now, we will 
discuss ways to find out \emph{how much} entanglement a given state has, once we know that it is entangled. 
This quantification is necessary, at least partly because  
entanglement is viewed as a resource in quantum information theory.
There are several complementary ways to quantify entanglement 
(see \cite{Bennett96b,Vedral97,DiVincenzo98,Laustsen03,Nielsen99,Vidal00,Jonathan99,Horodecki04,Horodecki01b,Plenio05b}
and references therein). 
We will present here three possible ways to do so.

\paragraph{Entanglement of formation}

Consider a bipartite pure state \(\left|\psi_{AB}\right\rangle\) shared between Alice and Bob.
It was shown by Bennett \emph{et al.} \cite{Bennett96a}, that given \(n E(\psi_{AB})\)  copies of the singlet state shared between Alice and Bob, 
they can by LOCC, transform them into \(n\) copies of the state \(\left|\psi_{AB}\right\rangle\), if \(n\) is large, 
where:
\begin{equation}
\label{eq_ent_pure}
E(\psi_{AB}) = S(\varrho_A)  = S(\varrho_B),
\end{equation}
with \(\varrho_A\) and \(\varrho_B\) being the local density matrices of \(\left|\psi_{AB}\right\rangle\).
Here \(S(\rho)\) is the von Neumann entropy of \(\rho\), given by: 
\[
 S(\rho) = - \mbox{tr} \rho \mbox{log}_{2} \rho.
\]
If the state \(\left|\psi_{AB}\right\rangle\) describes
a \(\mathbb{C}^{d_1} \otimes \mathbb{C}^{d_2}\) system, then  \(E\) ranges from \(0\) to \(\mbox{log}_2 d\), where
\(d= \mbox{min} \{d_1, d_2\}\). Note that \(E\) is vanishing for separable (pure) states.
 A state is called maximally entangled in \(\mathbb{C}^{d_1} \otimes \mathbb{C}^{d_2}\),
 if its \(E\) is \(\mbox{log}_2 d\). The singlet state is a 
maximally entangled state in \(\mathbb{C}^{2} \otimes \mathbb{C}^{2}\) and its entanglement is unity. This amount
of entanglement of a singlet is also called an {\it ebit}. A maximally entangled state in \(\mathbb{C}^{d_1} \otimes \mathbb{C}^{d_2}\) has
\(\mbox{log}_2 d \) ebits. An example of a maximally entangled state in \(\mathbb{C}^{d_1} \otimes \mathbb{C}^{d_2}\)
 is: 
\[ \frac{1}{\sqrt{d}} \sum_{i=1}^{d} \left|  \alpha_i \right\rangle_{A}  \otimes \left|  \beta_i \right\rangle_{B}, \]
where 
\(\left|  \alpha_i \right\rangle_{A}\)   (\(\left|  \beta_i \right\rangle_{B}\)) are orthogonal states in Alice's (Bob's) subsystem.
With the above terminology, we can therefore say that the state \(\left|\psi_{AB}\right\rangle\) has 
\(E(\psi_{AB})\) ebits. Since \(E(\psi_{AB})\) is the number of singlets required to prepare a copy of the state 
\(\left|\psi_{AB}\right\rangle\),  it  is called the ``entanglement of formation'' of 
\(\left|\psi_{AB}\right\rangle\). We are therefore using the amount of entanglement of the singlet state
as our unit in quantifying entanglement. 

For (bipartite) pure states, the entanglement of formation is essentially the only (asymptotic) measure of entanglement, as 
the singlet state can be locally obtained from a pure state at the same asymptotic rate \cite{Bennett96a}.

After obtaining the definition of entanglement of formation for pure states, let us now define this measure for mixed states 
\cite{Bennett96b}. 
Any mixed state can be expressed as a mixture (convex combination) of pure states. 
That is, a mixed state \(\varrho\) can always be expressed as: 
\[\varrho= \sum_i p_i \left|\psi_i \right\rangle \left\langle \psi_i \right|,\]
where \(\{p_i\}\) are a set of probabilities. The states \(\left|\psi_{i}\right\rangle\) are not
necessarily orthogonal.
It is then tempting to define the entanglement of formation of a mixed state
\[ \varrho_{AB} = \sum_i p_i \left|  \psi^i_{AB} \right\rangle \left\langle \psi^i_{AB} \right| ,\]
as the average 
\[\sum_i p_i E(\psi^i).\]
However, a mixed state can be decomposed into pure states in an infinite number of ways. The entanglement of formation
\(E_F\) of \(\varrho_{AB}\) is defined as the minimum over all such averages. Precisely, \cite{Bennett96b}:
\[ E_F(\varrho_{AB}) = \mbox{min} \sum_i p_i E(\psi^i_{AB}),\]
where the minimum is over all decompositions of \(\varrho_{AB}\) into pure states.

The procedure for calculating entanglement of formation of any state in 
\(\mathbb{C}^2 \otimes \mathbb{C}^2\) was given in Refs. \cite{Hill97, Wootters98}. 
For states in higher dimensions, the entanglement of formation (or even more difficult asymptotic entanglement cost
for many copies) has been calculated for only a few rare 
instances
using certain symmetries (see for instance Ref. \cite{Terhal00,Vidal02a,Horodecki03,Giedke03,Wolf04}).

\paragraph{Concurrence}

As stated before, the entanglement of formation of an arbitrary (possibly, mixed) state 
in \(\mathbb{C}^2 \otimes \mathbb{C}^2\)
has been calculated. It is given by: 
\[
E_F(\varrho_{AB}) = \mbox{H}\left( \frac{1+ \sqrt{1 - C^2}}{2} \right),
\]
where 
\[
C=C(\varrho) = \max\left\{0, \lambda_1 - \lambda_2 - \lambda_3 - \lambda_4 \right\},
\]
and \(H(\cdot)\) is the binary entropy function, defined for \(0\leq x \leq 1\) as 
\(H(x) = -x\log_2 x - (1-x)\log_2 (1-x)\). 
The \(\lambda_i\) are the eigenvalues (with \(\lambda_1\) being the greatest) of the Hermitian matrix
\(\left( \varrho^{\frac{1}{2}} \tilde{\varrho} \varrho^{\frac{1}{2}}  \right)^{\frac{1}{2}}\),
where \(\tilde{\varrho}= \sigma_y \otimes \sigma_y \varrho^* \sigma_y \otimes \sigma_y\),
with the complex conjugation over \(\varrho\) being taken in the \(\sigma_z\) eigenbasis.

In \(\mathbb{C}^2 \otimes \mathbb{C}^2\), the quantity \(C(\varrho_{AB})\), called the concurrence, 
is defined for arbitrary states, and  
moreover, the entanglement of formation and concurrence are monotonically nondecreasing 
functions of each other, with
both quantities ranging from 0 to 1.
Therefore, in  \(\mathbb{C}^2 \otimes \mathbb{C}^2\), we will also use the concurrence as a measure of entanglement.

\paragraph{Logarithmic negativity}

The partial transposition criterion can be used to define a useful measure of entanglement, that is 
easily computable. To define the measure, let us first introduce a quantity called negativity. 
 The negativity \(N(\varrho_{AB})\)
  of a bipartite state \(\varrho_{AB}\) is defined as the absolute value of the sum of the negative
eigenvalues of  \(\varrho_{AB}^{T_{A}}\).
The logarithmic negativity (LN)
is defined as \cite{Vidal02b}
\[
E_{N}(\varrho_{AB}) = \log_2 (2 N(\varrho_{AB}) + 1). 
\]
Note that for two qubit states, \(\varrho_{AB}^{T_{A}}\) has at most 
one negative eigenvalue \cite{Sanpera98}.
Moreover, it follows from our discussion in \ref{subsubsec-jog-biyog} that 
for states in \(\mathbb{C}^2 \otimes \mathbb{C}^2\) and in \(\mathbb{C}^2 \otimes \mathbb{C}^3\), 
a positive LN implies that the state is 
entangled, while 
\(E_{N} =0\) implies that the state is separable.

The logarithmic negativity has an operational interpretation in terms of another measure
related to the entanglement of formation \cite{Audenaert03}, and moreover, 
is an entanglement monotone under deterministic LOCC \cite{Plenio05a}. However, it is not convex.

\subsection{Entanglement and phase transitions}
\label{subsec-dusra-qi}

Over the past few years, the question whether quantum phase transitions (QPT) 
can be understood from the perspective 
of quantum information has been posed, and at least partially positively answered.

QPTs are nonanalyticities in the ground state energy of physical systems. A typical situation is 
the following (see e.g. \cite{Sachdev99}). Consider a physical system that is defined on some lattice, and that is
described by a Hamiltonian of the form 
\[
H = H_0 + \lambda H_1,
\]
where \([H_0, H_1] \ne 0\).
The parameter \(\lambda\) is to be visualized as an external parameter, which can be changed by the experimenter. 
In the limit of an infinite lattice, 
 if as the system crosses  a certain value of \(\lambda = \lambda_c\), the ground state has a qualitative change, and there is an 
 associated nonanalyticity in the ground state energy at \(\lambda = \lambda_c\), then the system is said to have a 
QPT at that point. Since the transition is in the ground state, so that the system is at zero temperature, 
the transition is driven by quantum fluctuations (and not by thermal fluctuations like in classical 
phase transitions), and this is the reason why it is called a ``quantum'' phase transition.

The study of entanglement in strongly correlated systems was initiated in e.g. 
\cite{Nielsen98, Preskill00, Wang01a, Wang01b, Wang01c, Zanardi02a, Zanardi02b,
Bose02, Wang02, Fu02, Wootters02a, Wootters02b, Meyer01, Coffman00, Dennison01, OConnors01, Gunlycke01, Arnesen01,
Bose03,Benjamin03}
(and references therein). It was realized by Osborne and Nielsen \cite{Osborne01, Osborne02} and Osterloh \emph{et al.}
\cite{Osterloh02} that the distinctly quantum phenomenon of QPTs can be related to the quantum phenomenon of 
entanglement. The scaling of entanglement as an infinite chain of spin-half particles, described by 
the XY model,  goes through a QPT was considered. Ideally, one would like to consider the scaling of the multiparticle 
entanglement of the whole infinite spin system in the ground state. However, the subject of multiparticle entanglement is not 
fully developed at present. As a way out, the authors in Refs. \cite{Osborne01, Osborne02, Osterloh02} considered the nearest neighbor entanglement 
 of the ground state of the spin chain. This required the consideration of entanglement of (mixed) states in 
\(\mathbb{C}^2 \otimes \mathbb{C}^2\), which is relatively well-developed. We consider this topic in subsection \ref{subsubsec-babajibon-qi}.

As we just commented, for higher number of parties, the theory of entanglement is not 
so well developed, even for pure states, at least not quantitatively. 
However, the entanglement of pure states in arbitrary \emph{bipartite} dimensions is 
quite well-understood, as we have discussed in  \ref{subsubsec-entanglement-measures}. Consequently, and since 
the ground state is a pure state, one may consider the scaling of entanglement of the ground state for a bipartite split consisting on a certain number of sites \(n\) and the rest of the lattice, as \(n\) grows. The entanglement of one site to the rest of the chain was 
considered in e.g. Ref. \cite{Zanardi02a, Zanardi02b, Osborne02}, and the scaling in this scenario was 
initiated by Vidal \emph{et al.} \cite{Vidal03b}. We discuss this topic in \ref{subsubsec-barishal-qi}.

Yet another type of scaling was considered by Verstraete \emph{et al.} \cite{Verstraete04a} by using the concept of 
``localizable entanglement'', 
which will be defined and then its scaling discussed in \ref{subsubsec-bardhaman-qi}. 

We omit several topics in the following, such as 
the topic of scaling of certain multipartite entanglement measures 
(see e.g. \cite{Roschilde04, Anfossi05, Roschilde05, Wei05, Oliveira06}, and references therein). 
There is also an interesting string of research that deals with criticalities in the Hubbard model (see e.g. 
\cite{Korepin04, Gu04, Anfossi05, Larsson05}, and references therein). 
Dorner \emph{et al.} \cite{Dorner03} consider a string of neutral atoms in a one dimensional beam splitter configuration, 
where the longitudinal motion is controlled by a moving optical lattice potential; they show that it is possible to 
create maximally entangled states in this setup, by crossing a QPT. Lambert \emph{et al.} \cite{Lambert04} 
considers an infinite collection of two-level atoms interacting via the Dicke Hamiltonian with a single bosonic mode, and 
describes the behaviour of entanglement when the system is near its QPT.

Most of the papers that deal with the scaling of entanglement near criticality deal with asymptotic measures of 
entanglement, or those related to such measures. The entanglement of formation (and concurrence), for 
example, is defined in terms of the the entropy of the reduced density matrix of a pure state, which has meaning only in the asymptotic regime. Exceptions include the work by Eisert and Cramer 
\cite{Eisert05}, which considers single copy entanglement of the ground state of critical quantum spin chains.
In the last part of this subsection, a possible way to use (bipartite) entanglement to 
observe a type of phase transitions (called dynamical phase transitions) in the evolved state
of a strongly correlated system is discussed \cite{SenDe05b}.

\subsubsection{Scaling of entanglement in the reduced density matrix}
\label{subsubsec-babajibon-qi}

The first system for which the scaling of entanglement near a QPT was considered is 
a system of spin-half particles on a chain, described by the XY Hamiltonian:
\[
H_{XY} = - \frac{J}{2}\sum_i \left[ (1 - \gamma) \sigma_i^x \sigma_{i+1}^x 
+ (1 + \gamma) \sigma_i^y \sigma_{i+1}^y \right] - h \sum_i \sigma_i^z,
\]
with  \(J>0\),  \(h >0\),  and  \(0 < \gamma \leq 1\). 
This system is known to undergo a QPT as the parameter \(\lambda = \frac{J}{2h}\) passes over 
\(\lambda = \lambda_c \equiv 1\) (see e.g. \cite{Lieb61, Pfeuty70, McCoy68, Barouch70,
Barouch71a, Barouch71b, Barouch71c, Sachdev99}). 

\begin{figure}[t]
\centering
\includegraphics[width=0.75\textwidth]{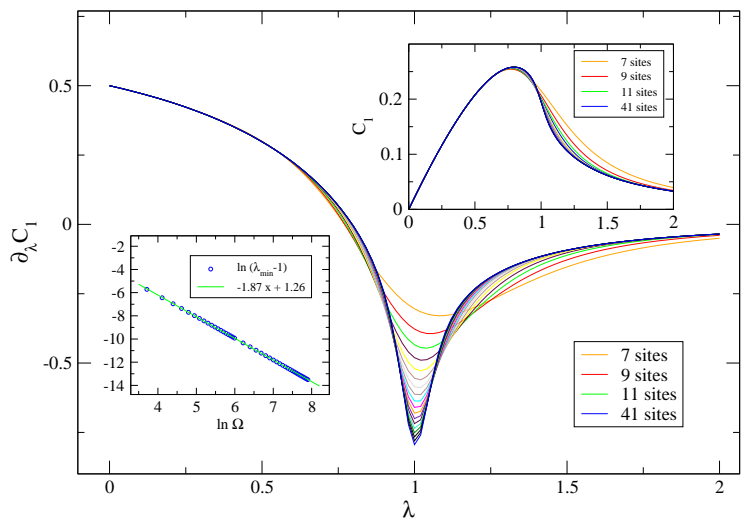}
\vspace*{0.5cm}
\caption{(Color online).Scaling of nearest neighbour entanglement of the ground state of 
the transverse Ising Hamiltonian on a chain (from \cite{Andreas_alada}). 
The inset at top right shows the nearest neighbour entanglement, while the main figure shows its derivative, plotted against 
the parameter \(\lambda\).
In the main figure (top right inset), say at \(\lambda =1\) (say at \(\lambda = 1.25\)), 
the curves from the top are respectively for 7, 9, 11, ..., 41 sites (the quality of the figure improves in the online version). Bottom left inset: The position \(\lambda_{\min}\) of the minimum of \(\partial_{\lambda} C_1\) in the main figure
changes with the total number of sites. The current inset shows the plot of the logarithm of \(\lambda_{\min} -1\) against 
the logarithm of the total number of sites (denoted in this figure as \(\ln \Omega\)).
}
\label{figure5.1}
\end{figure}

In Refs. \cite{Osborne01, Osborne02, Osterloh02}, it was proposed that this QPT may be 
analysed by looking at the behavior of the entanglement in the two site states
(the state obtained by tracing over all except two sites of the chain) of the ground state 
of the system, as it goes through the QPT  (see Fig. \ref{figure5.1}). In particular, in the case of the Ising Hamiltonian (\(\gamma = 1\))
on an infinite chain,
the \(\lambda\)-derivative of the nearest neighbor concurrence diverges as \cite{Osterloh02}:
\[
\partial_\lambda C_1 = \frac{8}{3 \pi^2}\ln|\lambda - \lambda_c|,
\]
(up to an additive constant) as \(\lambda\) approaches \(\lambda_c\) (See Fig. \ref{figure5.1}). The subindex 1 of \(C_1\)
indicates the fact that the \emph{nearest} neighbor entanglement is calculated. 

The Refs. \cite{Osborne01, Osborne02, Osterloh02} stimulated further
studies of  two-site entanglement  to characterise QPTs. 
Bose and Chattopadhyay \cite{Bose02} have considered it in the case of some frustrated spin models in 
one dimension, ladder, and two dimensions. Two-site entanglement in the XXZ chain was 
studied by Gu \emph{et al.} \cite{Gu03}. The Lipkin-Meshkov-Glick model has been considered in Refs.
\cite{Vidal04a, Vidal04b, Dusuel04}. Stauber and Guinea \cite{Stauber04} have looked at 
the entanglement of spins at the boundary of a spin chain 
governed by the transverse Ising model, as well as that of two spins coupled to a critical reservoir. 
The anisotropic ferromagnetic Heisenberg chain in the presence of domain walls 
was considered by Alcarez \emph{et al.} \cite{Alcarez04}.

Yang \cite{Yang05} obtained the following apparently 
counterintuitive result. An exactly solvable 
three-spin interaction Hamiltonian  was considered, and it was shown that there can appear a discontinuity 
of the first derivative of the two-site concurrence, which has no connection to any QPT (see also 
Refs. \cite{Vidal04a, Syljuasen03}). It was further shown that this discontinuity originates from finding the maximum 
in the definition of concurrence. Subsequently, Wu \emph{et al.} \cite{Wu04},
found necessary and sufficient conditions under which discontinuities in the two-site entanglement
(as quantified by concurrence, or negativity) can be an indicator of a QPT in physical systems governed 
by Hamiltonians containing only two-body interactions.

\subsubsection{Entanglement entropy: Scaling of spin block entanglement}
\label{subsubsec-barishal-qi}

A different approach than the two-site entanglement was considered by Vidal \emph{et al.} \cite{Vidal03b}.
The idea is to consider a block of spins and consider its entanglement with the rest of the system, for 
the ground state, as 
the block size increases. 
Since the ground state is a pure state, a good measure of
 entanglement of a block of spins with the rest of the system, is 
the entropy of the block of spins. 
This has been termed the entanglement entropy (EE) of the model. 
The models that were considered in Ref. \cite{Vidal03b} (see also \cite{Latorre04})
are the one dimensional XY and XXZ models, where the latter is described by the following Hamiltonian:
\[
H_{XXZ} = \sum_i \left( \sigma_i^x \sigma_{i+1}^x + \sigma_i^y \sigma_{i+1}^y 
+ \Delta \sigma_i^z \sigma_{i+1}^z - \lambda \sigma_i^z \right).
\]
It was shown that away from the criticalities, EE saturates to a constant as a 
function of the block size. However, near the criticalities, EE diverges logarithmically with block size. 
Similar behaviour was seen to be true for the Affleck-Kennedy-Lieb-Tasaki valence bond solid models by 
Fan \emph{et al.} \cite{Fan04}.

Techniques from conformal field theory were used by Korepin \cite{Korepin04}
for evaluating the EE in the limit of large block size for infinite 
spin chains.
Its \emph{et al.} \cite{Its05} have evaluated this limiting EE for an infinite 
chain governed by the XY model.

EE was also considered by Wellard and Or{\' u}s \cite{Wellard04} for the case of QPTs in antiferromagnetic (Ising) 
planar cubic lattices in the presence of homogeneous transverse and longitudinal magnetic fields, and 
by Latorre \emph{et al.} 
Ref. \cite{Latorre05} for the Lipkin-Meshkov-Glick model.

The relation between  entanglement entropy and area for the cubic harmonic lattice has been discussed in Refs.
\cite{Plenio05c, Cramer06} (for fermionic systems,
see \cite{Wolf06, Gioev06}) (cf. \cite{Verstraete06b, Riera06}).

In Ref. \cite{Dur05}, an unbounded increase of entanglement entropy was obtained for a lattice of spins that interact via some 
long-range Ising type interaction.

\subsubsection{Localizable entanglement and its scaling}
\label{subsubsec-bardhaman-qi}

For a given multiparty state, \(\varrho_{1,2, \ldots, N}\), of \(N\) parties, the localizable 
entanglement (LE) \cite{Verstraete04a} is the maximum average entanglement that can be made to be shared 
between two pre-determined 
parties (say, 1 and 2), by measurements at the rest of the parties. We still have to state which entanglement
measure we consider in finding the average entanglement. Similar considerations were used 
in the context of violation of Bell inequalities \cite{Bell64} in Refs. \cite{Popescu92, SenDe03a, SenDe03b}.
In Ref. \cite{Verstraete04a}, the LE of the ground state of 
certain spin-half systems were considered, where concurrence was used as the measure of entanglement 
in the definition of average entanglement.

Using the concept of LE, a related ``entanglement length'', \(\xi_E\), was defined in \cite{Verstraete04c}, 
which gives the
typical length scale (of a spin-half chain, e.g.) at which it is possible to create maximally entangled states
by performing measurements at the rest of the parties. Let us make the definition more specific in the 
case of a spin chain. Let \(E_{i,i+n}\) be the 
LE that can be made to share between the \(i\)th and the \((i+n)\)th spins of the chain. If \(E_{i,i+n}\) decays 
exponentially with \(n\), then the entanglement length of the chain is defined by: 
\[
\xi_E^{-1} = \lim_{n \to \infty} \left( \frac{-\ln E_{i,i+n}}{n} \right).
\]
If the decay of \(E_{i,i+n}\)  is non-exponential, then the entanglement length is infinite.
A similar definition was given by Aharonov \cite{Aharonov00}, considering a transition from quantum to classical physics. 
The definition of entanglement length parallels that of correlation length, for which if the correlation 
\(\left\langle O_i O_{i+n} \right\rangle\) between two observables \(O_i\) and \(O_{i+n}\) decays exponentially, then the 
correlation length \(\xi\) is defined by: 
\[
\xi^{-1} = \lim_{n \to \infty} \left( \frac{-\ln \left\langle O_i O_{i+n} \right\rangle}{n} \right).
\]

A diverging correlation length implies a diverging entanglement length. However, the converse is not true: 
for example, the gapped spin system governed by the modified 
Affleck-Kennedy-Lieb-Tasaki model, which has a finite correlation length, was proven  to have a diverging entanglement length by Verstraete 
\emph{et al.} \cite{Verstraete04c}.

\subsubsection{Critical behaviour in the evolved state}
\label{subsubsec-sundarban-qi}

Phase transitions are usually viewed for ground states, at least when considered from the perspective 
of quantum information theory. In contrast, Ref. \cite{SenDe05b} considers 
a criticality of nearest-neighbor entanglement in the \emph{evolution} of 
an infinite spin chain described by the asymmetric \(XY\) model in a time-dependent transverse field.

\begin{figure}[t]
\centering
\includegraphics[width=0.75\textwidth]{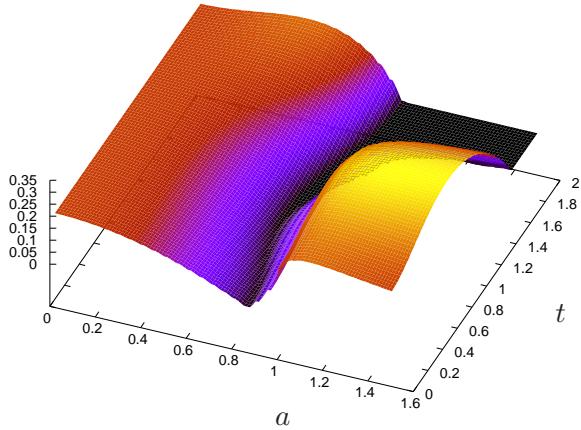}
\put(-150,7){\(a\)}
\put(-44,46){\(t\)}
\vspace*{0.5cm}
\caption{Dynamical phase transitions: a ``river" of separable states between the two regions of control 
parameter space where nearest neighbour entaglement is non-zero in an asymmetric XY model in a 
time-dependent transverse field.
(from  \cite{SenDe05b}). Entanglement (logarithmic negativity) is plotted against 
the initial transverse field (\(a\)) and (real) time (\(t\)).  
A similar behaviour is absent in magnetization \cite{SenDe05b}.}
\label{figure5.2}
\end{figure}

The  initial state of the evolution is taken to be the canonical equilibrium state 
at zero temperature (which, in this case, is the ground state). The transverse field is then suddenly turn off at time equal to zero.  
The system is thus initially disturbed, and its properties are then studied 
at later times. 
The nearest-neighbor entanglement (quantified by the logarithmic negativity) in the evolved state at a fixed time shows a 
criticality (which has been called a dynamical phase transition (DPT)) 
 with respect to the transverse external field. 
The  region of the  initial transverse field for 
  which the entanglement is nonvanishing (vanishing),  at a fixed time, is referred to as the 
  ``entangled phase'' (``separable phase'') (see Fig. \ref{figure5.2}).  
  
Interestingly, the 
nature of the DPT  depends on whether we 
are near, or far from the time of initial disturbance. Moreover, 
for values of the initial transverse field near the criticalities, as well as in the separable 
phase, and for short times, the nearest-neighbor entanglement 
shows \emph{nonmonotonicity} with respect to temperature. 
Accordingly, the criticalities are referred to 
as ``critical regions'', signifying that ``critical" effects persist for a small region around the critical value of
the transverse field.

The study of nonmonotonicity of entanglement with respect to temperature is 
interesting, as preservation of entanglement in a hostile environment is one of the 
main challenges in quantum computation and quantum information in general.
A common belief is that temperature is a form of noise, i.e. it destroys the subtle quantum correlations. 
However, this seems not to be universally true.

Nonmonotonicity of entanglement with respect to temperature was also 
found by Arnesen \emph{et al.} \cite{Arnesen01}, in the canonical equilibrium state of the 
one-dimensional isotropic antiferromagnetic Heisenberg model. 
Similar behavior was obtained by Scheel \emph{et al.} \cite{Scheel03}, in the Jaynes-Cummings model.
The idea that dissipation can assist generation of entanglement has also been put forward in Refs.
\cite{Plenio99, Bose99, Beige00, Horodecki01a, Plenio02, Braun02, Yi03}.

It should be mentioned here that the asymptotics of temperature correlations in the isotropic (i.e \(\gamma =0\))  $XY$ model was considered 
by Its \emph{et al.} \cite{Its93} (see also \cite{Vaidya78, Perk80, McCoy83a, McCoy83b}). 
Also, previous studies of the quantum dynamics of spin models after a rapid change of the field include 
Refs. \cite{McCoy68, Barouch70,
Barouch71a, Barouch71b, Barouch71c, Greiner02, Sengupta04, SenDe04}, while 
effects of a sudden switching of the interaction in arrays of oscillators were studied in Ref. \cite{Plenio04}.

\subsection{Quantum computing with lattice gases}
\label{subsec-teesra-qi}

The computers that have become a part of our daily life are very efficient and useful machines.
However, there are certain problems that may not be efficiently solved with such computers. 
For example, it is believed that the problem of finding prime factors of a given integer is exponentially inefficient on a classical 
computer, since the number of steps required to find the factors scales exponentially with the size of the input integer.

Classical computers follow the principles of classical physics, but it is possible to consider computers that are ruled by the laws of quantum mechanics. There are at least two important motivations for considering 
computing at a quantum level: (1) The classical laws of physics governing the functioning of classical computers are an 
approximation that suits our purpose for the current size of microchips. If size is going to keep 
decreasing, so that computational speed increases, quantum effects will unavoidably emerge; (2) In 1994, Shor discovered  
a quantum algorithm that can efficiently factorize integers into their prime factors \cite{Shor94}.

\begin{figure}[t]
\centering
\includegraphics[width=0.75\textwidth]{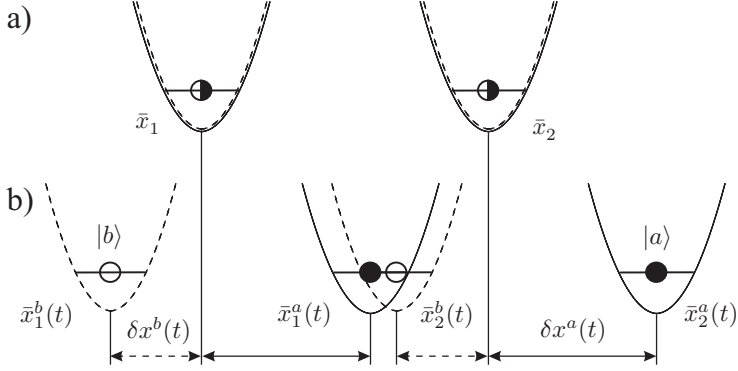}
\vspace*{0.5cm}
\caption{Schematic representation of the gate operation. Solid and dashed lines 
represent potentials felt by the atoms in the different internal states, respectively. By shifting 
the potentials, the pair of atoms in the middle experiences a collisional shift (from \cite{Jaksch99})}
\label{figure5.3}
\end{figure}

There are several proposals for implementing quantum computing \cite{Williams97}. A very promising one was put forward  by Cirac and Zoller in 1995 
that uses cold trapped ions interacting with laser beams \cite{Cirac95}. For the effects of environmental decoherence on such an ion trap quantum 
computer and possible ways to get over them, see e.g. 
\cite{Plenio96, Garg96, Wineland98, James98,  Poyatos98, King98}, and references therein. 
Following the Cirac-Zoller 
proposal, a fundamental quantum logic gate, the CNOT gate, 
was experimentally demonstrated by Monroe \emph{et al.} \cite{Monroe95}, where the two qubit state was encoded in the 
two internal states and two external states of a single trapped ion. 
For recent developments in ion trap computers, see \cite{Gulde03, Schmidt-Kaler03, Haeffner05, Wineland05, Korbicz06}. 
In particular, Ref. \cite{Schmidt-Kaler03} has realized the CNOT gate
by two \(^{40}\)Ca\(^+\) ions, held in a linear Paul trap. Jaksch \emph{et al.} \cite{Jaksch99} proposed to use atoms in an optical lattice for quantum computation.
They use cold controlled collisions
between two atoms in moving trap potentials for implementing the following two-qubit gate: 
\begin{eqnarray}
\label{tobildari-qi}
|a\rangle_1 |a\rangle_2 & \to & |a\rangle_1 |a\rangle_2, \nonumber \\
|a\rangle_1 |b\rangle_2 & \to & e^{-i\phi_{ab}}|a\rangle_1 |b\rangle_2, \nonumber \\
|b\rangle_1 |a\rangle_2 & \to & |b\rangle_1 |a\rangle_2, \nonumber \\
|b\rangle_1 |b\rangle_2 & \to & |b\rangle_1 |b\rangle_2.
\end{eqnarray}
The two qubits are encoded in the internal states \(|a\rangle\) and \(|b\rangle\) of the two atoms. The two atoms are labelled as 
1 and 2 and each atom is trapped in the ground state of a potential well. The phase \(\phi_{ab}\) will be a 
fixed number depending on the implementation. Initially, the potential wells
are at a 
distance 
that 
is sufficiently large distance so that there is no interaction between the atoms. They are then moved along certain trajectories 
toward each other, so that the atoms interact for a certain time, and 
then finally the wells are restored to their initial positions. The potential wells are 
moved adiabatically, so that the atoms 
remain in the respective ground states of the wells. 
The implementation requires an interaction that induces internal state dependent potentials, so that the atoms in different 
internal states feel different potentials (see Fig. \ref{figure5.3}).

The effective Hamiltonian governing the system is given by: 
\begin{eqnarray}
H= \sum_i \left[ \omega^a(t) a_i^\dagger a_i + \omega^b(t) b_i^\dagger b_i  + u^{aa}(t) a_i^\dagger a_i^\dagger a_i a_i
+ u^{bb}(t) b_i^\dagger b_i^\dagger b_i b_i\right] \nonumber \\ 
+ \sum_{ij} u^{ab}_{ij}(t) a_i^\dagger a_i b_j^\dagger b_j, \nonumber
\end{eqnarray}
\(a_i\) (\(a_i^\dagger\)) and \(b_i\) (\(b_i^\dagger\)) are the annihilation (creation) operators of the internal levels 
\(|a\rangle\) and \(|b\rangle\) for an atom 
in the ground state of the potential well at position \(i\). By assuming that (i) there is no sloshing motion of the atoms in the wells when the latter are moved, (ii) the potentials are moved 
so that the atoms remain in the ground state (adiabaticity condition), and (iii) the rms velocity of the atoms in the 
vibrational ground state is sufficiently small so that zero energy \(s\)-wave scattering approximation is valid, then it is possible 
to show that the internal states of the atoms transform in the way described in Eq. (\ref{tobildari-qi}).

There are several other papers that deal with quantum computation with atoms in optical lattices. 
In particular, Brennen \emph{et al.} \cite{Brennen99} have used dipole-dipole interactions to implement the CNOT gate. 
For further work, see e.g. \cite{Charron02, Lee05, Greiner02, Tian03, Pachos03, Dorner03, Sklarz02, Benjamin02,
Hemmerich99, Murg04, Derevianko04, Vala05, Duan03, Vollbrecht04, Christandl04, Jaksch98, 
Serafini06, Rabl03, Mompart03, Eckert02, 
Lukin01, Andersson01, Jaksch00, Garcia-Ripoll03, Brennen00, Schlosser01, Dumke02, Dur99, Folman02,
Calarco04, Ionicioiu02}.

\subsection{Generation of entanglement: The one-way quantum computer}
\label{subsec-chauthha-qi}

There has been several studies for generating entanglement in different physical systems. We want to 
discuss generation of entanglement from the perspective of performing quantum computation. We shall discuss entanglement generation in ordered and disordered lattices separately.

\subsubsection{The one-way quantum computer}
\label{subsec-chauthha-qi1}
The one-way quantum computer (also referred to as ``measurement-based quantum computation'')
was proposed by 
Raussendorf and Briegel  \cite{Raussendorf01a}, who showed that arbitrary 
quantum computation can be simulated via single-particle measurements on a specially prepared quantum state
termed  as ``cluster state''. Cluster states can be created efficiently in any ultracold atomic systems with
Ising-type interactions, and allow the possibility of preparing the system in a specific initial product state.
A proposal to implement cluster states uses four atomic levels trapped in appropriately tuned optical lattices \cite{Kay06}
(cf. \cite{Clark05, Christandl05}).

\begin{figure}[ht]
\centerline{
\unitlength=0.5mm
\begin{picture}(150,40)(0,0)
\thicklines
\put(10,25){\circle*{5}}
\put(10,25){\line(1,0){25}}
\put(35,25){\circle*{5}}
\put(35,25){\line(1,0){25}}
\put(60,25){\circle*{5}}
\put(60,25){\line(1,0){25}}
\put(85,25){\circle*{5}}
\put(35,25){\line(0,-1){25}}
\put(35,0){\circle*{5}}
\end{picture}
}
\caption{An example of a graph. In this case, it has
 five vertices and four edges. The vertices represent qubits, and the edges represent 
Ising interactions between them.}
\label{graph-er-udahoron-qi-fig} 
\end{figure}
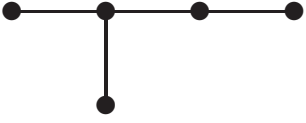

\begin{figure}[ht]
\centerline{
\unitlength=0.5mm
\begin{picture}(150,40)(0,0)
\thicklines
\put(10,25){\circle*{5}}
\put(10,25){\line(1,0){25}}
\put(35,25){\circle*{5}}
\put(35,25){\line(1,0){25}}
\put(60,25){\circle*{5}}
\put(60,25){\line(1,0){25}}
\put(85,25){\circle*{5}}
\put(85,25){\line(1,0){25}}
\put(110,25){\circle*{5}}
\put(4,15){\(\left|\psi_{in}\right\rangle\)}
\put(29,15){\(\left|+\right\rangle\)}
\put(54,15){\(\left|+\right\rangle\)}
\put(79,15){\(\left|+\right\rangle\)}
\put(104,15){\(\left|+\right\rangle\)}
\put(7,35){\(\sigma_x\)}
\put(32,35){\(\sigma_y\)}
\put(57,35){\(\sigma_y\)}
\put(82,35){\(\sigma_y\)}
\end{picture}
}
\caption{A graph to implement the Hadamard gate. 
}
\label{Hadamard-er-udahoron-qi-fig} 
\end{figure}
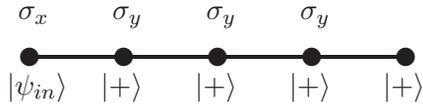

To define cluster states \cite{Raussendorf01b}, consider an arbitrary graph made up of \(N\) ``vertices'' (which will be identified 
with the qubits making up the quantum computer), and a certain number of ``edges'' (which will be 
identified with Ising interactions between the qubits) connecting the vertices 
(see Fig. \ref{graph-er-udahoron-qi-fig}). 
The terminology used here is  
from graph theory (see e.g. \cite{Deo74}), and thereby cluster states are also called ``graph states''. 
Suppose that the qubits are spin-half particles, and that each of them is prepared in the state 
\(|+\rangle = (|0\rangle + |1\rangle)/\sqrt{2}\), where \(|0\rangle\) and \(|1\rangle\) are 
respectively the spin up and down states in the \(z\) direction. 
We then apply an Ising interaction governed by the Hamiltonian 
\[
H = - J\sum \sigma_i^z \sigma_{j}^z,
\]
between the qubits, where the summation runs over all pairs \((i,j)\) of sites that 
are connected by an edge in the graph. 
After a certain finite time (\(=\frac{\pi \hbar}{4J}\)), the state 
\begin{equation}
\label{telari-qi}
\left|\phi_C\right\rangle = \prod CS_{ij} |+\rangle^{\otimes N}
\end{equation}
is created.
Here \(CS_{ij}\) is the controlled-PHASE gate, defined as 
\begin{eqnarray}
|0\rangle |0\rangle & \to & |0\rangle |0\rangle, \nonumber \\
|0\rangle |1\rangle & \to & |0\rangle |1\rangle, \nonumber \\
|1\rangle |0\rangle & \to & |1\rangle |0\rangle, \nonumber \\
|1\rangle |1\rangle & \to & i|1\rangle |1\rangle, \nonumber
\end{eqnarray}
acting on the spins \(i\) and \(j\).
The product on the right hand side of Eq. (\ref{telari-qi}), again runs 
over all pairs \((i,j)\) of sites that 
are connected by an edge in the graph. 
The state \(\left|\phi_C\right\rangle\) is called the  cluster state corresponding to the given 
graph. These states have several interesting properties. Perhaps the most interesting property is that 
for sufficiently large graphs, the corresponding cluster states can be used as a substrate for 
quantum computational tasks. 
More specifically, any quantum gate can be implemented with this resource,
simply by performing measurements on the different spins of a suitably chosen graph, where it is 
assumed that the choice of a measurement basis on a particular spin may depend on 
outcomes of previous measurements on other spins. As an example, 
the Hadamard gate (\(\mathbb{H}\)),
defined as 
\begin{eqnarray}
|0\rangle  & \to & (|0\rangle + |1\rangle)/\sqrt{2}, \nonumber \\
|1\rangle  & \to & (|0\rangle - |1\rangle)/\sqrt{2}, \nonumber
\end{eqnarray}
can be implemented in the following way. 
Suppose that at a certain point of computation, we obtain the state \(\left|\psi_{in}\right\rangle\),
which we want to transform into \(\mathbb{H}\left|\psi_{in}\right\rangle\). 
This can be realized by using a graph 
of five vertices in a chain as shown in Fig. \ref{Hadamard-er-udahoron-qi-fig} 
\cite{Raussendorf01a, Raussendorf02, Raussendorf03}. The input state is initially at the extreme left. 
The rest of the spins are initially in the state $|+ \rangle$, $\sigma_z|+ \rangle=|+ \rangle$. Ising interaction with nearest neighbor 
interactions (as depicted in the figure by the edges) for 
the requisite amount of time (equivalent to the controlled-PHASE gates, as described in the text) 
results in a (entangled) state of 
the five spins. Subsequently, measurements are performed on all spins except that on the extreme right,
in the eigenbasis of the operator
indicated over the corresponding vertices. 
This results in the state \(\mathbb{H}\left|\psi_{in}\right\rangle\) at the extreme right spin.
A five-spin graph and corresponding measurements implementing an arbitrary single-qubit gate, and 
a 15-spin graph implementing the CNOT gate is given in the same papers.

There is a large body of work in this direction that deal with several important issues ( see 
e.g. \cite{Gottesman97,Nielsen03,Leung01,Leung03,Hein04, Asano04,
VandenNest04a,VandenNest04b,Aliferis04,Aschauer05,Duan05,Childs05,Borhani05,
Hostens05,Nielsen05a,Walther05,Nielsen05b,VandenNest05a,Tame05,VandenNest05b,Chen06,Tame06,Anders06,Aliferis06, Kay06}).
In particular, D{\" u}r \emph{et al.} \cite{Dur05} deals with the entanglement properties of the time-evolved states in a model with 
long-range 
Ising-type interactions, in the thermodynamic limit.

\subsubsection{Disordered lattice}
\label{subsubsec-hazarchurasi}

Interestingly, it is also possible to perform
quantum computing in systems with
quenched disorder. 
This may sound contradictory, but we want to employ here the possibility of creating
\emph{controlled} disorder in atomic gases in optical
lattices. Here, by ``controlled'', we mean that a particular realization of disorder remains fixed for times much longer than 
over which we monitor the evolution of the system (the disorder is ``quenched'').
This allows, as discussed in Section \ref{sec_disorder_Konark}, 
to study Anderson and Bose glasses in a Bose gas \cite{Damski03},
or spin glasses with short range interactions  
in  Fermi-Bose, or Bose-Bose mixtures \cite{Sanpera04}.
Using linear chains of trapped ions \cite{Porras04}, or dipolar atomic gases \cite{Baranov02, Schmidt03}, it
is possible to realize complex spin systems with long range interactions
that may serve as a model for classical and quantum neural networks 
\cite{Pons} (cf. \cite{Cho05}).

Disordered systems 
offer at least two possible advantages for quantum computing. First, 
since they have typically a large number of different metastable (free) 
energy minima, 
such states might be used to store information distributed over the whole
system, similarly to classical neural network (NN) models \cite{Amit92}. 
The information is thus naturally stored in a redundant way, 
like in error correcting schemes \cite{Shor95,Steane96}. 
Second, in disordered systems with long range interactions, 
the stored information is robust: metastable states have quite large
basins of attraction in the thermodynamical sense.

The aim of using  
complex disordered systems is therefore to be able to perform distributed quantum computing
in a way that is resistant to noise.
In Refs. \cite{SenDe05c, SenDe05d},
the following questions in the \emph{non-distributed} case were addressed: \footnote{Since the submission 
of an earlier version of this review, the \emph{distributed} case has also been considered \cite{Pons,Braungardt}.}
(i) Can one generate
entanglement in such systems that would survive quenched averaging over long times? 
(ii)  Can one realize quantum gates with reasonable fidelity?
Both questions were answered affirmatively for both
short and long range disordered systems.
Here we present as an example,  the second question for the Hadamard gate \cite{SenDe05c}.

We assume that the computation is performed in a spin lattice created with cold ions in an optical lattice, of which particles 1 and 2 are part of. At a certain time, particle 1
is in an arbitrary state \(\left|\psi_{in}\right\rangle\), while 2 is in an eigenstate of $\sigma^{z}$ with eigenvalue $+1$ e.g. \(|+\rangle\).  
Then we  let 1 and 2
evolve according to the Edwards-Anderson Hamiltonian 
\[H_{E-A}
= - J \sigma_1^z \sigma_{2}^z
\] 
for a suitable duration of time. Here \(J\) is a random quenched coupling, that in the case 
of the Edwards-Anderson model, can be regarded a Gaussian random variable. After a suitable duration of 
evolution, a 
measurement is performed on particle 1 (in a suitable basis).
For the case when \(J\) is a Gaussian variable with mean= 5 and variance=1,
the state in particle 2 attains the Hadamard rotated 
state \(\mathbb{H}\left|\psi_{in}\right\rangle\), with quenched averaged
fidelity greater
than \(0.85\). One can increase such fidelity by increasing the 
number of spins, and employing measurements assisted by outcomes of previously performed measurements.
Note, that the fidelity 
of the Hadamard rotated state 
using the classical information obtained only from the
 measurement of particle
 \(1\), 
is only \(2/3\) \cite{Popescu94}.

\section{Summary}
\label{sec-summary-oma}

Summarizing, we have tried
to survey the state-of-the-art on the
active research field of ultracold atomic gases in optical lattices.
On one hand, we have aimed at giving a general outline of new topics 
and open problems that can be addressed with these systems. 
Our list, however, is by no means exhaustive. 
On the other hand, we have briefly reviewed some of the best established methods
and techniques to study strongly correlated systems in the framework of
ultracold gases. In spite of the fact that we have mostly focused 
on the connections of ultracold gases with condensed matter topics, 
we envision and hope that, in the next years, the frontiers 
of this field will greatly expand.

Finally, we hope we have at least partially convey our enthusiasm for 
ultracold gases physics. A concise version of our summary 
is contained  
in the  motto, which has in fact been used by one of us (M.L.)
at the end of his conference presentations and seminars. This motto should express a pure joy of curiosity and discovery
(even if it was re-discovery) that we and many others have experienced while working on ultracold atomic gases over the last years. 
This is the  most important message: Ultracold atom physics is FA-A-ANTASTIC!!!

\section{Acknowledgements}
\label{ackn}

This review is to a great extent based on ideas and discussions that we shared with our long lasting friends, close collaborators,
and colleagues: 
Jan Arlt, Alain Aspect, Nuri Barberan, Michael Baranov, Indrani Bose, Immanuel Bloch, Kai Bongs, 
Sibylle Braungardt, Dagmar Bru\ss, Ignacio Cirac, Eugene Demler,
Kai Eckert, Wolfgang Ertmer, Uli Everts, Henning Fehrmann, Massimo Inguscio, Jaros{\l}aw Korbicz, Anna Kubasiak,
Jonas Larson, Mikko Juhani Leskinen, 
Alem Mebrahtu, Chiara Menotti, Armand Niederberger, Belen Paredes,
Eugene Polzik,  Andreas Osterloh,  Klaus Osterloh, Carles Rod{\'o}, Oriol Romero-Isart, Robert Roth, Kazimierz Rz{\c a}zewski, 
Krzysztof Sacha, Laurent Sanchez-Palencia,
Luis Santos, Klaus Sengstock, Gora Shlyapnikov, Christian Trefzger,  Janek Wehr, Jakub Zakrzewski, {\L}ukasz Zawitkowski, and Peter Zoller.

We acknowledge support from the 
Deutsche
Forschungsgemeinschaft 
(SFB 407, SPP1078 and SPP1116), 
the Spanish Ministerio de Ciencia y Tecnolog{\' i}a 
grants BFM-2002-02588, FIS2005-04627, FIS2005-01497, FIS2005-01369 and Consolider Ingenio 2010 CSD2006-00019, the Alexander von Humboldt Foundation, 
the ESF Program QUDEDIS, the EU IP SCALA, and the U.S. Department of Energy.

\appendices

\section{Effective Hamiltonian to second order}
\label{effhamil}
The Bose Hubbard Hamiltonian or Fermi-Bose Hamiltonian 
can be split into $H=H_0+H_{int}$, where $H_0$ denotes  
the on-site Hamiltonian obtained in the limit of zero tunneling,
and the hopping term is denoted by $H_{int}$. 
The Hamiltonian $H_0$ can be easily diagonalised 
and has  well defined eigenstates, that are grouped in blocks (or manifolds). 
Each manifold is well separated from each other. Typically different
manifolds are separated by terms of the order of $U$, 
being $U$ the interaction between two bosons. 
We denote by  $P_{\alpha}$, the projector on each
block space, where $\alpha$ is the block index,
and the $i$-th state in any block is denoted
by $|\alpha, i\rangle$. Note that 
$P_{\alpha} H_0 P_{\beta}=0$ holds for $\alpha\neq\beta$.
The $H_{int}$ part of the Hamiltonian will 
introduce couplings between the bare block $\alpha$ and $\beta$, i.e.,
$P_{\alpha} H_{int} P_{\beta}$ can be different from zero for 
$\alpha\neq\beta$. Following ~\cite{Cohen92},
one can construct an effective Hamiltonian,
$H_{eff}$, from $H$ such that it describes the slow, low-energy
perturbation-induced tunneling {\it strictly within}
each manifold of the unperturbed block states and has the same eigenvalues as $H$. 
Tunneling processes between different manifolds are thus neglected.
We demand that the effective hamiltonian $H_{eff}$
fulfills that:
\begin{enumerate}
\item is by construction hermitian, 
with the same eigenvalues and the same degeneracies as $H$.
To achieve that, one defines $T:=e^{iS}$,
with $S$ hermitian, $S=S^{\dagger}$, and chosen
such that:
\begin{equation} H_{eff}=THT^{\dagger} ~. \end{equation}
\item $H_{eff}$ 
does not couple states from different  manifolds:
\begin{equation}
P_{\alpha} H_{eff}P_{\beta}=0,\quad\alpha
\neq\beta\quad ~. 
\end{equation}

\item As the first two conditions still allow for
an infinite number of unitary transformations (all
$UT$ are still possible, $U$ being any unitary
transformation acting only {\it within} the manifolds),
the following additional condition is imposed:
\begin{equation}P_\alpha S P_\alpha=0\quad\mbox{for any}
\hspace{0.1cm}\alpha .\end{equation}
\end{enumerate}

Expanding the first condition using the Baker-Hausdorff
formula, one obtains:
\begin{eqnarray}\lefteqn{\label{heff1}H_{eff}=H+\lsb
iS, H\rsb+\frac{1}{2!}\lsb iS, \lsb iS,
H\rsb\rsb}\hspace{1.0cm}\nonumber\\& &+\frac{1}{3!}\lsb iS, \lsb iS,
\lsb iS, H\rsb\rsb\rsb + \ldots \; .
\end{eqnarray}
Making a power-series ansatz for in the perturbative parameter $t$ 
$S$,
\begin{equation}\label{sexpand}S=t\Sone+
t^2\Stwo+t^3S_3+\ldots,\end{equation}
and employing $H=H_0+t H_{int}$ one obtains
from (\ref{heff1}) to second order

\begin{eqnarray}
H_{eff}=H_0+
t\overbrace{\lrb\lsb iS_1, H_0\rsb+H_{int}\rrb}
^{H_{eff}^1}\hspace{1.0cm}\nonumber\\
+t^2\overbrace{\lrb\lsb iS_2,
H_0\rsb+\lsb iS_1,\Hint\rsb+\frac{1}{2}\lsb
iS_1, \lsb iS_1, \HO\rsb\rsb\rrb}^{\Heff^2} ~.
\label{heff2}
\end{eqnarray}


This is a power series for $H_{eff}$, with its
moments denoted by $H_{eff}^1$, $H_{eff}^2$..
to first order, i.e. $H_{eff}=H_0+t H_{eff}^1$
and $S=t\Sone$. Using the second and third conditions,
as well as \Palpha\HO\Pbeta=0 and the
expression for $H_{eff}^1$ in (\ref{heff2}),
one finds:
\begin{equation}\label{sone1}\braai iS_1\ketbj
\lrb E_{\beta j}-E_{\alpha i}\rrb+\braai\Hint
\ketbj=0\end{equation}
\begin{equation}\label{sone2}
\Rightarrow\braai
iS_1\ketbj=\lgb \begin{array}{cr}\frac{\braai
\Hint\ketbj}{E_{\alph i}-E_{\beta j}}&\alph\neq
\beta\\0&\alph=\beta\end{array}\right. ~.
\end{equation}
Thus, the effective Hamiltonian within
the $\alpha$ -manifold, depends only on $H_{int}$ and not on \Sone, 
i.e., ${
\braai\Heff^1\ketaj=\braai\Hint\ketaj}$. 
A general result for any $n$ is that \ensm{\braai\Heff^n\ketaj} is independent of 
\ensm{S_n}. Based on the third condition, and on the observation that $S_n$ enters the
expression for \ensm{\Heff^n} only in the commutator
with \HO, which is diagonal in the manifold index.

Thus, when continuing to second order, the
term \ensm{\lsb iS_2, \HO\rsb} in the expression
for \ensm{\Heff^2} can be dropped. Of the two
remaining terms defining \ensm{\Heff^2} in
(\ref{heff2}), the second one can be simplified
by observing, that according to (\ref{sone1})
the operator \ensm{[iS_1, \HO]} is purely
non-diagonal in the manifold index, with values
opposite to those of the non-diagonal
part of $H_{int}$. Thus, \ensm{\frac{1}{2}\lsb
iS_1, \lsb iS_1, \HO\rsb\rsb=-\frac{1}{2}\lsb
iS_1, \Hint^{nd}\rsb}. Now inserting the identity
between the operators in the still untreated
second term in \ensm{\Heff^2}, \ensm{\lsb i\Sone, \Hint\rsb} ,
one sees that due to \Sone\ being non-diagonal
in \alph, again only the non-diagonal part
of \Hint\ can contribute: \ensm{\lsb iS_1, \Hint\rsb
=\lsb iS_1, \Hint^{nd}\rsb}. Therefore, one has:
\begin{equation}\label{heff3}\Heff^2=[ iS_1,
\Hint^{nd}]+\frac{1}{2}[ iS_1, \overbrace{
\lsb iS_1, \HO\rsb}^{-\Hint^{nd}}]=\frac{1}{2}
[iS_1, \Hint^{nd}] ~.
\end{equation}
Collecting all terms relevant for \braai\Heff
\ketaj\ to second order in $t$, and introducing
the notation \ensm{\Qalphai:=\sum_{k,\gamma\neq
\alph}\frac{|\gamma,k\rangle\langle\gamma,k|}
{E_{\gamma k}-E_{\alpha i}}}, one finds:

\begin{eqnarray}\label{heff4}\braai\Heff\ketaj
=E_{\alph i}\delta_{ij}+t\braai\Hint\ketaj \nonumber\\
-\frac{t^2}{2}\lrb\braai\Hint\lsb\Qalphai+
\Qalphaj\rsb\Hint\ketaj\rrb,
\end{eqnarray}
where the identity operator has been inserted
in the final expression for \ensm{\Heff^2} in
formula (\ref{heff3}), and then evaluated using
formula (\ref{sone2}), which naturally leads one
to define the operator \Qalphai\ as above.
Note that this construction can be generalized to
arbitrary orders in $t$ in a straightforward
manner, as detailed in \cite{Cohen92}.

\section{Size of the occupation-reduced Hilbert space}
\label{appendix_S}
We derive here Eq. (\ref{HSK}), i.e., an expression for the size of the Hilbert space
in the system subjected to the restriction that at most 
$K$ bosons can populate a single lattice site. We assume that 
there are $N$ bosons placed in $M$ lattice sites, therefore
$K\le N$ and $K M\ge N$. By definition
${\cal HS}(N,M)|_K= \sum_{n_1=0}^K\cdots\sum_{n_M=0}^K\delta_{n_1+\cdots+n_M,N}$.
Using the following Kronecker delta representation, 
$\delta_{l,m}=\frac{1}{2\pi}\int_{-\pi}^\pi{\rm d}\varphi e^{i(l-m)\varphi}$, one gets
$${\cal HS}(N,M)|_K=\frac{1}{2\pi}\int_{-\pi}^\pi{\rm d}\varphi e^{-iN\varphi}
\left(\frac{e^{i(K+1)\varphi}-1}{e^{i\varphi}-1}\right)^M=\frac{1}{N!}\frac{d^N}{dz^N}
\left.\left(\frac{z^{K+1}-1}{z-1}\right)^M\right|_{z=0},$$
where the right hand side is obtained after integration 
on a unit circle $z=\exp(i\varphi)$ in the complex plane. Then, using the 
Leibnitz formula one obtains 
$${\cal HS}(N,M)|_K= \frac{1}{N!}\sum_{n=0}^N 
\left(
\begin{array}{c}
N \\ n
\end{array}
\right)
\left.\frac{d^{N-n}}{dz^{N-n}}(z^{K+1}-1)^M\frac{d^n}{dz^n}
\left(\frac{1}{z-1}\right)^M\right|_{z=0},
$$
which can be further reduced to (\ref{HSK}).



\end{document}